\documentclass[iop]{emulateapj}
\usepackage{graphicx, amsmath,natbib, longtable}
\usepackage[usenames]{color}  \shorttitle{The COS-Halos Survey} \shortauthors{}

\newcommand{\noprint}[1]{}

\providecommand\scription[2]{\scriptsize#1$\;${\scriptsize\uppercase\expandafter{\romannumeral
#2}}\relax}%
\providecommand{\Lya}{\ensuremath{\mbox{Ly}\alpha}}
\providecommand{\lya}{\ensuremath{\mbox{Ly}\alpha}}
\providecommand{\Lyb}{\ensuremath{\mbox{Ly}\beta}}
   \providecommand{\HI}{\ensuremath{\mbox{\ion{H}{1}}}}
\providecommand{\OVI}{\ensuremath{\mbox{\ion{O}{6}}}}
\providecommand{\kms}{\,\ensuremath{\rm{km\,s}^{-1}}}

\providecommand{\scHI}{\ensuremath{\mbox{\scription{H}{1}}}}
\providecommand{\Rvir}{\ensuremath{R_{\rm vir}}}
\providecommand{\NHI}{\ensuremath{N_{\scHI}}}

\providecommand{\kpc}{\,\ensuremath{\mbox{kpc}}}

\def\spose#1{\hbox to 0pt{#1\hss}} \def\simlt{\mathrel{\spose{\lower
3pt\hbox{$\mathchar"218$}}
     \raise 2.0pt\hbox{$\mathchar"13C$}}}
\def\simgt{\mathrel{\spose{\lower 3pt\hbox{$\mathchar"218$}}
     \raise 2.0pt\hbox{$\mathchar"13E$}}}

\providecommand{\Msun}{\,\ensuremath{\mbox{M}_{\odot}}}

\begin{document}

\title{The COS-Halos Survey: Rationale, Design, and A Census of Circumgalactic Neutral Hydrogen\altaffilmark{1}} 
\author{
Jason Tumlinson\altaffilmark{2},
Christopher Thom\altaffilmark{2}, 
Jessica K. Werk\altaffilmark{3}, 
J. Xavier Prochaska\altaffilmark{3}, 
Todd M. Tripp\altaffilmark{4}, 
Neal Katz\altaffilmark{4}, 
Romeel Dav{\'e}\altaffilmark{5,6}, 
Benjamin D. Oppenheimer\altaffilmark{7,8},  
Joseph D. Meiring\altaffilmark{4}, 
Amanda Brady Ford\altaffilmark{5}, 
John M. O'Meara\altaffilmark{9}, 
Molly S. Peeples\altaffilmark{10,2}, 
Kenneth R. Sembach\altaffilmark{2}, 
David H. Weinberg\altaffilmark{11}
}

\altaffiltext{1}{Based on observations made with the NASA/ESA Hubble
Space Telescope, obtained at
  the Space Telescope Science Institute, which is operated by the
  Association of Universities for Research in Astronomy, Inc., under
  NASA contract NAS 5-26555. These observations are associated with
  program GO11598.}
\altaffiltext{2}{Space Telescope Science Institute, Baltimore, MD}
\altaffiltext{3}{UCO/Lick Observatory, University of California, Santa Cruz, CA} 
\altaffiltext{4}{Department of Astronomy, University of Massachusetts, Amherst, MA} 
\altaffiltext{5}{Steward Observatory, University of Arizona, Tucson, AZ} 
\altaffiltext{6}{University of the Western Cape, South African Astronomical Observatories, and African Institute for Mathematical Sciences, Cape Town, South Africa}
\altaffiltext{7}{Leiden Observatory, Leiden University, the Netherlands}
\altaffiltext{8}{CASA, Department of Astrophysical and Planetary Sciences, University of Colorado, Boulder, CO 80309, USA} 
\altaffiltext{9}{Department of Chemistry and Physics, Saint Michael's College, Colchester, VT}
\altaffiltext{10}{Center for Galaxy Evolution, University of California Los Angeles, Los Angeles, CA} 
\altaffiltext{11}{Department of Astronomy, The Ohio State University, Columbus, OH}

\begin{abstract} 
We present the design and methods of the 
COS-Halos survey, a systematic investigation of the gaseous halos of 
44 $z = 0.15-0.35$ galaxies using background QSOs observed with the 
Cosmic Origins Spectrograph aboard the {\it Hubble Space Telescope}. 
This survey has yielded 39 spectra of $z_{em} \simeq 0.5$ QSOs with
S/N $\sim$10-15 per resolution element. 
The QSO sightlines pass within 150 physical kpc of the galaxies, which span early and late types over stellar mass $\log M_* / M_{\odot}= 9.5 - 11.5$. 
We find that the CGM exhibits strong \HI, averaging $\simeq1$ \AA\ in \lya\ equivalent width out to 150 kpc, 
with 100\% covering fraction for star-forming galaxies and 75\% covering for passive galaxies. We find good agreement in column 
densities 
between this survey and previous studies over similar range of impact parameter. There is weak evidence for a difference 
between early- and late-type galaxies in the strength and distribution of \HI. Kinematics indicate that the detected material is  bound to the host galaxy, such that 
$\gtrsim 90$\% of the detected column density is confined within $\pm 200$ \kms\ of the galaxies.  
This material generally exists well below the halo virial temperatures at $T \lesssim 10^5$ K. We evaluate a number of possible 
origin scenarios for the detected material,  and in the end favor a simple model in which the bulk of the detected 
\HI\ arises in a bound, cool, low-density photoionized diffuse medium that is 
generic to all $L^*$ galaxies and may harbor a total gaseous 
mass comparable to galactic stellar masses. 
\end{abstract}

\keywords{ galaxies: halos, formation --- quasars: absorption lines ---
intergalactic medium}


\section{Introduction and Motivations} \label{intro-section}

The means by which galaxies acquire their gas, process it into stars, and 
expel it as energetic feedback have assumed central importance in the modern picture of
galaxy formation.  Solutions to important puzzles such as the galactic
``missing baryons'' problem, the mass-metallicity relation, and the
color-magnitude bimodality must involve the flows of gas that cycle
through the intergalactic medium (IGM), circumgalactic medium (CGM),
and interstellar medium (ISM) during galaxy evolution. Those parts of
galaxies that are readily visible in emission -- stars and the ISM gas
they illuminate -- constitute the outcome of the flows from the IGM and
CGM, not those flows themselves. At least in the IGM and CGM, these flows
are difficult to observe directly because the gas is diffuse and spread
over large regions of space. Fortunately, absorption-line techniques can
access physical tracers at the relevant densities with extremely high sensitivity.

This paper describes the properties of a new survey (``COS-Halos'') of
the CGM gas surrounding a sample of $L \sim L^*$ galaxies in the low-redshift
Universe using the Cosmic Origins Spectrograph (COS) aboard the {\it
Hubble Space Telescope}. The primary motivation for the COS-Halos survey
is to examine the content of the CGM and to better understand its role
in galaxy formation. The design of the survey leverages the large advance
in UV spectroscopic sensitivity offered by COS \citep{green-etal-12-COS}
to address the CGM with a larger sample and better control over galaxy
populations than was possible with earlier instruments and selection
techniques. As a result of the improvement in sensitivity, it has become 
feasible to choose a sample of galaxies with particular properties and to observe their halo gas in a commonly
studied suite of hydrogen and metal-line diagnostics. The
major scientific motivations for this survey are:

{\it Galaxy Accretion -- Hot, Cold, and Multiphase:} The
question of how galaxies acquire their gas dates back at least to
\cite{1978MNRAS.183..341W}, who posited that gas entering a halo shock
heats to roughly the virial temperature ($T \sim 10^6$ K for a Milky
Way-sized galaxy), before cooling and collapsing to form the central
galaxy. This basic picture has been modified in many ways and has reached
its fullest development in detailed hydrodynamical simulations that track
gas infall, cooling, star formation, and feedback self-consistently within
a cosmological context. When these details are included as faithfully
as possible in simulations, a theoretical picture emerges that is more
complex than the traditional picture in which all gas shock-heats to
the halo virial temperature before cooling and falling, more or less spherically, into the galaxy. 

The ``bimodal accretion" scenario posits two primary modes by which gas can
accrete. Galaxies with $M_* \gtrsim 5 \times 10^{10}$ M$_{\odot}$ reside 
in halos massive enough at $z \sim 0$ to accrete through the ``hot mode'', with gas shock-heated near the halo virial radius ($R_{\rm vir}$) to the halo virial temperature ($T_{\rm vir}
\sim 10^6$ K). This accretion mode closely resembles the earlier shocked-accretion 
spherical inflow scenario. 
For galaxies at $M_* \lesssim 5 \times 10^{10}$ M$_{\odot}$, rapid radiative cooling removes
the pressure needed to maintain the shock and gas can penetrate far inside
$R_{\rm vir}$ without heating above  $\sim 10^{4-5}$ K as it accretes along
narrow filaments \citep{Keres:2005gba, db06}. This is the Òcold modeÓ of
accretion. The predicted transition halo mass between the hot and cold modes
of accretion lies near the observed transition between high-mass, red,
bulgy galaxies and low-mass, blue, disky galaxies, so it is tempting to
associate the two phenomena and thus explain the bimodality of galaxy
colors in terms of gas accretion. Recent simulations based on new 
treatments of hydrodynamics have questioned the details of cold accretion 
predictions \citep{2012MNRAS.424.2999S,2013MNRAS.429.3341B}, but in any case it is important that observations provide tests of any proposed accretion scenarios. 

The ÒmultiphaseÓ accretion scenario
\citep{1996ApJ...469..589M, 2004MNRAS.355..694M} instead posits the existence
of the hot, diffuse gaseous halo left over from virialization or a major
merger, and then considers the behavior of density fluctuations that
cool within this hot medium. In contrast to the canonical picture in which hot gas within 
$R_{\rm vir}$ cools from the inside out (starting at small radii where the densities are highest 
and cooling times are shortest), in the multiphase model cooling proceeds unevenly and $T
\sim 10^4$ K fragments form in rough pressure equilibrium with the hot
diffuse halo. These cooled fragments then spiral in and form the primary means
by which gas accretes to the central galaxy.

\begin{figure*}[!t] 
\begin{center} 
\plotone{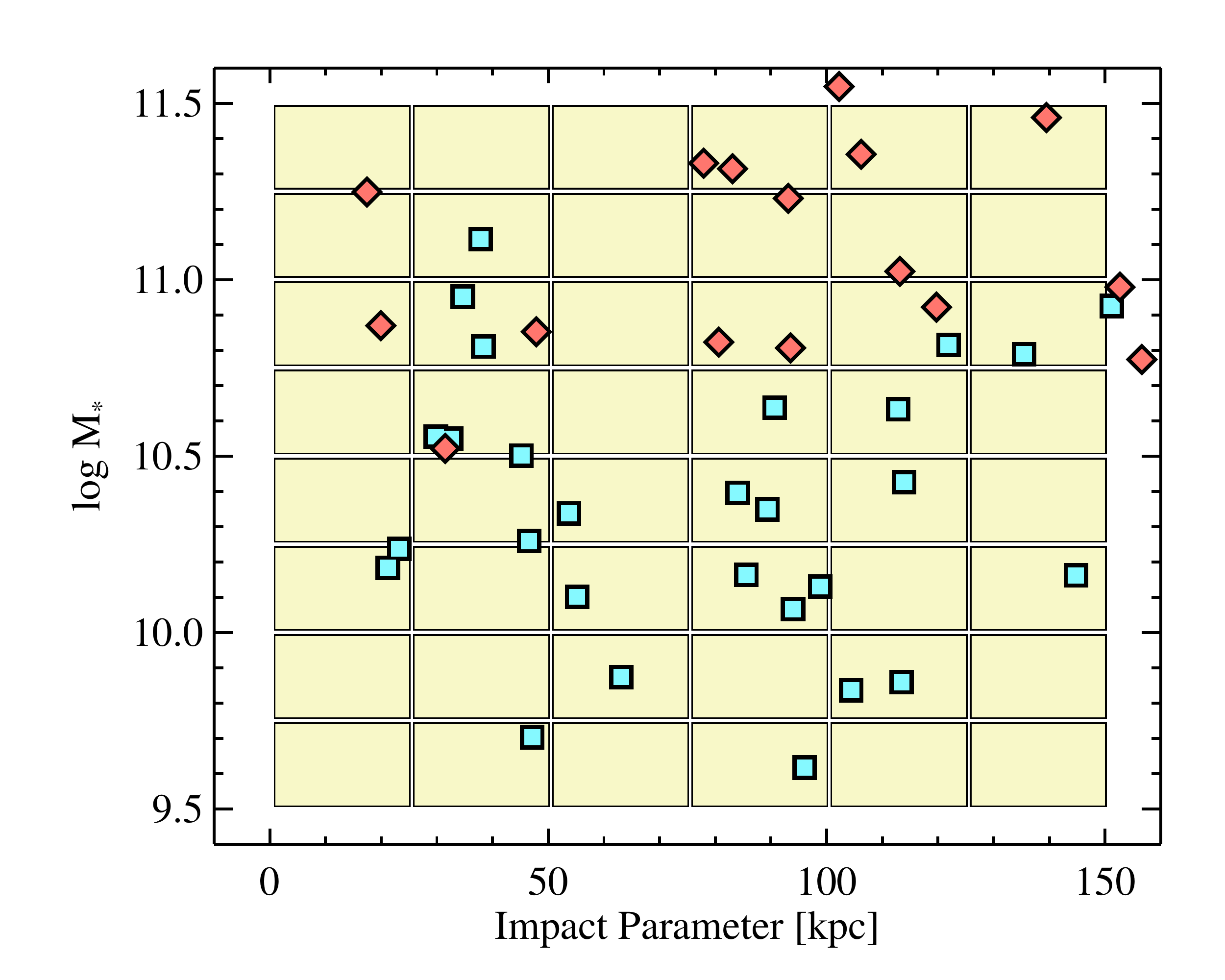} 
\label{binfig}
\end{center} 
\caption{The original COS-Halos selection distributed 43 
galaxies in 25 bins of stellar mass and impact parameter 
between $\log M_* / M_{\odot}= 10$ and 11 as evenly as possible. 
The confirmed spectroscopic redshifts, as well as target substitutions and bonus galaxies, 
brought the sample to those 44 galaxies shown, which populate the 48 bins in 
this parameter space as shown. The stellar masses assume a Salpeter
IMF, 
and the impact parameters
are given in physical kpc at the galaxy spectroscopic redshift $z_{\rm sys}$. The symbol coding
of blue squares for star-forming galaxies and red diamonds for passive
galaxies are used throughout all subsequent figures.} \end{figure*}

This ``multiphase'' picture
draws support, and COS-Halos is also motivated by, the 
widespread detection of ionized gas and colder clouds within the hot medium surrounding the Milky Way, as traced by
the population of high-velocity clouds \citep[HVCs;][]{Sembach:2003dra,
Fox:2006kga, Collins:2003bla}. The widespread HVC O VI was interpreted by \cite{Sembach:2003dra} as collisionally ionized material at interfaces between cooler infalling clouds seen as the classical HVC complexes and a hot ($\sim 10^7$ K), extended ($\gtrsim 70$ kpc) Galactic
halo. In this model, the HVC O VI is an indirect indicator of the hot Galactic corona 
gas rather than a major halo component in its own right, at least within $\sim 10$ kpc 
of the Galactic disk \citep{2012ApJ...759...97H}. 
We were motivated to design 
COS-Halos in part to provide some comparisons to the MW halo gas by exploiting 
the very different viewing geometry possible for external galaxies. In particular, the FUSE 
results on Galactic halo \ion{O}{6} \cite{Sembach:2003dra} was a key motivation for our choice to 
design COS-Halos around galaxies with sufficiently high redshift for COS to measure their \ion{O}{6}. 

All of these considerations about galaxy accretion motivated us to design a survey
that would examine CGM gas over a range of galaxy stellar masses spanning the
expected transition from cold to hot accretion ($\sim 3-5 \times 10^{10}$
M$_{\odot}$, or $L \sim L^*$ galaxies), and to use ionization diagnostics
that could be compared directly to Milky Way O VI absorption 
and to the samples of absorbers in which O VI is detected in association 
with diffuse HI and sometimes other metals \cite[e.g., ][]{2008ApJS..177...39T, Thom:2008gma}. 

{\it Mass, Physical Phases, and Metal Content of the CGM and IGM:}
The COS-Halos Survey is also motivated by a desire to assess the total
mass of gas in the gaseous halos of galaxies. Measuring the mass of the
CGM addresses at least three key problems in galaxy evolution. First,
this mass reservoir is a potential source of fuel for star formation in
galaxies, which generally have short gas consumption times compared to
the duration of their star formation histories. Second, the
mass in the CGM may help to explain the deficiency of galactic baryon
budgets with respect to their dark matter halos, if a significant
budget of baryons resides in the CGM. Third, the baryonic mass
in the CGM may be a significant reservoir of {\it cosmic} baryons,
which are undercounted at low redshift
\citep{1992MNRAS.258P..14P, 1998ApJ...503..518F, 1999ApJ...514....1C, prochaska-etal-11-OVI-HI} 
but which may reside in diffuse ionized gas within galaxy halos
\citep{2007ARA&A..45..221B}. All these problems can be addressed by a
survey that can estimate the total quantities of neutral and ionized gas
within $R_{vir}$ of low-redshift galaxies. These considerations drive
the survey toward a design that gives good constraints on the total
gas column densities and ionization corrections derived by the best
available metal-line ionization diagnostics, for galaxies over a range
of stellar mass (as a proxy for dark matter halo mass). Such a survey can
also assess the CGM mass as a function of gas temperature and/or density,
if appropriate diagnostics are available. It is important to understand
the physical state of the gas to avoid double-counting 
baryons already in the census, such as might occur if they are highly
overlapping with the metal-enriched, mostly photoionized absorbers in the
\Lya\ or O VI phases of the IGM \citep{Thom:2008gma,
2008ApJS..177...39T}. A full accounting of the CGM mass by physical phase
is thus an important long-term goal of the COS-Halos survey. Finally,
the COS-Halos survey is also motivated by a desire to assess the heavy
element content of the CGM and to find out how far from galaxies metals have  
propagated. Metal transport is important both as a factor in Galactic
chemical evolution and as a tracer of feedback by galaxies into their
surroundings, which in turn alters their evolution. 

Papers addressing the main COS-Halos survey so far include 
the \cite{werk-etal-12-galaxies} compilation of galaxy spectroscopy, the
\cite{tumlinson-etal-11-OVI-statistics} study of \ion{O}{6} bimodality
in galaxy halos, the \cite{2012ApJ...758L..41T} study of \HI\ in
early type galaxies, and the \cite{2013ApJS..204...17W} empirical description 
of the CGM as seen in metal lines. These studies collectively show the power of the
COS-Halos dataset to reveal the properties of gas surrounding galaxies and
its relationship to the properties of those galaxies. The results of these main survey
papers will be summarized in connection with results below. 

Data from COS-Halos has also been used for several
additional investigations of gas within and around galaxies in the low-redshift
universe, apart from the main survey. \cite{tumlinson-etal-11-J1009-LLS} 
reported the detection of a strong intervening \ion{O}{6} absorber associated with a galaxy
toward J1009+0713, and found it to be a complex, multiphase system
associated with a nearby star forming galaxy that likely contributes to
the ionization of the detected absorption. \cite{thom-etal-11-J0943-OVI}
reported the detection of a metal-poor cloud in close association
with a nearby star-forming galaxy that resembles the
expected properties of cold, metal-poor accretion onto star-forming
galaxies. \cite{Meiring2011} presented the first survey of low-redshift
DLAs using COS, which were analyzed for metallicities and relative
abundances by \cite{2012ApJ...744...93B}. \cite{2012MNRAS.424.2896L}
combined a portion of data from COS-Halos with other COS programs
to produce an unbiased estimate of the covering fraction of high-velocity
ionized gas in the Milky Way. Finally, using data from a combination 
of COS programs with published absorbers, \cite{2013ApJ...770..138L} 
have investigated the bimodal distribution of metallicities in Lyman-limit 
systems (LLSs) presumably tracing CGM gas. 

This paper presents both the general design and execution of COS-Halos 
and the resulting census of \HI\ detected near the targeted galaxies. The 
paper is organized as follows. Section~\ref{design-section} describes the design 
features of the program and how they meet the scientific goals described 
in this introduction. Section~\ref{data-section} covers the data collection and 
analysis, concentrating on the COS data; the reader interested in the 
full details of the complementary ground-based spectroscopic survey is referred
to \cite{werk-etal-12-galaxies}. Section~\ref{results-section} presents the basic 
empirical characterization of the \HI\ near the survey galaxies, in terms of
absorption strength, kinematics, and correlations with galaxy properties. Section~\ref{other_studies_section} compares
these results to prior studies of \HI\ near galaxies. Section~\ref{CGM_section} examines 
possible origins for the detected \HI, including gas inside the targeted
halos and gas outside in the IGM, and various specific sites of origin such as satellite 
galaxies or galaxy groups. Section~\ref{summary-section} summarizes our major conclusions. 

Throughout our analysis we adopted a
cosmology specified by WMAP3 ($\Omega _m = 0.238$, $\Omega _{\Lambda} = 0.762$, 
$H_0 = 73.2$ \kms\ Mpc$^{-1}$, $\Omega _b = 0.0416$). 
Distances and galaxy virial radii are given in proper coordinates. 



\section{The Design of COS-Halos}
\label{design-section}

\subsection{QSO and Galaxy Selection}

We began our selection with the SDSS DR5 catalog of quasars \citep{2007AJ....134..102S}. 
This catalog was cross-matched with the GALEX GR3 photometric catalog\footnote{\cite{2005ApJ...619L...1M}, http://galex.stsci.edu} to assign FUV
and NUV magnitudes to each SDSS QSO. Objects with multi-epoch detections
in the GALEX data had their fluxes averaged, weighted by their respective 
errors, for a final value. We selected for QSOs with $z_{\rm em} < 1$ to minimize contamination by
foreground LLSs  that would have masked absorption
by the targeted galaxies at $z \lesssim 0.3$. 
We included in our search only QSOs bright enough (GALEX FUV
$\lesssim 18.5$) for COS to obtain S/N $\sim  10-12$ in 5 orbits with
the medium-resolution gratings. We also avoided QSOs with $> 1$ \AA\ 
\ion{Mg}{2} absorbers at $z > 0.4$ based on published catalogs \citep{2006ApJ...639..766P} and visual inspection of 
optical QSO spectra. This was done to avoid losing a large range of QSO spectrum to Lyman limit systems. This criterion selects against absorber systems at $z \gtrsim 0.4$, well above the redshifts of interest for the prime sample. This screening benefits our own scientific goals, but we emphasize that the COS-Halos dataset is strongly biased against any absorbers that would exhibit strong \ion{Mg}{2} at $z \gtrsim 0.4$, such as metal-enriched LLSs and DLAs, and therefore should not be used to derive quantities such as redshift number-density ($dN$/$dz$) where unbiased sightlines are necessary.  We did not apply any other selection criteria to the QSOs themselves. The final list of QSOs observed appears in Table~\ref{qso_table}.


\subsection{Galaxy Catalog and Selection}

COS-Halos departs from the typical selection technique used for QSO 
absorption-line 
studies by choosing galaxies in advance of absorbers but without
secure spectroscopic redshifts. Most previous studies have either
selected galaxies at known (usually very low) redshifts \citep[e.g.,][]{1996ApJ...464..141B}, 
obtained spectroscopic redshifts prior to analyzing absorber data \citep[e.g.,][]{2011ApJS..193...28P},   
performed post-facto galaxy surveys after the absorber data were obtained \cite[e.g.,][]{chen-etal-01-Lya-imaging,1998ApJ...508..200T}, 
or compiled pre-existing galaxy catalogs surrounding the QSOs for which absorption data had already been 
obtained or was expected to be obtained \cite[e.g.,][]{2002ApJ...565..720P, wakker-savage-09-OVI-HI-lowz}. Rather than adopt any
of these techniques, COS-Halos chose galaxies in advance, but based on
SDSS {\em photometric} redshifts. Selection based on photometric redshifts vastly increases the number
of galaxies available, since the photometric survey (DR5 at the
time of our planning in 2008) contained tens of millions of galaxies but
only 675,000 of these were observed spectroscopically. Furthermore the SDSS spectroscopic survey is concentrated at $z \leq 0.1$, too low to place \ion{O}{6} in the COS band ($\lambda > 1150$ \AA). Using the photometric redshifts allows a selection to push beyond $z \sim 0.1$, and makes close pairings between galaxies and QSOs much easier to find, especially at the low projected angular separations needed to probe the CGM inside $\sim 50$ kpc. The tradeoff is that secure spectroscopic redshifts must be obtained subsequently. However, this places COS-Halos in no worse position than every survey that relies on post-facto galaxy surveys with no prior knowledge of galaxies in the field. With relative errors of $\sigma _z / z \lesssim 0.2$ \citep{2008ApJ...674..768O},  the photometric redshifts were adequate to ensure that the key ionization diagnostics of CGM gas, particularly \ion{H}{1} and \ion{O}{6}, still fall within the $1150-1800$ \AA\ range of the COS M gratings. Thus the photometric redshifts allowed us to select good candidate pairs from a large pool with little additional risk.

\begin{figure*}[!t] \begin{center} 
\plotone{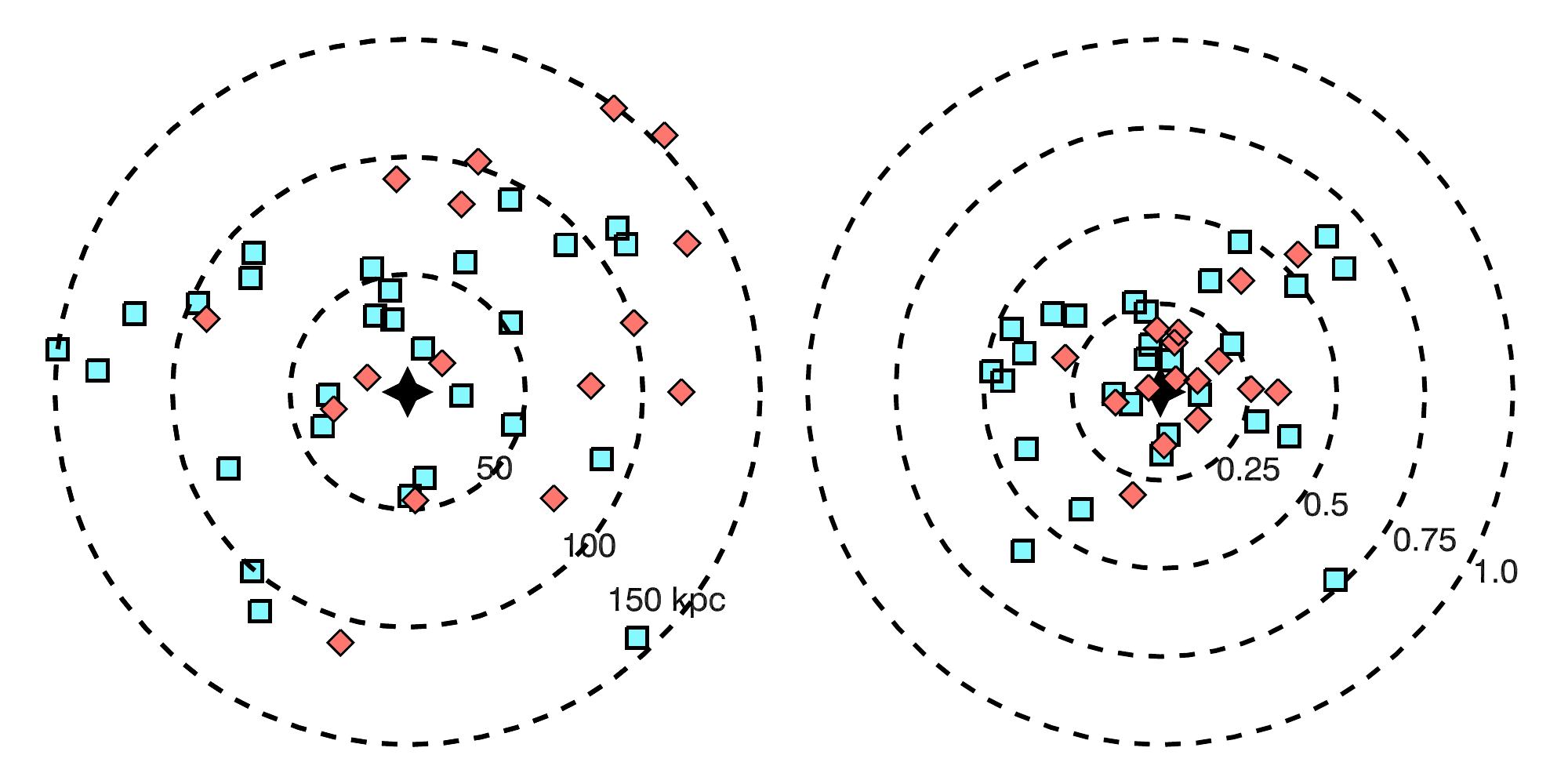}
\end{center} 
\caption{Target figure showing
the distribution of QSOs on the sky with respect to the target
galaxies (shifted to the center). Star-forming galaxies are shown in 
blue, and passively evolving galaxies in red. 
At left the radial coordinate is physical kpc at the galaxy redshift, 
at right this coordinate is translated to the fraction of galaxy's 
virial radius, $R_{\rm vir}$, at which the sightline intercepts 
the halo. No knowledge of galaxy disk orientation or inclination 
with respect to the sightline is implied here. \label{targfig} } 
\end{figure*}

We began our galaxy searches by obtaining a list of all SDSS galaxies (photoObj.type = 3) within 8$'$ ($\simeq$1 Mpc at $z = 0.1$) of every QSO that met the FUV magnitude and redshift selection criteria described above. The dereddened $ugriz$ magnitudes of these galaxies were then $k$-corrected \citep{2007AJ....133..734B} based on the photometric redshift ($z_3$, which agreed best with the spectroscopic redshifts when both were available). We adopted a cosmology specified by WMAP3 ($\Omega _m = 0.238$, $\Omega _{\Lambda} = 0.762$, $H_0 = 73.2$ \kms\ Mpc$^{-1}$, $\Omega _b = 0.0416$). From these $k$-corrected magnitudes we derived estimated stellar masses using the relation of \cite{McIntosh:2008kqa}. Galaxy impact parameters to the sightline were computed in physical kpc at $z_3$ in the adopted cosmology. All these galaxy quantities entered our master catalog of galaxies from which to choose pairings with QSOs. 

\begin{figure*}[!t] 
\begin{center} 
\plotone{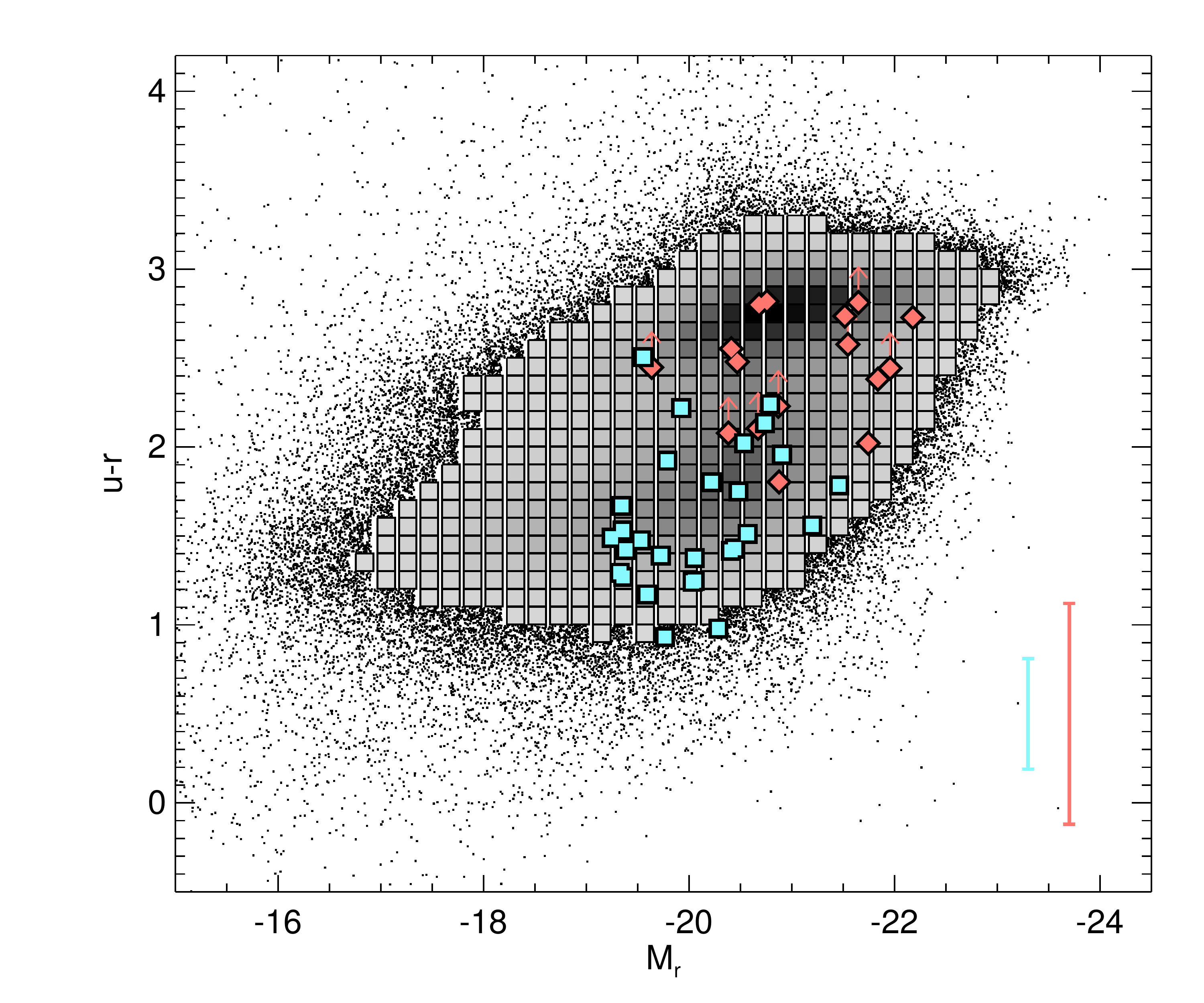}
\end{center} 
\caption{The COS-Halos sample in a color-magnitude diagram using the $u-r$ color and the absolute $r$-band magnitude, both from dereddened, $k$-corrected SDSS photometry. Color coding is the same as before. The background distribution of galaxy number densities is derived from the NYU Value Added Galaxy Catalog \citep{2005AJ....129.2562B}. The bars at lower left show the mean error in the colors for the star-forming ($\pm 0.3$ mag) and passive subsamples ($\pm 0.6$ mag). Roughly half of the passive galaxies have uncertain intrinsic colors with errors up to 1-2 magnitudes as a consequence of their non-detection in the SDSS $u$ band; these are plotted with lower limit arrows.
\label{cmdfig}} 
\end{figure*}

We then binned this subset of galaxies by mass over $\log M_* \simeq 10 - 11$ and by 
impact parameter over  $\rho < 150$ kpc, in 25 bins of $\Delta M_* = 0.2$ and $\Delta \rho
= 30$ kpc (see Figure~\ref{binfig}). At the time of selection, these quantities were estimated based on photometric 
redshifts (Figure~\ref{binfig} adopts the final spectroscopic redshifts). We required that the galaxies satisfy the condition $0.11 +
\sigma _{z} < z_3 < z_{\rm QSO} - \sigma _{z}$ to ensure that \OVI\
was redshifted into the COS FUV band, and to reduce the chance that
the galaxy would turn out to lie at the same redshift as the QSO. We
attempted to place 2 galaxies in each bin, to cover this parameter space
as evenly as possible. All else being equal, we chose pairings with 
brighter QSOs to minimize the observing time required to build a sample. This 
led to a selected sample of 43 ``target'' galaxies, which was then modified by 
additions and subtractions (see below) into a sample of the 44 galaxies shown in Figures~\ref{targfig} and \ref{cmdfig} and listed in the first part of Table 2.  
The technique of building an aggregate map of galaxies, each probed by a single 
sightline, is visualized in Figure~\ref{targfig}, which shifts the target galaxy to the center
and displays the QSOs distributed properly in impact parameter and
position angle. Figure~\ref{cmdfig} shows the galaxies in a color magnitude 
diagram along with the parent galaxy population from the NYU-VAGC \citep{2005AJ....129.2562B}. 

\subsection{Target and Bonus Galaxies}

While collecting spectra on the ``target'' galaxies -- those that
were originally selected -- we also obtained spectra on photometrically-selected
galaxies near the sightline that met the criteria of stellar
mass and impact parameter but different $z3$. Galaxies that ended up
with $z \simeq z_{\rm QSO}$ were discarded, but those with acceptable
redshifts are included in our analysis as ``bonus'' galaxies. Note
that these galaxies were included without regard to absorption along
the sightline, as generally their redshifts were obtained before the COS data, 
so they fulfill the same requirements of prior galaxy
selection as the original ``targets''. The galaxies in our sample are
detailed in Table 2. The listed properties are drawn
from \cite{werk-etal-12-galaxies} directly and are listed here for
reference in later tables. 

\subsection{Galaxies Omitted from the Main Sample}

Four of the galaxies originally targeted turned out to have spectroscopic 
redshifts of $z \lesssim 0.1$, placing their \ion{O}{6} out of the COS band. 
In these four cases the galaxies also have significantly lower luminosity and/or
stellar mass than we intended to include in the sample. These sub-$L^*$
galaxies have been excised from the main sample that is analyzed 
further here. Another originally-selected galaxy (J1553$+$3548 97\_30) has not had its spectroscopic redshift measured securely, so we cannot include it in the sample. 
These five cases are listed in Table~\ref{galaxy_table} for completeness. 

\subsection{Galaxy Neighbors and Environment \label{environment_section}}

The properties of gas in galaxy halos, and even the galaxies themselves, can be influenced by galaxies nearby in group environments or the same regions of the large-scale structure. Thus quantifying galaxy environment can be an important component of studies of their star formation, evolution, and surrounding gas. 

The COS-Halos galaxies were selected based on SDSS photometric redshifts, so this survey does not feature the ready recovery of environment information enjoyed by surveys done with multi-object spectroscopy before or after the absorption line data is obtained. However, to mitigate the effects of multiple galaxies at the same redshift, during the selection of targeted galaxies we preferred candidate galaxies without photometric redshift coincidences within the same $\sim 1$ Mpc search radius used to pick the candidate targets. That is, sightlines with two photo-$z$ candidates at similar $z$ (within about 1 $\sigma _z$) were not chosen. This choice introduces a bias into the nearby environment of these galaxies, in favor of isolation over group membership with other galaxies. This bias acts mainly against other $\sim L^*$ or brighter galaxies, as fainter galaxies at the same redshifts either have much larger photo-$z$ errors or drop out of the SDSS photometry altogether. This bias against close neighbors is difficult to quantify because it can only really be assessed with a complete spectroscopic redshift survey and comparisons to control samples to which different selection criteria have been applied. However,  because we did not aggressively omit all possible coincidences and photometric redshift errors are not perfectly well-behaved, and because we did not apply any such screening to the bonus galaxies, redshift coincidences did arise during the spectroscopic followup stage of the survey. 

The full list of galaxies for which we obtained spectroscopy is described in \cite{werk-etal-12-galaxies}. In a few fields the followup spectroscopy
identified more than one galaxy at the redshift of the pre-selected,
targeted galaxy. In these cases, we have taken the most massive
(equivalently the brightest in SDSS $r$) as the adopted ``canonical'' galaxy for
analysis purposes. There are only two cases for which choosing the closest galaxy rather than the most massive would change the type of the galaxy and therefore affect the comparisons we do in later sections. The galaxy 270$\_$40 toward J2257$+$1340 is a passive galaxy with two smaller, star-forming galaxies slightly nearer the sightline. This system is not detected in \HI, so the change would slightly affect the \HI\ detection rates examined below (see \S~4.1). The damped system associated with galaxy 110$\_$35 toward J0928$+$6025 has a less massive star forming galaxy nearer the sightline, which would result in two, rather than just one, of the three damped systems being associated with star-forming galaxies. These minor ambiguities do not affect the larger conclusions reached below.

We have not yet performed complete spectroscopic surveys in these fields, so we cannot say anything more about their large-scale environment, near neighbors, or possible satellites without knowledge of exact redshifts. To address this issue with the available information, \cite{werk-etal-12-galaxies} searched the fields surrounding the targeted galaxies for photometric redshift candidates and then quantified neighboring galaxies by (1) counting the number of galaxies within 5 Mpc, and (2) identifying the distance to the nearest neighbor. This same search was performed for a set of 500 SDSS control galaxies with the same range of $r$-band absolute magnitudes and redshift as COS-Halos. Figure 8 of \cite{werk-etal-12-galaxies} shows the results of these environment tests. For all but a few COS-Halos galaxies the nearest-neighbor candidate at similar luminosity is at $> 1$ Mpc projected separation, and the median nearest neighbor distance is 2.5 Mpc (2.7 Mpc in the control sample, a statistically insignificant difference). There is also no significant difference in the counts of galaxies within 5 Mpc for COS-Halos and the control sample. On this basis, \cite{werk-etal-12-galaxies} concluded that there was no evidence that the COS-Halos galaxies are unusual in their large-scale (1-5 Mpc) environments. The canonical galaxy in each field is the most luminous galaxy within 300 kpc of the QSO sightline at its redshift and in most cases there are no likely $L \sim L^*$ photo-$z$ candidates within 1 Mpc. We also believe that the large-scale environments ($\lesssim 5$ Mpc) of these galaxies are not unusual. We will revisit this issue when considering a group gas origin for the detected halo gas (\S~\ref{group_gas_section}). 

\begin{figure*}[!t] 
\begin{center} 
\plotone{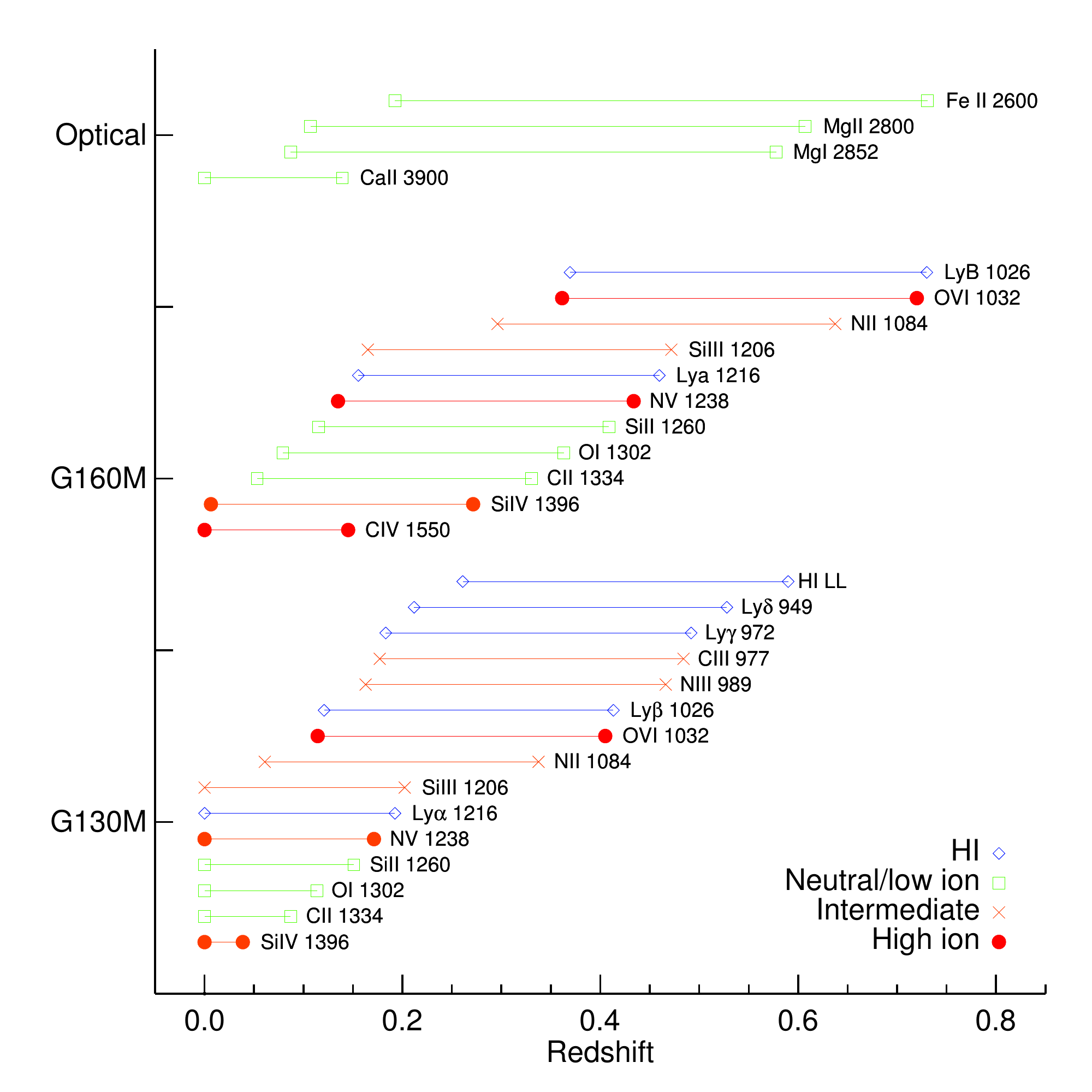}
\end{center} 
\caption{COS-Halos was designed with galaxies at $z = 0.15 - 0.35$ to place \OVI\ and other key ionization
diagnostics on the COS FUV G130M and G160M gratings. \label{lineplan_fig}} 
\end{figure*}


\section{Data Collection and Analysis}
\label{data-section}

\subsection{COS Data Reduction} 




\begin{figure*}[!t] 
\begin{center} 
\epsscale{0.6}
\plotone{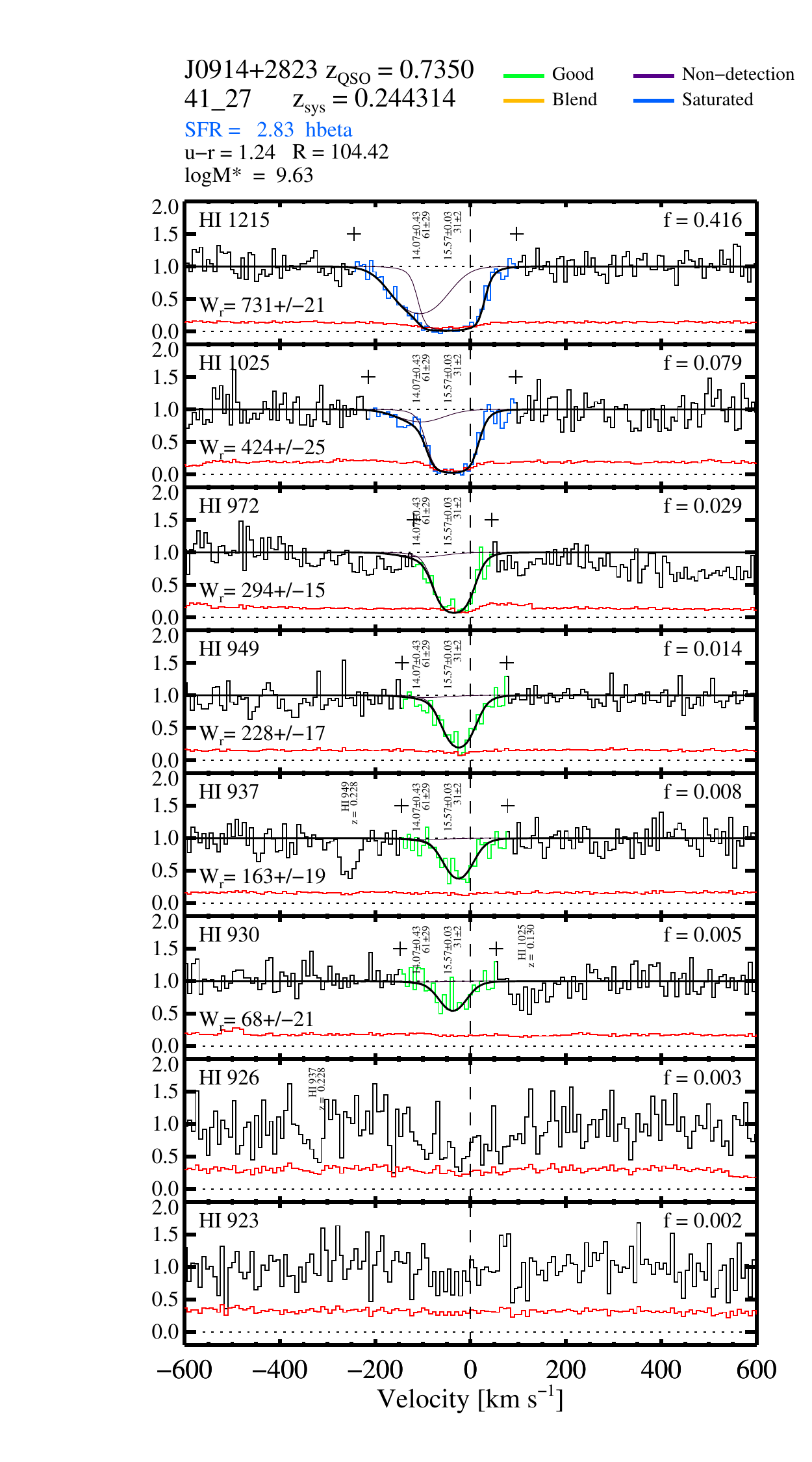}
\end{center} 
\caption{A stack plot illustrating our data
and its analysis. These panels show \HI\ absorption in the
rest frame of the targeted galaxy 41\_27 toward J0914+2823. All data is shifted to
place the rest-frame of the galaxy systemic velocity at $v = 0$. Galaxy
quantities are shown at the top. Line quality flags are listed at the
top and color-coded in the individual panels, where $W_r$ measurements
are also displayed. Lines identified with other absorption line systems
are marked with their species and redshift in rotated text. The profile 
fits to individual components are shown in thin purple lines and their 
fitted column density \NHI\ and Doppler $b$ are given above positioned 
at the component's central velocity. The composite profile fit is shown in the 
heavy black line. Profile fits have been convolved with the nominal 
COS LSFs for comparison to the data. Plots like these for each system 
are provided in the appendix and available in the electronic journal. 
\label{stack_fig}} \end{figure*}


The COS observations are detailed in Table~\ref{qso_table}. 
We planned for a uniform set of COS data obtained with both of the FUV
medium resolution gratings, G130M and G160M \citep{Dixon:2010tv, green-etal-12-COS}. We
used the FUV channel detectors in \verb1TAGFLASH1 mode. We tuned the
central wavelengths (\verb1CENWAVE1s) to avoid placing the 14 \AA\ gap
between FUV detector segments at the position of \OVI\ for the targeted galaxies, based on $z_{\rm sys}$, and to ensure that \OVI\ would be covered at the short-wavelength end of the G130M setting. Generally 
two \verb1CENWAVE1s were used to ensure complete wavelength coverage from 1140-1800 \AA,  though for a few QSOs only one position was used to ensure that all the exposure 
time went into covering lines expected to be near the edge of the recorded spectrum. The balance
between the time allocated to G130M and G160M was set to ensure S/N =
8 per resolution element or better over $1150 - 1700$ \AA. 

The COS data were obtained from MAST\footnote{http://archive.stsci.edu} and uniformly processed by CALCOS
(v2.12) with standard parameters and reference files. We
performed our own co-additions to merge exposures obtained with different
\verb1CENWAVE1s and the two gratings. This was the same method
used by \cite{Meiring2011}, \cite{tumlinson-etal-11-J1009-LLS} and
\cite{thom-etal-11-J0943-OVI}. This procedure operates on the gross counts
vectors stored in the CALCOS \verb2x1d2 output files, and tracks the
count rates in each raw pixel; each COS resolution
element at $R \sim$ 18,000 covers $\simeq 16$ \kms\ and is sampled by six raw pixels. 
We track counts and count rates so that
variances can be computed rigorously in the Poisson limit at low count
rates using the tables of \cite{1986ApJ...303..336G}. Exposures taken at
the same grating and \verb1CENWAVE1 were added first. This coadd was then
summed with exposures in the same grating at different \verb1CENWAVE1s,
followed by a sum of the two grating spectra to
produce a single 1D trace from 1150 - 1800 \AA. At each stage of the
co-addition, exposures were shifted in wavelength by steps of 1 raw pixel (1/6 resel) 
by aligning common Milky Way interstellar lines (e.g. Si II, C II, Al II) in velocity space. These
alignments ensure that small zeropoint shifts in the wavelength solution
from exposure to exposure are mitigated as much as possible.

The photocathode grid wires lying above the COS microchannel plates cast
shadows onto the detector and are the main source of fixed-pattern
noise in our data. Smaller fluctuations caused by the microchannel plate
pores generally do not appear at the S/N ratios of our data. However, the
grid wires are easily visible in our raw data and must be corrected. There 
are other fixed-pattern noise features that must also be removed. 
We adopted flat-field reference files prepared and communicated to us by D. Massa
at STScI and filtered for high-frequency noise by E. Jenkins. These 1D
files allow us to correct the shadowed pixels by modifying the 
effective exposure time and count rate in each pixel prior to coadding
it with the others. The resulting 1D, flat-corrected summed spectra
were then binned by 3 raw pixels to give final science-grade
spectra with $\sim 2$ bins and S/N $\sim 10-12$
per COS resolution element (FWHM $\simeq 18$ \kms). 
These 2-bins-per-resel spectra are in units of wavelength and counts per second and are used in all further analysis.

\subsection{COS Data Analysis}

Our absorption-line analysis begins with the optimally binned
one-dimensional count-rate spectra described above. The goal of the absorption-line
analysis procedure is to identify and measure a set of common
lines associated with the sample galaxies. The design of COS-Halos
introduces a significant simplification over the usual requirement of
identifying every line in a QSO spectrum, because we have measured the
galaxy systematic redshifts $z_{sys}$ to high precision ($\sigma _{specz} \sim
30$ km s$^{-1}$ in the rest-frame) and can focus on common lines at predictable places in 
observed wavelength. A subset of the common lineset
appears in Figure~\ref{lineplan_fig} to illustrate the redshift ranges over
which they are available in the COS-Halos data, including the
optical data described further in \S~3.4. This figure also motivates the design choice to focus on 
galaxies at $z \simeq 0.1 - 0.3$. Of course the availability of a given line depends not only 
on redshift but on local S/N and the contaminating presence of other lines, 
so that each line in the common set must be identified and evaluated 
on its own to judge detection. 

We developed a semi-automated procedure for identifying and measuring
lines that automatically extracts and processes slices of spectrum
around each of the lines in the standard set. This procedure is used 
here and in the complementary analysis of metal lines in COS-Halos \citep{2013ApJS..204...17W}.
Each slice covers $\pm 500$ bins (each bin is 3 raw pixels spanning approximately 1/2 of a resolution element) 
on either side of the systemic galaxy
redshift, whether the corresponding line is detected or not. These slices are then 
independently continuum-normalized using fifth-order Legendre polynomials
and trimmed to $\pm 1500$ \kms\ of $v = 0$. Lines of a single element
are grouped and plotted together in rest-frame velocity space, as shown in
the example in Figure~\ref{stack_fig}. The complete set of stack plots is 
available as Figure set 5 in the electronic journal. 

These uniform plot sets allow for quick assessments of which lines from a given element and species should be measured. These visual inspections determine only whether or not a 
particular line will be flagged for measurement; they do not decide formal detections or limits. 
They also rely on ``sightline'' linelists, or catalogs of absorption lines from systems 
at other redshifts in the sightline, to 
facilitate the identification of nearby lines blended with 
absorption from the target. 
The process is then repeated to measure equivalent widths
and apparent column densities over velocity intervals
specified manually in the system linelist by an inspection of the stack
plots. This generates rest-frame equivalent widths $W_r$
for each line, and assigns data quality flags based
on S/N, blending, and saturation. We calculated apparent column densities
using the method of \cite{Savage:1991ka}, which converts the normalized 
flux spectrum into an optical depth and then to an apparent column density 
prior to integrating over the desired velocity range. 
Detections are automatically flagged as 
such by a requirement for $\geq 3\sigma$ significance (not the visual inspection used
to flag the line for measurement).   Undetected lines are recorded
as upper limits based on the local S/N, usually integrated over $\pm 50$ \kms; the 
errors and limits in the stack plots and Tables 4 and 5 are $1 \sigma$ values, and we generally 
adopt $2 \sigma$ limits in analysis and interpretive plots. Any line that drops below 
10\% transmission relative to the normalized continuum is automatically flagged as 
saturated and treated as a lower limit for column density measurement; lines can 
also be manually flagged as saturated based on more subjective judgement, flagged 
as blended based on detailed inspection of possible contamination, or left out 
entirely if the data quality is poor, continuum normalization is inadequate, or 
for any other reason prohibiting a clean measurement. Even lines that are
not used generally appear in the stack plots for completeness. The directly integrated 
apparent column densities (replaced with the \lya\ profile fitting for damped systems) 
are used in almost all of our subsequent analysis. Tests performed with the 
profile-fitted column densities are called out as such throughout the text. 

After this process, the resulting rest-frame 
equivalent width $W_r$, directly integrated column density $N$, velocity ranges, and data quality flags are stored in a table of line
results for each system for later analysis. Adopted column densities are computed
using an average of uncontaminated, unsaturated Lyman lines weighed by their 
relative errors. For systems in which all the detected lines are saturated, 
the adopted column density is a lower limit set by the measurement for the weakest (e.g. most sensitive) available line. At the end of the process the plots are then regenerated to incorporate 
all the resulting information about column densities, errors, flags, and profile fits as 
shown in Figure~\ref{stack_fig}. 

Our automated analysis routines searched for every HI Lyman series
line that was covered in the COS data. The number of available lines
increases as the galaxy redshift increases and shorter-wavelength Lyman
lines redshift onto the detector. \Lya\ always falls on the detector, but
in a few cases is blended with extraneous absorption and cannot be measured. 
The basic \HI\ results are listed in Tables~4 and 5, where we list the system labels 
($=$ QSO name and galaxy name), the
velocity width for integration of the \Lya\ profile (or \Lyb\ where \Lya\
is damped or contaminated). Table~4 gives rest-frame equivalent
widths $W_r$ for the first six Lyman series lines. Table~5
gives the full results including line-profile fitting. Full results for all lines are available in our electronic tables and stack plots. 

We have attempted to thoroughly identify and measure all absorption appearing 
within $\pm 600$ \kms\ of the targeted galaxies, and to associate it convincingly with either the target galaxy itself or an 
intervening absorber at another redshift. These identifications for unrelated lines are 
marked in the stack plots. Some weaker lines ($\lesssim 100$ m\AA), 
particularly those that appear only in the \Lya\ region of the targeted absorber, 
cannot be allocated conclusively to intervening systems and cannot be confirmed as \Lya\ components 
near the targeted system because they are too weak for Ly$\beta$ to be expected, or because there is 
blending, low S/N,  or no wavelength coverage. We have chosen to allocate these lines to the targeted 
absorber if there is no reliable alternate identification at a different redshift, even if the line cannot be confirmed with Ly$\beta$; these lines are called out 
in Table 5 with the flag ``Ly$\alpha$ only''. 

The following analysis uses the rest-frame equivalent widths $W_r$
where appropriate, but sometimes also the best estimate or limit
on the \HI\ column density, \NHI,\ derived from fitting the profiles. We obtain the
\NHI\ measurement or limit with one of three methods depending on the
column density of the system expressed in the bitwise flags ``Adopt" and ``Method"
in Table~5.  ``Adopt'' is set to 1 for good
measurements of $N_{\rm HI}$, 4 for non-detections that give upper
limits, and 8 for saturated systems that give lower limits. Damped
(DLA) or nearly damped (subDLA) absorbers ($\log$ \NHI\ $\simeq 18 -
20$) have their \NHI\ derived by profile fitting to the \Lya\ profiles
(Adopt = 1, Method = 3). Undetected absorbers ($\log$ \NHI\ $\lesssim
14$) have their upper limits on \lya\ and \NHI\ derived from the
directly integrated noise over $\pm 50$ km s$^{-1}$ around $z_{sys}$
(Adopt = 4, Method = 1 for direct integration). The plots that follow
include these $2\sigma$ upper limits, while the tables specify the
$1 \sigma$ values. Finally, cases with intermediate column densities have their
\NHI\ estimated by direct integration of the line profiles (Method =
1) in apparent column density. If the highest available line is weak
and unsaturated, the resulting measurement is considered a measurement
(Adopt = 1). If the highest available line shows evidence of saturation,
the resulting \NHI\ is considered a lower limit (Adopt = 4).

In addition to measurements of line equivalent widths
and integrated column densities, for most systems we use line-profile fitting to resolve
the kinematic components of detected absorption and to estimate column
densities that take into account profile shapes and saturation. This
procedure uses the line slices generated by the automated pipeline 
and performs Voigt-profile fits to derive the column density $N$,
the Doppler width $b$, and the velocity offset $v$ for each component. 
When multiple transitions for a given species are available, the same $N$, $b$, and
$v$ parameters are applied to all transitions and optimized jointly. The number of
components fitted to a given profile or set of profiles is determined by
visual inspection of the data. We do not attempt to fit every transition
of every species and we ignore strong blended profiles. 
Non-detections are used only when they provide significant constraints 
on the column density to the high end, such as when when a stronger transition 
is detected and a weaker transition of the same species is not. 
User judgment is required to decide the number
and placement of unrelated nuisance absorption lines from other redshifts
that blend with the lines of interest. When multiple transitions of the
same ion are fitted simultaneously (such as more than one line from the
Lyman series), lines appearing on different regions of the COS detector
can experience small shifts (usually $\lesssim$ 5-10 \kms\ but in some
rare cases up to a full $\sim 20$ \kms\ resolution element) owing to errors in the COS
geometric distortion and/or wavelength solutions. To allow for these
small instrumental shifts, our multi-line fitting allows for small
velocity shifts relative to the strongest line (here, generally \lya)
as nuisance parameters whose fitted values are then ignored. 
Model intrinsic profiles are constructed with nuisance lines and velocity
shifts applied. These intrinsic model profiles are then convolved with the
COS line-spread function (LSF) as given at the nearest observed-wavelength 
grid point in the compilation by \cite{2009cos..rept....1G}. 

The \verb1MPFIT1
software\footnote{http://cow.physics.wisc.edu/$\sim$craigm/idl/fitting.html}
is used to do the optimization of the fit and to generate errors near
the best fit point, which are stored in the corresponding line slice
along with individual and total model profiles. These are formal
errors, computed from parameter covariance matrices derived within
\verb1MPFIT1. These fits are subject to several sources of error 
affecting components at different column densities and $b$-values. 
For systems with strongly saturated Lyman series lines that
do not yet exhibit damping wings ($\log$\NHI\ $\sim 16.5 - 18.5$), these
formal errors likely underestimate the true uncertainty in the fitted
column densities. In this column density range single-component absorbers
are on the flat part of the curve of growth, so that errors in \NHI\ and
$b$ are correlated. Furthermore, we generally cannot discern individual
component structure in these strongly saturated profiles (examples are
J1016$+$4706 359\_16 and J1322$+$4645 349\_11), so only one component
can be fitted, typically resulting in high \NHI\ and a single broad $b$
value. Thus, in the analysis that follows we typically use the more
conservative lower limits to \NHI\ derived from apparent optical depth
measurements of the highest available Lyman line. Assignment of a single
component most likely overestimates the true $b$-value in these cases,
so the formally derived $b$ can be considered reasonable upper limits
for the characterization of gas kinematics. It can also be difficult 
to recognize a broad and shallow component (as would be expected for \HI\ in hot gas at $> 10^5$ K) when it is superimposed on profiles with complex and strong narrow components.
Moreover, even when evidence of broad components is present, its interpretation is often ambiguous \citep[e.g. Figures 33-35 in][]{2008ApJS..177...39T}.  In some systems broad features
arising in continuum fluctuations or noisy, blended weak components 
can appear as a broad single profile giving very large $b$-values in the formal fits. 
Component fits that are considered uncertain because of saturation, poor data quality, or unreliable 
parameters are flagged as such in Table 5 and 
identified in figures. 

\subsection{Errors, Biases and Problems} \label{problems_section}

The COS-Halos database includes information from many sources: SDSS
photometry, Keck and Magellan spectroscopy, and COS spectroscopy. To assess the
robustness and statistical significance of our results, it is important
to budget for both statistical and systematic errors in the measured
and derived quantities. Our error budgets for galaxies and absorbers
are displayed in Table~3.

\begin{figure}[!t] 
\begin{center} 
\epsscale{1.2}
\plotone{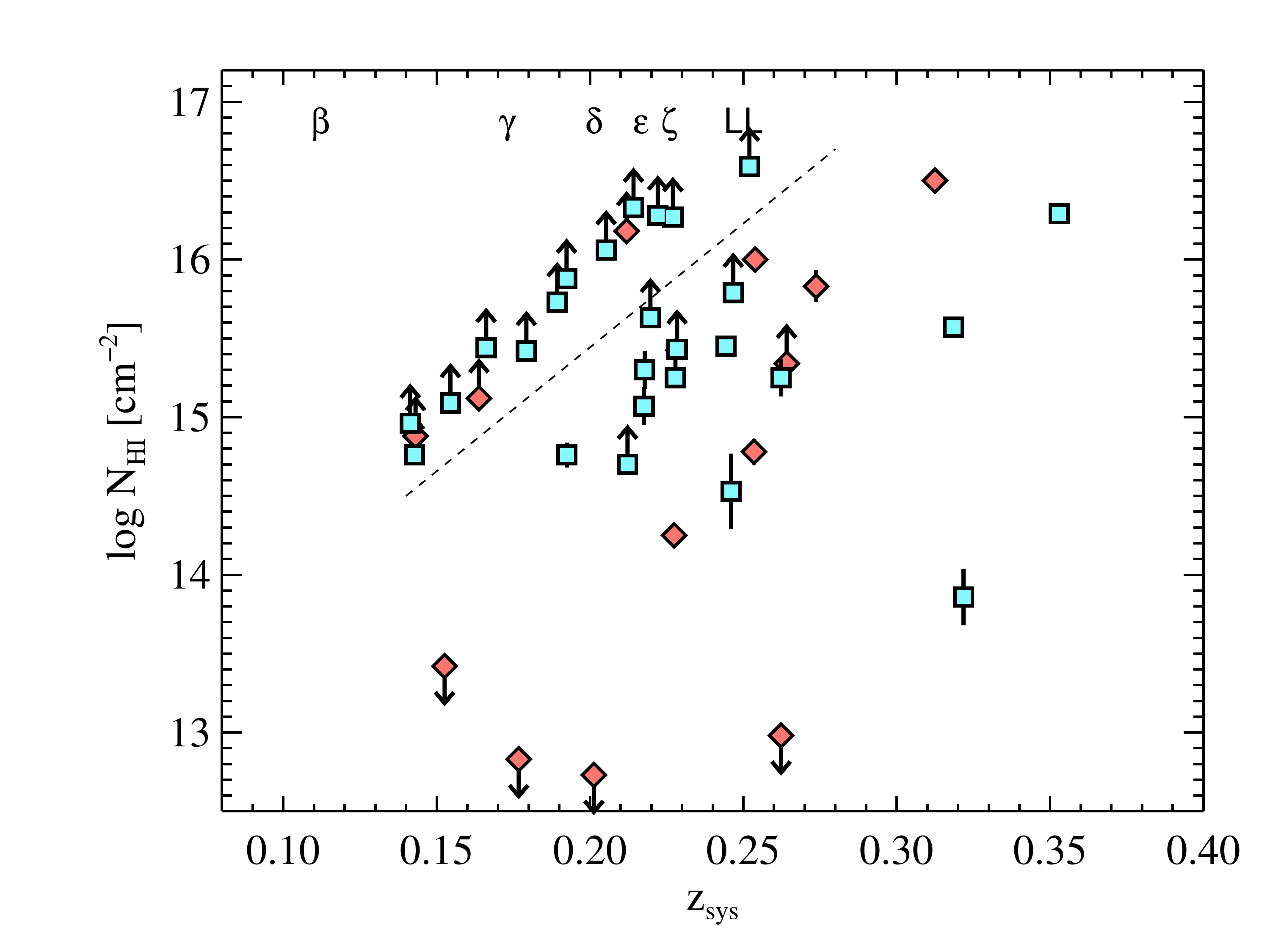}
\end{center} 
\caption{An illustration of the effects of saturation in the COS-Halos sample using adopted  
\NHI\ column densities versus redshift. For systems with upward pointing arrows we can infer only a lower limit to
the column density of \HI\ because of saturation. The \HI\ limit depends mainly on the redshift,
through the strength of the highest available Lyman series line. The dashed line 
roughly indicates the boundary of the saturation region. 
Thus the \HI\ limit increases with redshift as weaker lines appear on the
G130M detector at $\lambda \gtrsim 1140$ \AA. The redshift at which 
each line appears on the COS detector is marked by the corresponding Greek letter. 
The lower limits below the 
indicative saturation line are screened by higher-redshift Lyman limit
systems, so that the weakest line they have available is not determined
by their redshift. 
\label{saturation_fig} } 
\end{figure}

The saturation of the Lyman series lines, even the intrinsically weak
higher series lines, is a significant problem affecting a large portion of
our systems. The saturation effect means that the lower limit we adopt
for \NHI\ depends predominantly on the highest available Lyman line, which
depends in turn on the system redshift (given a fixed short-wavelength
cutoff in the COS detector). This strong saturation vs. redshift effect
is seen in Figure~\ref{saturation_fig},
where the points with lower limits follow a clear trend with redshift.  
As redshift increases, 
higher Lyman series lines become available and the minimum reliable 
column density estimate increases as well. The dashed line in the figure roughly
indicates the region where column densities are high enough to 
be saturated in the highest available line. Note that because saturation 
flags are set based on local factors such as S/N and line-profile 
shape, there is no one-to-one correspondence between saturation and 
column density and this line is just an approximation.
Symbols without arrows are systems for
which the \HI\ is weak enough to give a clean measurement\footnote{Four
systems outside the saturation region possess lower limits because their
highest Lyman series lines are screened by foreground Lyman limit systems
or other absorption at a different redshift, so the highest Lyman line for
which we can obtain a measurement is not accurately reflected by their
redshift alone.}.

Component structure, both resolved and unresolved, is an important issue
affecting the analysis and interpretation of our results. Multiple
components of absorption are often seen in the weaker Lyman series
profiles. In many of these cases the components are fully blended, at least
at COS resolution, and so inseparable in the \Lya\ profile. In many cases components are seen in \Lya\ that
are too weak to be detected in the higher transitions. The simultaneous
fitting of multiple Lyman series lines accounts for this effect by using
component information in the higher lines and then self-consistently
modeling the stronger transitions. But even this improvement imperfectly
captures the true underlying component structure, which usually can only
be seen in the weaker lines. Components that contribute to \Lya\ but not,
e.g. Ly$\beta$, will not always be included in the model. In most of our tests
below we consider the {\it total} \NHI\ measured from either the total
$W_r$ or by fitted components, but it must be recalled that
these are only approximations to the true \NHI\ and its distribution into
components. In our interpretations below we use metrics that are robust
to these considerations, where possible.

\subsection{Optical Data Analysis}

Comparing the properties of CGM gas to the properties of the host galaxies
requires robust redshifts, masses, star formation rates, colors, and
metallicities for the targeted galaxies. To measure these quantities
we obtained medium resolution optical spectroscopy using the LRIS
spectrograph at Keck and the MagE spectrograph at Magellan. The details
of data collection, reduction, and analysis for the COS-Halos optical
data are presented in \cite{werk-etal-12-galaxies}. Here we describe
only those aspects of the measurements that are necessary to evaluate the
CGM properties of the galaxies. The quantities derived from optical
spectroscopy are listed in Table~2 and their errors in Table 3. 

The high quality of the SDSS photometry and the ground-based spectroscopy
mean that the most significant errors in Table~3
affecting galaxy quantities are not statistical errors from noise in the
data. Rather, the most important galaxy uncertainties are systematic:
the calibration of the stellar mass derivations, including the IMF (10),
the SFR (11), and the calibration of the metallicity scale (13). These
errors are consistent with the literature on studies
of similar type.

The galaxy spectroscopic redshifts $z_{\rm sys}$ and the
systemic velocity scales below were derived from
a cross-correlation of the available nebular emission and/or stellar
absorption lines. The 25 \kms\ error in the galaxy systemic velocity
is a predominantly systematic error caused by instrumental effects such
as flexure. The galaxy colors
(e.g. $u-r$) and stellar masses $M_{*}$ were derived from the 5-band
SDSS photometry using the template-fitting approach implemented in the
{\it kcorrect} code \citep{2007AJ....133..734B} and the measured $z_{\rm
sys}$. Errors in color are from the underlying SDSS photometry. Systematic
errors from the mass-to-light ratio and IMF dominate the $\pm 0.2$ dex
error in $M_*$. Star formation rates are estimated from the detected
nebular emission lines or limited by their absence, with errors up
to $\pm 50$\%. For passive galaxies the SFR is given as a 2$\sigma$ upper
limit. Errors on combined quantities, such as the specific star formation
rate $sSFR = SFR/M_{*}$, are obtained from quadrature sums of the basic
terms as specified in Table~3. 

We compute halo masses $M_{\rm halo}$ by interpolating along the abundance-matching 
relation of \cite{2010ApJ...710..903M} at the stellar mass determined by {\it kcorrect} from the SDSS 
$ugriz$ photometry of the galaxy. We then determine the virial radius, \Rvir, with the relation 
\begin{equation} 
\Rvir\ = (3 M_{\rm halo}  / \Delta_{\rm vir}  \rho_{\rm crit}  4\pi )^{1/3}, 
\end{equation} 
where $\rho_{\rm crit}$ is the critical density at the spectroscopically-determined redshift 
of the galaxy and $\Delta _{\rm vir} = 200$. Accounting for systematic errors in the 
$M_*$ estimates (Table 3) and the scatter and uncertainty in the $M_{\rm halo} - M_*$ relation, we adopt 
uncertainty of 50\% on \Rvir. 

We have also observed most of these target QSOs with Keck HIRES to measure the
near-UV and optical-band absorption from neutrals and low ions such as
\ion{Fe}{2} and \ion{Mg}{2}. These data are described in \cite{2013ApJS..204...17W}. 

\section{The HI CGM As Characterized by COS-Halos}
\label{results-section}

The COS-Halos data measure the incidence, strength, and kinematics
of \HI\ and metal-line absorption surrounding galaxies as a function
of galaxy stellar mass and out to impact parameter $ \simeq
150\kpc$. We have already published results from the COS-Halos survey
of \OVI\ absorption surrounding galaxies, and concluded that \OVI\
traces a significant reservoir of metals in a highly ionized
CGM \citep{tumlinson-etal-11-OVI-statistics}. An analysis of the
COS-Halos sample of lower metallic ions (e.g., CII/III/IV, SiII/III/IV) has been 
published separately by \cite{2013ApJS..204...17W}. The remainder of this
paper is focused on the COS-Halos survey of \HI\ absorption in the sample.

This section presents the key COS-Halos results on \ion{H}{1} surrounding galaxies. 
In \S~\ref{strong_subsection}, we examine the
strength of the  \HI\ absorption with equivalent widths and column
densities. Section~\ref{bound_subsection} examines the kinematics
of the detected \HI\ with respect to the galaxies.
Section~\ref{cold_subsection} examines the internal kinematics and linewidths of the
detected gas. We compare
the star-forming and passive galaxy subsamples to one another in all these sections (and later in 
\S~\ref{redblue_subsection}).  These results are compared to other studies in \S~\ref{other_studies_section} and to interpretive models in \S~\ref{CGM_section}. 
We summarize our chief
results and remark on open questions in \S~\ref{summary-section}.

\begin{figure*}[!t] 
\begin{center}
\epsscale{1.0}
\plottwo{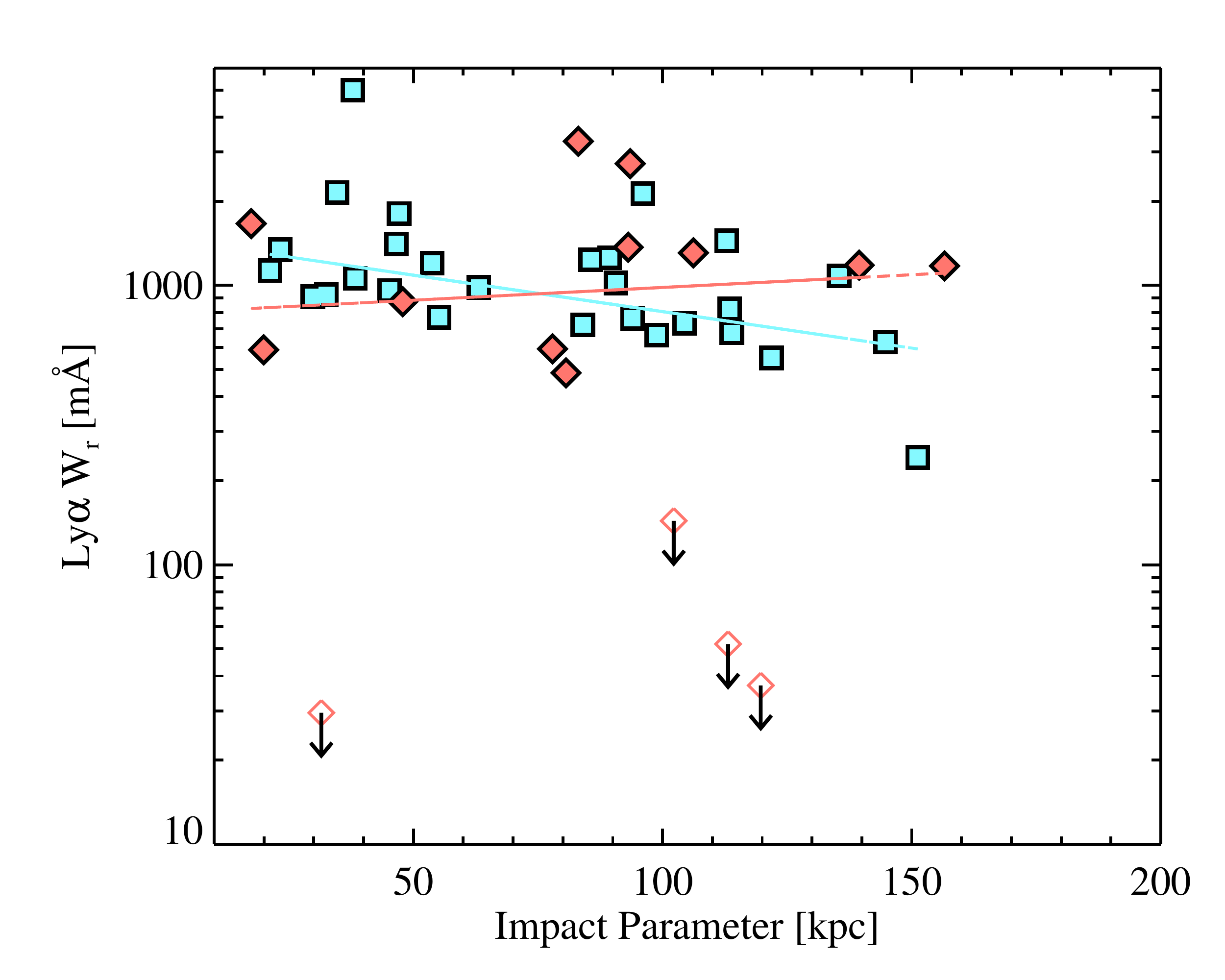}{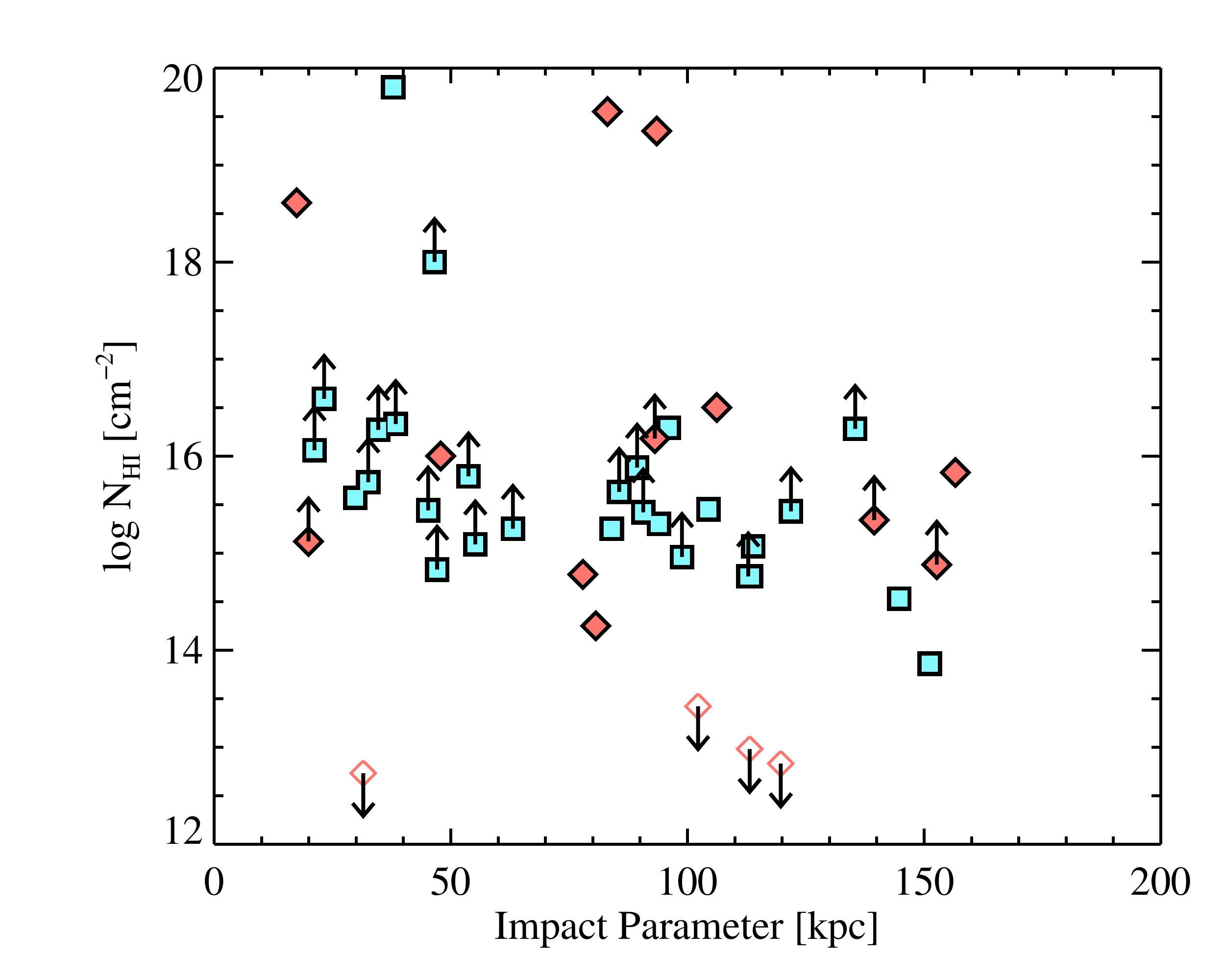}
\plottwo{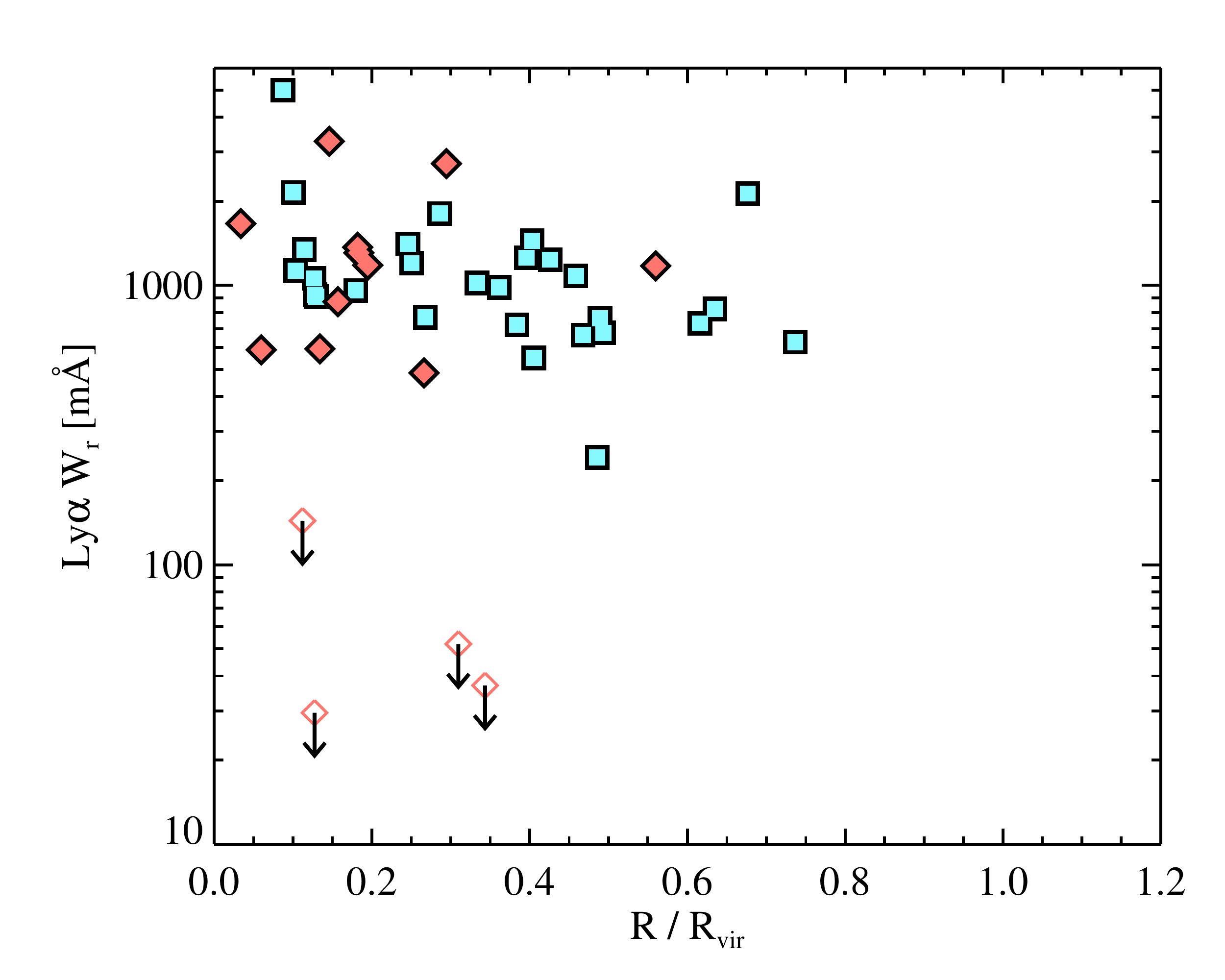}{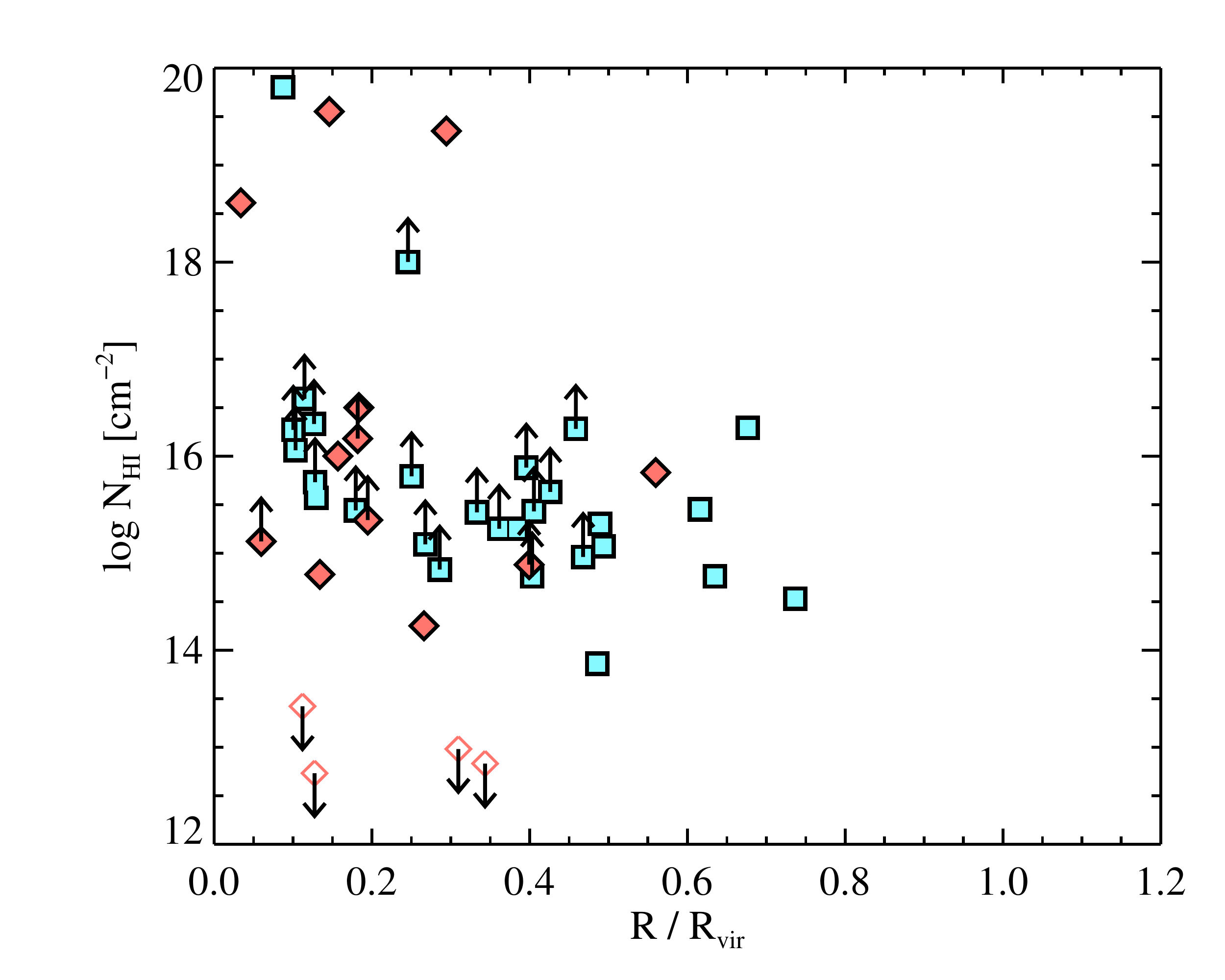} 
\end{center} 
\caption{\HI\ \Lya\ rest-frame equivalent width $W_r$ (left) and column density \NHI\ (right) versus impact parameter (top) and \Rvir\ (bottom). Typical errors are of the same order as the symbol size. The left panels show two fits to the (non-damped) detections in the star-forming and passive subsamples. The slopes are marginally different; the apparent strengthening of the red line is still consistent with a flat slope given the small sample size. \label{NHI_rho_fig} } \vspace{0.3in} 
\end{figure*}


\subsection{Strong HI: Equivalent Widths and Column Densities} 
\label{strong_subsection}

The COS-Halos survey allows us to assess the quantity of \HI\ near
galaxies, and to compare these quantities and trends with galaxy type,
mass, and impact parameter. We perform these comparisons in terms of either \lya\
equivalent width (in the rest frame of the absorber, $W_r$), or in
terms of column density \NHI\ (for complete systems, or by component). Both
quantities carry useful information, but offer different advantages
and suffer different shortcomings. Considering \lya\ $W_r$ only restricts
the comparisons to well-defined measurements with no major systematic
error and a small degree of censoring owing to non-detections, at the
cost of losing the more physical information contained in \NHI. Conversely,
considering column densities allows for the calculation of important physical
quantities such as the absorber mass and size, but greatly increases the degree
of censoring owing to saturation, which converts many well-measured equivalent widths 
into lower limits on \NHI. In the analysis that follows, we endeavor 
to use the quantity best suited for the test in question, but these limitations 
must be borne in mind.

The complete sample of \lya\ $W_r$ appears in Figure~\ref{NHI_rho_fig},
coded by galaxy type, and plotted versus galaxy impact parameter (top panels) 
and \Rvir\ (bottom panels). 
Three features in this figure are notable. First, there is a high
degree of overlap between the red and blue subsamples in terms of $W_r$ alone, with
most values scattered between 500 and 2000 m\AA, indicating a 
strong overlap between the HI properties of star-forming and passive galaxies \citep{2012ApJ...758L..41T}.
Second, we have detected
three damped systems in the main sample; these are visible as the three
strongest $W_r$ points. Third, the four non-detections marked here are
all passively evolving galaxies\footnote{There is one system, associated 
with galaxy J0943+0531 216\_61, which is detected strongly in Ly$\beta$ 
but has its Ly$\alpha$ profile blended with the Ly$\beta$ profile from another system 
at $z = 0.356$ \citep{thom-etal-11-J0943-OVI}. This system counts as a
detection of \HI\ in association with a passive galaxy, but it does not appear 
in any $W_r$ analysis based on Ly$\alpha$. We use Ly$\beta$ to estimate its kinematic extent.}.
These four non-detections in the passive galaxies constitute a suggestive hint of a difference between the two kinds of galaxies.
The gap between the main trend above $\sim 500$ m\AA\ and the lower points
in Figure~\ref{NHI_rho_fig} indicate this possible difference. To assess this, we use the Wilson score interval to estimate the underlying binomial hit rate above a $W_r$ threshold of 200 m\AA. The 28 of 28 star
forming galaxies yield $f_{\rm hit} = 96 \pm 4$\%, while the 12 of 16
passive galaxies give $f_{\rm hit} = 70 \pm 20$\% (95\% confidence limit)\footnote{There is ambiguity for the system 270$\_$40 toward J2257+1340, where two small star-forming galaxies appear to lie closer to the sightline than the canonical passive galaxy. Were we to adopt a ``closest" rather than ``most massive'' galaxy association rule (\S~2.4), these ``hit rates'' would be 28/29 for star-forming galaxies and 12/15 for passive galaxies, well within the errors quoted for these subsamples.}.
There is some indication simply in the hit rates that ETGs show \HI\ less
frequently, but uncertainty related to the sample size prevents a strong conclusion.

Simple linear fits to $\log W_r$ versus impact parameter are shown in Figure~\ref{NHI_rho_fig}, excluding non-detections and damped systems. The star-forming subsample shows a trend with slope $d\log W_r / dR = -0.0026 \pm 0.0008$. The passive subsample shows a slight strengthening at larger impact parameter, but this is not statistically significant: the slope is $d\log W_r / dR = 0.0009 \pm 0.0015$. This mild inconsistency between the two subsamples could still be a result of small sample size - the two strong detections at $\sim 150$ kpc are the chief cause of the flattened slope in the passive subsample. 

If we consider
the ``hit rate'' of damped systems, we find that $f_{\rm damped} =
9 \pm 8$\% (1 in 28) for the star-forming galaxies and $f_{\rm damped}
= 20 \pm 16$\% (2 in 16) for the passive galaxies (95\% confidence). 
This comparison is made even more
ambiguous by the presence of a star-forming galaxy at the same redshift
as the canonical (and more massive) passive galaxy J0925+4004 196\_22, with only a $15$ kpc separation, 
and by the presence of two star-forming galaxies at the same redshift as 
the canonical passive galaxy J0928$+$6025 110$\_$35. Thus we cannot draw strong 
conclusions based on these small samples, but the damped systems are plausibly 
associated with galaxies of either type.


We can use one- and two-sample nonparametric statistics to assess the
differences between the star-forming and passive subsamples in terms of
\lya\ $W_r$. We use the Kaplan-Meier estimator \citep{Feigelson:1985by}
to derive the mean $W_r$ and the error in the mean. If we include all
28 star-forming and 15 passive galaxies in these tests (216\_61 has no clean Ly$\alpha$ measurement), we find $\langle
W_r \rangle < 1033 \pm 250$ m\AA\  for the passive galaxies and $1200 \pm
260$ m\AA\ for the star-forming galaxies (one-sided censoring makes the
passive value an upper limit). Taking the two subsamples in a two-sided
KS test\footnote{In this test, the passive upper limits are taken as
values - the presence of all these on one side of the full distribution
means their exact values between those plotted and zero do not affect
the cumulative probability distribution from which the KS statistic is
computed.}, we find that we can reject the null hypothesis that the two
subsamples are drawn from the same parent distribution at 93\% confidence
($D = 0.395$). However, a closer look shows that this test is affected
by the presence of the four passive upper limits on one side and by
the damped systems on the other. If we exclude the damped systems, the
probability of null rejection increases to 96\%, with mean values $730
\pm 160$ m\AA\ and $970 \pm 70$ m\AA\ for passive and star-forming. So, 
apart from the damped systems, there is a  
suggestive indication of a difference between the two subsamples. 

\begin{figure*}[!t] \begin{center}
\epsscale{1.0}
\plottwo{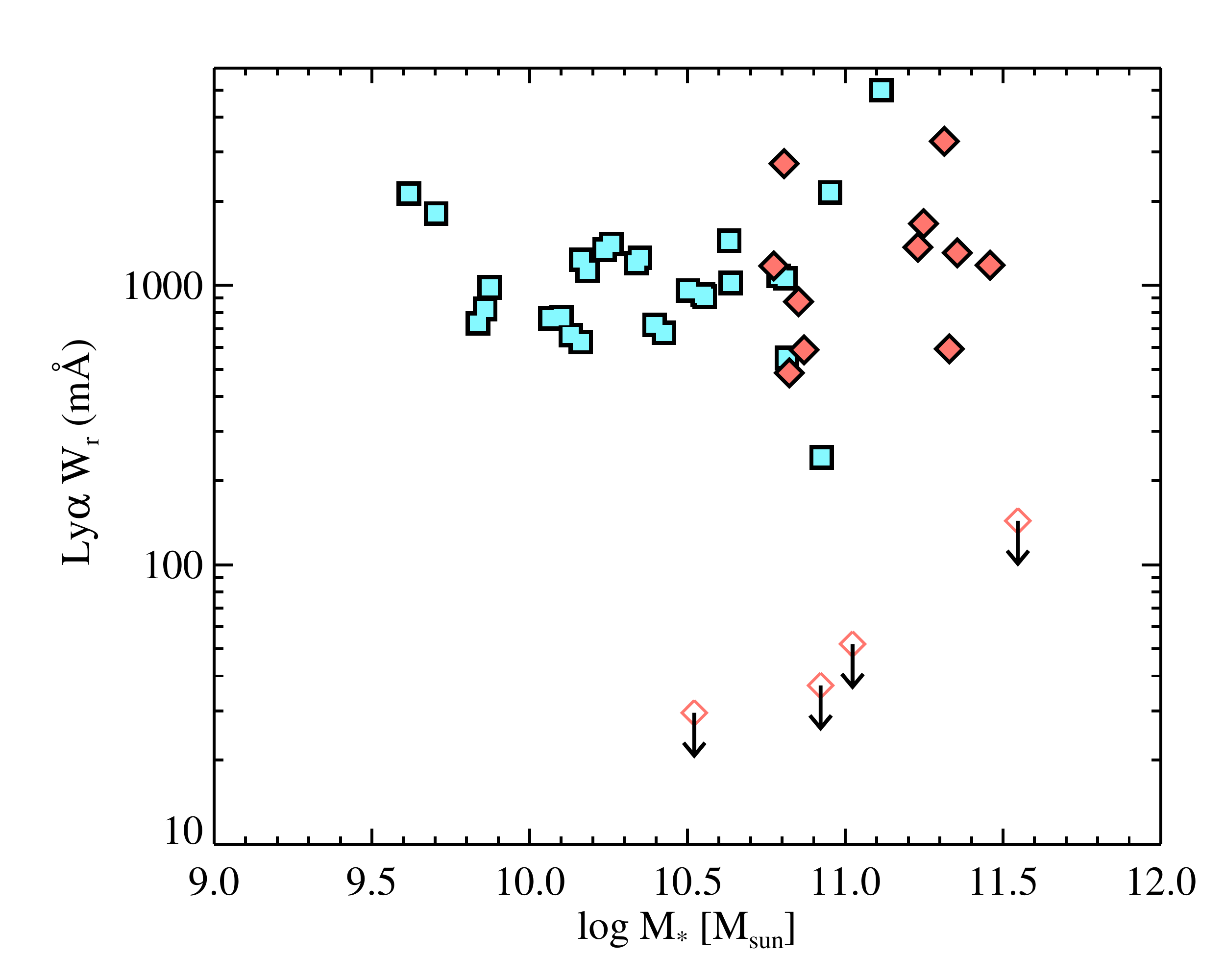}{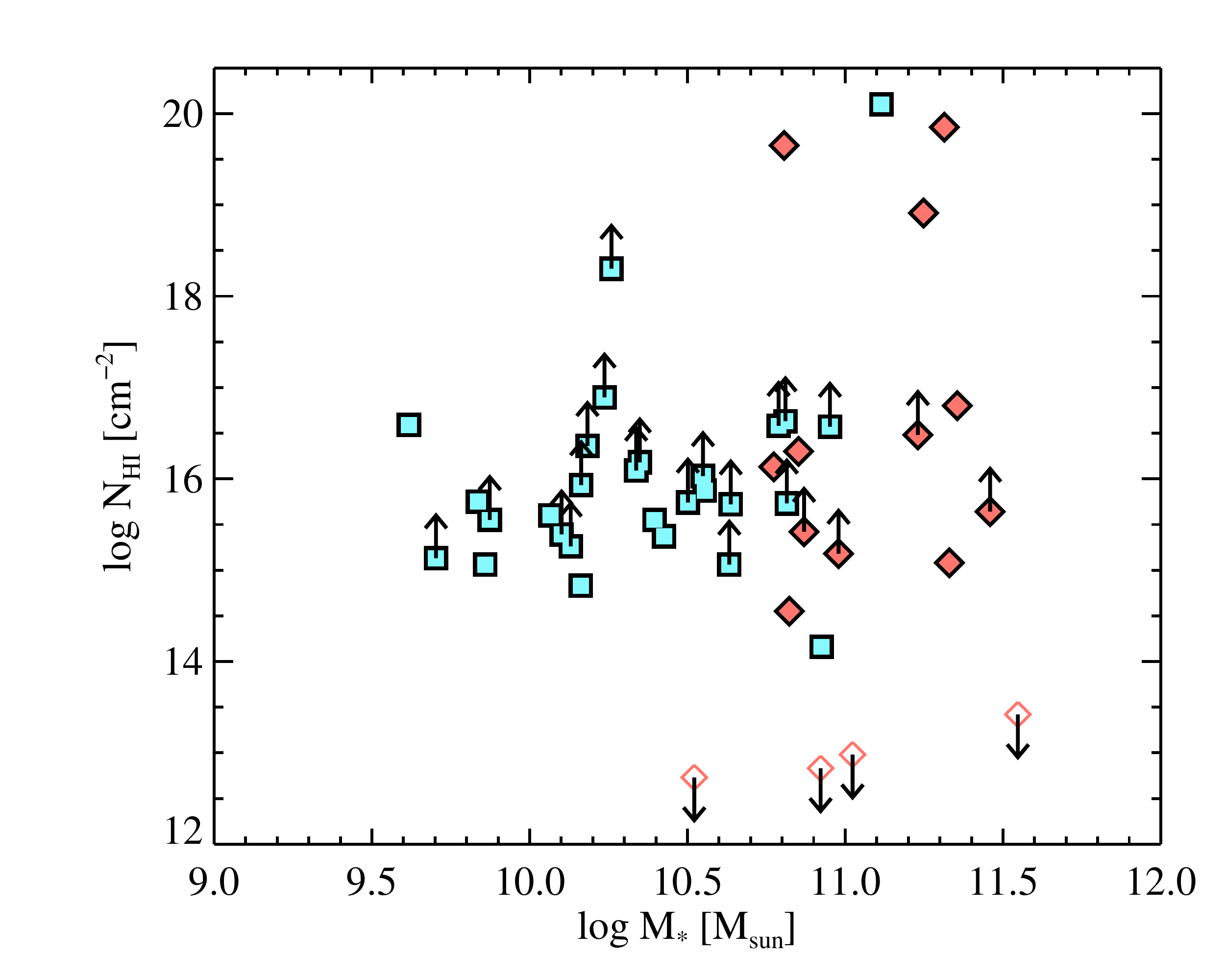} 
\end{center} 
\caption{\HI\ \Lya\ rest-frame equivalent width $W_r$ (left)
and column density \NHI\ (right) versus stellar mass $M_*$. \label{NHI_Mstar_fig}} \end{figure*}

However, it appears that a detection or non-detection of \HI\ is almost an
either-or proposition in the passive subsample, with a distribution 
of \lya\ strengths in the detections that is similar to the star-forming subsample 
and well-separated from the upper limits.  It may be that \HI\ around passive
galaxies, when it occurs, is very similar in strength to the ubiquitous
detections in the star-forming subsample. 
To assess this case, we repeated
the tests above, this time excluding {\it both} the damped systems and
the upper limits in the passive subsample. In this case, we can only
reject the same-parent null hypothesis at 62\% confidence, and we find
that the mean $W_r = 1030 \pm 140$ m\AA\ for passive and a slightly
lower $W_r = 970 \pm 70$ m\AA\ for the star-forming galaxies. While
the non-detection of \HI\ in 25\% of the passive galaxies does
indicate a difference in the hit rates, we have no reason to conclude
that, when it occurs, it is any weaker than in star-forming galaxies.

In light of all these comparisons, we conclude that there is only weak
evidence of a difference in the frequency of \HI\ absorption surrounding
star-forming and passive galaxies. Once we remove the small number of non-detections, the two
subsamples are not significantly different in terms of \HI\ strength. Because
the column densities \NHI\ contain both left and right censoring (both
upper limits for non-detections and lower limits caused by saturation),
we have not attempted these non-parametric statistics for \NHI. It remains
possible that there is a difference in the intrinsic \NHI\ distributions
of the two subsamples that is masked by saturation. In other words,
it is possible that one of the subsamples has a systematically higher
average \NHI\ (or shallower power law distribution of column density, $f_{\rm HI}$), 
which has gone undetected because of saturation effects on the flat part of 
the curve-of-growth. Another confusing factor is that the lower limits on saturated 
\NHI\ depend mainly on the unrelated redshift of the system through the 
availability of the Lyman series lines (see Figure~\ref{saturation_fig}). 
Thus the actual values could be higher, on average, than the formal 
lower limits for one of the subsamples and still go unseen. 

\begin{figure*}[!t] \begin{center}
\epsscale{1.0}
\plottwo{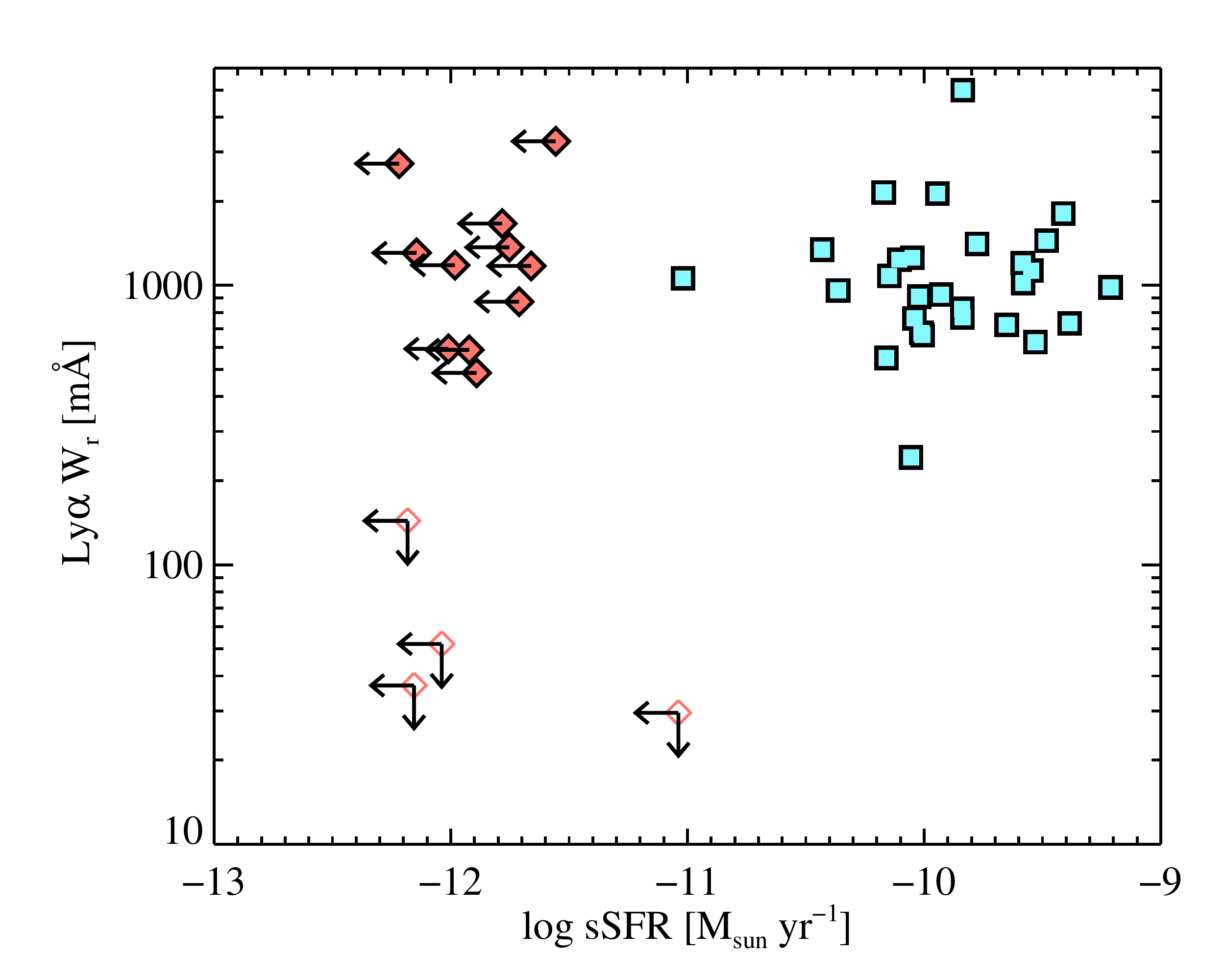}{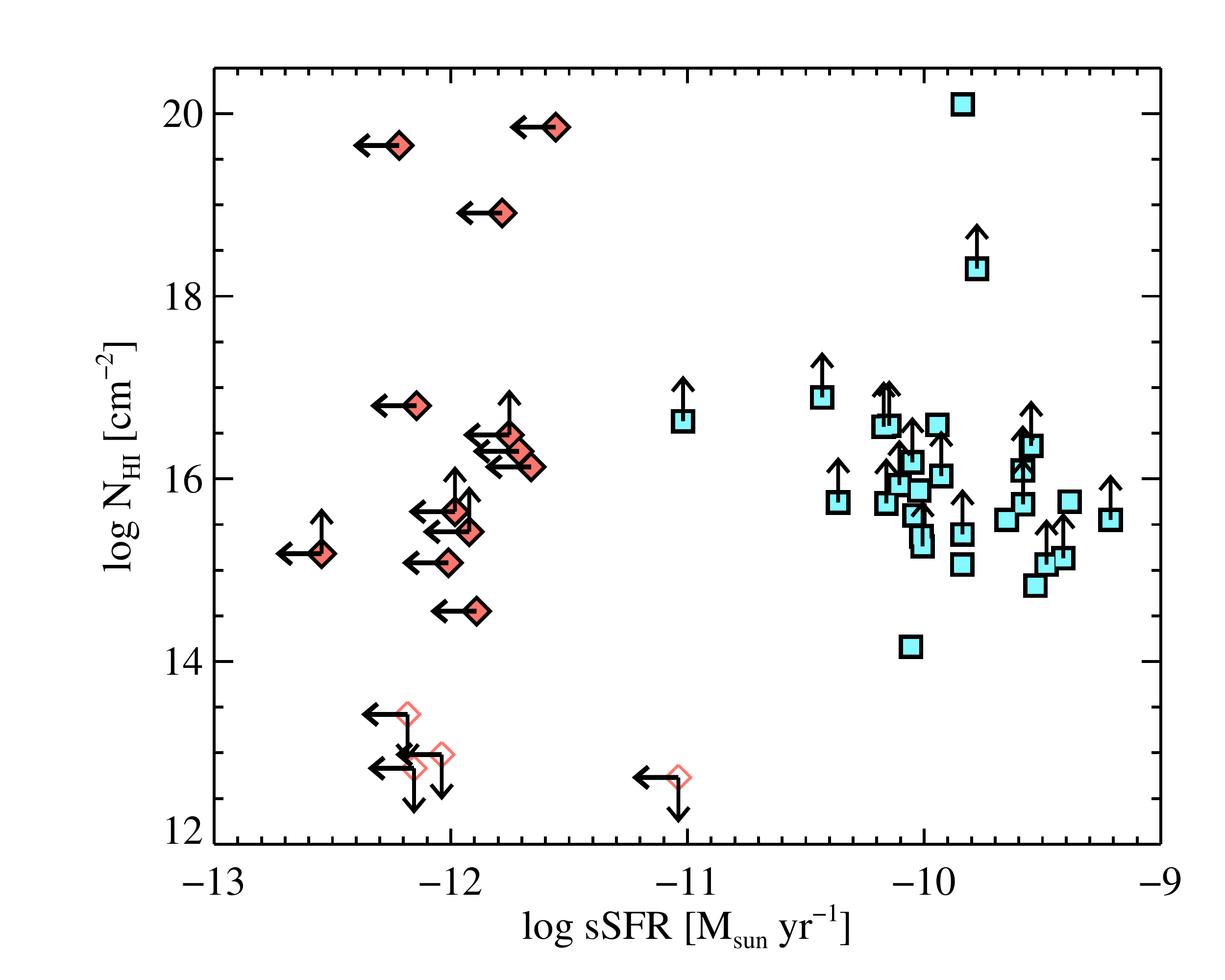}
\end{center} \caption{Left: \lya\ equivalent widths versus sSFR. Right:
\HI\ column density versus sSFR.  \label{NHI_sSFR_fig} } \end{figure*}

We can also examine the properties of the detected \NHI\ in comparison to
the stellar mass $M_*$ and star formation rate. Figure~\ref{NHI_Mstar_fig}
shows $W_r$ and \NHI\ versus $M_*$ and Figure~\ref{NHI_sSFR_fig}  shows
these in comparison to the specific star formation rate (sSFR $\equiv$ SFR/M$_*$). Here the slight
preference of the low values for the passive galaxies can be readily seen,
but we have already judged it to be of marginal statistical significance. 
There are no apparent trends of absorption strength with either stellar
mass or sSFR. From the right panel of Figure~\ref{NHI_Mstar_fig} it appears that all three damped systems have $\log M*
> 10.5$. While the number is small, we speculate that this reflects
the increasing chance of the sightline through a massive galaxy halo
passing near a gas-rich satellite that might give rise to a damped
absorber (the target galaxies themselves are $\gtrsim 40$ kpc from the
sightline, so that the target themselves are less likely to host the damped
absorption). Conversely, the damped systems found serendipitously 
in the COS-Halos data \citep{2012ApJ...744...93B} do not appear to have luminous galaxies
nearby. This suggests that at least some DLAs are associated with sub-$L^*$ or 
low surface-brightness galaxies.  It is also notable that the increased frequency of damped systems occurs in the same range of stellar 
mass as the four non-detections. Thus the scatter in \NHI\ in this region is much 
larger than below $\log M* / M_{\odot} = 10.5$. We speculate that this large scatter reflects a greater diversity of 
origins for the detected gas, in contrast with the lower mass range where the 
detected gas follows a narrower trend. 
However, as demonstrated above these sample sizes are too
small to draw a robust conclusions; more data above $\log M_* \sim
10.5$ is needed to assess whether damped absorbers and non-detections have higher covering
fractions in these halos as the host mass increases, and to identify these trends with 
particular origin scenarios.

\subsection{HI Kinematics}

\label{bound_subsection}

\begin{figure*}[!t] \begin{center}
\epsscale{1.0}
\plottwo{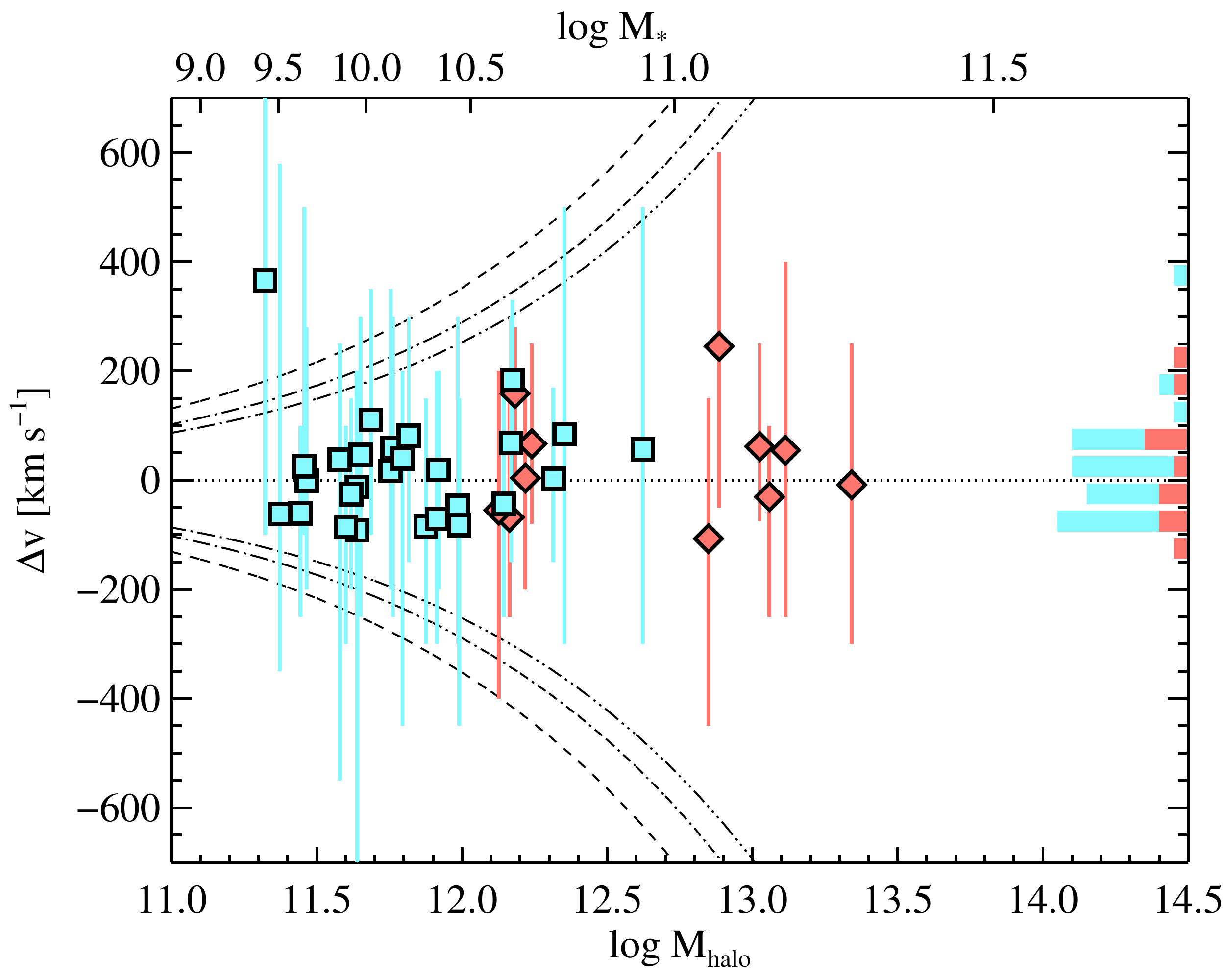}{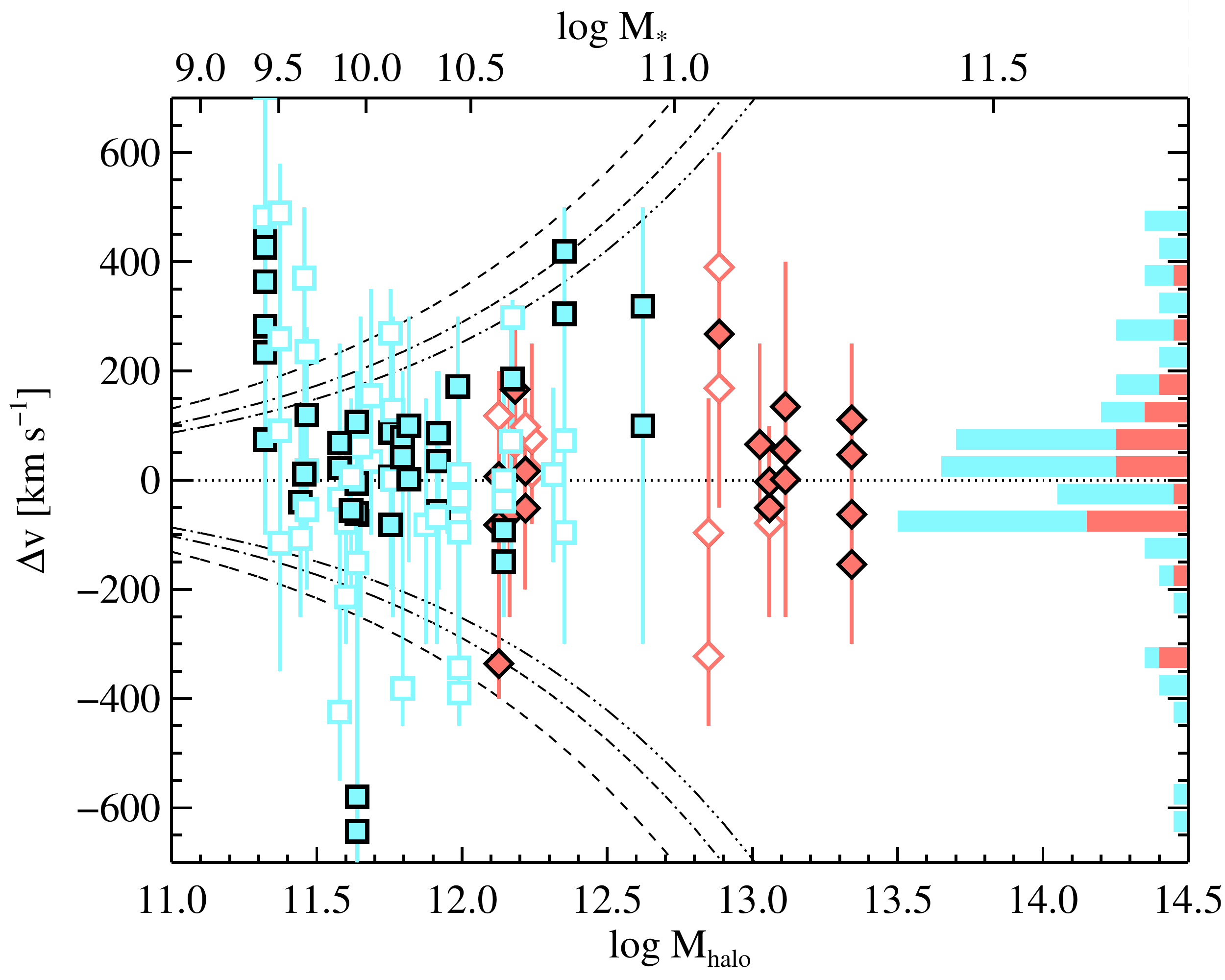}
\plottwo{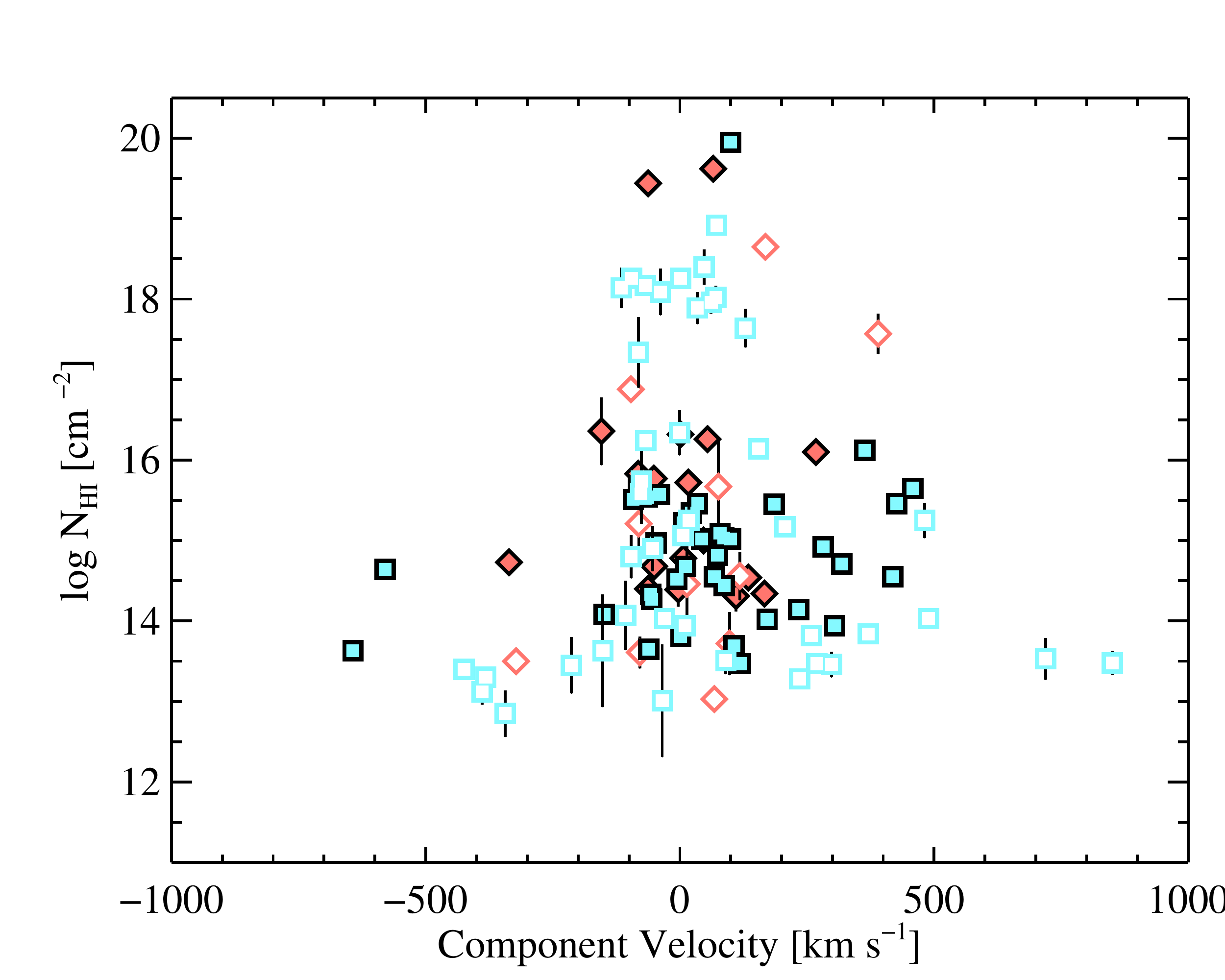}{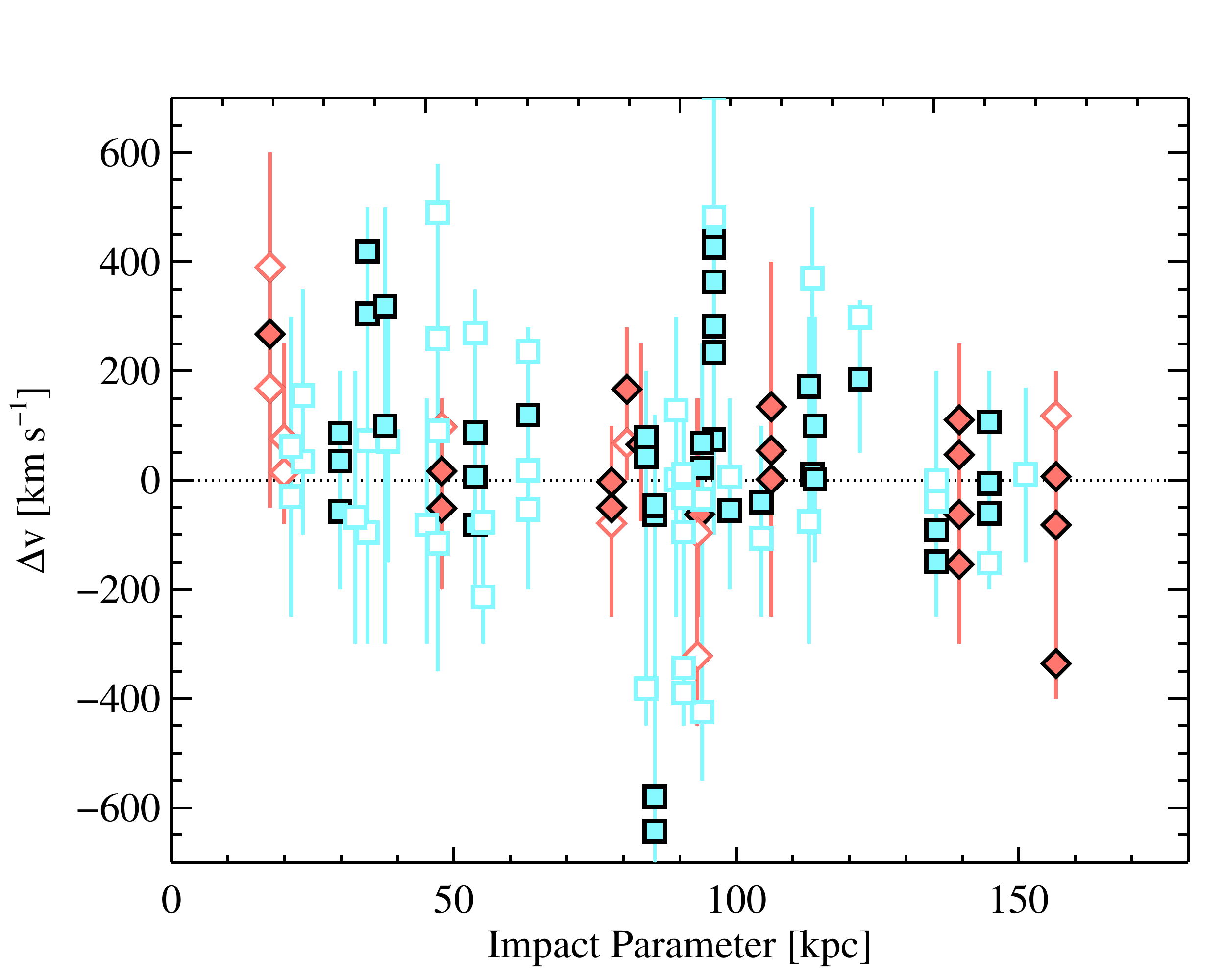}
\end{center} 
\caption{Measures of \HI\ kinematics in COS-Halos. Top left:
Velocity ranges (full width at zero optical depth) with respect to
the galaxy systemic velocity. The points mark the velocity centroid of the total absorption profile (the first moment of equivalent width). 
The three pairs of dashed curves are the halo escape
velocities $v_{\rm esc}$ from 50, 100, and 150 kpc, from the inside
out. Velocity ranges are given for \Lyb\ for absorbers where \lya\ is damped or blended. Top right: A similar plot with the same total velocity
ranges but with the points giving velocity component centroids from
fitting rather than the centroid of the total absorption profile. In
both panels the histograms at right show the distributions of the two
subsamples. Components flagged as uncertain are plotted as 
open symbols. Bottom left: Fitted component column densities 
versus centroid velocity, showing the concentration of the  strongest \HI\ at $\pm 200$ \kms. Bottom right: \HI\ kinematics versus sightline impact parameter. As in the panel at top right, the extent of the solid lines marks the full velocity width at zero optical depth, while the symbols mark the centroids of fitted components.  
\label{Lya_kinematics_fig}} 
\end{figure*}

In addition to line strengths and column densities, we can also examine
the kinematics of the \HI\ absorption and its distribution across
the galaxy subsamples. Measures of absorber kinematics with respect to galaxies appear in
Figure~\ref{Lya_kinematics_fig}. First, we examine $\Delta v$, the full velocity
extent of the detected absorption at zero optical depth (marked as solid lines in the figures), using the
velocity integration ranges of \lya\ as the measure. The full velocity width includes 
contributions from both thermal broadening and bulk flow, so it is only a proxy for 
constraining the maximum kinematic extent
of the absorbing gas\footnote{For systems with $\log$ \NHI\ $\gtrsim 18.5$, the \lya\ width reflects damping wings, not kinematic broadening. Also, galaxy 216\_61 toward J0943+0531 is detected in Ly$\beta$ but contaminated at Ly$\alpha$. For those systems, we substitute the full width at zero optical depth of Ly$\beta$ for this test.}. We also use the velocity centroid of the detected absorption, measured as the first moment of equivalent width. Recall that these velocities are measured in the rest-frame of the galaxy systemic redshift, with $z_{\rm sys}$ identified with $v = 0$. Here we find that most of the  detected \HI\ absorption is located within $\pm 200$ \kms\ of the galaxy. There are exceptions, with some systems showing a total extent of $\gtrsim 500$ \kms, including such notable systems as the ``cold accretion'' absorber toward J0943+0531 \citep{thom-etal-11-J0943-OVI}. There are also some strong components in other systems and a few weaker ``Ly$\alpha$ only'' components (see Table 5) at $|v| > 200$ \kms. However, statistically the data argues for concentration within $\pm 200$ \kms; all but two centroids out of 40 detected systems lie within $\pm 200$ \kms. This concentration is seen clearly in the other three panels of Figure~\ref{Lya_kinematics_fig}, which use the velocity centroids of fitted components. In the upper right panel, we see that, by number, 74\% of all detected components lie within $\pm 200$ \kms\ of the systemic velocity and 81\% are within $\pm 300$ \kms. By contrast, only 9\% of detected components lie between $400$ and $600$ \kms\ away from the systemic velocity.  

The lower left panel shows the column-density dependence of the component velocity distribution. Here we see that component fits implying high column density concentrate more strongly at low velocity than weaker components. While the column densities of the strongest components are made uncertain by saturation, there is no systematic effect preventing high $N_{\rm HI}$ components from appearing at higher velocities, so this is a real effect. Counting by column density produces an even stronger concentration than counting by number: 90\% of the total fitted column density lies within $\pm 100$ \kms, while 99.8\% lies within $\pm 200$ \kms\ and only 0.2\% is outside that range (this excludes the damped systems, which all have $v < 100$ \kms\ and so would only reinforce the trend toward low velocities, but would unfairly dominate the total column density).  It is also notable that, while there are weak $\log N_{\rm HI} \lesssim 15$ components at all relative velocities, the high-velocity components  tend strongly to have these lower column densities. The lower right panel of the figure shows that the full velocity range of the absorption and the velocity centroids of fitted components do not exhibit any clear trends with impact parameter (the full width at zero optical depth is still marked in solid lines). 

These velocities are of the same order as the velocity dispersions of the galaxies, and so are consistent with a gravitational association between the galaxies and the gas. Figure~\ref{Lya_kinematics_fig} also compares these velocity ranges (top left panel) and the velocities of fitted components (top right panel) to the inferred escape velocity of the host halos, with $M_{\rm halo}$ converted from $M_*$ using the relation of \cite{2010ApJ...710..903M}. We find that only a small portion of the absorption lies outside the inferred $v_{\rm esc}$, especially above $\log M_{\rm halo} \simeq 12$. Thus there is scant evidence that the bulk of the detected \HI\ is able to escape the host galaxy. It is possible that there is escaping material outside $\pm 200$ \kms\ that is too weak to detect at the S/N ratio of our survey, or that substantial material escaped at earlier epochs of galaxy evolution, but the bulk of the detected \HI\ (and, by inference, the mass budget that it traces) is not escaping at the time we observe it.  These results could mean that the CGM we detect is essentially bound to the host galaxies. Furthermore, its kinematics have no apparent dependence on galaxy type.  

 Finally, we note that the full linewidths of the absorbers as shown in Figure \ref{Lya_kinematics_fig} generally range over $\pm 100-200$ \kms\ and center near zero; because they depend weakly on \NHI, the total profile linewidth is a poor measure of total column density or internal kinematics and should be treated cautiously in low-resolution or low-S/N data. 

\begin{figure*}[] \begin{center}
\epsscale{1.0}
\plottwo{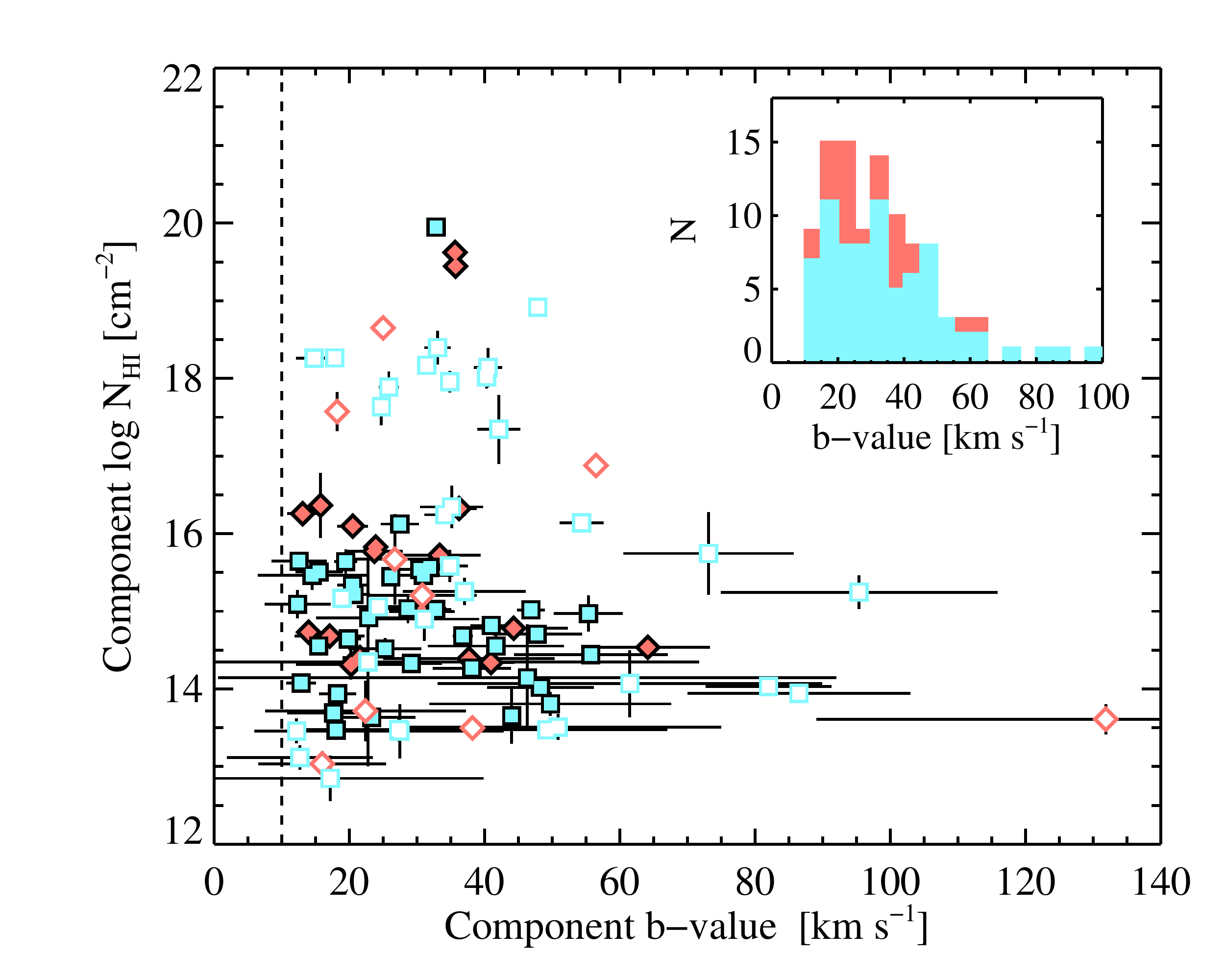}{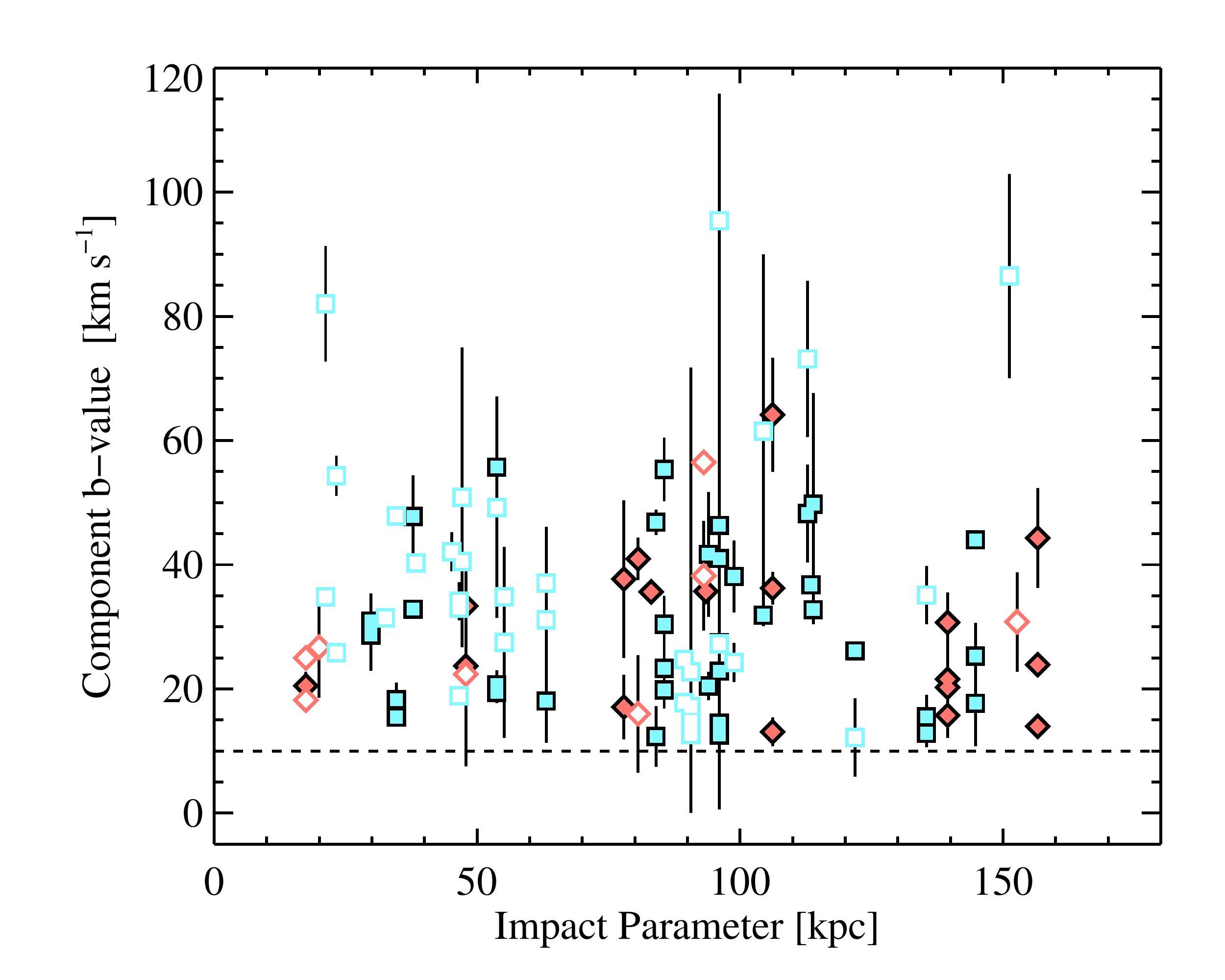}
\end{center} 
\caption{Left: fitted component Doppler $b$-values versus \NHI\ for profile-fitted
components. Systems with uncertain fits owing to saturation, 
ambiguous placement of components, or data quality issues are flagged 
as such in Table~5 and plotted with open symbols here. 
Generally speaking, component column densities are uncertain 
above $\log$ \NHI\ $\simeq 16$ owing to saturation in even 
the highest available Lyman lines and remain so until until damping wings 
reappear at $\log$ \NHI\ $\simeq 18.5$. The main uncertainty in the fitted 
$b$-values arises from properly assigning the number of components in a 
strong profile; given limited resolution, S/N, and blending it is likely 
that the number of components is underestimated and so the fitted $b$ 
here are overestimates. 
The components with 
$b \gtrsim 80$ \kms\ are not robustly constrained at the S/N of 
the data. Some weaker components, $\log$ \NHI\ $\lesssim 14$, are considered 
uncertain because of noise or possible contamination. 
The inset histogram shows the distribution of the Doppler $b$
for both subsamples together. Right: Fitted component linewidths versus
impact parameter with the same symbol coding. 
\label{component_bval_rho} }
\end{figure*}

\subsection{HI Linewidths and Temperatures}

\label{cold_subsection}

Another important test is the internal kinematics of the detected
absorption as represented by the Doppler $b$ parameters of the fitted
components. Figure~\ref{component_bval_rho} shows the fitted \NHI\ and $b$
for all components without grouping them into their host systems. We
see that almost all the absorption lies in systems with $b < 60$ \kms,
while more than half (by component number, not column density) has $b \lesssim 40$
\kms. We can interpret these Doppler $b$ parameters as robust upper limits
to the temperature, if we adopt $b = \sqrt{2kT/m_H}$.  Thus almost
all the detected \HI\ has $T \lesssim 220,000$ K (60 \kms) and most have
$T \lesssim 100,000$ K (40 \kms). These temperatures are significantly
below the implied virial temperatures of the host halos, which are
1 million K or more for galaxies with $M_{\rm halo} \gtrsim 10^{12}$
\Msun. 

There are two reasons why these values are robust upper limits
to $b$ and $T$. First, our method for fitting decomposes the observed
profiles into the minimum number of components necessary to achieve a
good fit; the actual number of components could be larger, which would
reduce the typical $b$ and thus the implied temperatures. Second, the
values quoted above assume that the observed line broadening is purely
thermal, which maximizes the implied temperatures. With non-thermal
motions of 30 \kms, the implied temperatures for $b = 40$ and 60 \kms\
drop to 40,000 and 160,000 K, respectively\footnote{The $R \sim 18-20,000$  resolution of COS means that we cannot reliably fit 
linewidths below $b \simeq 10$ \kms. Fits to broader lines can be affected by the non-Gaussian 
wings of the COS LSF, for which we include a detailed model}. Finally, recall that components 
with uncertain fits owing to saturation, component placement, or poor 
significance are marked with open symbols in the plots.

Figure~\ref{component_bval_rho} also shows that there is no discernible
difference in line widths between the star-forming and passive
subsamples. A KS test on the red and blue distributions in the inset
histogram returns a KS statistic of $D = 0.164$ and a null rejection probability $P =
0.64$ for 26 red and 73 blue components. We conclude there is no evidence
for a difference in line widths in the two subsamples. This corresponds
to no significant difference in the inferred temperature limits, and no
difference in the temperatures themselves if the relative contributions
of non-thermal broadening are the same.

\begin{figure*}[!t] \begin{center} 
\epsscale{0.68}
\plotone{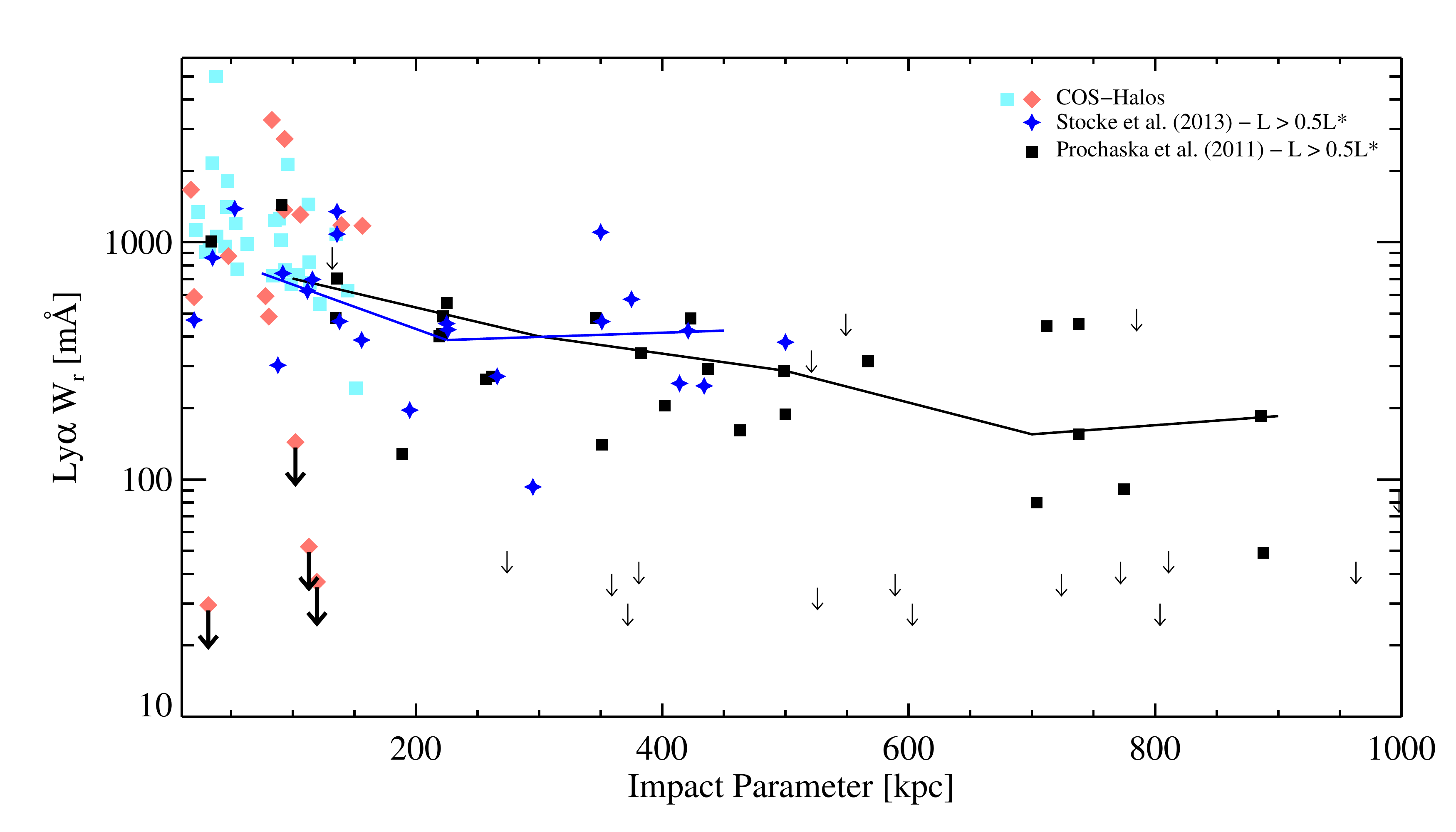} 
\plotone{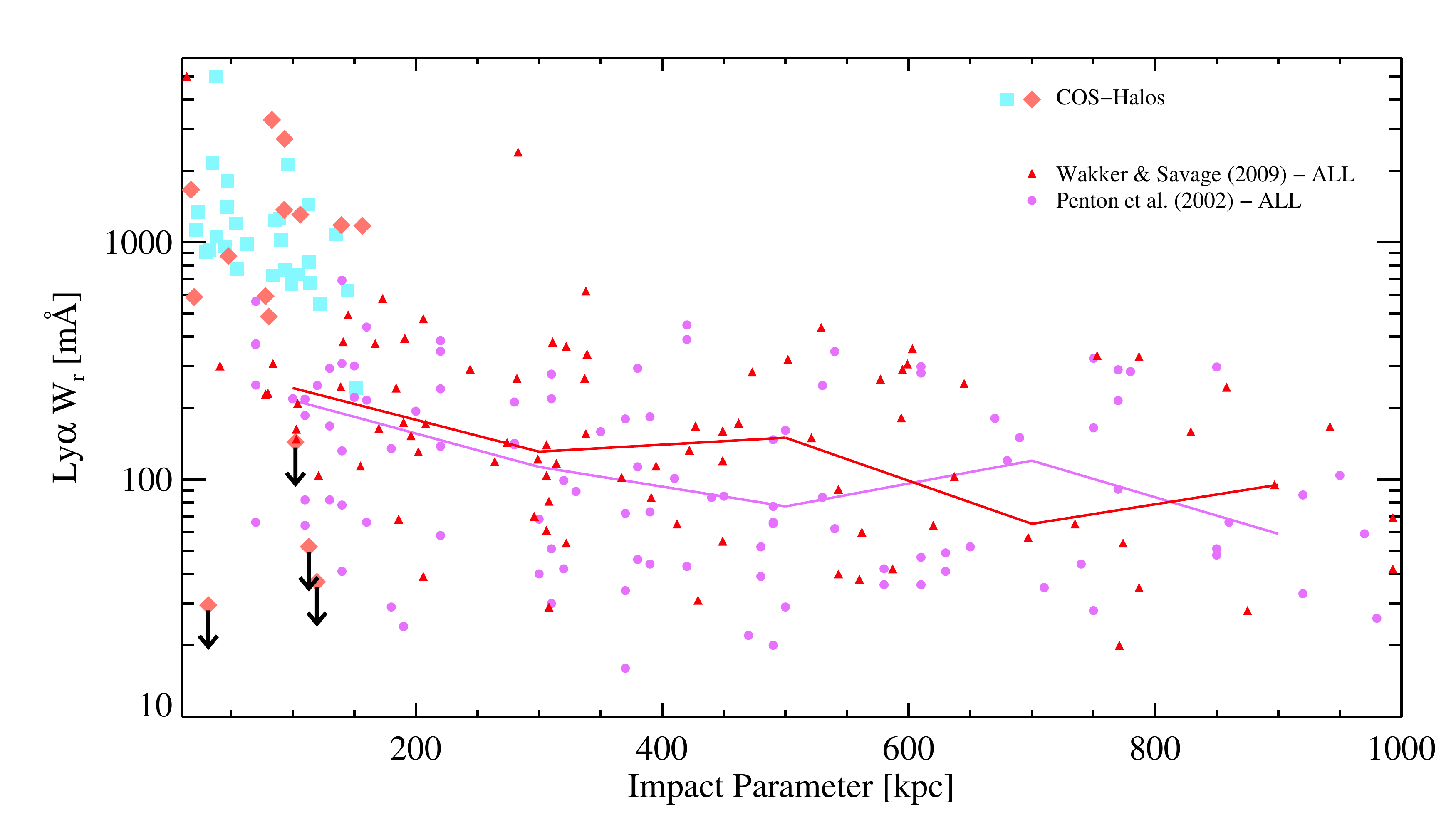} 
\plotone{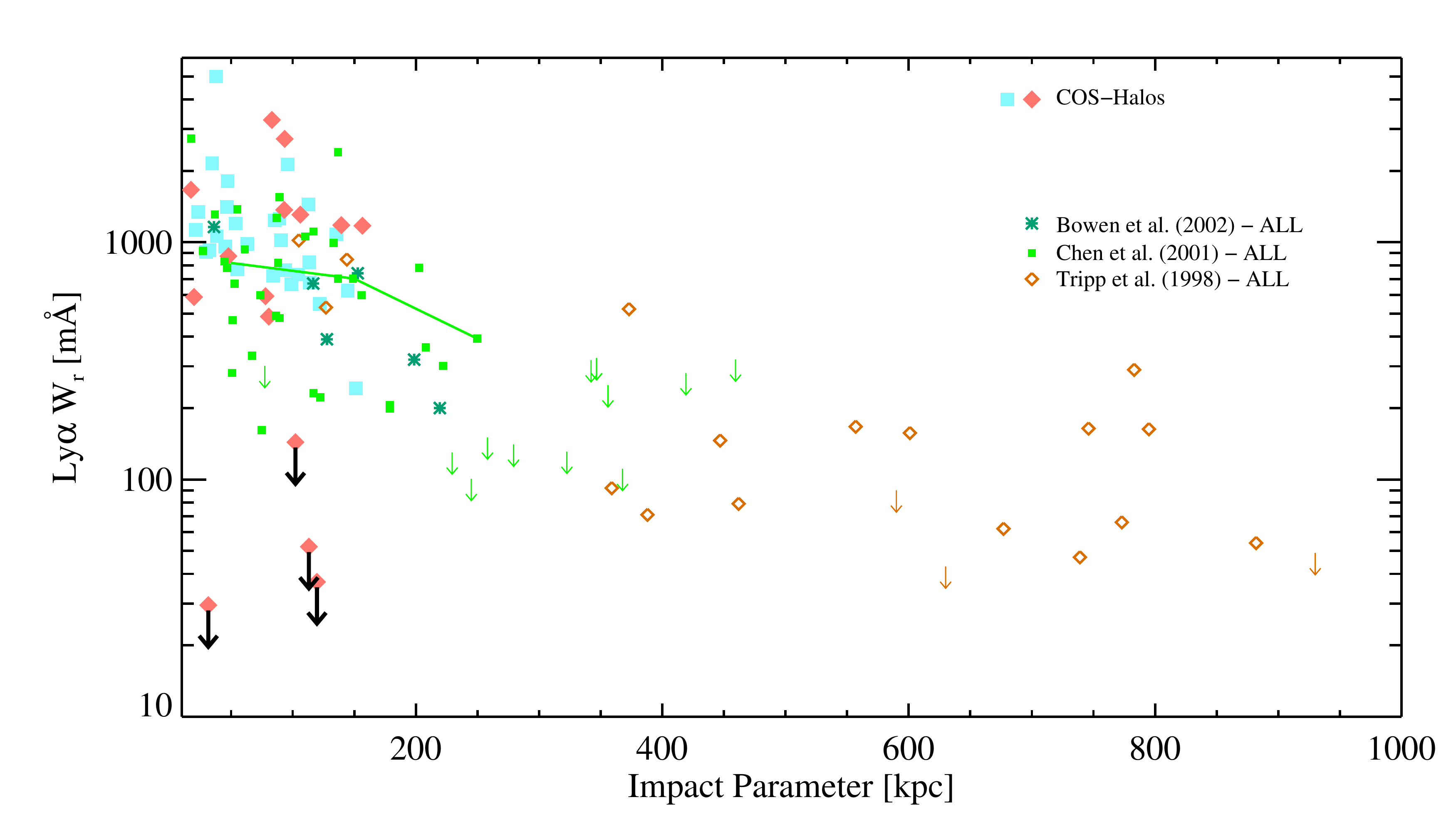}
\end{center} 
\caption{A comparison of HI strength with respect to galaxies from COS-Halos and seven previous studies. The points from \cite{1998ApJ...508..200T}, \cite{chen-etal-01-Lya-imaging}, and \cite{prochaska-etal-11-OVI-HI} include upper limit arrows for galaxies in their surveys without detected absorption at the level indicated. The \cite{prochaska-etal-11-OVI-HI} and \cite{2013ApJ...763..148S} samples have been cut for $L > 0.5 L^*$ to approximate the selection criteria for COS-Halos.  When present, the solid trend lines show the median $W_{\rm Lya}$ for detections in the indicated bins for samples large enough to give meaningful binned results (5 points per 200 kpc bin). 
\label{lya_rho_compare_fig} } 
\vspace{0.2in} 
\end{figure*}

\begin{figure*} 
\begin{center}
\epsscale{1.0}
\plotone{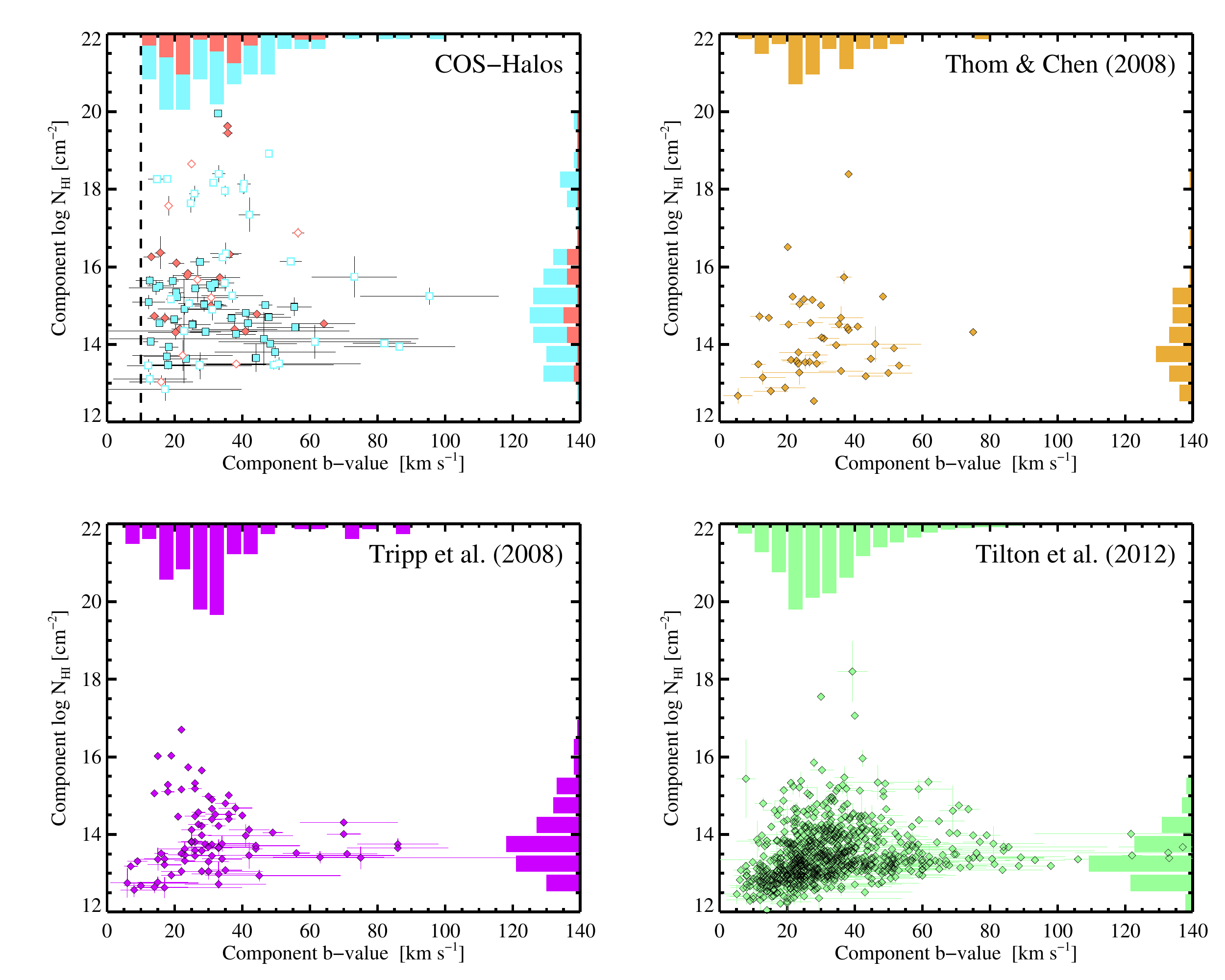}
\end{center} 
\caption{Four samples of \NHI\ vs. component
Doppler $b$ parameters for CGM and IGM absorption. For COS-Halos,
the star-forming and passive galaxies are shown with symbol coding
as in all previous plots. The three blind intervening samples from
\cite[Table 3]{2008ApJS..177...39T}, \cite[Table 3]{Thom:2008gma}, and
\cite[Table 5]{2012ApJ...759..112T} are plotted separately from their tables
without modification. COS-Halos components flagged as saturated and/or uncertain 
are marked with open symbols. These column densities should be treated
cautiously, though the given $b$-values can be considered reasonable
upper limits to the true $b$.  \label{compare-NHI-bval-fig}} 
\end{figure*}

\section{COS-Halos Compared to Previous Studies of IGM/CGM HI and Galaxies}
\label{other_studies_section} 

\subsection{Correlation with Radius: Comparisons of HI Strength}

Surveys of \HI\ absorbers in the spectra of QSOs have been pursued fruitfully 
with every generation of spectrographic capability
on HST, starting with the Key Project on QSO absorption-line systems using FOS
\citep{1993ApJS...87....1B, 1998ApJS..118....1J} and the first pioneering 
high-resolution studies with GHRS by \cite{1991ApJ...377L..21M}. 
Investigators using
data from the Key Project and follow-up galaxy
redshift surveys firmly vindicated the early prediction by \cite{1969ApJ...156L..63B}
that diffuse gas in galaxy halos would imprint absorption lines in the spectra of
distant QSOs.  This work established the existence of an extended, diffuse
gaseous medium surrounding galaxies out to hundreds of kpc, traced
mainly by \lya\ \citep{1993ApJ...419..524M, 1995ApJ...442..538L}. Further studies that
added images of the galaxy fields and additional spectroscopy were able to
examine the relation of this gaseous halo to galaxy type and mass (proxied
by luminosity).  \cite{1998ApJ...498...77C} reported that the physical
extent of the \lya\ CGM scaled slowly with $B$ and $K$ band luminosity,
such that more massive galaxies have more extended gaseous halos. 
A contemporaneous study by \cite{1998ApJ...508..200T} used more 
sensitive absorption-line data to examine the galaxy correlations of weaker 
Ly$\alpha$ absorbers with a smaller sample of galaxies, finding that even 
absorbers with $<100$ m\AA\ are not randomly distributed with respect to 
galaxies.

In a series of papers based on GHRS/G160M data, the Colorado IGM group examined a set of $\sim 100$ \Lya\ absorbers \citep{2000ApJS..130..121P} 
and studied their correlations with galaxies from published surveys, finding that the stronger absorbers cluster more weakly with galaxies 
than galaxies cluster with themselves, while the weakest absorbers are 
even more randomly distributed and so do not lie in individual 
galaxy halos but arise instead in the large-scale structure of filaments \citep{2002ApJ...565..720P}.  This tendency for the strongest \HI\ to appear near galaxies 
and for weaker \HI\ to arise in larger, more diffuse structures such as cosmic web filaments in also seen in hydrodynamical simulations at low redshift \citep{1999ApJ...511..521D}
and at high redshift \citep{2003ApJ...594...75K}, making it clear that even absorption that appears to lie very near galaxies ($\lesssim 100$ kpc) may in fact arise 
further out in extended large-scale structures ($\gtrsim 1$ Mpc), particularly at low column densities. 
Taking all low-redshift \HI\ clouds together, \cite{2000ApJ...544..150P} estimated 
that \lya\ clouds contain approximately 20\% of the cosmic baryons at low redshift. Their later 
study using STIS extended this analysis to weaker systems and brought the baryonic mass fraction up to 29\% of the cosmic value \citep{2004ApJS..152...29P}. 

Using cross-correlation methods on a single sightline, \cite{2005ApJ...629L..25C} confirmed that even weak \Lya\ 
correlates with galaxies out to $\sim 1$ Mpc, particularly for star-forming galaxies.
With STIS, \cite{2002ApJ...580..169B} further characterized
the \lya\ surrounding galaxies by pre-selecting galaxies prior to obtaining the 
absorption data, as COS-Halos has done. \cite{prochaska-etal-11-OVI-HI}
followed up absorber samples from HST and FUSE after compiling galaxy redshifts 
for these sightlines, but did not select the sightlines themselves based on 
the availability of foreground galaxies. They reported a strong \HI-traced CGM out 
to $\simeq 300$ kpc, with no strong dependence on galaxy type but with some 
preference of the highest $W_r$ systems to be
associated with galaxies at $L \gtrsim L^*$. Surveying a large compilation of 
data on galaxies of $z < 0.017$ and absorbers from HST and FUSE,
\cite{wakker-savage-09-OVI-HI-lowz} confirmed this basic conclusion. Thus
the existence of a diffuse, ionized medium surrounding galaxies,
with an extent weakly depending on galaxy properties, was well established
before the advent of COS. Recently, \cite{2013ApJ...763..148S} have 
revisited this issue with a new sample of targeted galaxies and serendipitous 
QSO-galaxy pairs drawn from earlier studies. They report that CGM clouds 
within $\sim 0.4$ \Rvir\ of galaxies tend to be be ``warm'' -- at photoionized temperatures -- metal enriched ($Z \sim 0.1 - 1Z_{\odot}$), and probably bound. 

Our goal now is to compare the COS-Halos results to these prior studies to test for similarities 
and differences in the gas surrounding galaxies, mainly as a function of impact parameter. 
Because of its selection technique, COS-Halos covers a range of impact parameter that is lower than the typical radius
probed by the previous studies of this size. Our sample is about twice the size of \cite{chen-etal-01-Lya-imaging} out to 150 kpc, while
\cite{prochaska-etal-11-OVI-HI} contains only three $L \sim L^*$ galaxies with $\rho \lesssim 150$ kpc.
Thus COS-Halos is well-positioned to study the dependence of the CGM on galaxy type and/or mass, while it is less
useful for probing the outer regions of the CGM ($R \gtrsim 0.5R_{\rm vir}$). To test for evolution as a function of 
impact parameter, we need samples at $\gtrsim 300$ kpc, for which many of these other 
studies are well-suited. 

However, this goal presents the immediate problem that these various studies used a wide range of 
selection techniques and analysis methods. The COS-Halos galaxies (15-160 kpc) were selected for study without any 
knowledge of the absorption; this is true as well of the pioneering study of six targeted galaxies by \cite{2002ApJ...580..169B} and 
of the ``targeted'' portion of the COS sample of \cite{2013ApJ...763..148S}. The other surveys are ``blind'', 
in the sense that they are based on absorber data from sightlines that were not chosen to probe particular 
galaxies, and the galaxy data was compiled separately and then correlated with the absorbers. 
We emphasize that these studies are heterogeneous in many respects; they 
use different criteria for absorber and galaxy selection, different sources of 
galaxy redshifts and photometry from custom follow-up 
spectroscopy (e.g. Chen and Prochaska) to literature galaxy catalogs (e.g., Penton and Wakker). 
Most importantly for comparison with COS-Halos, each survey has used different rules  
for associating galaxies and absorbers in physical and velocity space. Moreover, as
many of these surveys are based on the common body of HST spectroscopic data, 
they overlap in the sightlines and redshift intervals that they cover, 
and so have many individual absorber / galaxy pairs in common. They are thus 
not independent surveys in terms of the objects studied. Rather than attempt the daunting task of
reconciling all these disparate studies, or combining overlapping samples, instead we take these prior studies as 
independent empirical characterizations of the \ion{H}{1} gas surrounding galaxies as seen by different groups using 
different techniques and analysis criteria at different times.  

Figure~\ref{lya_rho_compare_fig} compares the COS-Halos results to these
other studies in terms of \lya\ equivalent width $W_r$ versus the impact
parameter at $< 1$ Mpc. Upper limits for galaxies near the sightlines with no 
corresponding absorption are shown for \cite{1998ApJ...508..200T}, 
\cite{chen-etal-01-Lya-imaging}, and \cite{prochaska-etal-11-OVI-HI} and COS-Halos.

There is a clear trend for \lya\ detections to
increase in strength within $\sim 200$ kpc, while outside that
impact parameter range the detections are weaker, with large scatter. This
trend is consistent with the conclusion that absorption found within 150 kpc
of galaxies by COS-Halos is associated with the selected galaxies and
not (generically) with gas at larger distances away from the galaxies
(see \S~\ref{gas_outside_rvir_subsection} below for more on this point).
This interpretation is supported by the kinematic associations of strong \HI\ with galaxies
presented above in \S~\ref{bound_subsection}. 
Within the region they have in common, COS-Halos and the samples 
from Bowen, Chen, and Prochaska agree well considering only detection 
strength. 

The samples from COS-Halos, Bowen, Stocke, and Prochaska provide 
the closest possible comparisons, since they all use galaxies predominantly 
selected or cataloged prior to the absorber data. For these samples
the evolution from $\lesssim 150$ kpc to $\gtrsim$ 300 kpc is clear. 
Over the range 100-200 kpc there is apparent disagreement in the median 
equivalent widths between COS-Halos on the one hand and 
Wakker \& Savage and Penton on the other; the latter two agree with each other well. 
This may be caused by a difference in the selection technique or the typically lower
luminosity of the galaxies in the latter two samples. The lower 
average redshifts in the latter two samples may also play a role if the CGM
evolves to lower column density at lower redshift. 

Statistical tests of the observed trends with impact parameter provide another 
argument for the containment of the detected \HI\ within the
virial radius. As such a test, we compare our empirical distributions of $W_r$
against those of \cite{prochaska-etal-11-OVI-HI}, which was cut for $L >
L^*$ galaxies and contains mostly points outside outside 200 kpc (see Figure 12). 
Accounting for censoring in a two-sample generalized Wilcoxon test, we find that we can reject at $>99.7$\% confidence 
the null hypothesis that the two $W_r$ distributions are drawn from the 
same parent population ($P = 0.003$ or less). If we restrict the \cite{prochaska-etal-11-OVI-HI} 
sample to $\rho < 500$ kpc, this confidence drops to $> 99$\% but is 
still highly significant. These probabilities are $< 10^{-5}$ for the Wakker \& Savage 
and Penton samples, allowing us to reject the null hypothesis at high confidence. 
Alternatively, if we compare to the Chen, Bowen, and Stocke samples by restricting them 
to the $< 160$ kpc region surveyed by COS-Halos, we find that the rejection probabilities 
drop to 60-70\%, indicating highly overlapping samples and no significant evidence 
of disagreement in \lya\ strength. Thus the surveys concentrated at $< 200$ kpc agree with 
one another well, but not with those surveys concentrated outside $200$ kpc. 
Based on these results we conclude that 
there is significant evidence of evolution in \ion{H}{1} between the $< 200$ kpc 
region near galaxies and further out. 
On the basis of these 
simple models and empirical characterizations with respect to impact 
parameter, we conclude that the \HI\ detected by COS-Halos and these other surveys is concentrated 
within the $\sim 300$ kpc region surrounding galaxies. However, there remains 
a strong likelihood that the equivalent widths observed in COS-Halos 
receive a contribution from foreground and/or background gas at 
$\gtrsim 300 - 1000 $ kpc, up to the $\simeq 100-300$ m\AA\ levels seen at these impact 
parameters in other samples with sightlines further away from galaxies. 

Gas surrounding galaxies at high redshift has also been observed in the spectra 
of background QSOs using optical telescopes on the ground. A recent example is 
the Keck Baryon Structure Survey \citep[KBSS;][]{Steidel:2010go, 2012ApJ...750...67R}. 
In their characterization of the \HI\ surrounding 886 galaxies with $z \sim 2.3$ (25 of which are 
probed at $\leq 200$ physical kpc), 
\cite{2012ApJ...750...67R} show that the strongest absorbers, $\log$ \NHI\ $\sim 15-17$ 
concentrate within 200 kpc (physical) and $\pm 300$ \kms\ of their targeted galaxies. 
Comparing against the low-redshift samples, they find that the covering fractions at $z \sim 2$ 
are similar to those measured within 300 kpc for limits $\log$ \NHI\ $\gtrsim 14$, with 
$f_c = 81 \pm 6$\% in the KBSS sample ($z \sim 2.3$), 72\% for \cite{chen-etal-01-Lya-imaging} ($z = 0.1 - 0.9$) 
and 70\% for \cite{2011ApJS..193...28P} ($z < 0.4$).  The overall 91\% detection frequency in the 
COS-Halos sample is consistent with these values, though higher likely because our survey region 
is restricted to 150 kpc. Removing the COS-Halos passive galaxies from the comparison, we find a 100\% detection rate 
above 10$^{14}$ cm$^{-2}$, in good agreement with the $92 \pm 5$\% rate inside 200 kpc. Thus we conclude that 
the overall column density distribution and covering fractions of \HI\ surrounding galaxies at the different 
epochs are broadly similar. We do not, however, reproduce the KBSS finding that \HI\ component 
linewidths increase slightly inside 300 kpc (their Figure 28 vs. our Figure 11. This difference may 
result from the lower spectral resolution of the COS data, the smaller physical radius over which 
our data span, or to a real difference in the internal kinematics and/or temperature of circumgalactic gas 
at two different cosmic epochs. KBSS does not include detailed comparisons to galaxy properties such as mass and SFR, but the small number of $z > 2$ CGM absorbers studied in connection with detailed galaxy properties reveal CGM accretion at this early cosmic epoch \cite[e.g.,][]{2013Sci...341...50B, Crighton:2013tm}.   We note also that the \HI\ envelopes surrounding mainstream galaxies at all these  redshifts are substantially weaker than those detected around $z \sim 2$ QSOs by \cite{2013ApJ...762L..19P}. We leave a more thorough and detailed comparison to be made later, 
when the galaxy properties for the KBSS fields have been published and we can compare against the 
COS-Halos findings with respect to galaxy mass and type. 

\subsection{Column Density and Temperature Comparisons}

Because COS-Halos typically covers Lyman series lines above \Lyb\
with moderate resolution ($\sim 20$ \kms), we can address the
internal kinematics of the CGM gas in a more rigorous manner than
previous studies that focused on \lya\ alone. As discussed above, line-profile broadening
takes two basic forms: thermal broadening that scales as $T^{1/2}$, and
non-thermal broadening from, e.g. turbulence, bulk flows, or gravitational
motions. If we can assess these separately, we can address whether
temperature or other physical quantities differ in the CGM. To do this,
we compare the line-profile fits from COS-Halos to the ``blind'' samples of
\cite{2008ApJS..177...39T}, \cite{Thom:2008gma}, and \cite{2012ApJ...759..112T}. 
These three studies analyze (highly overlapping) samples of intervening \HI\ absorbers from HST
and FUSE data. All have tabulated line-profile fits based on Lyman series
lines for each absorber. Figure~\ref{compare-NHI-bval-fig} shows \NHI\
versus $b$ for COS-Halos and these three samples. The blind comparison
samples have no statistically significant differences between them in
terms of b-value distribution; considered pairwise, the two-sided KS
tests yield $P$ values in the range $0.15 - 0.92$. It is important
that the comparison samples were blind sample of absorbers obtained without 
prior knowledge of galaxy coincidences, and have
a large degree of overlap with one another. They effectively provide a measure of how
different the \NHI\ and $b$ value distributions of the same parent
sample can differ depending on the sample selection and treatment of fitting by different
investigators\footnote{We use Tilton's consensus values from their Table 5, which attempted
to reconcile disparate measurements in this sense.}. Indeed, there is
no statistically significant difference between them. If we compare the
COS-Halos $b$-value distribution to these samples, we find that its KS
probabilities with respect to each are in the range 0.4-0.92; in other
words, there is no evidence that the COS-Halos CGM distribution is more
different from the blind IGM samples than they are from each other.
We find no evidence that there is significant evolution in internal kinematics and 
the implied temperatures inside and outside the 150 kpc region of our survey. 

Considered together with the \HI\ equivalent width and kinematics
shown above, we interpret these line broadening
results as follows. There is a trend in the \HI\
equivalent width at low impact parameter
(Figure~\ref{NHI_rho_fig}). Because the
typical \lya\ line is strongly saturated, this trend is manifested
primarily as a broadening of the profile of total \HI\ absorption,
but component structure is usually not resolved in \lya. For those
systems where we can use higher Lyman series lines to separate blended
components in a profile fit, the fitted Doppler $b$ parameters, which
imply upper limits on temperatures, fall in a distribution that is not
significantly different from blind samples of intervening absorbers. 
The COS-Halos data indicate that that the broadening of \lya\ absorption
near galaxies owes more to the piling up of multiple absorption
components and/or increases in total column density than to an apparent increase in the temperature of CGM gas
\cite[cf.][]{wakker-savage-09-OVI-HI-lowz}. Indeed,
given the inability of COS to resolve any components at $b \lesssim 10$ \kms, 
and the severe blending problems that impede the proper counting of 
components, it is possible that gas near galaxies is {\em colder} on average
than further out in the IGM but that the effect is masked by blending and saturation 
and goes unseen. It seems that ``cold'' IGM/CGM
clouds, wherever they are found, have roughly the same temperatures that
are not determined by their locations with respect to galaxies. 

However we emphasize that using \ion{H}{1} as a tracer of CGM gas temperatures
presents a biased view, as it preferentially detects material at $T \lesssim 10^5$ K 
where the neutral fraction of \HI\ is high enough to yield detected, or even strong, 
absorption. 
There may be a hotter and/or more photo-ionzed CGM 
component, such as that traced by high ions like \ion{O}{6} or \ion{Ne}{8}, that contributes 
little to the observed \HI. 
It is possible that some broad, weak \HI\ absorption, which could trace
a component of the CGM nearer to the expected halo virial temperatures, has
gone undetected in our data. \Lya\ lines with $T = 0.5 - 1$ million K and purely
thermal broadening will have $b = 90-130$ \kms, and detections of such
``Broad Ly$\alpha$ Absorbers'' (BLAs) suggest that they may depart
from the unabsorbed continuum by only 10-15\%, even in the line center
\citep{2011ApJ...731...14S, 2011ApJ...743..180S}. While some 
broad \lya\ components are distinctly evident and separated from 
nearby narrow components \cite[e.g.,][]{2001ApJ...563..724T,2006A&A...445..827R}, 
in many cases such absorbers could be hidden within the much stronger, 
narrower absorption that we detect, which is spread over $\sim 200-400$ \kms\ and is often line-black saturated. Even BLA features with wings exposed at the edges of stronger,
narrow absorption could easily go undetected at the typical S/N ratio of
our data. This is clearly seen in the four $b \geq 80$ \kms\ components in Table 5; the 
presence of these is required for formally good fits, but we regard their presence 
as truly broad \Lya\ absorbers as highly ambiguous - they could be continuum fluctuations, 
blended narrower components seen at low S/N, or other kinds of artifacts. We cannot 
confirm this with Ly$\beta$, so we must regard these profiles as unconfirmed. They are, 
however, potentially good candidates for higher S/N observations to determine 
their shape and kinematics. The widespread detection of a cold CGM does not preclude 
the hidden presence of a hot CGM traced by undetected BLAs. Gas at temperatures 
significantly above 1 million K would likely be undetectable in the UV 
under even the best circumstances.

\subsection{HI Properties by Galaxy Type}
\label{redblue_subsection}

COS-Halos was deliberately designed with both star-forming and passive
galaxies in the sample to permit comparisons between these two populations
of galaxies. The final sample has 16 passive or early-type galaxies
and 28 star-forming galaxies. The
chief goal of this design was to investigate an expected transition in
the gas properties of galaxy halos around the stellar mass where they
transition in star forming properties, to see if halo gas plays some
role in the quenching or continued suppression of star formation.

On the basis of the COS-Halos measurements analyzed so far, the evidence
for a change in the behavior of gas halos between star-forming and passive
galaxies is mixed. \cite{tumlinson-etal-11-OVI-statistics} reported
the bimodality of highly ionized gas traced by \ion{O}{6}, with
a lower rate of detection (30\% vs.~90\%) and weaker absorption in the
COS-Halos ETG sample. 
However, as reported by \cite{2012ApJ...758L..41T}, there is much less difference
in the presence or strength of cool gas traced by \HI\ in the star-forming
and passive samples. There are also modest differences in the covering fraction 
of low-ionization gas between the star-forming and passive subsamples
\cite{2013ApJS..204...17W}.
Our sample of ETGs, combined with our measures
of galaxy/absorber and internal gas kinematics, clearly indicates that gas
with temperatures well below the halo virial temperature is common in
the vicinity of passive galaxies \citep{2012ApJ...758L..41T} with a 75\% rate of
detection even down to low column density limits. This is only a weak 
rejection of the null hypothesis that the two galaxy subsamples 
exhibit a similar equivalent width distribution. Considering detections only, there is
no evidence that the CGM \HI\ around passive galaxies is
significantly weaker or stronger than for star-forming galaxies. These
findings led \cite{2012ApJ...758L..41T} to conclude that the CGM of ETGs could
harbor a cool, photoionized mass of $\sim 10^{10}$ M$_{\odot}$ or more.

Figures 7$-$11 reinforce these results with detailed comparisons between the
line kinematics and internal component structure of absorption in the two
subsamples. Figure~\ref{Lya_kinematics_fig} shows that the kinematics
of the \HI\ with respect to the galaxy systemic velocity are not
significantly different (a two sided KS test rejects the null hypothesis
at only 10\% confidence), indicating that both sets of absorbers could be
effectively bound to their galaxies. Figure~\ref{component_bval_rho}
shows similar distributions of internal kinematics from linewidths
and a lack of evolution with impact parameter. Thus the strong, cold, and
bound medium traced by HI does not vary significantly from star-forming
to passive galaxies, except in those 25\% of cases where the latter
do not show it. 

In light of the COS-Halos results on \HI\  strength, kinematics, and
comparisons to galaxy properties, we conclude that a possibly bound, 
cool CGM is a generic feature of $L \sim L^*$ galaxies independent of 
their type. We have only statistically modest suggestions 
that passive galaxies exhibit \HI\ less often than 
their star-forming counterparts. A larger sample of early type galaxies 
could address this suggestion with better statistics. We have robust 
evidence that the detected CGM has temperatures of $T \lesssim 100,000$ 
K, well below the expected virial temperatures of the halos that host 
these absorbers, but we cannot rule out the presence of some hot 
material that would go undetected. Finally, the relative velocities 
indicate that the detected gas is consistent with being bound to the 
host galaxy. 
None of these results shows a strong dependence on galaxy type, so 
we conclude that a cold, bound CGM is a generic feature of 
galaxies near $L^*$.


\section{The State and Origins of Circumgalactic \HI}
\label{CGM_section}

The COS-Halos measurements of \HI\ are intended to characterize empirically the distribution and content of gas surrounding $L^*$ galaxies. In this section, we use our results and the previous studies to address some open questions about the origins of this gas. Is it really a CGM? What is its source - the CGM, or the surrounding large-scale structure? How does it relate to satellites, disks, accretion, or feedback? What is its fate - has it recently escaped its host galaxy and hence is flowing out, or is it accreting to fuel star formation?

We examine two basic classes of explanation for the detected \HI. The basic division between the two classes depends on whether the gas is outside the galaxy's halo, or ``inside'' the galaxy's halo. For the purposes of this discussion, we define ``inside'' the halo to be inside the virial radius $R_{\rm vir}$. 

First, in \S~\ref{gas_outside_rvir_subsection} we consider the class of explanation in which the gas resides in the large-scale structure of the ``cosmic web'', and not directly in
the galactic halos that we have targeted at $R \leq R_{\rm vir}$.  Possible origins outside include IGM ``filament'' gas near the galaxy but outside the halo (\S~\ref{filaments}), gas residing in a group environment of which the target galaxy is a member (\S~\ref{group_gas_section}), and finally gas arising from the halos of other galaxies nearby that are not the target, including gas driven out in winds or gas residing in nearby ``minihalos'', or small halos in the same large-scale structure (LSS) filament as the target (\S~\ref{othergalaxies}).

Then in \S~\ref{gas_inside_rvir_subsection} we consider the class of origins that lie within \Rvir, including three basic origins: gas arising in galaxy disks themselves (\S~\ref{disk_section}), gas bound to or stripped from satellites (\S~\ref{satellite_section}), and gas distributed through a diffuse medium in a true ``circumgalactic medium'' of clouds and diffuse gas (\S~\ref{cgm_section}). In the first case, the gas can arise directly in the galaxy disk ISM, or in ionized disks extending beyond the stars.  In the second case, gas related to satellites can arise directly in their interstellar medium, in dark-matter bound, starless ``minihalos'', in gas ejected  from satellites by their own star formation, or in material stripped by  tidal forces or ram pressure from an ambient medium surrounding the host. 

``True'' CGM gas can arise in the small neutral fraction of an otherwise hot
halo (near the virial temperature), denser clouds cooled from and
pressure-confined by this hot medium \citep{1996ApJ...469..589M, 2004MNRAS.355..694M},  
gas accreting in a ``cold mode'' below $T_{vir}$, gas accreting 
from the IGM in a ``hot mode'' (i.e. cooling flows), supernova winds or galactic
fountain gas ejected from the host galaxy, or interface regions between
cold clouds from one of these other sources and the expected hot corona.

Our goal in investigating these possible origins is not to conclusively 
rule out or confirm a particular unique explanation, as it is generally true 
that contributions from multiple sources are compatible with the data as well as any single 
scenario. Rather, our goal is 
to identify the properties that a particular model origin would have 
if adopted alone, and thereby to assess the relative viability of the 
various scenarios and their relative possible contributions. The purpose of these 
simple models is to aid our own interpretations of the data, and those of others, and 
possibly to guide followup observations. These simple models are best understood 
on that limited basis, not as unique or conclusive explanations of the empirical results, which 
speak for themselves. 

\begin{figure*}[!t] \begin{center}
\epsscale{1.0}
\plotone{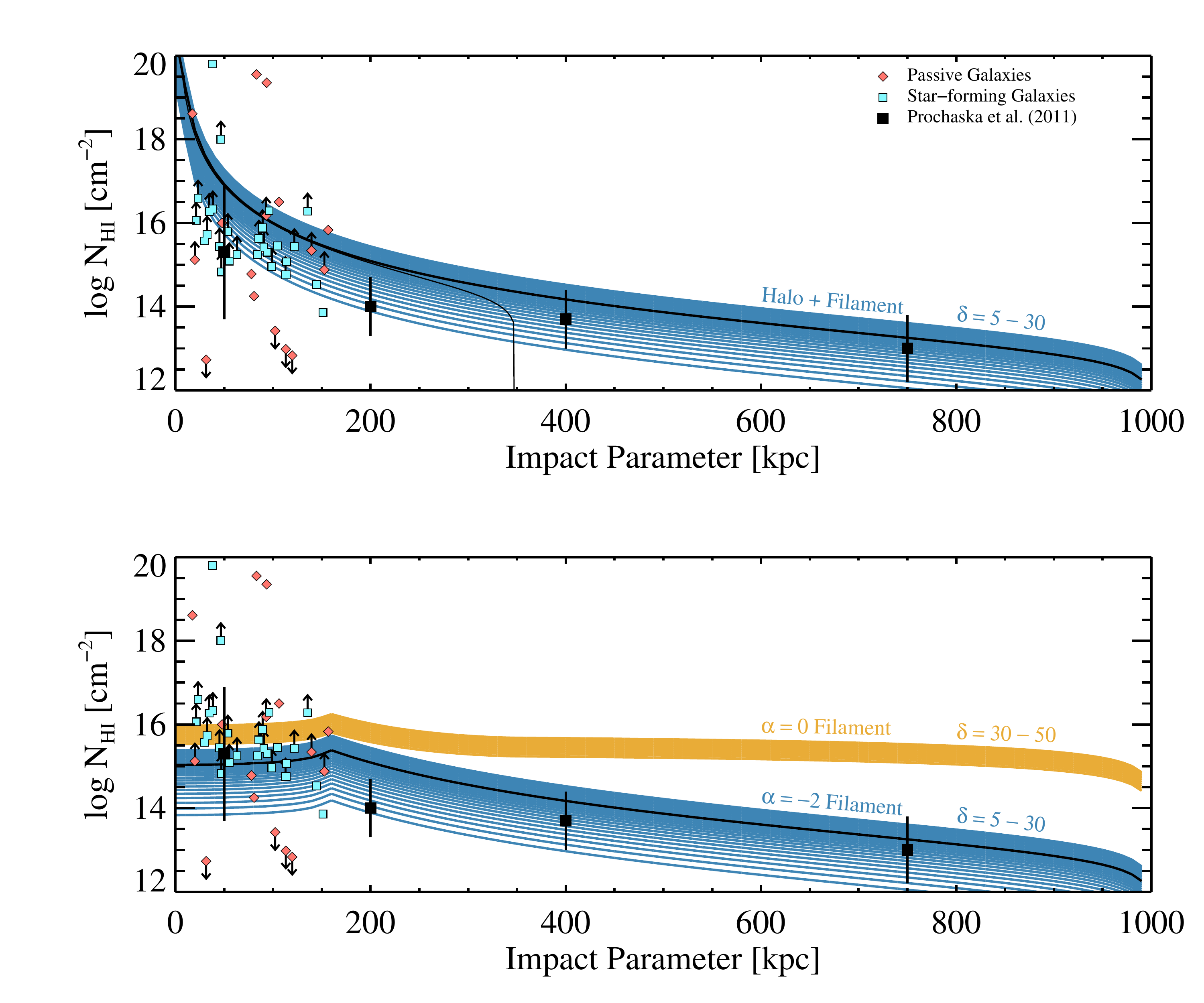} 
\end{center} 
\caption{Simple models of halo and IGM gas compared to \HI\ data. The COS-Halos data are plotted as before. The black squares mark the median column density and r.m.s.~scatter for all galaxies in \cite{prochaska-etal-11-OVI-HI}. In the upper panel, the blue colored region marks the model density profile with $\alpha = -2$ described in the text normalized to overdensity $\delta = 5$ to $30$ at \Rvir\ with constant temperature $T = $ 30,000 K. The black lines mark the curve with $\delta = 20$, and then repeat it excising all gas at $>$ \Rvir\ (thin black curve). The simple model which extends gas to 1 Mpc with declining density (e.g. the thick black curve) provides a reasonable description of the trend with impact parameter. In the lower panel, we repeat the same $\alpha = -2$ models with the spherical region inside 150 kpc entirely removed. The orange curves increase the range of  $\delta = 30-50$ with the halo and again omit $<150$ kpc.  This model, which matches the COS-Halos detections by including a diffuse, extended background outside 150 kpc, overproduces absorption far away from galaxies. The best fitting models are those in which gas density increases inside \Rvir. 
\label{nhi_rho_simple_fig} } 
\end{figure*}

\subsection{A Simple Model} 
\label{simplemodelintro}

In the sections to follow, we will consider whether the detected \HI\ can
arise outside the virial radius $R_{\rm vir}$ (either from the IGM or from
other galaxies), inside \Rvir\ but not in a diffuse CGM (from satellites or extended
galaxy disks), or whether it is most likely a true diffuse CGM. To assist
with these tests, we have built a toy model of gas halos with very simple
parameterizations that describe the geometry and ionization state of
the gas. This model will be used below to assess the viability
of the various origin hypotheses, so we describe its details here.

This ``simple halo'' model includes a spherically symmetric density
profile running with physical radius $R$ specified by a power-law
of particle density, 
\begin{equation} 
n_H = n_0 R^{\alpha},
\end{equation}  
where $\alpha$ is the
power law slope and the density is normalized to cosmic overdensity
\begin{equation} 
\delta_0 = n_0 / \bar{n}
\end{equation} 
at the given \Rvir. 
We take 
\begin{equation} 
\bar{n} = (3H^2 / 8 \pi G) \Omega_b / m_H \times (1+z)^3 = 4.3 \times 10^{-7} {\rm cm}^{-3} 
\end{equation} 
at $z = 0.2$. 
At each $R$, we calculate the
ionization fraction of \HI\ given the local density, a fixed parametric 
temperature $T$, and the extragalactic ionizing background computed at
 $z = 0.2$ \citep{2001cghr.confE..64H}. The ionization model assumes that the gas is optically thin and is in thermal 
and ionization equilibrium, the latter set by recombinations in balance 
with ionizations from photons and collisions with electrons and protons 
in CIE at the parametric temperature. The ionization tables are given 
in terms of local density and temperature and return the ionization fraction 
for a range of particular observable ions; here we focus on the \HI.

The spherical halo can be surrounded by either a medium with a density
that continues to decline in density as $R^{\alpha}$, or by a fixed
medium of constant parametric density $\delta _{\rm back}$ over  
$R_{\rm vir} \leq R \leq 1$ Mpc. This external medium can represent a larger
structure, such as a filament of large scale structure or intragroup medium, in which the
model halo is embedded.

Once the density profile and ionization structure of the model halo
are calculated, column densities are integrated along mock lines
of sight through the medium at a range of impact parameters to
compare to the COS-Halos data. These column densities are calculated by
line-integrals along chords running through
the spherical density profile. Finally, total masses of ionized gas are
calculated for various regions in physical radius or in projection.

Naturally these simple models leave out important physical details such
as density fluctuations, non-equilibrium ionization, and velocity fields
among other potentially relevant effects. However, they are not intended
to faithfully represent all the rich detail of the gas giving rise to
the detected absorption. These models are useful for testing whether
simple configurations and assumptions can explain the data or not, and to
help derive simple estimates of the properties of the detected gas under
certain assumptions. Their results should be considered in
light of these significant limitations and modest goals.

\subsection{Gas outside \Rvir} 
\label{gas_outside_rvir_subsection}

\subsubsection{IGM Filaments\label{filaments}}

First we examine the possibility that the detected gas lies
outside \Rvir\ in extended filaments of gas associated with large-scale
structure in which the COS-Halos galaxies are embedded
Figure~\ref{nhi_rho_simple_fig} shows the results of this
simple model for this case. The plot assumes $\alpha = -2$ and $T = 30,000$ K. In addition to the COS-Halos data plotted as in all
figures above, the black points show the mean and RMS column densities from
\cite{prochaska-etal-11-OVI-HI} as a characterization of HI surrounding $\sim L^*$ galaxies 
at $\gtrsim 300$ kpc. In the upper panel, the blue curves show \NHI\ versus
impact parameter for the full model, with the $\alpha = -2$ density profile normalized to $\delta _0 = 5 - 30$ at \Rvir\ = 350 kpc. 
In this model, the steeply declining density profile adopted inside \Rvir\ is extended (arbitrarily) 
out to 1 Mpc to represent a filament of decreasing density. 
Here, the ``observed'' column density \NHI\ does not decline as rapidly as $n_{\rm H}$, 
because the column density is obtained by a line integral over large pathlengths 
that compensate for the declining density. The range of adopted normalizations 
$\delta = 5 - 30$ is reasonable for gas near the virial radius. 
The thick black curve 
marks a full model with $\delta _0 = 20$; the thin black curve truncates this model at 350 kpc, 
showing that the column densities outside 300 kpc are produced predominantly by the 
``filament'' and those inside predominantly by the ``halo''. 
The total masses in the spherical structure out to 1 Mpc are $0.6 - 4 \times 10^{11}$ \Msun. 
This model adequately reproduces the data over a broad range of impact parameter by extending 
gaseous structures to high density at low radius (the ``CGM'') and low density at high radius (the ``filament'').

The lower panel of Figure~\ref{nhi_rho_simple_fig} shows the behavior of models containing 
only this simple model filament, in two forms. The blue curves repeat the ``halo+filament'' ($\alpha = -2$, $\delta _0 = 5-30$)
models from the upper panel, but this time excises all the gas inside the 150 kpc sphere covered 
by our data prior to the line integral to obtain \NHI. This model expresses the (extreme) scenario in which all the gas
``observed'' at $< 150$ kpc impact parameter actually sits at $> 150$ kpc radius from the center of the system. 
These ``halo deleted''  models clearly fall short of the COS-Halos detections at $\leq 150$ kpc, though 
they still provide a reasonable match to the data points at larger impact 
parameters. By deleting the gas from the region surveyed by COS-Halos, 
they fail to reproduce the trend to higher \NHI\ observed in the two samples 
of data.

To assess whether a more extreme ``filament" model may be able to restore 
the fit when gas inside 150 kpc is deleted, the solid orange curves in the lower panel assume 
that the density $\delta _0$ at \Rvir\ is continued 
at a constant value out to 1 Mpc; this is a ``constant-density'' filament\footnote{The 
``filament'' must have $\delta _0$ less than or equal to the halo
at the virial radius, by the definition of virtualized halos.}. Once 
again gas is deleted inside 150 kpc. To better match the data in the COS-Halos 
region with this ``constant-density'' filament, we must increase the range 
$\delta _0 = 5 - 30$, as in the lighter curves, to $\delta _0 = 30 - 50$.  This
model, which matches the COS-Halos detections by including a diffuse, very
extended background but omitting the halo, overproduces absorption far away from galaxies (the 
implied mass within 1 Mpc is $1-2 \times 10^{12}$ \Msun). Thus, ``filament" gas can help explain 
the data at $>$\Rvir\, but fits best if the density continues to decline outside the halo 
(top panel). More extreme models which attempt to explain the data by 
either (1) deleting gas at  $< 150$ kpc from a filament with a declining density profile (top), or 
(2) including only a constant-density filament at $> 150$ kpc (bottom) both 
fail to recover the trend of \HI\ strength seen in the data.  This examination 
of these simple models leads us to a simple explanation: 
that the detected \HI\ is not likely to be caused simply by large-scale structure 
filament gas, with no contribution from within the halo itself. It remains possible that 
some smaller portion of the detected material arises from such large-scale structures, but 
 this modeling indicates that the contribution is of order $\log$ \NHI\ $\sim 14-15$ or less. 
More detailed 
examinations of these models and the behavior of \HI\ gas within 1 Mpc of 
galaxies is deferred for a later study. 

\subsubsection{Intra-group Gas \label{group_gas_section} }

It is also possible that gas observed at low impact parameters near galaxies
arises outside $R_{vir}$ but within a galaxy group. Such ``intragroup''
gas has been claimed in some circumstances where groups are evident
and gas alignment with any particular galaxy is not obvious \citep[e.g.,][]{2003ApJ...594L.107S, 2006MNRAS.367..139A}. We do not have the data necessary to perform
a fair test of this hypothesis with COS-Halos, for two reasons (see also \S~\ref{environment_section}). First,
our original selection of candidates actively selected against galaxies
with coincident photometic redshifts, because the goal was to obtain a
sample of isolated L$^*$ galaxies to minimize confusion. Nevertheless
some galaxies turned out to have neighbors at the same redshift owing to
imperfect selection and photometric redshift errors. Second, we have not
obtained redshifts for all galaxies in these fields, and so we cannot
make robust statements about neighbors and possible group membership
in all cases.

Despite this original selection against groups and the
heterogeneous redshift database, we do know of a few targets galaxies
that are likely in groups, defined as those with one or more nearby
galaxies at the same redshift with similar luminosity. Six of our galaxies
meet these criteria \citep{werk-etal-12-galaxies},
and their detections in \HI\ show a wide degree of diversity. Two exhibit
subDLA systems (J0925$+$4004 galaxy 196$\_$22 and J0928$+$6025 galaxy
110$\_$35), three show strong \HI\ with multiple components (J0910$+$1014
galaxy 242$\_$34, J0820$+$2334 galaxy 260$\_$17, J1133$+$0327 galaxy
110$\_$5), and one is a non-detection \HI\ (J2257$+$1340  galaxy
270$\_$40). This diversity means that we cannot make firm statements
about group gas origins in those COS-Halos galaxies for which group
membership is known. The nearest-neighbor analysis 
indicates that their large-scale environments ($1-5$ Mpc) are not unusual for 
galaxies at this luminosity \citep{werk-etal-12-galaxies}. 

However, despite the lack of detailed knowledge of group membership for the 
COS-Halos galaxies, we regard it as unlikely that group gas is a dominant  
cause of the detected absorption in \HI\ because of the 
challenging set of observational facts that a group model must explain. 
First, the gas we detect is {\it apparently} within $R_{vir}$ for the selected galaxies. Second, the 
relative velocities are well within the expected escape velocities and tend to 
be centered on the systemic redshift of the targeted galaxies (\S~\ref{bound_subsection} and Figure \ref{Lya_kinematics_fig}); even in poor groups the 
velocity dispersions of a few hundred \kms\ would not be expected to show such  
a trend. Third, we find a nearly unity covering fraction of gas near $L^*$ galaxies; this would imply a nearly unity covering fraction in an intra-group medium if it were not associated with 
particular galaxies. Fourth, there are relatively clean trends of \HI\ strength with 
impact parameter, which would not necessarily be expected if the gas reside in 
an intragroup medium but is not associated with particular galaxies. 

To prove that this gas actually arises in an intra-group medium instead of inside 
the virial radius, the statistical associations of the gas properties with group 
membership must be as strong as or stronger than correlations with the 
nearest galaxy individually. Evidence for this would include findings that 
group membership causes a systematic change in gas properties - 
stronger, weaker, more extended, hotter, etc.  This would require a 
sample of galaxies within groups and a control sample of galaxies that are not  in groups. 
It must also be shown that the apparent change in absorption properties in a group sample is not caused simply by the superposition of gas  inside the $R_{vir}$ of group members, as has been claimed for, e.g. 
\ion{Mg}{2}-traced CGM gas by \cite{2011ApJ...743...10B}. No such controlled 
experiment has yet been done for \HI\ and other UV-band ions. Should 
such a sample of selected group sightlines exist, a subset of COS-Halos fields 
containing galaxies outside groups could serve as a control sample once their 
spectroscopic followup is 
complete. 

\subsubsection{Other Nearby Galaxies \label{othergalaxies}}

Apart from galaxies nearby our targets that share the same group-scale dark matter
halo, it is possible that some of the detected gas is associated
instead with interloper L* galaxies nearby that create chance projections
along the sightline (less luminous satellites are considered separately
below). This material could be halo gas within, ejected from, or bound
to the satellites of the neighbor. Interloping galaxies could lie within a few
Mpc in the foreground or background of our sample galaxies, far enough
for significant Hubble flow velocities with respect to our targets, and
still have peculiar velocities move them back into chance coincidences
in velocity space.

Chance coincidences such as this are extremely difficult to rule out
conclusively, particularly for individual cases, but are 
disfavored by our selection technique and by our knowledge of the
fields. Even though we have not completely surveyed all these galaxy
fields to identify all possible interlopers, the comments above regarding
possible group membership cover all the cases where neighboring $L^*$
galaxies at the same redshift were identified. No other targeted systems
have massive galaxies closer to the sightline than the target and at
similar redshift, as a consequence of active selection for isolated $L^*$
galaxies. Any possible interlopers would need to have $L \ll L*$ or impact 
parameter $\gg R_{\rm vir}$, thus begging the question of how \ion{H}{1} is 
distributed around galaxies. Also,
any interlopers would not be distributed almost evenly in impact parameter
to the sightline as the targeted galaxies are (by selection). The
absorption they contributed would then be drawn preferentially from
larger radii
and would not be expected to produce the clean trends with
target impact parameter that are shown in Figures~\ref{NHI_rho_fig},
\ref{lya_rho_compare_fig}, and \ref{nhi_rho_simple_fig}. That is,
absorbers drawn from well-behaved relations around randomly distributed
galaxies are not expected to recover well-behaved relations around these
carefully-selected galaxies. The contribution of interlopers in velocity
space should be just as likely to lie outside the escape velocity of the
targeted galaxy as within it, not to give the tight kinematic correlations seen
in Figure~\ref{Lya_kinematics_fig}. Given the suppression of interlopers
in our selection and subsequent redshift screening, and these qualitative
considerations about the observed trends, we regard it as unlikely that
neighboring L$^*$ galaxies contribute significantly to the detected
absorption around our targeted galaxies. Finally, we note that attempting 
to construct a quantitative model for the contribution of
interloper $L^*$ galaxies (as we will do for satellites below) would beg 
the question of how diffuse \HI\
is distributed in space and velocity around $L^*$ galaxies. A full treatment 
of this issue would require deeper, complete galaxy surveys in 
these fields. 

\begin{figure*}[!t] \begin{center}
\plottwo{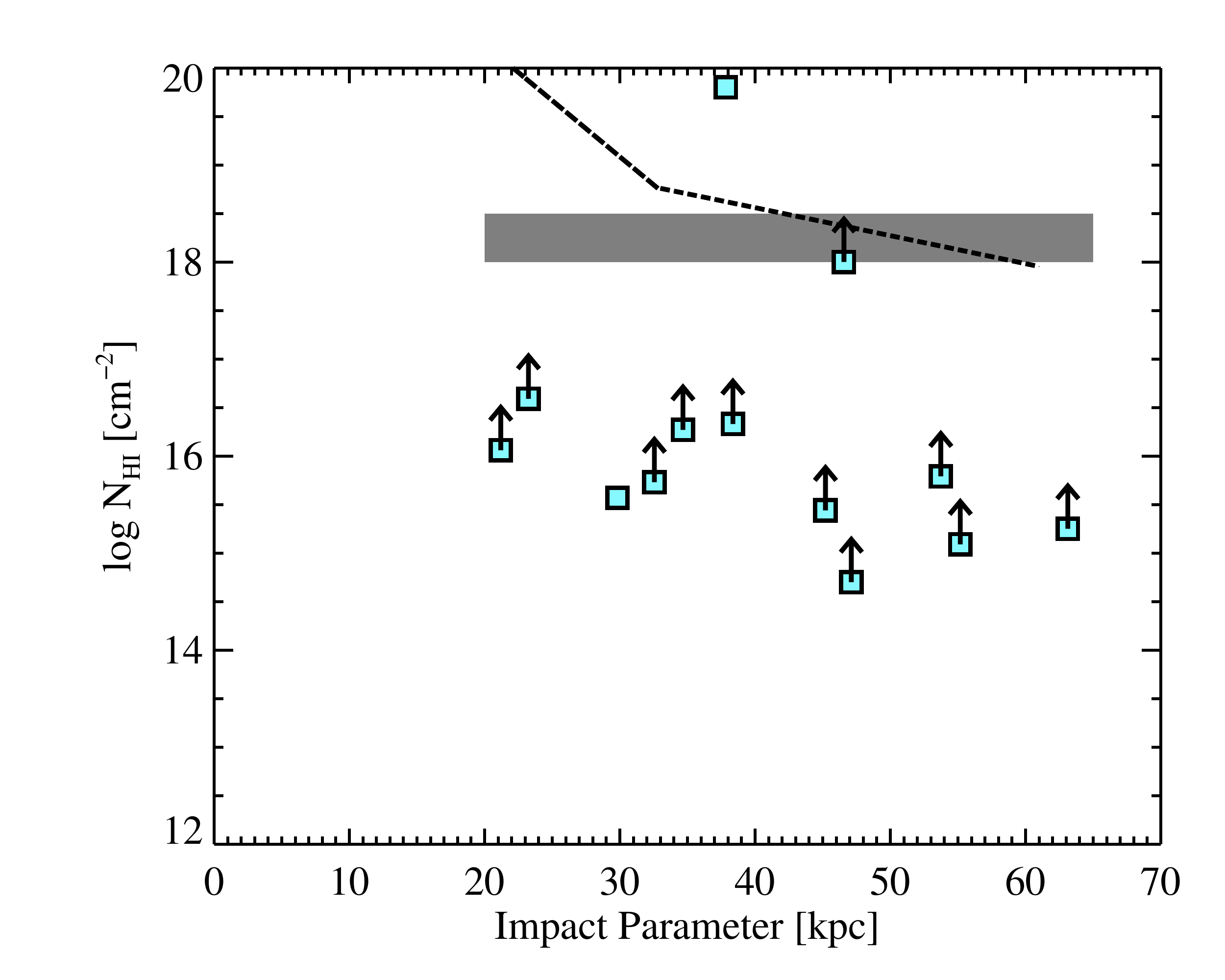}{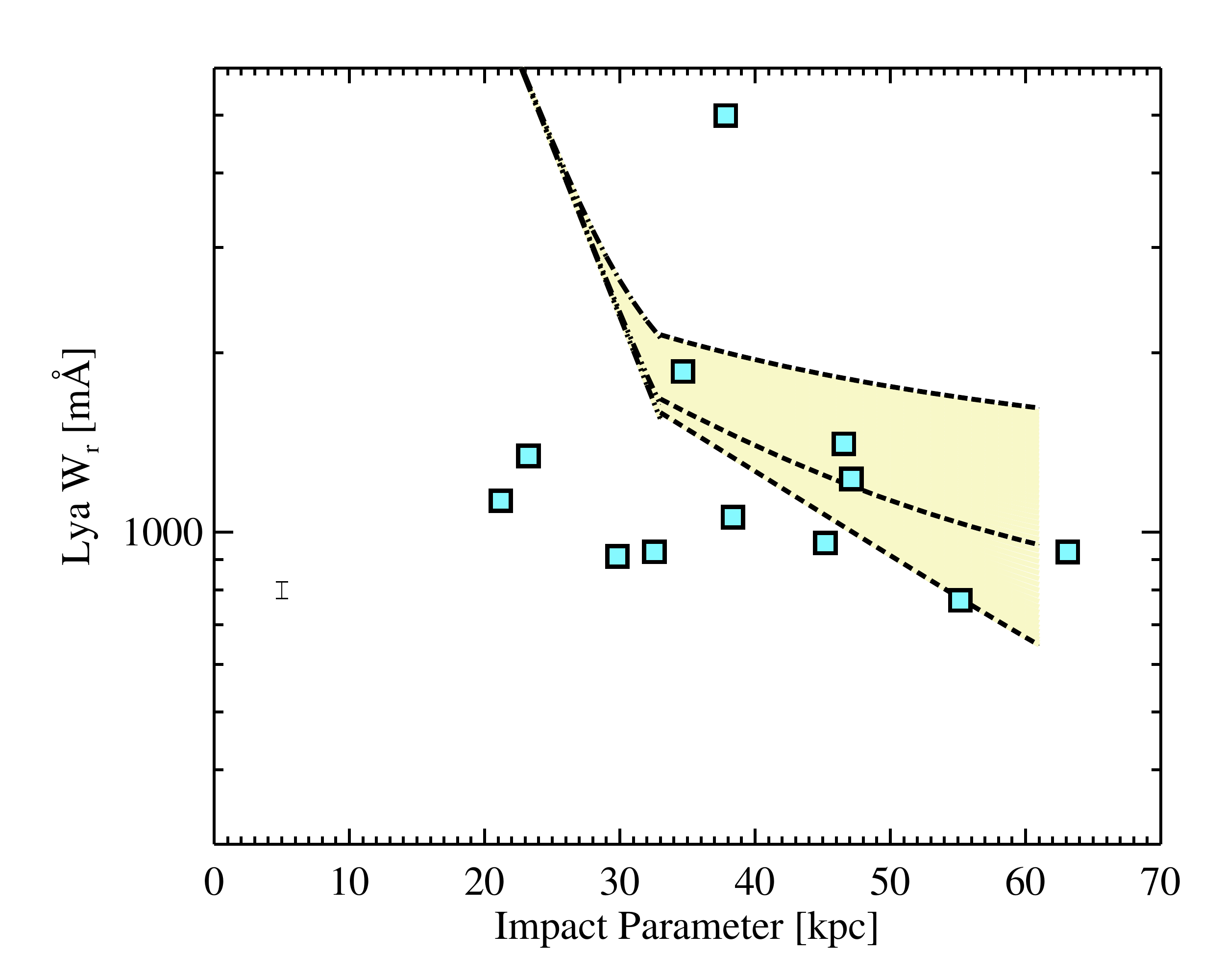}
\end{center} 
\caption{Simple models of the extended Milky Way disk
compared to COS-Halos \HI\ data. The COS-Halos data are plotted as before, omitting 
passive galaxies that should lack extended disks. 
The broken profile shows the disk and ``high velocity dispersion'' (HVD)
profiles for the MW from \cite{2008A&A...487..951K}, which are good fits
to the observed Milky Way surface density. The COS-Halos points are 
plotted with lower limit arrows where appropriate. These points are constrained 
to have column densities less the shaded region with $\log$ \NHI $\simeq 18 - 18.5$ 
by the absence of damping wings. Thus the COS-Halos points generally do not 
match the model profiles inside $\sim 50$ kpc. 
At right this the model profile has been
converted to $W_r$ for \lya\ assuming a single velocity component with $b
= 10$, $30$, and 60 \kms, from bottom to top. One damped system is present at the top - the others 
are $> 80$ kpc from the targeted systems. The typical error bar on these equivalent widths 
is smaller than the adopted symbol size, as  shown by the representative value at left in the panel. 
\label{nhi_rho_disk_fig}} \end{figure*}

\subsection{Gas inside \Rvir}
\label{gas_inside_rvir_subsection}

\subsubsection{Extended Galactic Disk Gas \label{disk_section}}

 An obvious possible explanation for \HI\ inside \Rvir\ is gas arising
 in the interstellar medium of the galaxy itself. This can
 be either the familiar, predominantly neutral inner disk or extended
 disks kept ionized by internal or external sources. Using 21 cm \HI\
 surveys of the Milky Way as a template for star-forming disk galaxies
 \citep{2009ARA&A..47...27K}, we expect that gas-rich disks themselves
 should imprint a damped \lya\ system (DLA; $\log$\NHI\ $\geq$ 20.3)
 or at least a strong LLS (sLLS; $\log$\NHI $\gtrsim 18$), but only
 within $\lesssim 35$ kpc. COS-Halos has 4 star-forming galaxies
 at $ < 35$ kpc and 12 at $< 60$ kpc impact parameters. We do not expect strong \HI\
 absorption from gaseous disks in our passive subsample.

In making these comparisons, we remain in the spirit of simple models
from above and so we adopt the fitted MW surface density profile from
\cite{2008A&A...487..951K} as a fiducial gas disk. This profile follows an
exponential profile out to 35 kpc, $\Sigma_{\rm HI}$ = 30
$\exp ((R-R_{\odot})/R_s)$  \Msun\ pc$^{-2}$, with $R_s = 3.75$ kpc. 
Outside $\sim 35$
kpc, the profile flattens to a shallower dependence on galactocentric
radius. \cite{2008A&A...487..951K}
treat this as an extension of the main disk, but one which may consist
of numerous small clouds or a turbulent medium with high velocity
dispersion. This part of the profile assumes a model instead of
being a direct fit to the emission data. Figure~\ref{nhi_rho_disk_fig}
shows the two components of the MW profile compared to COS-Halos data
inside 70 kpc. The main hindrance to direct comparison is the substantial
fraction of saturated systems with \NHI\ lower limits. Therefore, we
perform these comparisons as before in both \NHI\ and \lya\ equivalent
width. We note that the saturation effects manifested as lower limits in
Figure~\ref{nhi_rho_disk_fig} are no longer an issue above $\log$ \NHI\
$\simeq 18 - 18.5$ (grey box), where damping wings usually appear; their absence from
the observed profile can loosely constrain \NHI\ to $\lesssim 18.5$, above
which robust measurements can be derived from fitting the damping wings.

It is notable that the strong damped or sub-damped absorbers ($\log$
\NHI\ $\gtrsim 19$) that would be expected inside $\sim 30$ kpc are not
evident in both panels of Figure~\ref{nhi_rho_disk_fig}. Because of
their geometry, disks can be missed in cases where the target galaxy
appears near edge-on with respect to the QSO sightline. We do not know
the disk sizes or inclinations of our star-forming subsample, so any
contribution of disks to COS-Halos would be that of typical gas disks in
$\sim L^*$ galaxies that are randomly oriented and inclined with
respect to the line of sight. In such a sample, we would expect that
the disk absorption would still cover approximately half of the area out to
some impact parameter on the sky if it fills the disk out to that same
radius in the galaxy. Thus the absence of any clearly damped, disk-line
absorption inside $\sim 30$ kpc is somewhat puzzling, though with small
numbers it may still be attributable to the accidents of random viewing geometry 
or to the fact that the typical stellar mass for COS-Halos galaxies is slightly smaller 
than the Milky Way. 

The outer portion of the MW surface density also seems to over-predict
the absorption seen in COS-Halos. This effect is best seen in
the right panel of Figure~\ref{nhi_rho_disk_fig}, where we have
converted the MW column density versus $R$ into an equivalent width for
single-component absorption using a curve of growth with $b = 10, 30$,
and $60$ \kms\ from bottom to top. The two larger velocity dispersions
are more characteristic of the fitted components in COS-Halos (see
Figure~\ref{component_bval_rho}). Even these profiles exceed the data
points from COS-Halos, suggesting that the MW profile at $R > 30$ kpc, if
is an extended disk, does not match up with external galaxies. However,
this disagreement does not necessarily imply that COS-Halos does not
detect ionized, extended disks in some cases. Observed \HI\ disks can
have quite sharp edges induced by photoionization from an external
ionizing background  \citep{1993ApJ...414...41M, 1994ApJ...423..196D}, with the location
of the edge depending on where the total density profile effectively
becomes optically thin. Also, an extended photoionized disk could be
difficult to distinguish from a more diffuse halo medium in general;
the latter might be expected to continue the galaxy's rotation curve
while the former might not. Since we lack measurements of the  galaxy
inclination and orientation with respect to the line of sight, and also
any information about their rotation curves, we cannot yet perform the
relevant tests. We, therefore, conclude that COS-Halos likely probes the
region of space where the disk transitions to general halo gas, but we
cannot cleanly separate them with the present dataset. We leave a more
sophisticated analysis of these ideas to future work.

\subsubsection{Gas in and from Satellite Galaxies\label{satellite_section}}

It is possible that some fraction of the detected gas is bound to,
or has been recently stripped from, satellite galaxies surrounding the
targeted $L^*$ galaxies whether bound to them or not. We would like to
assess how much of a contribution satellite galaxies can make to the
observed column density and absorption profiles as a population. It is
straightforward to assess the possible contribution of gas {\it bound}
to satellites within the well-specified structure-formation model with
simple assumptions about gas inside the satellites. It is more difficult
to assess the contributions of gas stripped from satellites. We will do
the former first, and see if that gives any insights into the budgets
of stripped gas.

First, let us recall the observational information to be explained:
We detect \NHI\  $\gtrsim 10^{14}$ cm$^{-2}$ with nearly unity
covering fraction at all impact parameters $ < 150$ kpc, for both
galaxy types. The covering fractions at $>10^{15}$ cm$^{-2}$ are $\gtrsim
0.5$. The kinematic spread of the absorption is equally important as a
constraint. The detected absorption is usually distributed into a few
resolved components that appear to be $b \sim 20-40$ km s$^{-1}$ (Figure~\ref{component_bval_rho}). The
range of centroid velocities for the identifiable components is roughly
$\pm 100$ \kms\ (see Figure~\ref{Lya_kinematics_fig}). The absorption
beyond that out to the typical edges of the full profiles near $\pm$200
\kms\ is partially caused by  broadening of individual components
(e.g. curve-of-growth effects, whether thermal or non-thermal). Given
the 20 \kms\ resolution of COS, there is a strong possibility of narrow
unresolved components inside the saturated profiles and the distinct
possibility that what appear as single $b \sim 20-40$ \kms\ components
are actually composed of narrower blended components \citep[indeed this 
is often indicated by component structure in metal ions observed at higher 
resolution, e.g. \ion{Mg}{2};][]{2013ApJS..204...17W}. However, any model
must still match the total kinematic extent of the detected absorption,
which we take to be $\pm 100$ \kms\ from the typical range of component
centroids. The high covering fractions and broad kinematic extent of the
detected \ion{H}{1} jointly provide robust data that any model involving
satellites must match.

We regard the three high-column density sub-DLA \HI\ systems ($\log$ \NHI\
$= 19.4 - 19.9$) as the most likely of all the COS-Halos systems to
arise in the bound ISM of satellites galaxies, though extended \HI\ disks
and high-column ``HVC-like" origins are also possible. These systems were
previously analyzed by \cite{Meiring2011} and \cite{2012ApJ...744...93B},
who found them to have modestly sub-solar to super-solar metallicities,
[Z/H] $\sim -0.4$ to +0.3. The lower of these metallicities are consistent
with luminous dwarf satellites, while the higher metallicities may indicate an extended
galactic disk (but see above). In all three cases, we have not confirmed any galaxy redshifts
closer to the sightline than the targeted $L^*$ galaxies. 

In the Milky Way system, $\sim 17$\% of the sky is covered by 21-cm HVC gas at 
$\log$ \NHI\ $\gtrsim 18.5$ \citep{1991A&A...250..499W} and 37\% is covered at 
$\log$ \NHI\ $\gtrsim 17.9$ \citep{1995ApJ...447..642M}. The majority of this area 
is covered by the HVC cloud complexes, which are 
not known to be affiliated with particular satellites or stellar populations. The 
areal covering factor contributed by the prominent Magellanic Stream 
is only $\sim 5$\%; and only $\lesssim 1$\% for the bound ISM of the 
Clouds themselves, even though their collective mass far exceed that of 
the HVC complexes. Thus if we take the MW system as a template
(and ignoring the obvious differences in viewing geometry) we expect that
$\lesssim 5-10$\% of COS-Halos systems should show $\log$ \NHI\ $\gtrsim
19.5$, with most of the covering fraction arising in HVC-like gas clouds without stars
and a still smaller fraction from gas bound to larger satellites. These expectations are borne out
in \cite{2010ApJ...716.1263G}, who found that satellites of halos
near the upper end of the COS-Halos range ($M_{\rm halo} \sim 10^{13}$ M$_{\odot}$)
should give a covering fraction of roughly 3\% or less inside 150 kpc,
using strong absorption ($> 1$ \AA) by \ion{Mg}{2} as the proxy for
strong absorption by cold gas. These results appear consistent with
the model of \cite{2013A&A...550A..87H}, who translated the  covering 
fractions of Milky Way HVC low-ions to an external viewing geometry and 
found that this HVC population would yield a \ion{Mg}{2} covering 
fraction (for $>300$ m\AA\ absorption) of 20\% out to 60 kpc. 
Two of the three damped systems 
in COS-Halos have $W_r > 1$ \AA\ in \ion{Mg}{2} \citep{2013ApJS..204...17W}. 
While we cannot
definitely conclude that these damped systems arise in the ISM of 
satellites, their properties and hit rate are consistent with this explanation.

\begin{figure}[!t] \begin{center}
\epsscale{1.2}
\plotone{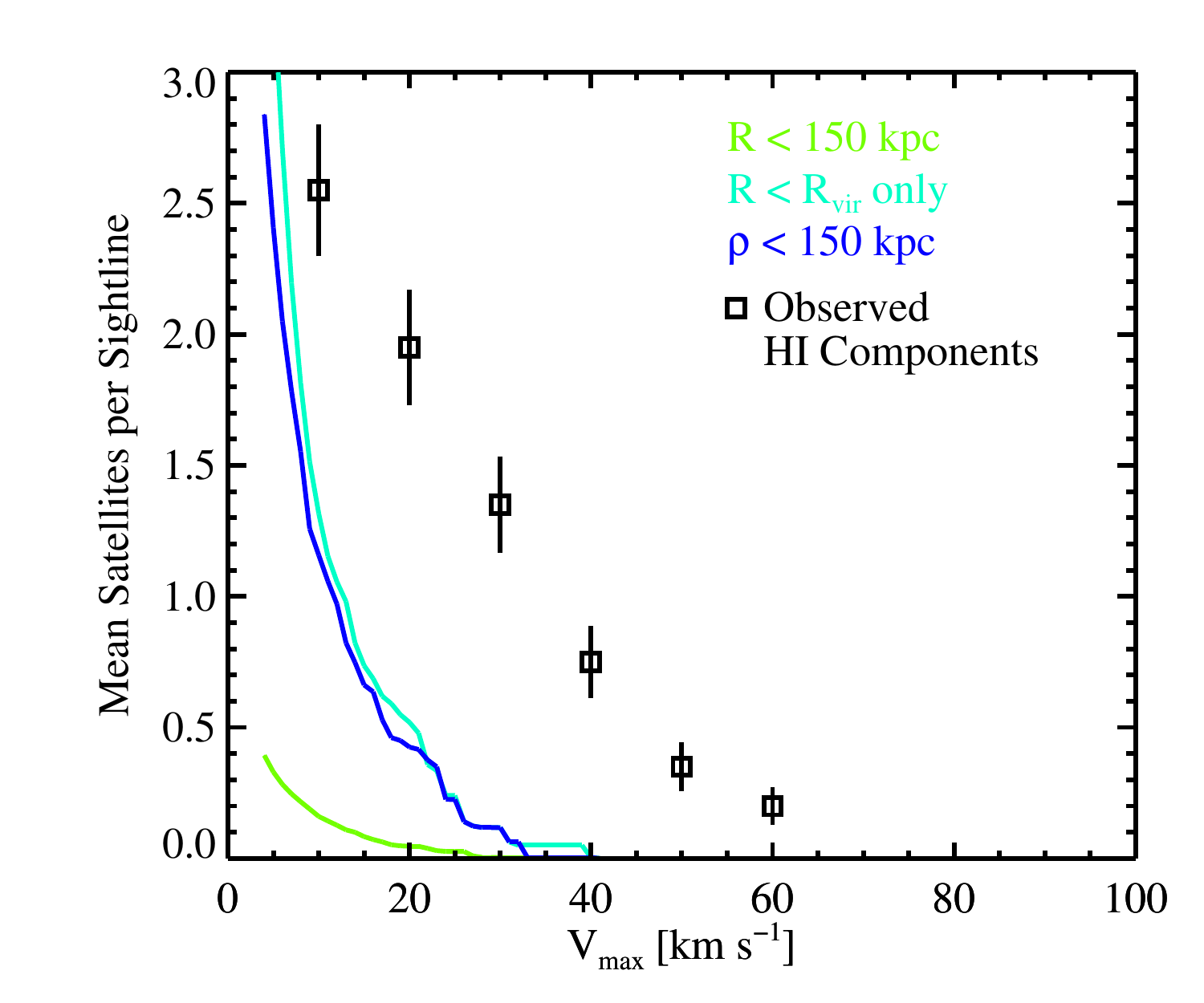} 
\end{center} 
\caption{Mean number of satellites encountered by mock sightlines through the 
Via Lactea halo compared with the number of detected \HI\ components
 in the COS-Halos data. The solid curves show the cumulative (mean) number 
of satellites expected per sightline, integrated down to the $v_{\rm max}$ given on the 
$x$ axis. The model VLII subhalos
with \HI\ at unity covering fraction out to their tidal radius
$R_{tidal}$. The mean number of satellites (or ``covering fraction'') can exceed unity because we allow for  
multiple components or satellites along single lines of sight. The data points use the profile-fitting results shown in
Figure~\ref{Lya_kinematics_fig} to estimate the covering fraction of
detected HI components as a function of doppler $b$ parameter. We assume 
that the doppler $b$ parameters map directly to $V_{\rm max}$ for satellites; this is merely an 
approximation. Even under these
generous assumptions the subhalos fail to explain more than a minority
of the detected \HI.  The three model curves show the VLII  results for all subhalos within 
150 kpc physical radius of the host (light green), all subhalos within \Rvir\ of the host (cyan) 
and, in blue, all subhalos that lie within a {\it projected} 150 kpc impact parameter of the host, whatever their physical radius 
(as viewed from a randomly chosen orientation). The latter model occupies a cylindrical volume that best 
approximates the COS-Halos viewing geometry. 
\label{vl_plot} } 
\end{figure}

For the lower column density COS-Halos systems ($\log$ \NHI\ $< 19$),
the possibility of satellites contributing significantly to the detected
\HI\ absorption exists but is more difficult to assess. The chief difficulty
is our ignorance of the mass and extent of ionized gas (below the 21 cm
detection threshold) surrounding satellites of varying mass. Building
such a detailed model would have to assume a density profile of
\HI, around lower-luminosity satellites, where we have little if any empirical guidance. 
So, instead of building a detailed model of \HI\ surrounding dwarf
satellites, and then computing their covering fraction with the survey
region, we attempt to work out the maximum possible cross section of
gas bound to satellites from a simple but physically motivated picture
of dark-matter substructure. The maximum possible number of satellites
which could contribute gas to the detected absorption is limited to
the number of dark-matter subhalos in host halos of $\log M_{\rm halo} =
11.5 - 13$. The number density of subhalos, each characterized by its
maximum circular velocity $v_{\rm max}$, increases as $v_{max}^{-3}$,
so most of the available subhalos are ``minihalos'' that may or may
not contain stars or ionized gas. 

\begin{figure*}[!t] \begin{center}
\plotone{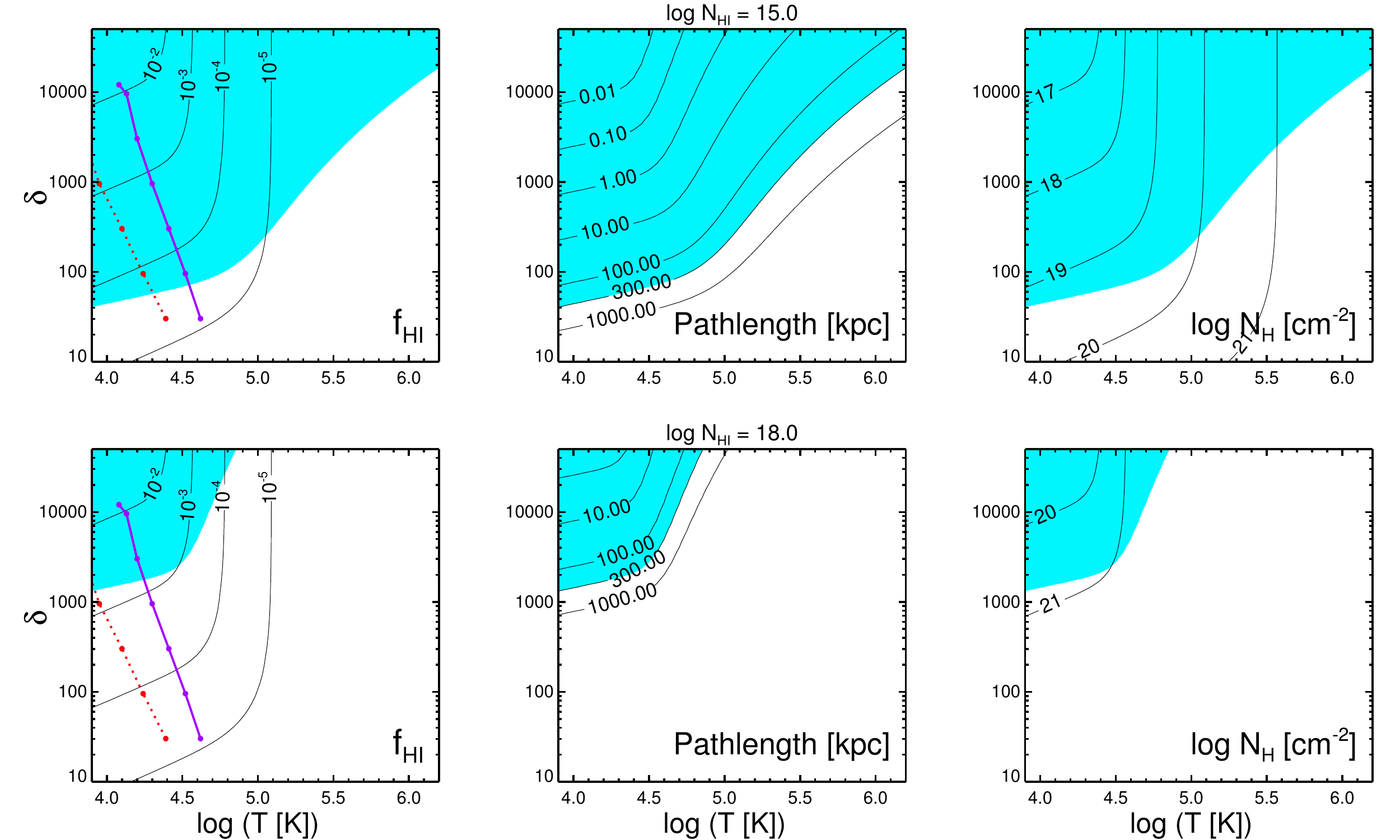} 
\end{center} 
\caption{ Basic ionization models for two fiducial column densities that bound the COS-Halos detections, $\log$ \NHI\ =
15 and 18. The two parameters are temperature $T$ and density $\delta = \rho /
\bar{\rho}$. From left to right in each row, the contours mark constant values 
of \ion{H}{1} ionization fraction $f_{\rm HI}$, pathlength required to yield the 
nominal \NHI, and the total H column density $N_{\rm H}$ through the medium. 
The shaded region with path length $< 300$ kpc is repeated in
each of the three panels in a row to show what combinations of parameters
fits within halos. The filled-circle curves in the $f_{\rm HI}$ panels mark the locus of temperature equilibrium (where 
photo-heating balances radiative cooling) for solar metallicity (dashed red) and 0.1 solar metallcity (solid purple). 
\label{modelh1_fig} } 
\end{figure*}

For concrete estimates, we use a subhalo catalog from the DM-only
simulation of an $L^*$-like Via Lactea II halo by \cite{Diemand:2007hb},
which resolves subhalos down to $v_{\rm max} = 4$ \kms. We make two
additional assumptions. First, that subhalos cannot host gas with
an internal velocity dispersion that exceeds their own: that is,
they cannot contribute to the observed gas profiles a velocity width
that significantly exceeds their own $v_{\rm max}$ (we assume that these velocities map directly to the 
doppler $b$ parameter of fitted components). Second, that they
cannot hold onto gas that falls outside their own instantaneous tidal radius $R_{\rm tidal}$, 
as given by the VLII catalogs. We compute the cross section for absorption 
of all the VLII subhalos by assuming that they all have unity covering fraction
of gas with $\log$ \NHI\ $> 10^{15}$ cm$^{-2}$ inside their own tidal
radius. We include the small contribution of subhalos outside $R_{vir}$
of the VLII host. The results of this very simple model are shown in
Figure~\ref{vl_plot}. Here, we also take at face value the distribution
of profile-fitting b-parameters in Figure~\ref{Lya_kinematics_fig} as an
{\it approximation} to the distribution of detected components versus
line broadening. The mean number of components per sightline is just above
2.5 at to the lowest limit of the reliable linewidths (10 \kms). 

If we assign detectable gas extending to $R_{\rm tidal}$ for  
all VLII subhalos inside an ``impact parameter'' of 150 kpc from the host, the
mean number of satellites per sightlines (equivalent to the 
areal covering factor, but allowed to exceed unity) of subhalos down to $v_{\rm max} = 20$ \kms\ is 0.5,
but the mean number of components in the data is already well in excess of unity. This
velocity is important as the value below which Local Group satellites
 do not contain detectable \HI, so using halos below this value
in this model presumes that such halos can retain bound, ionized
gas at the low levels detected in COS-Halos. The VLII curve
does not reach unity unless we use all subhalos down to $v_{\rm max} =
10$ \kms, and it does not reach 2 until we include all subhalos down to
$v_{max} = 6$ \kms. In other words, to match the COS-Halos data we must allow for
unity covering fraction of \HI\ in small subhalos that are not known to retain
gas at all. Note that this problem only gets worse if we account for the finite
resolution of COS and posit that the detected components may conceal 
narrower unresolved components. Narrower components can arise in subhalos of
lower $v_{max}$, but they are then more numerous in the data, forcing the
points in Figure~\ref{vl_plot} to shift up while the model curves do not
(e.g there would typically be $> 2$ per sightline if they are narrow and
unresolved). Thus, under assumptions that maximize the cross sections of
satellite DM halos (including minihalos), and conservatively estimating
the number counts of detected gas components, subhalos fail to match
the data by a large factor.

\begin{figure*}[!t] 
\begin{center}
\plotone{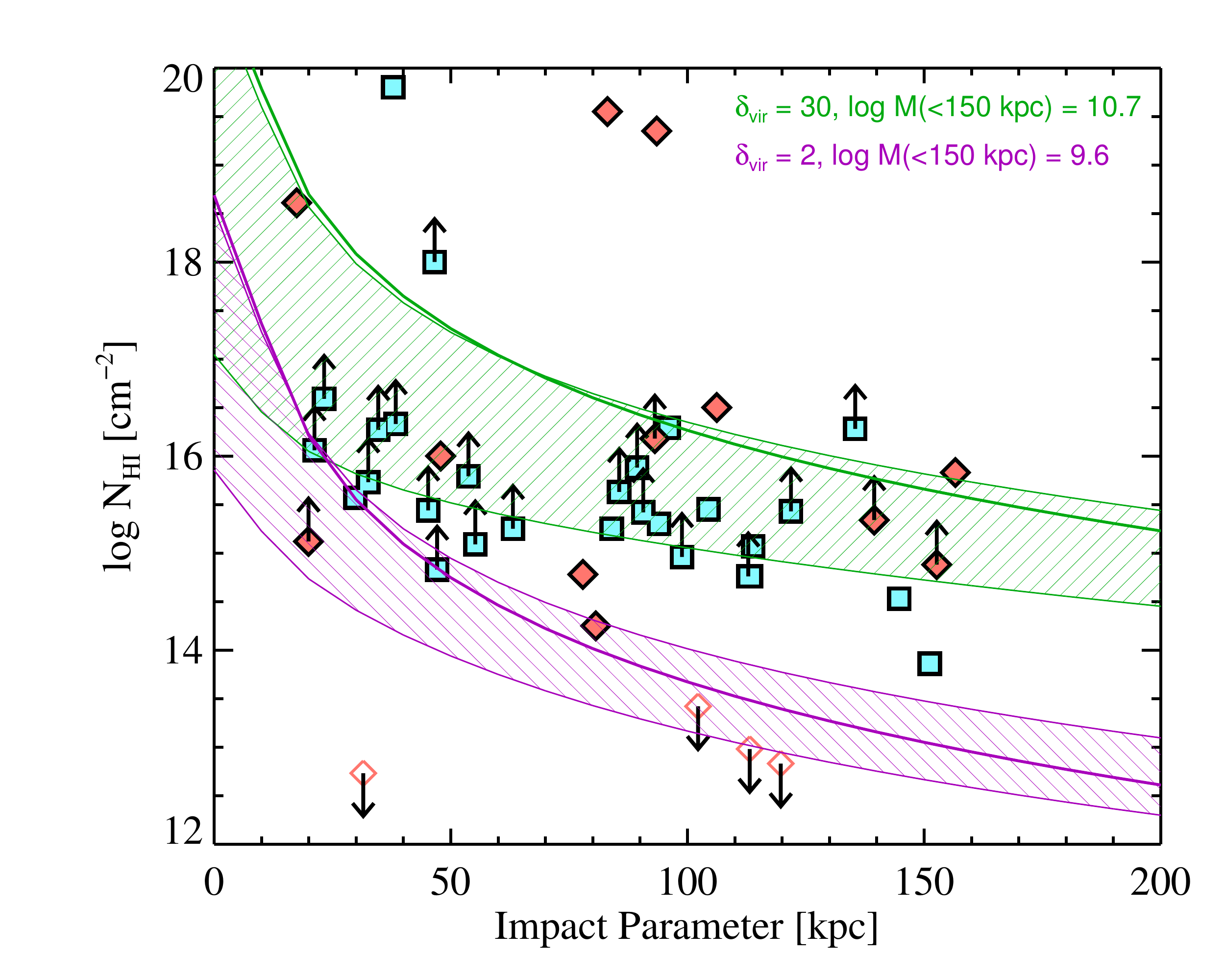}
\end{center} 
\caption{ Figure~\ref{NHI_rho_fig} repeated with simple
physical models included. The green and purple curve families show
the regions covered by models with $\delta _0 = 30$ and 2,
respectively. All models have $R_{\rm vir} = 350$ kpc and $\alpha =
-2$. In each shaded region, models with constant temperature $T =
10^4$ K define the upper bound and $T = 10^5$ K the lower bound. The
intermediate solid heavy curves in each set assume the same density profile 
but include the density-dependent equilibrium  temperature shown in Figure 17 (for solar metallicity). 
\label{real_CGM_fig} } 
\end{figure*}

In short, we find that even if we generously assign all DM subhalos with
\ion{H}{1} gas at the detected level out to their tidal radii, their
projected area and kinematic widths are not sufficient to explain the
strong, broad absorption surrounding the COS-Halos galaxies. Of course,
the satellite frequencies inferred from this exercise are not 
negligible (though the assumptions deliberately maximize them), so it
remains possible that individual systems trace gas bound to satellites. It
might be that a significant minority of the components and/or systems
arise in gas bound to satellites, but if that fraction exceeds about
one-half of all systems then our simple models imply that low-mass
galaxies retain small portions of ionized gas that is undetected by 21
cm measurements. Proving or disproving this hypothesis in single cases
would be very difficult if not impossible given only the \HI\ data
and lacking deeper images and spectroscopy of these fields. For now, we 
regard gas bound to satellites as CGM gas of interest
that ultimately contributes to the mass budget of the host galaxy like any
other CGM component. The origins of CGM gas in satellites is a possibly fruitful
line of research that could be addressed in the COS-Halos data on metal
lines, in new spectroscopic data on dwarf galaxies from our
Cycle 18 HST program (PID 12248, ``COS-Dwarfs''), and in the context of
highly resolved numerical simulations of galaxy halos that could assess
the relative contributions to the CGM of gas bound to satellites and
gas arising in the diffuse CGM.

Gas stripping from satellites surely contributes to the CGM (viz. the
Magellanic Stream) but the contributions of recently stripped material to
the COS-Halos data is even more difficult to assess quantitatively than
the possibility of gas bound to satellites. Stripping could increase the
\HI\ cross sections of small satellites, but to conserve mass it must then
lower the typical column density. We do not have quantitative constraints
on either cross-sections or column densities for real satellites at $\log
N_{\rm HI} < 10^{18}$ cm$^{-2}$. Note that even if we were to crudely account for 
tidal stripping effects by arbitrarily
assigning gas out to $2R_{tidal}$ in the simple model above, the expected
number of components would still not match the data without the contribution of
small subhalos that are not known to contain gas. This is another area
where insight from numerical simulations would be helpful.

\subsubsection{The CGM: Diffuse Gas within \Rvir \label{cgm_section}}

Having evaluated many sources of the detected \HI\ absorption from
outside the host halo, and from disks and satellites within it, we now
turn to examining the properties that this medium has if it is a true
``circumgalactic medium'': diffuse gas surrounding the
galaxies that is not directly bound to satellites or arising in the
IGM. This CGM might include flows on their way into the galaxy, ejecta
on their way out, gas stripped long before from satellites by tidal forces or 
ram pressure, material that is being heated by active feedback
or material cooling and falling in from the IGM. Perhaps the real CGM
contains gas from all these sources.

COS-Halos has generated a rich dataset of multiphase ions that can be used
to examine the ionization state, metallicity, kinematics, and origins
of this gas using a range of diagnostic lines. An empirical analysis of the
metal line survey is available in \cite{2013ApJS..204...17W}. The \HI\ by
itself provides too little information on which to draw conclusions about
the true state and origins of the CGM gas, but we can undertake simple
models to get order-of-magnitude estimates for the properties of the CGM
gas that might exist around these targeted halos. These models are rather
limited in the amount of detail they can capture and still be constrained
usefully by the \HI\ data alone. Among the things we do not know are the density
and temperature of the absorbing material, its distribution throughout
the halo, the degree of clumping in space or along the line of sight, the
thermal history of the gas, or its internal kinematics and bulk flows. In
light of all these missing elements, we instead have only two
modest aims: (1) to show that a diffuse ionized medium can reproduce the
observed column densities and trends with galaxy properties and still
fit within the spatial extent of halos, and (2)
to estimate the properties -- density, temperature and implied masses --
for simple parameterizations of diffuse CGM that match the data.

The first kind of ionization modeling is designed simply to show that
a diffuse ionized medium can reproduce the column densities we observe
for plausible physical conditions and still fit within the physical
extent of the relevant dark matter halos. To do this we model column
densities from a uniform diffuse medium of constant temperature $T$ and constant density $\rho$,  
expressed in a ratio to the cosmic mean baryon density
as $\delta _0 = n / \bar{n}$, where $\bar{n} = 4.2\times 10^{-7}$ cm$^{-3}$. This optically thin medium is exposed to the
extragalactic ionizing background scaled to $z = 0.2$ \citep{2001cghr.confE..64H}. 
In addition to photoionizations we also include temperature-dependent ionization from collisions in pure CIE. Thus
we have a two-parameter space, as shown in Figure~\ref{modelh1_fig}.
The two rows show results for a fiducial column density \NHI\ $= 10^{15}$
cm$^{-2}$ (top) and \NHI\ $= 10^{18}$ cm$^{-2}$ (bottom); these two limits
bound the region of our (non-damped) detections. The three columns are
(left to right) contours of constant \HI\ neutral fraction $f_{HI}$, pathlength 
required to achieve the nominal \NHI, and the total $N_{\rm H}$,
for each possible combination of $\delta _0$ and $T$. The light
blue shaded region shows where the implied pathlength is 300 kpc or less,
and is repeated in all panels to show allowable values of the parameters
by this criterion. 

The range of plausible temperatures in this space is further constrained 
by the line-broadening observed in COS-Halos systems. The fitted line widths 
constrain most of the detected gas to $T \lesssim 10^5$ K ($b \leq 40$ \kms), 
or roughly $f_{\rm HI} \gtrsim 10^{-5}$. Basic considerations of temperature 
equilibrium (red and purple curves in Figure~\ref{modelh1_fig}) and cooling also argue for temperatures of $10^{4-5}$ K. 
Cooling and photo-heating timescales also argue for temperatures to be $10^{4-5}$ K, 
near the locus of points indicating thermal equilibrium (red and purple curves in Figure~\ref{modelh1_fig})
At  $T = 10^5$ K and $\delta = 1000$, the cooling time is only 20 Myr at 
metallicity $Z_{\odot}$ and 65 Myr at $0.1 Z_{\odot}$, and scale down inversely 
with $\delta$. On the equilibrium locus, radiative cooling balances photo-heating exactly and the 
cooling time is effectively infinite. Gas that takes excursions away from 
the equilibrium curves will not remain there for long. Gas with the observed 
kinematics ($b \simeq 15- 40$ \kms) could exist at or near the equilibrium 
curves for much longer times, remaining stable long enough to be the 
most commonly detected component  of the CGM. A hot medium 
with $T \sim 10^6$ K could persist for long times if the density is low 
($\delta \sim 10-100$), but would then have neutral fractions and \HI\ columns and 
linewidths that could easily evade detection; the absence of such 
detections is not evidence against the existence of hot halos surrounding 
galaxies. These models show that a cool, photoionized medium a few hundred 
kpc in extent at moderate overdensity, with a moderate degree of clumping, can 
plausibly recover the observed column densities and line-kinematics in a physically plausible 
ionization scenario.

The second part of this simple analysis combines these basic elements
and revisits the more prescriptive halo models illustrated above
(Section~\ref{simplemodelintro} and Figure~\ref{nhi_rho_simple_fig}). Here
we assume that the gas is distributed according to a smooth power
law density profile, $\rho \propto R^{\alpha}$ normalized to $\delta _0$
at $R_{vir}$ and $\alpha = -2$. The temperature is again held fixed with $R$ but varied
in families of models. The same ionization tables are used as in the
sightlne analysis just above. The results of these models are shown
in Figure~\ref{real_CGM_fig}. Here we show two families of models,
with $\delta _{0} = 2$ (purple) and $\delta _0 = 30$ (green). The lower
envelope of each shaded region corresponds to $\log T = 5$, the upper
envelope to $\log T = 4$. The heavy curves in each set take the same 
density profile but apply the density-dependent equilibrium temperature 
from Figure 17 (at solar metallicity) to set the temperature as a function 
of local density. The higher curves are better matches to the
COS-Halos data, which is roughly the expected overdensity at 
\Rvir\ for halos at this mass scale ($\delta _0 \sim 30$). The implied mass
for these models is $M \simeq 5 \times 10^{10}$ \Msun\ out to 150 kpc 
in physical radius, comparable to 
the stellar masses, and apparently valid for both galaxy types. 
These models are too simple and our
data too meager to account for clumping, non-equilibrium ionization,
aspherical distributions, or any more complicated physics, but they do
show that there are plausible models for a diffuse ionizing CGM around
these galaxies and that such a medium has a significant total mass.


\section{Summary} 
\label{summary-section}

COS-Halos has characterized the diffuse gas near $L \sim L^*$ galaxies
using a new sample of QSO/galaxy pairs selected specifically for this
purpose. This survey spans both star-forming and passive galaxies 
with sightlines ranging at projected separations up to 150 kpc. This
paper has presented the detailed properties of the survey design and
the procedures followed in the collection and processing of the data. We
have also presented results of the COS-Halos census of \HI\ surrounding
these $\sim L^*$ galaxies. The key findings of the \HI\ survey are:

\begin{itemize} 

\item[1.] With detection limits at rest-frame equivalent
width $W_r \sim 30 - 50$ m\AA, or $\log$ \NHI\ $\sim 13$, neutral H
is detected 100\% of the time around star-forming galaxies and 75\%
of the time around passive galaxies (Figure~\ref{NHI_rho_fig})
within impact parameters of 150 kpc (physical). These
detections are stronger than those typically found more than
$\gtrsim 300$ kpc from galaxies, indicating that high-column 
density circumgalactic material is associated with the targeted galaxy 
at high statistical significance; weaker absorption is more broadly 
distributed and may not be associated directly with galaxies. We find generally 
good agreement between our sample and the prior studies that have 
examined sightlines within 200 kpc of galaxies \citep{chen-etal-01-Lya-imaging,2002ApJ...580..169B,2002ApJ...565..720P,wakker-savage-09-OVI-HI-lowz,2013ApJ...763..148S}. 

\item[2.] As reported by
\cite{2012ApJ...758L..41T}, there is modest but not conclusive evidence
for a difference in the CGM properties of the star-forming and passive
subsamples. COS-Halos shows four non-detections in the passive sample of
16 galaxies, but only (strong) detections in the 28 $L^*$ star-forming
galaxies (Figures~\ref{NHI_rho_fig}-\ref{NHI_sSFR_fig}). However, the
\HI\ strengths are similar for the detections in the two subsamples when
non-detections are excluded. 
The CGM gas mass implied by these 
measurements are similar for the two sub-populations (\S~6.3.3 and Figure~\ref{cgm_section}).   
Thus we conclude that even passive galaxies
are associated with strong \HI\ absorption and CGM mass, though possibly at a lower
frequency than star-forming galaxies. 

\item[3.] Considering relative
velocities between galaxies and their associated absorption, we find
that most of the detected material is within approximately $\pm 200$
\kms\ of the galaxy systemic velocity (Figure~\ref{Lya_kinematics_fig}). This
velocity range includes 74\% of fitted components by number and $>99$\% 
of the total column density of fitted components. Strong \HI\ ($\log N_{\rm HI} 
\gtrsim 16$) occurs within this range 90\% of the time. This
range is generally within the expected escape velocity of the galaxies as
calculated from their inferred dark-matter halo masses.   Conversely, weaker ($\log N_{\rm HI} \lesssim 15$)
components are seen at all relative velocities out to more than $\pm 
500$ \kms. Thus we conclude
that the detected strong \HI\ is most likely bound gravitationally to
the nearby galaxy, while weaker components seen at any velocity may 
be associated with extended large scale structures or nearby galaxies 
in addition to the targeted galaxies. 

\item[4.] Using line-profile fits to decompose
the observed profiles into resolved components, we find that the
line widths range over $b = 10 - 40$ \kms, with a few broader lines
(Figure~\ref{Lya_kinematics_fig}). These line widths indicate that
most of the detected column density (and, perhaps, inferred mass)
lies at temperatures of $T \lesssim 10^5$ K, far less than would be
expected for shock-heated gas in virialized halos of $\log M_{\rm halo}
\simeq 12-13$. A substantial
quantity of hot ($\sim 10^6$ K) gas could be present in these halos
and remain unseen owing to the strong presence of the cooler material.

\item[5.] The observed trend of \HI\ strength with impact parameter
(Figure~\ref{NHI_rho_fig}), the tight kinematic correlation with
galaxy systemic velocity (Figure~\ref{Lya_kinematics_fig}), and the
concentration of \HI\ near the galaxies with respect to results of
blind surveys out to $\sim 1$ Mpc (Figure~\ref{lya_rho_compare_fig}) lead us to conclude
that the detected material does not arise in the nearby IGM, in other
galaxies, or otherwise far away from the targeted galaxies. The simplest
explanation for these findings is that the detected gas is directly associated 
with the targeted galaxies, and probably gravitationally bound
to them.  

\item[6.] Comparing our line strength and width measurements
to blind \HI\ surveys in the literature, we find a strong indication that \HI\ {\em column densities} -- and perhaps the number of absorbing clouds -- evolve as sightlines
get nearer to galaxies, but there is no evidence that linewidths do so
(Figures~\ref{Lya_kinematics_fig} and \ref{compare-NHI-bval-fig}). We
interpret this lack of evolution in the linewidths as an indication
that the bulk of \HI\ absorption arises in gas with temperatures $T
\lesssim 10^5$ K regardless of location. Even in galaxy halos where higher
temperatures from shock-heating in virialization and/or feedback might be
expected, significant amounts of cold gas remain.   

\item[7.] Because of
our poor knowledge of the gaseous outskirts of galaxies, it is difficult
to constrain the direct contribution of gas bound to the satellites of
the targeted galaxies. However, a simple analysis based on dark-matter
substructure counts indicates that to explain the column densities and
kinematic extent of the detected absorption would require gas to be
commonly associated with very small subhalos ($\lesssim 10$ \kms) that
are typical of dwarf spheroidal satellites not known to retain gas at
the observed column densities. We conclude that gas directly bound to
satellites may contribute to the detected absorption but is not likely
to be the primary source.  \end{itemize}

The picture that emerges from these findings is of a diffuse, cool
CGM surrounding nearly all galaxies at $\sim L^*$, regardless of
type. This CGM is composed mainly of gas at temperatures
expected for low densities in photoionizing conditions. Its internal
motions may be turbulent (adding some non-thermal broadening) but
its bulk flows are insufficient to unbind it from the galaxy. This
medium exists around both star-forming and passive galaxies, though
the latter may possess a lower volume filling factor of cold gas that
projects a lower areal covering fraction owing to generally higher
halo gas temperatures or gas removal during the transition to passive
evolution. The
ionization correction that should be applied to these values of \NHI\ are
the critical factor in obtaining total gas surface density measurements
for the CGM; ionization factions cannot be measured with \HI\ itself but
can be inferred from associated metal lines from species over a range of
ionization potential.  An empirical characterization of the metal-lines in the COS-Halos survey 
has been presented recently by \cite{2013ApJS..204...17W}.  A forthcoming paper will 
present a combined analysis of the \HI\ and metals in terms of ionization models 
and physical interpretations. Other followup studies will consider the relationship
between \HI, the low-ionization metals, and the O VI results presented
by \cite{tumlinson-etal-11-OVI-statistics}. 
These measurements are only one piece of the CGM puzzle,
but as the \HI\ traces the dominant component of the gas, the hydrogen,
these measurements provide a critical basis for our planned studies of
the ionization state, metallicity, and mass of the detected CGM. 

\acknowledgments

Support for program GO11598 was provided by NASA through a grant from the
Space Telescope Science Institute, which is operated by the Association
of Universities for Research in Astronomy, Inc., under NASA contract NAS
5-26555. Some of the data presented herein were obtained at the W.M. Keck
Observatory, which is operated as a scientific partnership among the
California Institute of Technology, the University of California and the
National Aeronautics and Space Administration. TMT appreciates support
from NSF grant AST-0908334. MSP acknowledges support from the Southern
California Center for Galaxy Evolution, a multi-campus research program
funded by the University of California Office of Research. Thanks to
Derck Massa for delivering the COS flats and to Ed Jenkins
for processing them for our use. Thanks also to Shelley Meyett and Parviz
Ghavamian for ensuring the success of the COS observations.

{\it Facilities:} \facility{HST (COS)}, \facility{Keck (LRIS, HIRES),
\facility{Magellan}}.

\bibliographystyle{apj}
\bibliography{ms}

\clearpage 

\begin{deluxetable}{lccccccl}
\tablecolumns{7}
\tabletypesize{\scriptsize}
\tablenum{1}
\tablewidth{0pt}
\tablecaption{COS-Halos QSO Sample}
\tablehead{\colhead{QSO Name}&
\colhead{$\alpha$ [J2000]}&
\colhead{$\delta$ [J2000]}&
\colhead{$z_{em}$}&
\colhead{$m_{FUV}$}&
\colhead{$t_{G130M}$ [s]}&
\colhead{$t_{G160M}$ [s]}&
\colhead{Visit\tablenotemark{a}}
}
\startdata
      SDSS J004222.29$-$103743.8  & $   10.593$ & $  -10.629 $ &    0.424   &     17.42   &    2448   &  2781   &    05   \\
      SDSS J022614.46$+$001529.7  & $   36.560$ & $    0.258 $ &    0.615   &     17.87   &    5826   &  4361   &    26   A5
                              \\
      SDSS J040148.98$-$054056.5  & $   60.454$ & $   -5.682 $ &    0.570   &     18.33   &    5377   &  5912   &    22   \\
      SDSS J080359.23$+$433258.4  & $  120.997$ & $   43.550 $ &    0.449   &     18.41   &    5207   &  6110   &    09   \\
      SDSS J082024.21$+$233450.4  & $  125.101$ & $   23.581 $ &    0.470   &     18.37   &    5035   &  5926   &    17   \\
      SDSS J091029.75$+$101413.6  & $  137.624$ & $   10.237 $ &    0.463   &     18.72   &    4913   &  8699   &    18   \\
      SDSS J091440.38$+$282330.6  & $  138.668$ & $   28.392 $ &    0.735   &     18.76   &    5093   &  8735   &    21   \\
      SDSS J092554.70$+$400414.1  & $  141.478$ & $   40.071 $ &    0.471   &     18.14   &    3765   &  4303   &    24   \\
\enddata
\label{qso_table}
\tablenotetext{a}{These visit labels can be used to find full details of the observations in MAST and in the Phase II file stored at http://www.stsci.edu/hst/phase2-public/11598.apt.}
\end{deluxetable}

\begin{deluxetable}{lrrrrrrrrr}
\tablecolumns{9}
\tabletypesize{\scriptsize}
\tablenum{2}
\tablewidth{0pt}
\tablecaption{COS-Halos Galaxy Sample} 
\tablehead{\colhead{QSO Name}&
\colhead{Galaxy\tablenotemark{a}}&
\colhead{T/B\tablenotemark{b}}&
\colhead{$z_{sys}$\tablenotemark{c}}&
\colhead{$M_r$\tablenotemark{c}}&
\colhead{$\log M_*$\tablenotemark{c}}&
\colhead{SFR\tablenotemark{c}}&
\colhead{$R$\tablenotemark{c}}&
\colhead{$R_{\rm vir}$}\\
\colhead{}&
\colhead{}&
\colhead{}&
\colhead{}&
\colhead{[mag]}&
\colhead{[$M_\odot$]}&
\colhead{[$M_\odot$ yr$^{-1}$]}&
\colhead{[kpc]}&
\colhead{[kpc]}
}
\startdata
 J0226$+$0015 & 268\_22  &      T &  0.22744 & $-20.38$ & $  10.8$ &   $  < 0.09$ &    80 &   303 \\ 
 J0401$-$0540 &  67\_24  &      T &  0.21969 & $-20.05$ & $  10.2$ &   $    1.14$ &    85 &   200 \\ 
 J0803$+$4332 & 306\_20  &      T &  0.25347 & $-21.65$ & $  11.3$ &   $  < 0.21$ &    77 &   581 \\ 
 J0910$+$1014 &  34\_46  &      B &  0.14274 & $-20.57$ & $  10.6$ &   $   14.12$ &   112 &   279 \\ 
 J0910$+$1014 & 242\_34  &      B &  0.26412 & $-21.96$ & $  11.5$ &   $  < 0.30$ &   139 &   716 \\ 
 J0914$+$2823 &  41\_27  &      T &  0.24431 & $-20.03$ & $   9.8$ &   $    2.83$ &   104 &   169 \\ 
 J0925$+$4004 & 196\_22  &      B &  0.24745 & $-21.55$ & $  11.3$ &   $  < 0.57$ &    83 &   569 \\ 
 J0928$+$6025 & 110\_35  &      T &  0.15400 & $-20.41$ & $  10.8$ &   $  < 0.04$ &    93 &   317 \\
\enddata
\label{galaxy_table}
\tablenotetext{a}{We label the galaxies by the position angle with respect to the QSO, N through E, and with the angular separation in arcsec.}
\tablenotetext{b}{Indicates whether the galaxy was a pre-selected Target (T) or a bonus galaxy (B) meeting the same criteria.}
\tablenotetext{c}{Galaxy redshift $z_{sys}$, absolute magnitude $M_r$, stellar mass $M_*$, star formation rate (SFR), and impact parameter $R$ are drawn from \cite{werk-etal-12-galaxies}.}
\tablenotetext{d}{Photometric redshift. This galaxy does not have a secure spectroscopic redshift.}
\end{deluxetable}

\clearpage 
\begin{deluxetable}{lccccl}
\tablecolumns{6}
\tabletypesize{\scriptsize}
\tablenum{3}
\tablewidth{0pt}
\tablecaption{COS-Halos \ion{O}{6} Sample Error Budgets}
\tablehead{
\colhead{Category}&
\colhead{Ref \#}&
\colhead{Quantity} &
\colhead{Statistical}&
\colhead{Systematic}&
\colhead{Comments}
}
\startdata
Absorbers       &1&  $v_{sys}$ zeropoint      &      \nodata        &       $\pm$25 \kms\                   &       directly from $z_{sys}$ of target galaxy                 \\
                &2&          "                &      \nodata        &       $\pm$2-3 pix = 6-8 \kms\    &       COS $\lambda$ zeropoint\tablenotemark{b}                \\
                &3&          "                &      \nodata        &       $\pm$10 \kms                    &       derived from doublet shifts\tablenotemark{c} \\
                &4&  $v_c$ total              &      \nodata        &       $\pm$30 \kms                    &       root-square-sum of (1) and (2)+(3)                  \\
           	&5&  $N$ (integrated)	      &    0.05-0.2 dex     &       $\simeq$0.1 dex\tablenotemark{a}&       systematic is continuum                    \\
           	&6&  $N$ (fitted)             &  $\lesssim 0.2$ dex &       depends on saturation           &       see Section \ref{problems_section}   \\
                &7&  $b$                      &  \nodata  &               from profile fits               &       " \\ 
\hline     \\
\enddata
\label{error_table}
\tablenotetext{b}{\cite{Oliveira:2010wz} derived the COS on-orbit wavelength solution and then compared the results with higher-resolution STIS data on the same targets. The COS and STIS wavelength zeropoints agree to within 3 raw pixels on average, equivalent to 1/2 COS resolution element. }
\tablenotetext{c}{As part of the study of CGM \ion{O}{6} in COS-Halos, \cite{tumlinson-etal-11-OVI-statistics}  applied small shifts to the 1037 line to align it with 1031 in cases where the COS wavelength solution has errors in the first and higher order terms. In the 29 systems where both lines of the doublet were detected, 12 cases required a non-zero shift, with a median of 10 \kms. Twenty-five of the 29 cases (86\%) were found to have 10 \kms\ or less of shift. We therefore adopt 10 \kms\ as the uncertainty associated with the first-order and higher terms of the COS wavelength solution. For propagation into other quantities, this term is added to the zeropoint error in a straight sum (terms 2 and 3), and then root-square-summed with (1) to derive the total error in $v_c$.}
\end{deluxetable}

\begin{deluxetable}{lrccccccccc}
\tablecolumns{11}
\tabletypesize{\scriptsize}
\tablenum{4}
\tablewidth{0pt}
\tablecaption{COS-Halos HI Line Measurements} 
\tablehead{\colhead{QSO Name}&
\colhead{Galaxy}&
\colhead{$z_{sys}$}&
\colhead{$v_{min}$,$v_{max}$\tablenotemark{a}}&
\colhead{$W_r$(Ly$\alpha$)}\tablenotemark{b}&
\colhead{$W_r$(Ly$\beta$)}&
\colhead{$W_r$(Ly$\gamma$)}&
\colhead{$W_r$(Ly$\delta$)}&
\colhead{$W_r$(Ly$\epsilon$)}&
\colhead{$W_r$(Ly$\zeta$)}\\
\colhead{}&
\colhead{}&
\colhead{}&
\colhead{[km s$^{-1}$]}&
\colhead{[m\AA]}&
\colhead{[m\AA]}&
\colhead{[m\AA]}&
\colhead{[m\AA]}&
\colhead{[m\AA]}&
\colhead{[m\AA]}
}
\startdata
 J0226$+$0015  &  268\_22  &  0.22744  &  $    0,  280$   &     $486\pm20$  &        \nodata  &        \nodata  &        \nodata  &        \nodata  &        \nodata  \\
 J0401$-$0540  &   67\_24  &  0.21969  &  $ -700,  120$   &    $1234\pm34$  &        \nodata  &     $463\pm39$  &     $290\pm25$  &        \nodata  &        \nodata  \\
 J0803$+$4332  &  306\_20  &  0.25347  &  $ -250,  100$   &     $592\pm26$  &     $277\pm18$  &        \nodata  &        \nodata  &        \nodata  &        \nodata  \\
 J0910$+$1014  &   34\_46  &  0.14274  &  $ -300,  300$   &    $1444\pm48$  &    $427\pm144$  &        \nodata  &        \nodata  &        \nodata  &        \nodata  \\
 J0910$+$1014  &  242\_34  &  0.26412  &  $ -300,  250$   &    $1180\pm36$  &     $802\pm56$  &     $278\pm70$  &        \nodata  &     $170\pm83$  &     $101\pm83$  \\
 J0914$+$2823  &   41\_27  &  0.24431  &  $ -250,  100$   &     $731\pm21$  &     $424\pm25$  &     $294\pm15$  &     $228\pm17$  &     $163\pm19$  &      $68\pm21$  \\
 J0925$+$4004  &  196\_22  &  0.24745  &  $  -75,  250$   &    $3276\pm64$  &     $896\pm18$  &        \nodata  &     $649\pm34$  &     $610\pm41$  &        \nodata  \\
 J0928$+$6025  &  110\_35  &  0.15400  &  $ -250,  150$   &    $2726\pm77$  &     $828\pm47$  &        \nodata  &        \nodata  &        \nodata  &        \nodata  \\
\enddata
\label{line_table}
\tablenotetext{a}{This velocity range is generally given for \Lya, but replaced with the values for Ly$\beta$ where \Lya\ is damped, blended, or otherwise unavailable.}
\tablenotetext{b}{Limits on $W_r$ are 1$\sigma$.}
\end{deluxetable}
 


\LongTables
\clearpage
\begin{deluxetable}{lcccccccccc}
\tablecolumns{12}
\tabletypesize{\scriptsize}
\tablenum{5}
\tablewidth{0pt}
\tablecaption{The COS-Halos \ion{H}{1} Sample} 
\tablehead{\colhead{Sightline}&
\colhead{Galaxy} &
\colhead{$z_{sys}$}&
\multicolumn{4}{c}{Adopted Values\tablenotemark{b}}&
\multicolumn{3}{c}{Profile fitting} &
\colhead{Notes\tablenotemark{c}}\\
\colhead{} &
\cline{3-6}  \cline{8-10}\\
\colhead{} &
\colhead{} &
\colhead{} &
\colhead{$v_{min},v_{max}$}&
\colhead{Adopt}&
\colhead{$\log$ \NHI}&
\colhead{Method}&
\colhead{$\log$ \NHI}&
\colhead{$b$}&
\colhead{$v$}&
\colhead{}
\\
\colhead{} &
\colhead{} &
\colhead{} &
\colhead{km s$^{-1}$} &
\colhead{} &
\colhead{cm$^{-2}$} &
\colhead{} &
\colhead{cm$^{-2}$}&
\colhead{km s$^{-1}$}&
\colhead{km s$^{-1}$}&
\colhead{}
}
\startdata
 J0226$+$0015 & 268\_22  &   0.22744 & $     0,   280$ & $     1$ & $   14.25 \pm  0.03$ &   1 & \nodata & \nodata & \nodata \\
          &          &            &          &          &                 &      &     $13.03\pm 0.16$ &      $   15\pm   9$ &      $   67\pm   5$ &             Unc  \\
          &          &            &          &          &                 &      &     $14.34\pm 0.06$ &      $   40\pm   3$ &      $  166\pm   1$ &                  \\
 J0401$-$0540 &  67\_24  &   0.21969 & $  -700,   120$ & $     8$ & $   15.63 \pm  0.05$ &   1 & \nodata & \nodata & \nodata \\
          &          &            &          &          &                 &      &     $13.63\pm 0.07$ &      $   23\pm   6$ &      $ -642\pm   5$ &                  \\
          &          &            &          &          &                 &      &     $14.64\pm 0.08$ &      $   19\pm   1$ &      $ -579\pm   2$ &                  \\
          &          &            &          &          &                 &      &     $15.54\pm 0.09$ &      $   30\pm   4$ &      $  -63\pm   5$ &                  \\
          &          &            &          &          &                 &      &     $14.97\pm 0.24$ &      $   55\pm   5$ &      $  -46\pm   3$ &                  \\
\enddata
\tablenotetext{a}{Columns are: QSO sightline, galaxy name, systemic redshift, minimum and maximum limits of direct integration, flag indicating a column density measurement (1$=$measurement, 4$=$upper limit, 8$=$lower limit), adopted column density and error, flag indicating method for column density estimate (1$=$direct integration, 3$=$ fitted damping wings, 4$=$Lyman limit), and three parameters for the profile fits. }
\tablenotetext{b}{Adopted column densities in this column are derived from direct integrations (Method = 1) over the apparent optical depth profile \citep{Savage:1991ka}, except the damped systems for which profile fitted values are always used (Method = 3) and the LLS toward J1009+0713 where the column density was derived from the Lyman limit (Method = 4).} 
\tablenotetext{c}{Components labeled ``Unc" have uncertain fits with large errors on parameters, or a doubtful identification as real absorption. Those labeled ``Broad" have fits demanding an unusually broad linewidth, which may be spurious. ``Sat" marks strongly saturated components with uncertain column densities. Fits derived from Ly$\alpha$ only are so indicated.} 
\label{component_table}
\end{deluxetable}
\clearpage

\clearpage
\begin{figure*}[!h]
\begin{center}
\epsscale{0.70}
\plotone{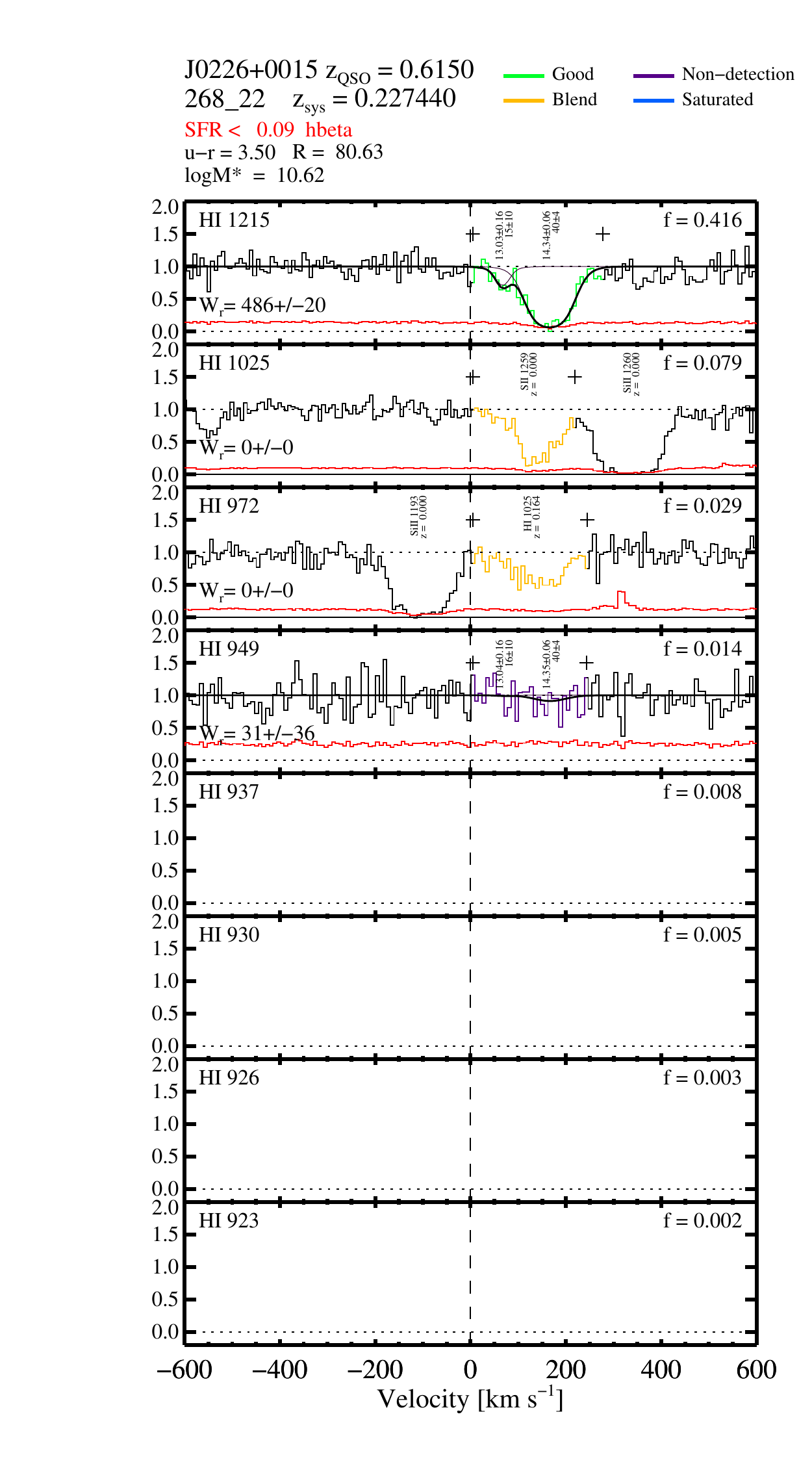}
\label{test_fig}
\end{center}
\caption{Hydrogen stack plot for system 268\_22 toward J0226$+$0015.}
\end{figure*}

\clearpage
\begin{figure*}[!h]
\begin{center}
\epsscale{0.70}
\plotone{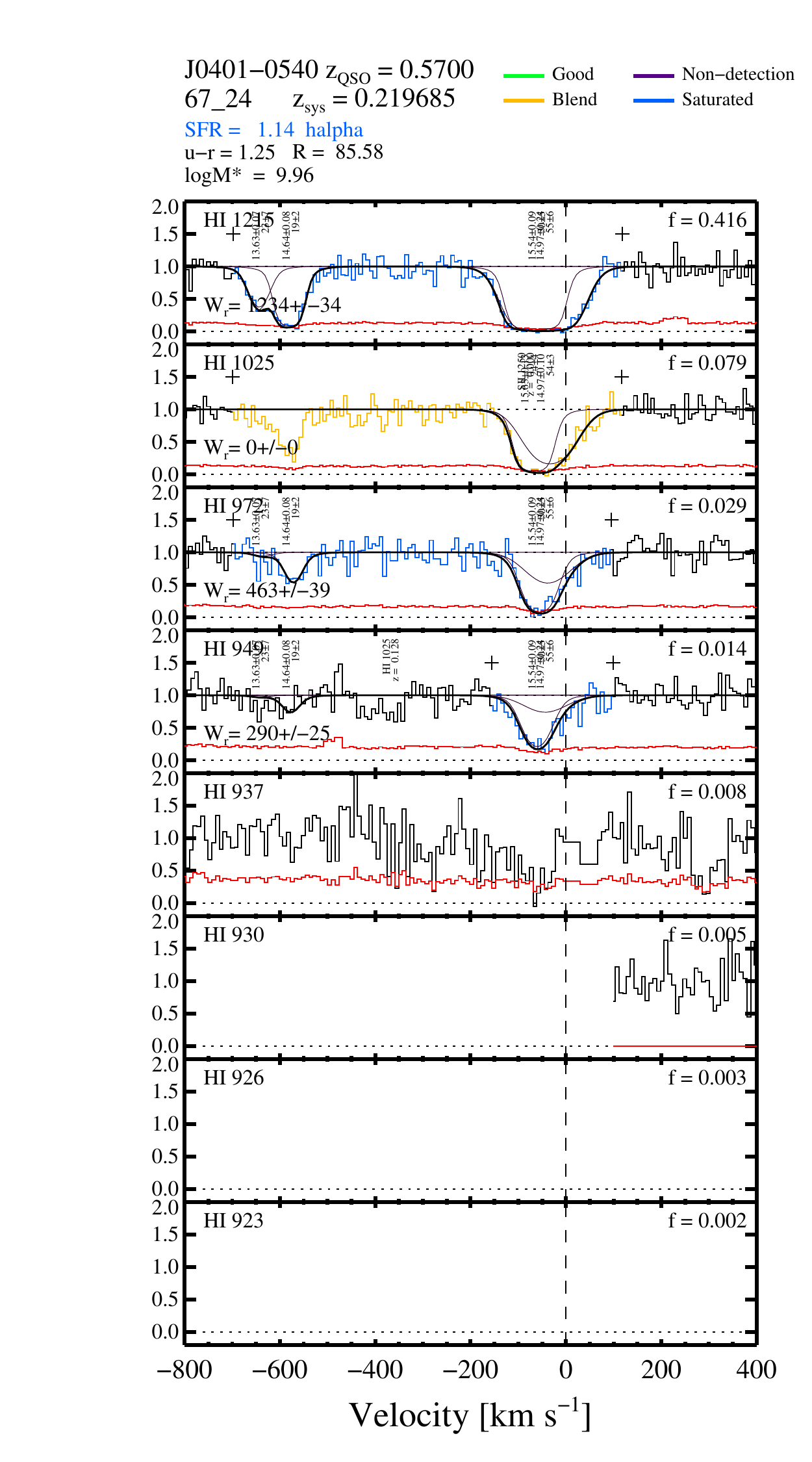}
\label{test_fig}
\end{center}
\caption{Hydrogen stack plot for system 67\_24 toward J0401$-$0540.}
\end{figure*}

\clearpage
\begin{figure*}[!h]
\begin{center}
\epsscale{0.70}
\plotone{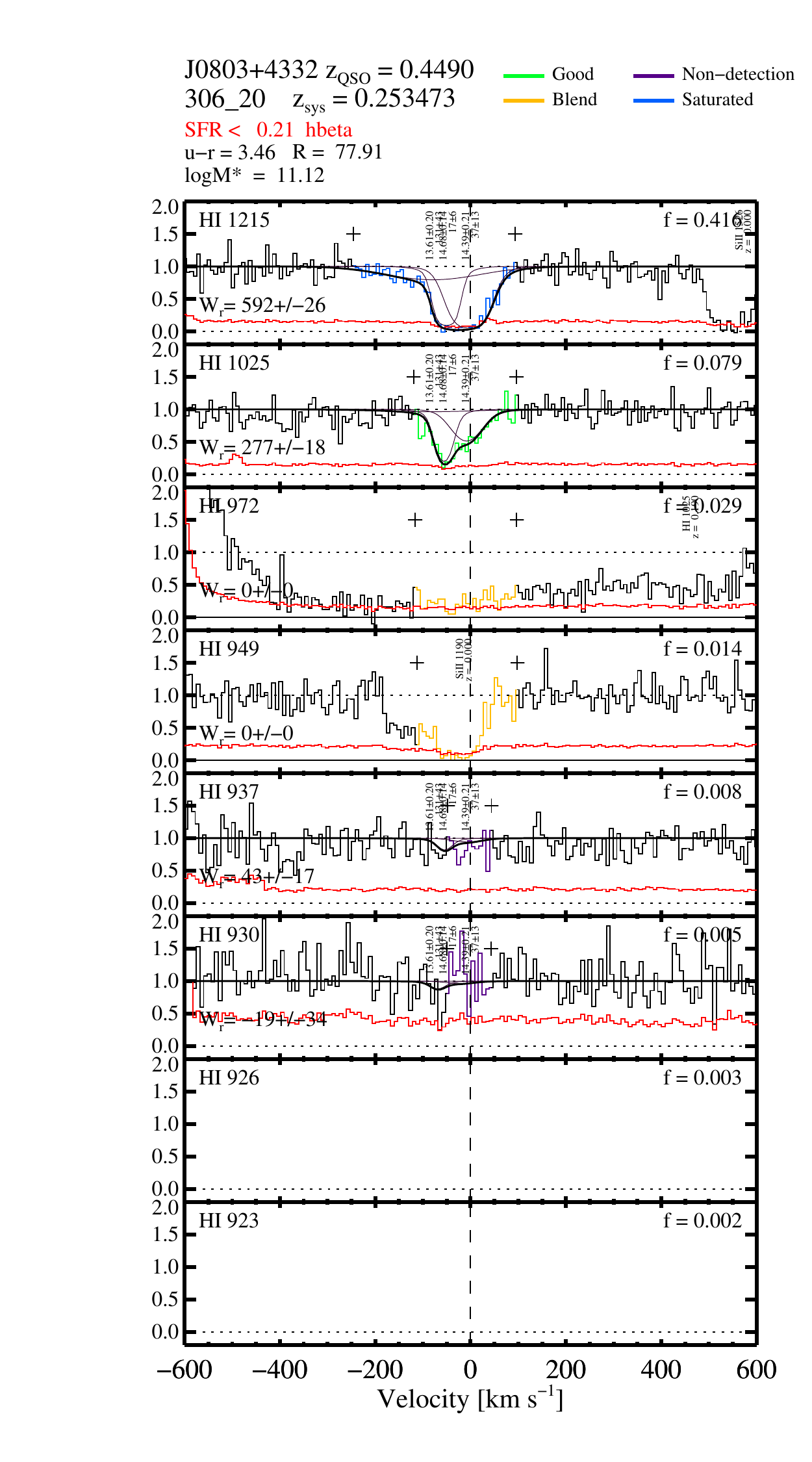}
\label{test_fig}
\end{center}
\caption{Hydrogen stack plot for system 306\_20 toward J0803$+$4332.}
\end{figure*}

\clearpage
\begin{figure*}[!h]
\begin{center}
\epsscale{0.70}
\plotone{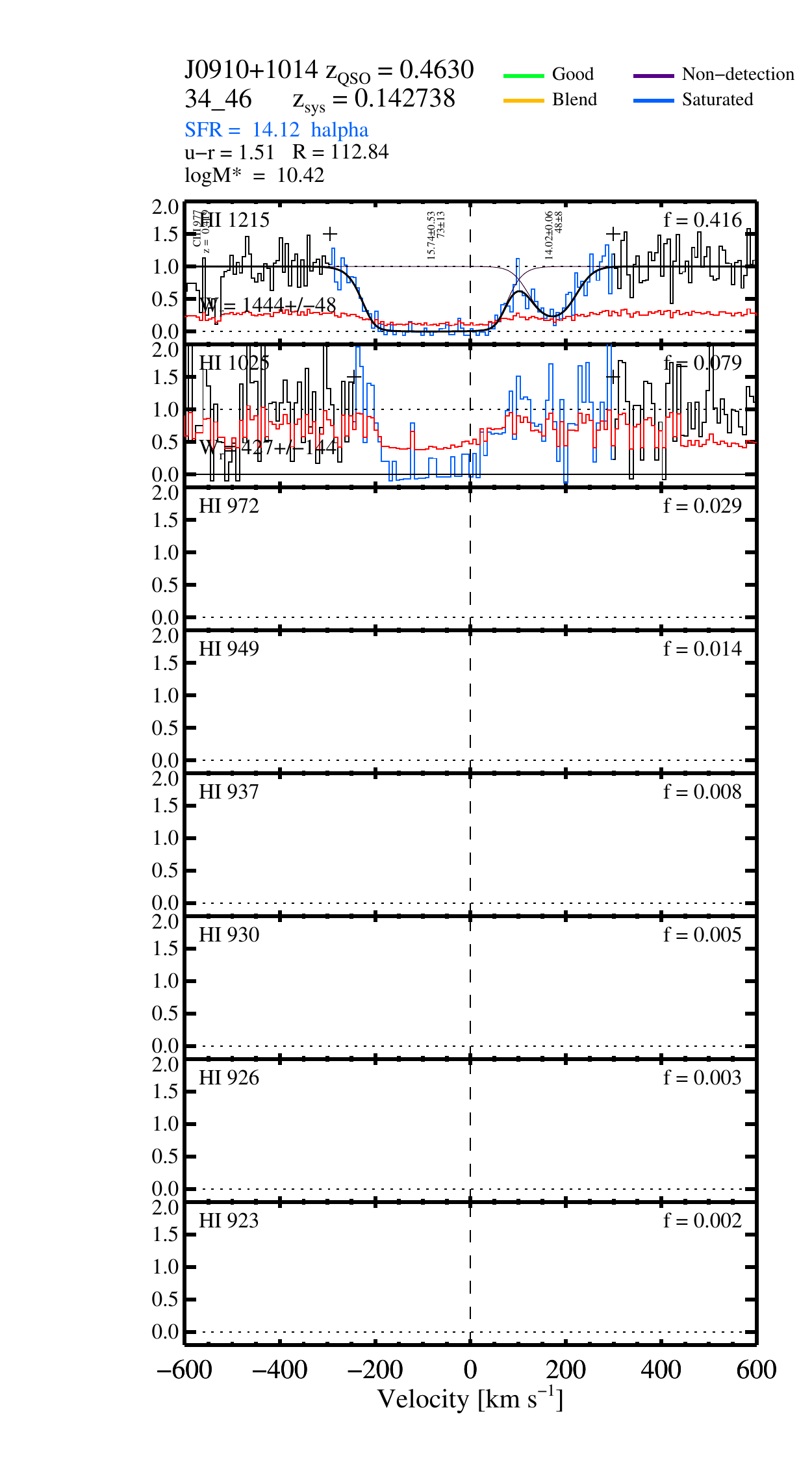}
\label{test_fig}
\end{center}
\caption{Hydrogen stack plot for system 34\_46 toward J0910$+$1014.}
\end{figure*}

\clearpage
\begin{figure*}[!h]
\begin{center}
\epsscale{0.70}
\plotone{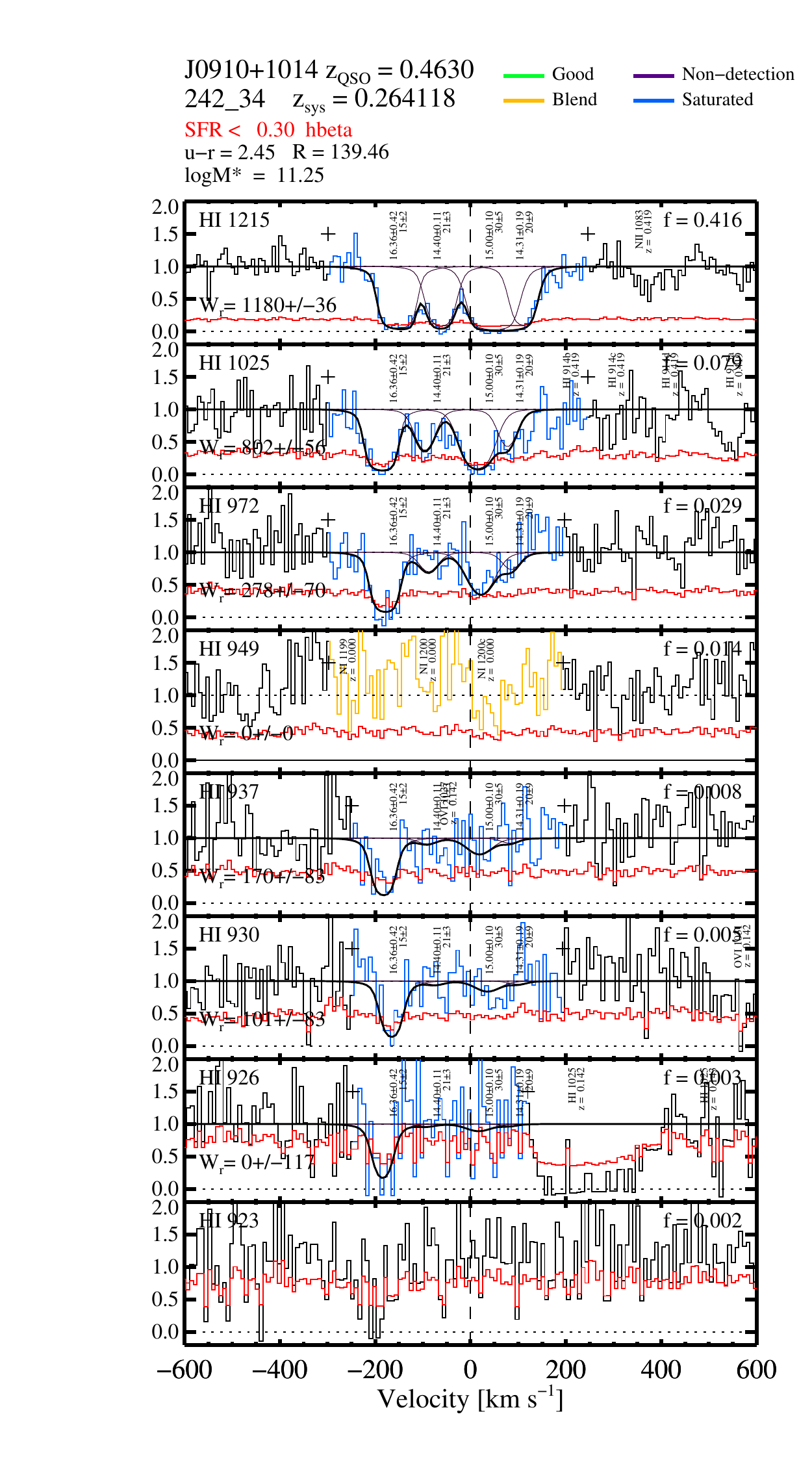}
\label{test_fig}
\end{center}
\caption{Hydrogen stack plot for system 242\_34 toward J0910$+$1014.}
\end{figure*}

\clearpage
\begin{figure*}[!h]
\begin{center}
\epsscale{0.70}
\plotone{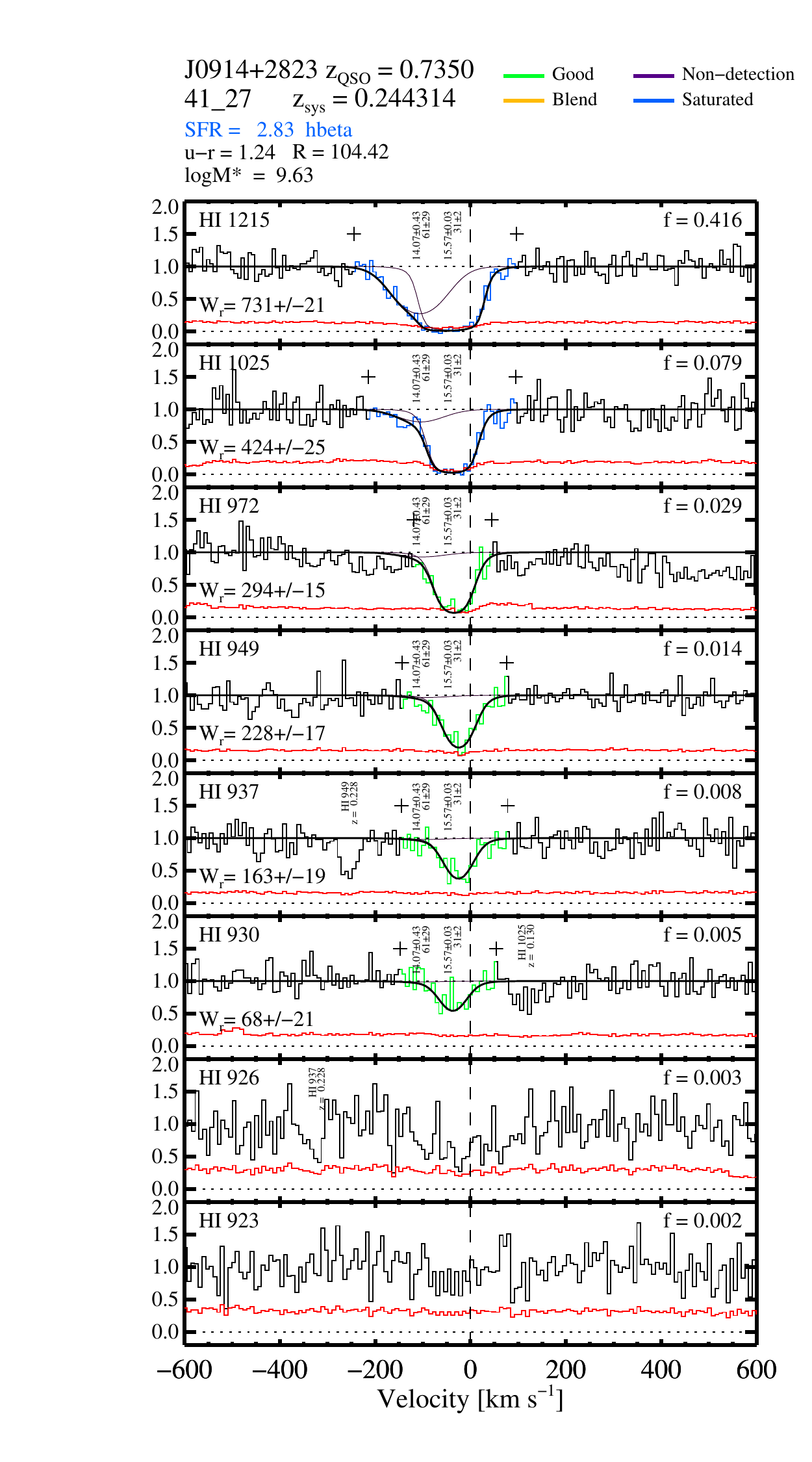}
\label{test_fig}
\end{center}
\caption{Hydrogen stack plot for system 41\_27 toward J0914$+$2823.}
\end{figure*}

\clearpage
\begin{figure*}[!h]
\begin{center}
\epsscale{0.70}
\plotone{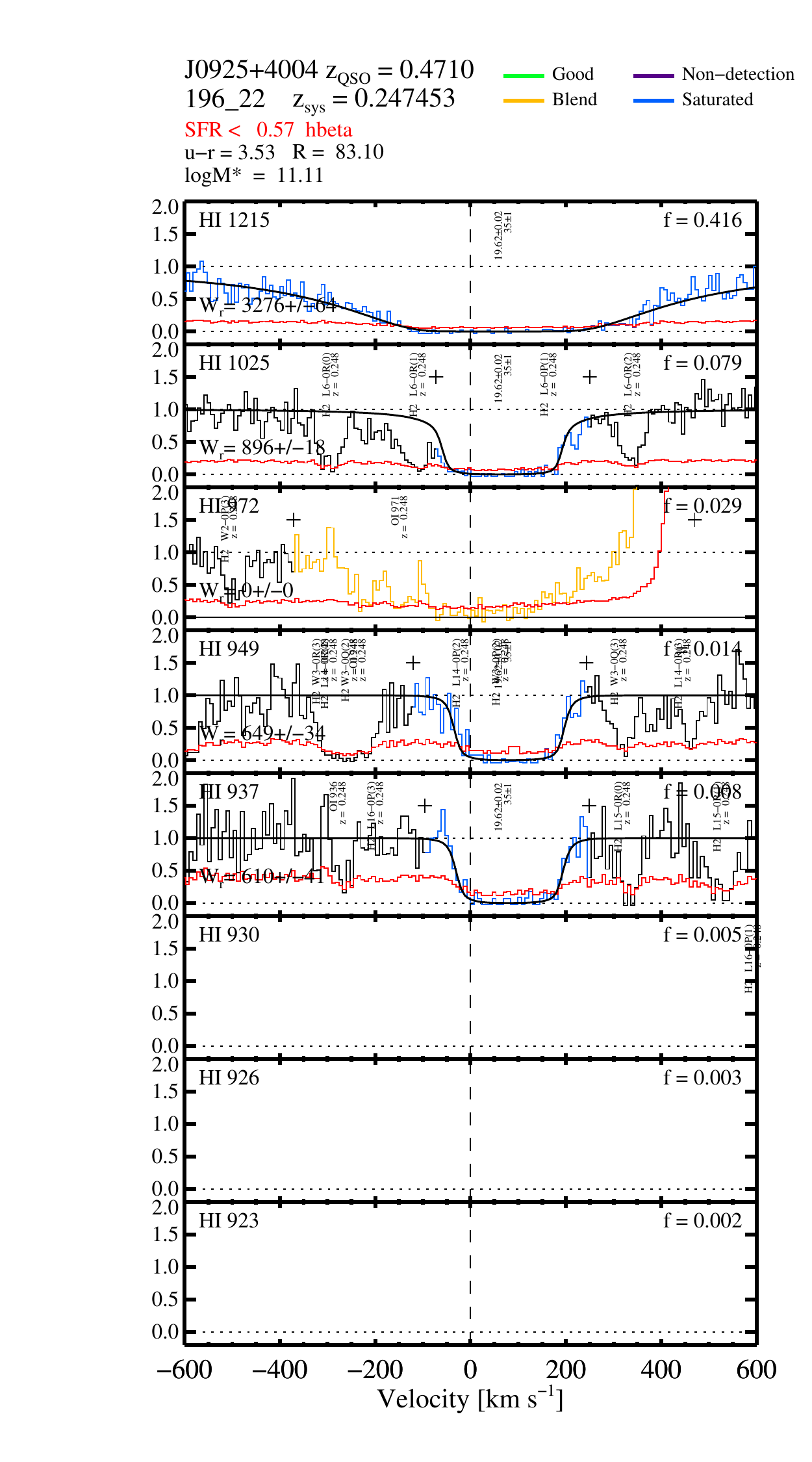}
\label{test_fig}
\end{center}
\caption{Hydrogen stack plot for system 196\_22 toward J0925$+$4004.}
\end{figure*}

\clearpage
\begin{figure*}[!h]
\begin{center}
\epsscale{0.70}
\plotone{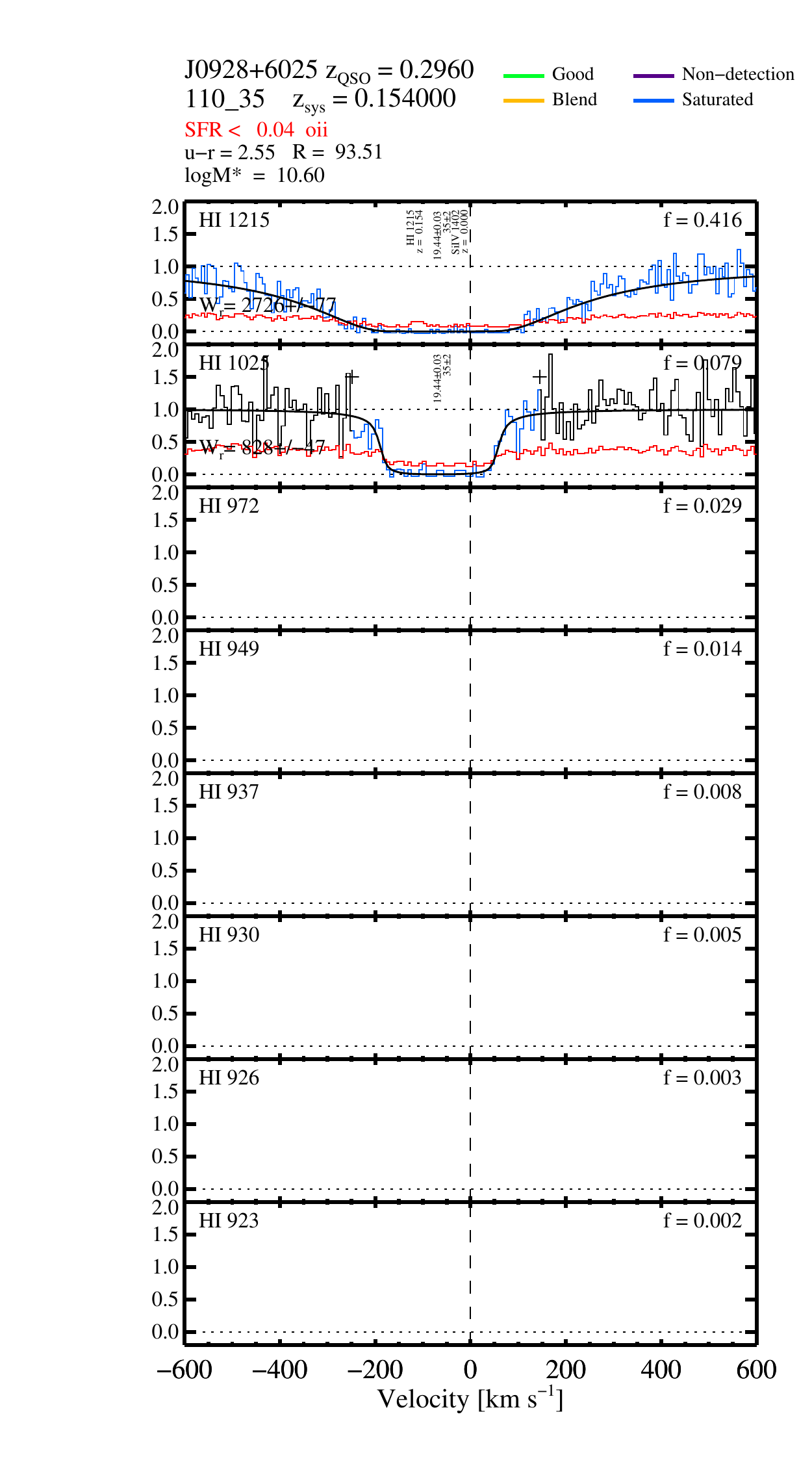}
\label{test_fig}
\end{center}
\caption{Hydrogen stack plot for system 110\_35 toward J0928$+$6025.}
\end{figure*}

\clearpage
\begin{figure*}[!h]
\begin{center}
\epsscale{0.70}
\plotone{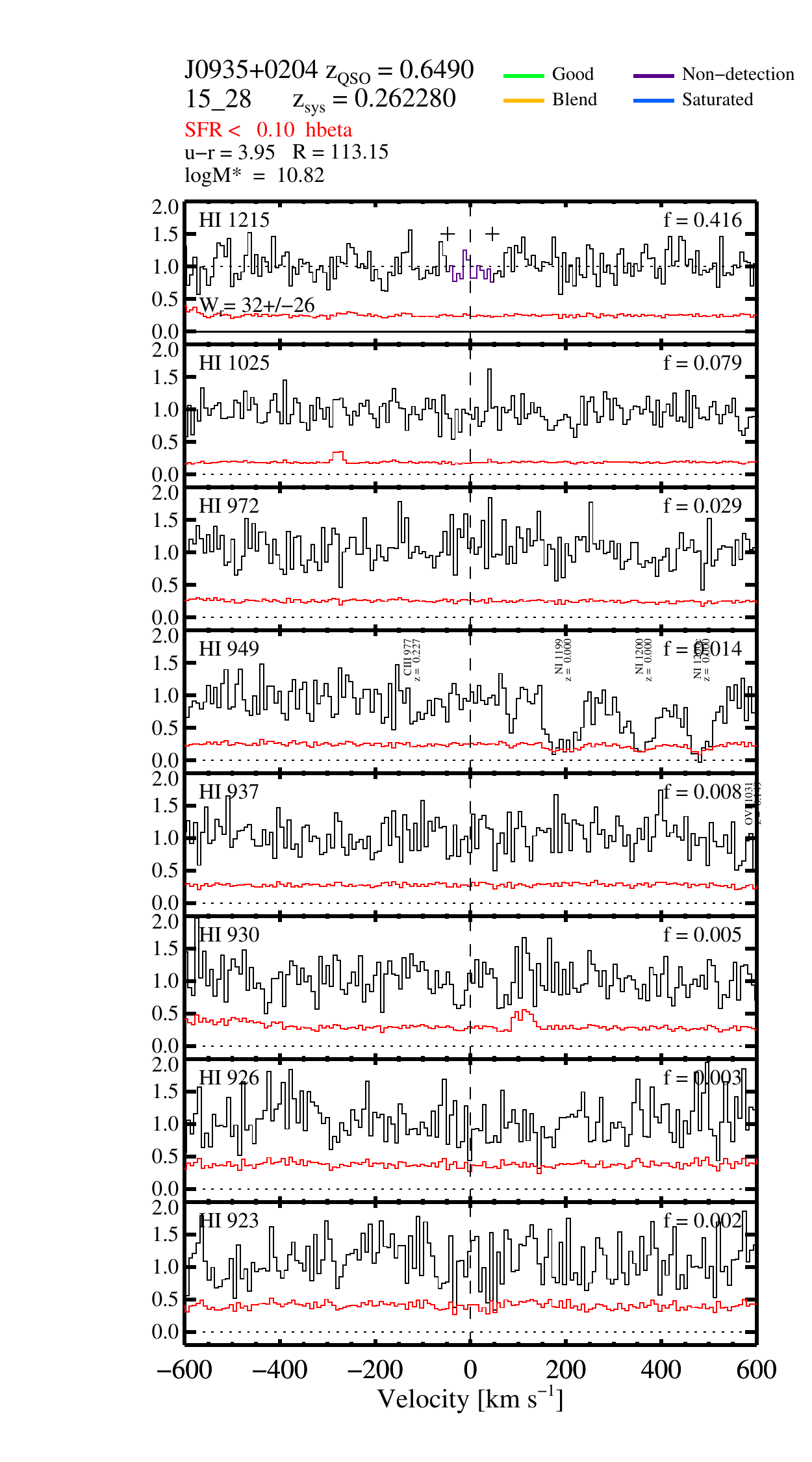}
\label{test_fig}
\end{center}
\caption{Hydrogen stack plot for system 15\_28 toward J0935$+$0204.}
\end{figure*}

\clearpage
\begin{figure*}[!h]
\begin{center}
\epsscale{0.70}
\plotone{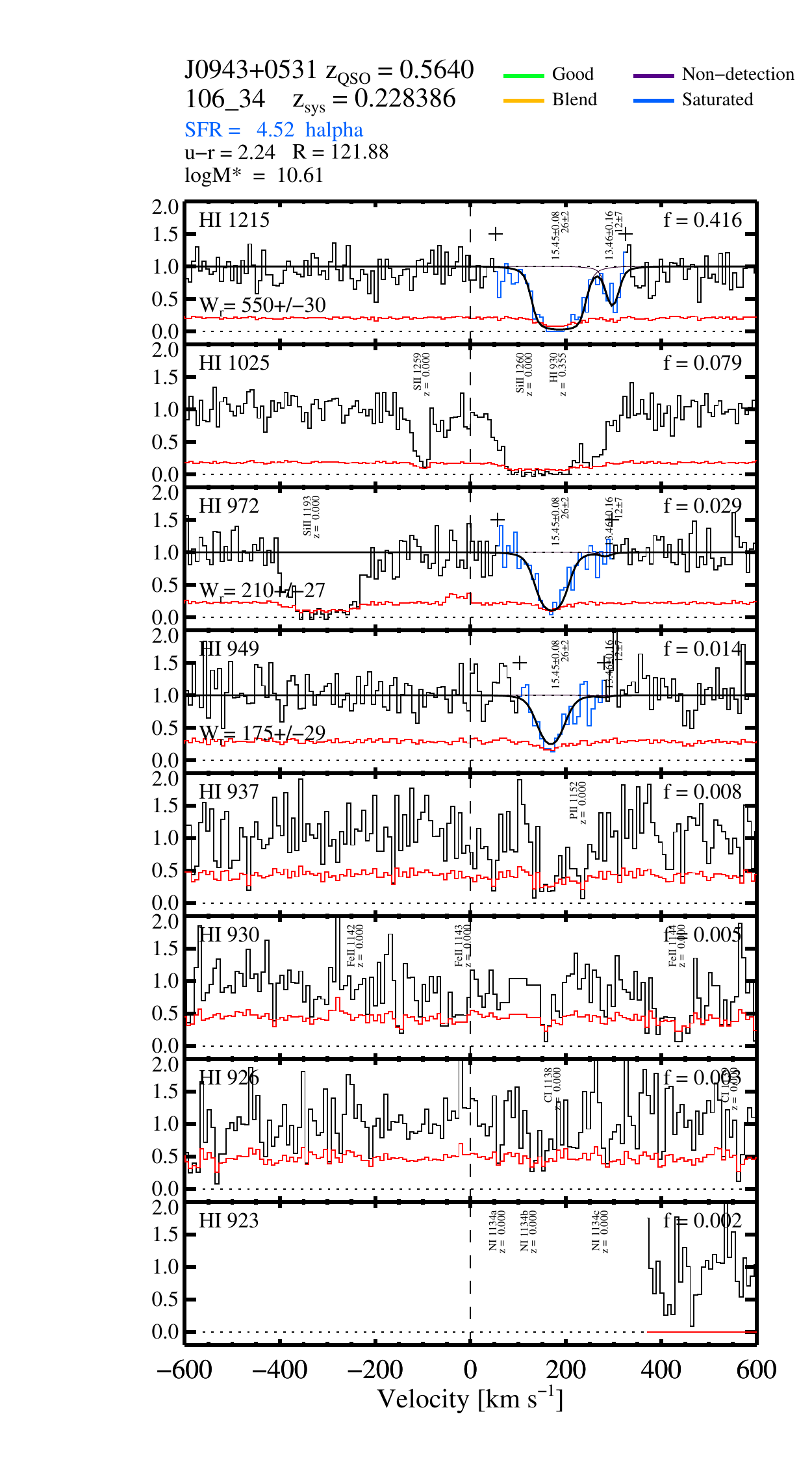}
\label{test_fig}
\end{center}
\caption{Hydrogen stack plot for system 106\_34 toward J0943$+$0531.}
\end{figure*}

\clearpage
\begin{figure*}[!h]
\begin{center}
\epsscale{0.70}
\plotone{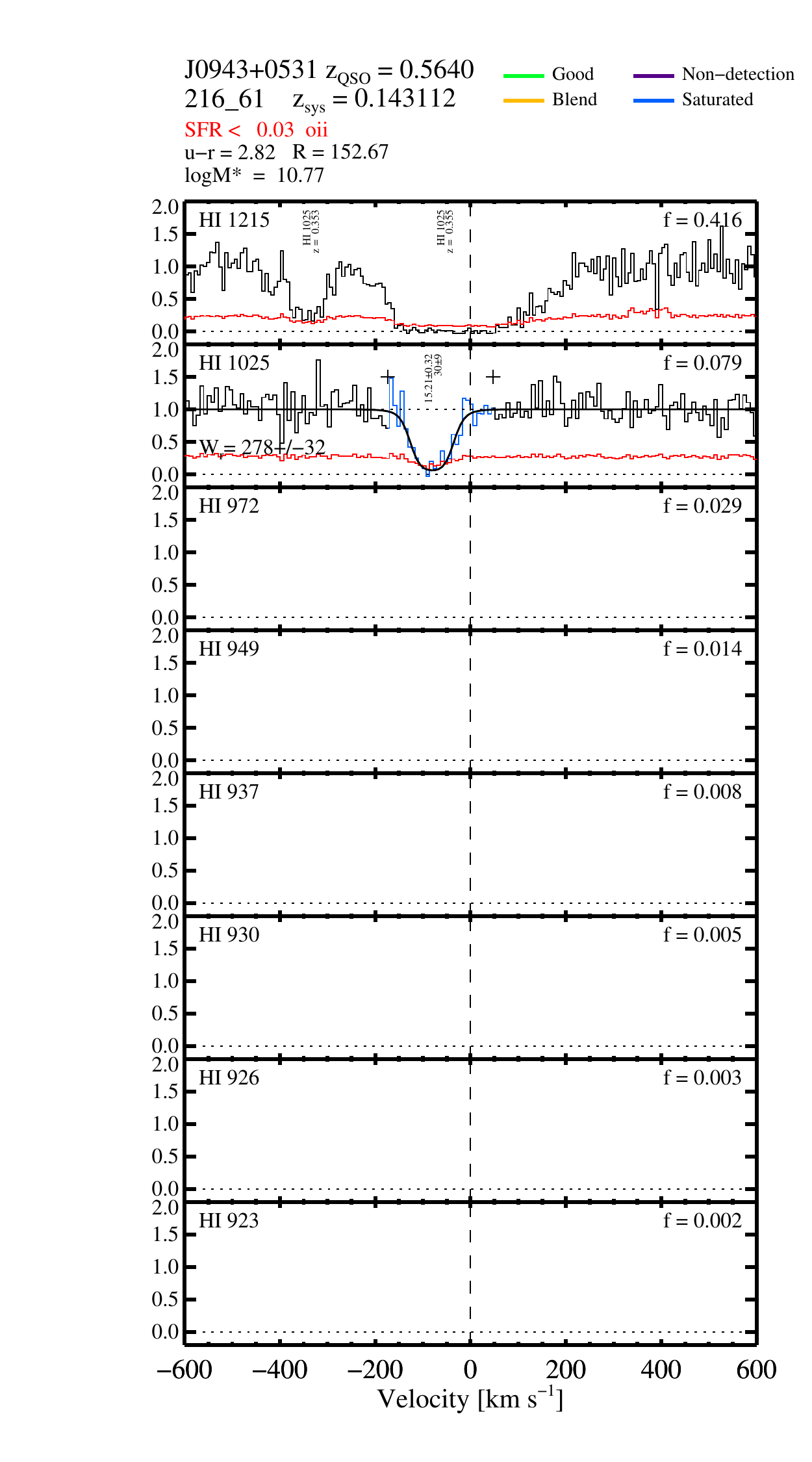}
\label{test_fig}
\end{center}
\caption{Hydrogen stack plot for system 216\_61 toward J0943$+$0531.}
\end{figure*}

\clearpage
\begin{figure*}[!h]
\begin{center}
\epsscale{0.70}
\plotone{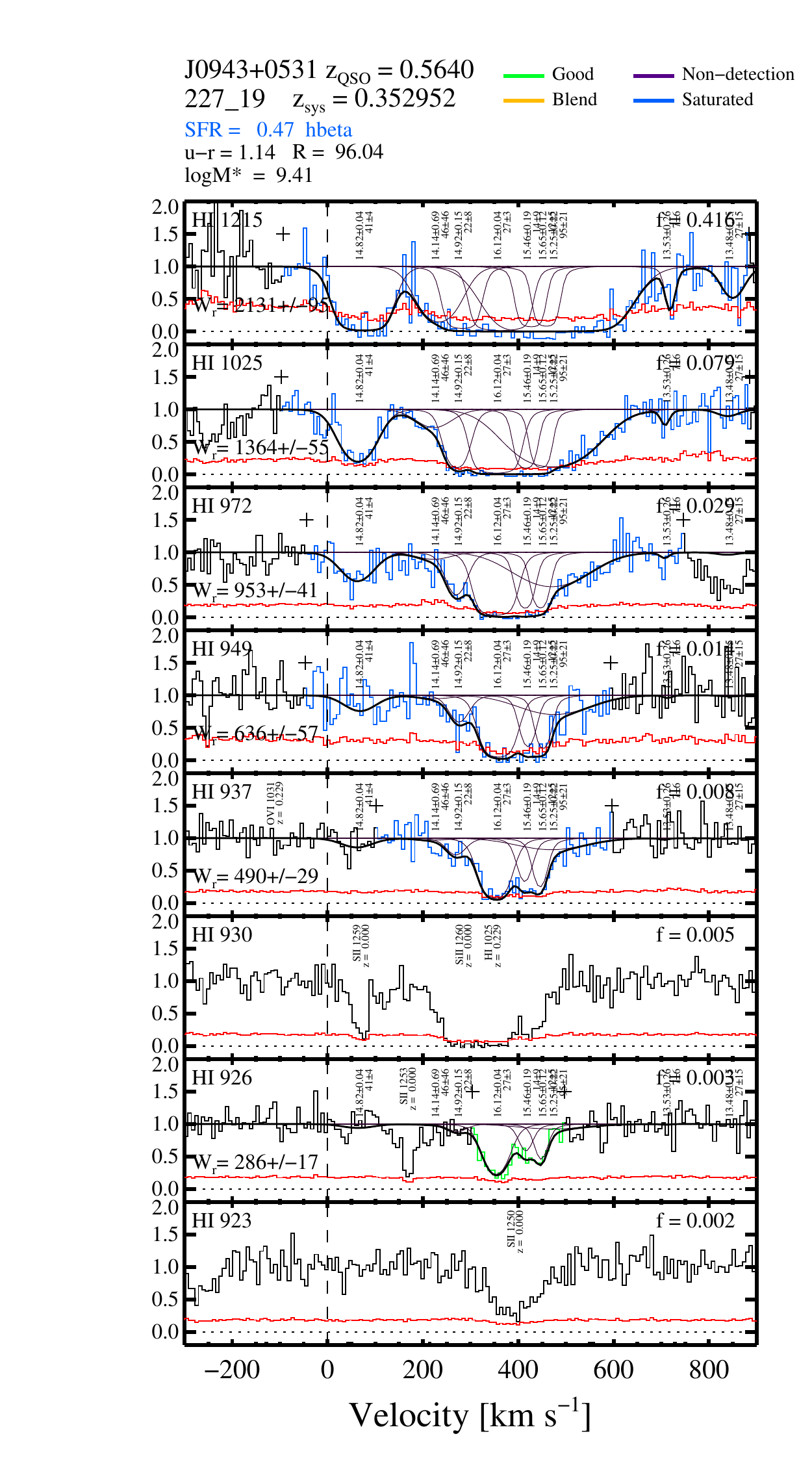}
\label{test_fig}
\end{center}
\caption{Hydrogen stack plot for system 227\_19 toward J0943$+$0531.}
\end{figure*}
\clearpage 
\begin{figure*}[!h]
\begin{center}
\epsscale{0.70}
\plotone{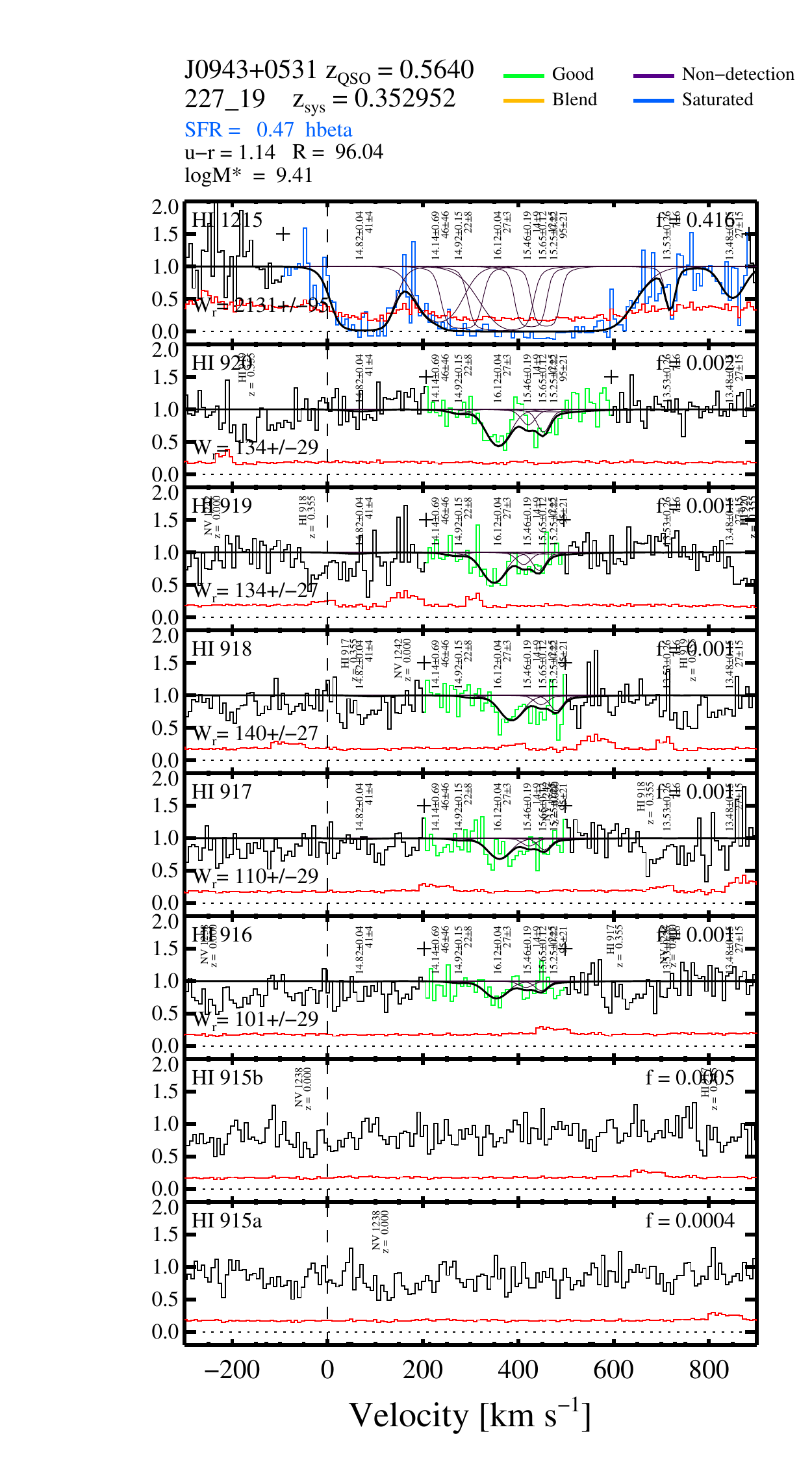}
\label{test_fig}
\end{center}
\caption{Hydrogen stack plot for the higher Lyman lines in system 227\_19 toward J0943$+$0531.}
\end{figure*}

\clearpage
\begin{figure*}[!h]
\begin{center}
\epsscale{0.70}
\plotone{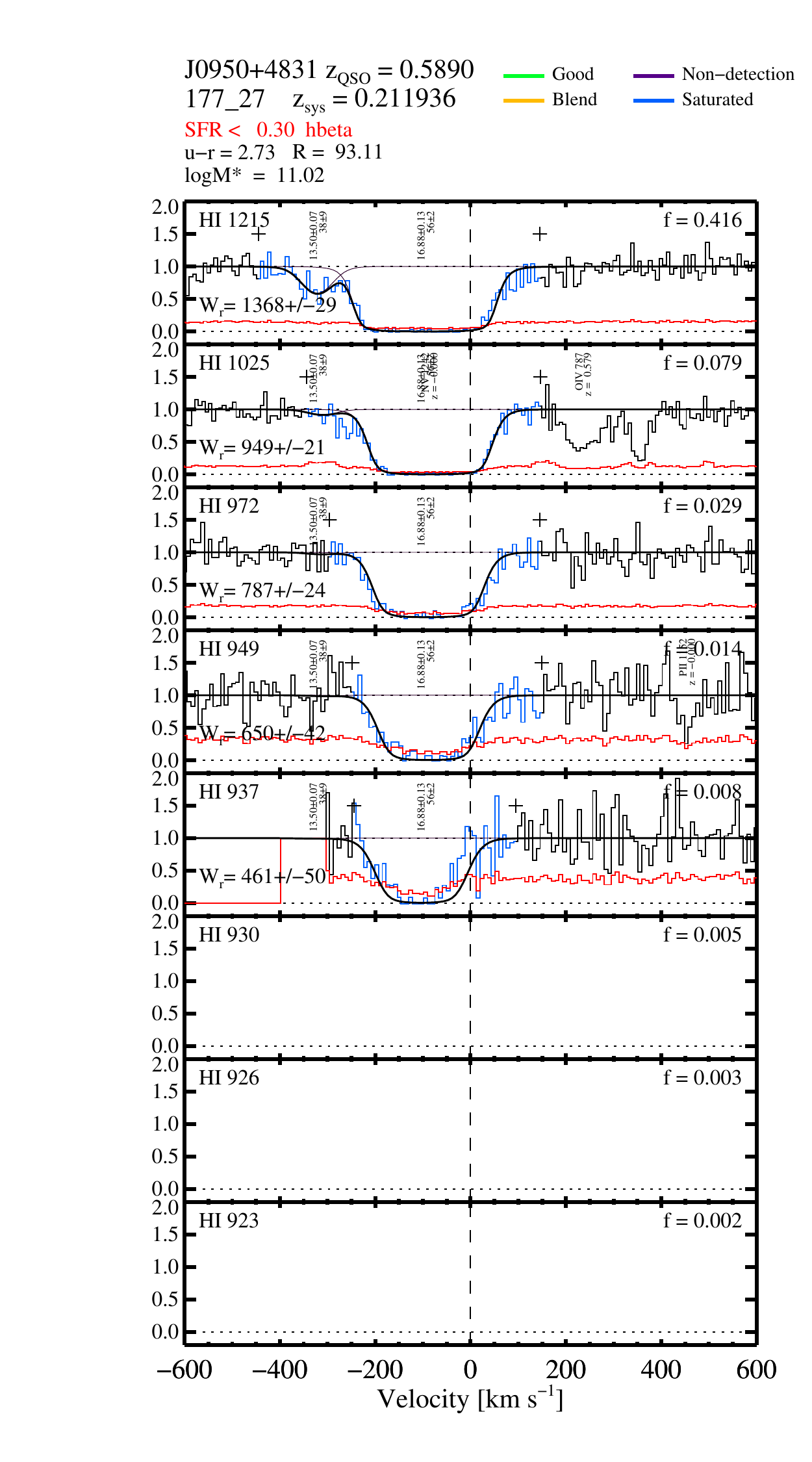}
\label{test_fig}
\end{center}
\caption{Hydrogen stack plot for system 177\_27 toward J0950$+$4831.}
\end{figure*}

\clearpage
\begin{figure*}[!h]
\begin{center}
\epsscale{0.70}
\plotone{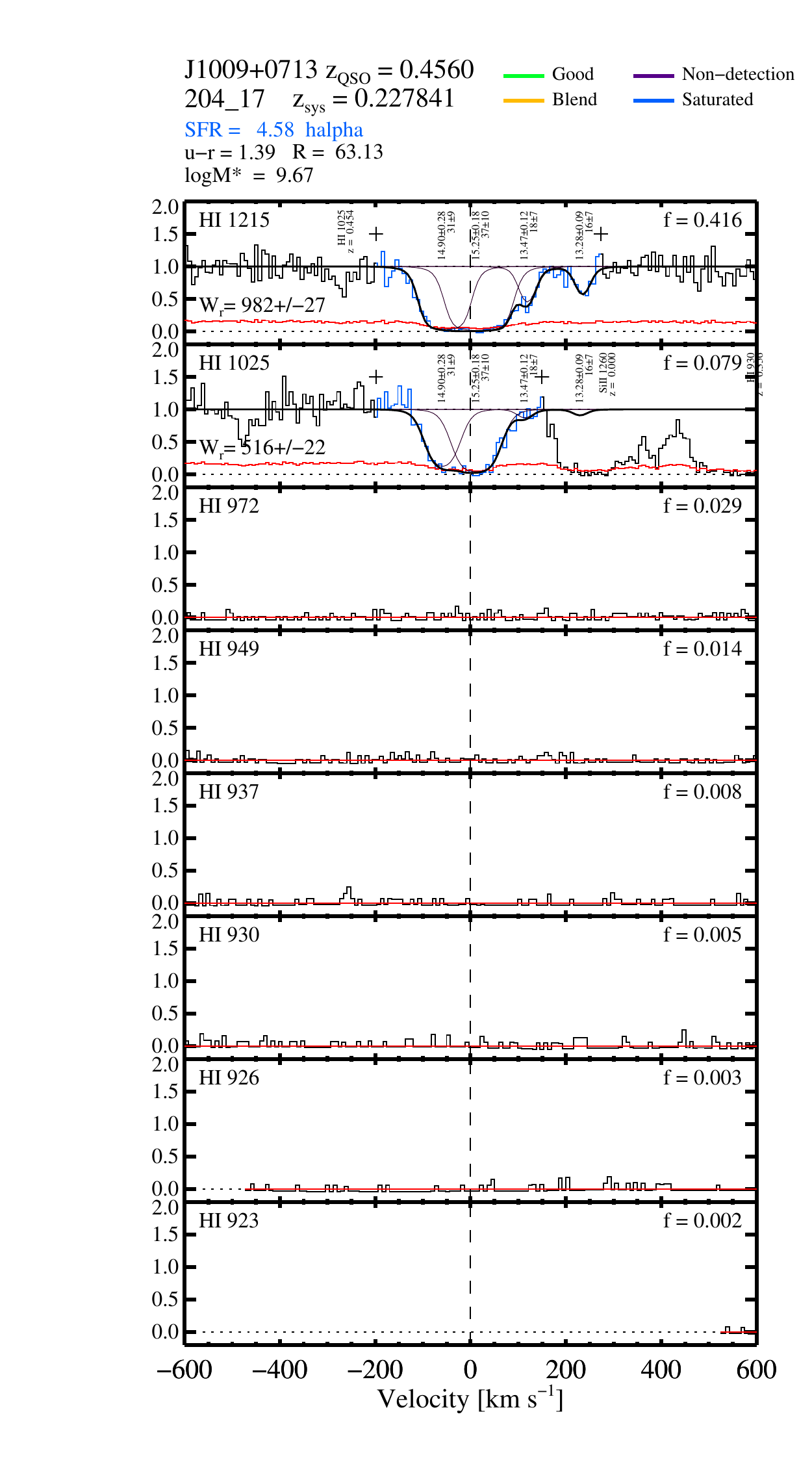} 
\label{test_fig}
\end{center}
\caption{Hydrogen stack plot for system 204\_17 toward J1009$+$0713.}
\end{figure*}

\clearpage
\begin{figure*}[!h]
\begin{center}
\epsscale{0.70}
\plotone{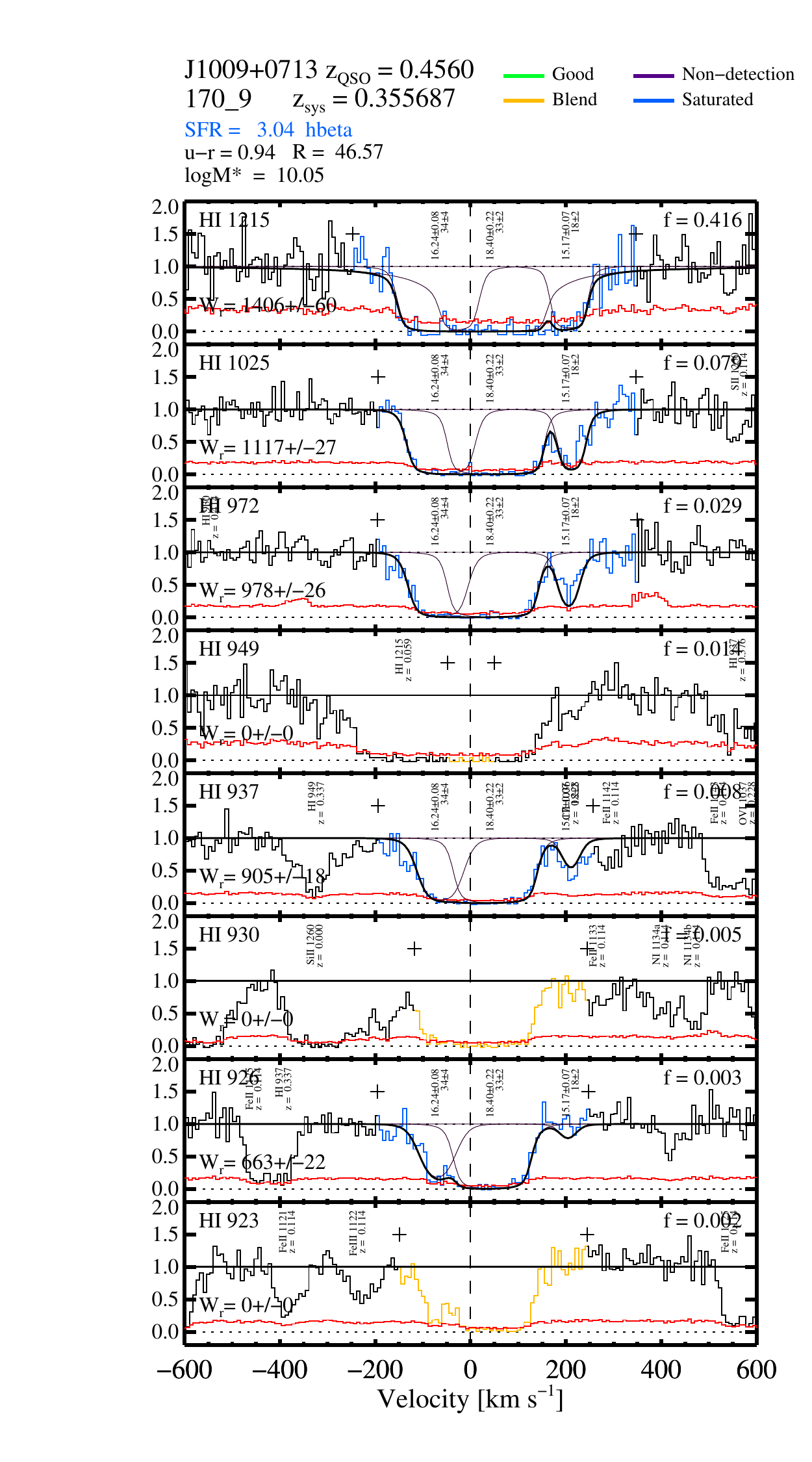} 
\label{test_fig}
\end{center}
\caption{Hydrogen stack plot for system 170\_9 toward J1009$+$0713.}
\end{figure*}
\clearpage 
\begin{figure*}[!h]
\begin{center}
\epsscale{0.70}
\plotone{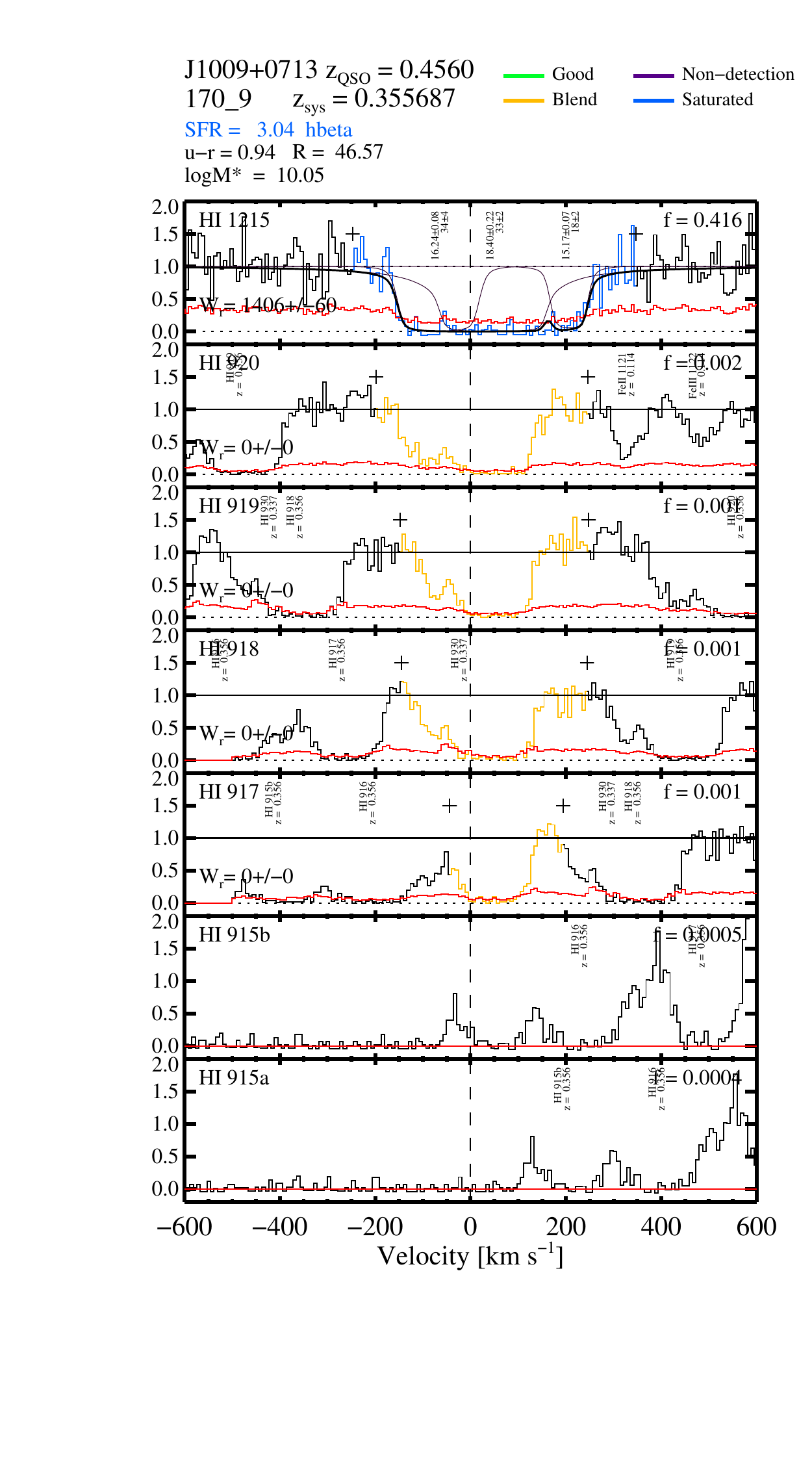} 
\label{test_fig}
\end{center}
\caption{Hydrogen stack plot for the higher Lyman lines in system 170\_9 toward J1009$+$0713.}
\end{figure*}

\clearpage
\begin{figure*}[!h]
\begin{center}
\epsscale{0.70}
\plotone{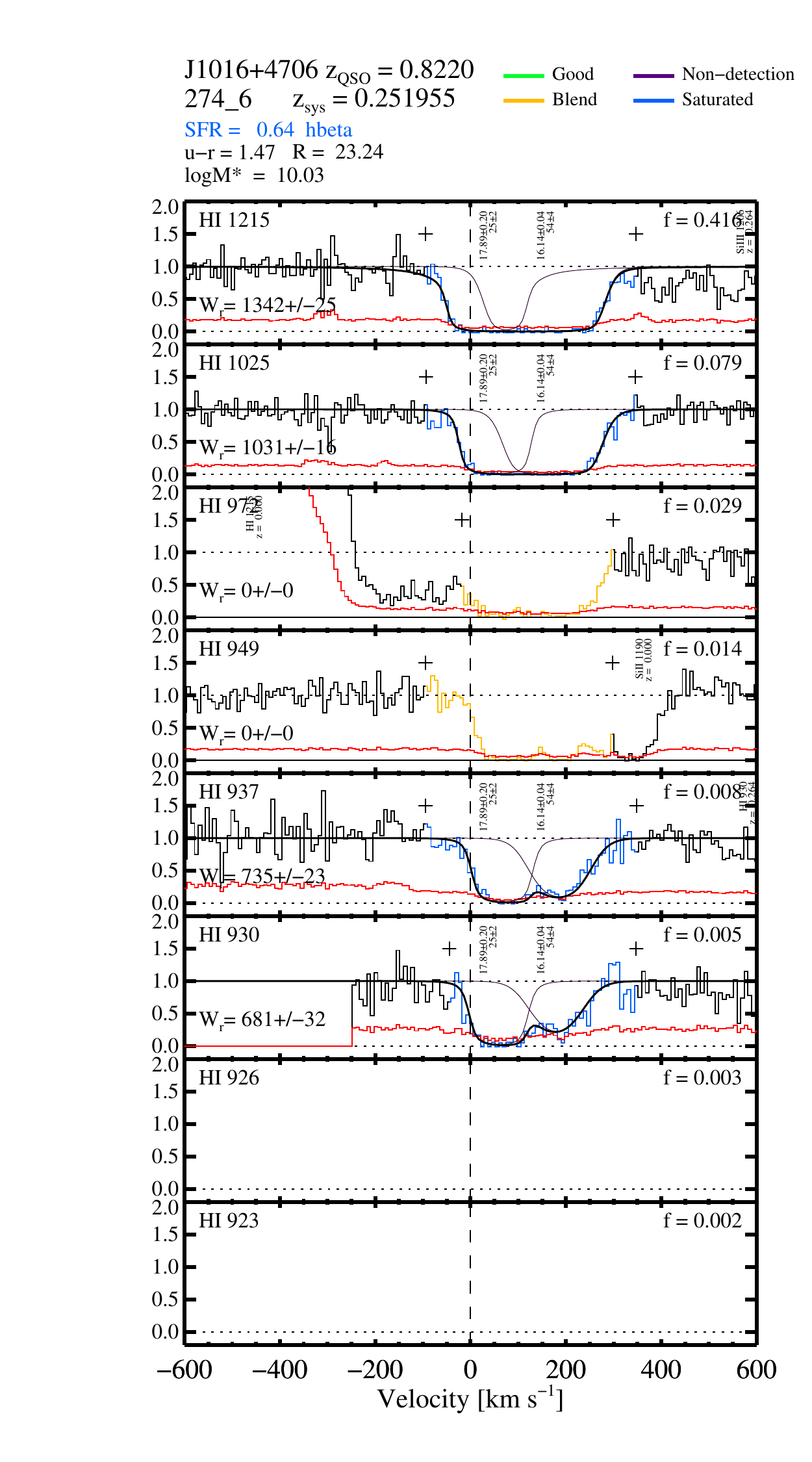}
\label{test_fig}
\end{center}
\caption{Hydrogen stack plot for system 274\_6 toward J1016$+$4706.}
\end{figure*}

\clearpage
\begin{figure*}[!h]
\begin{center}
\epsscale{0.70}
\plotone{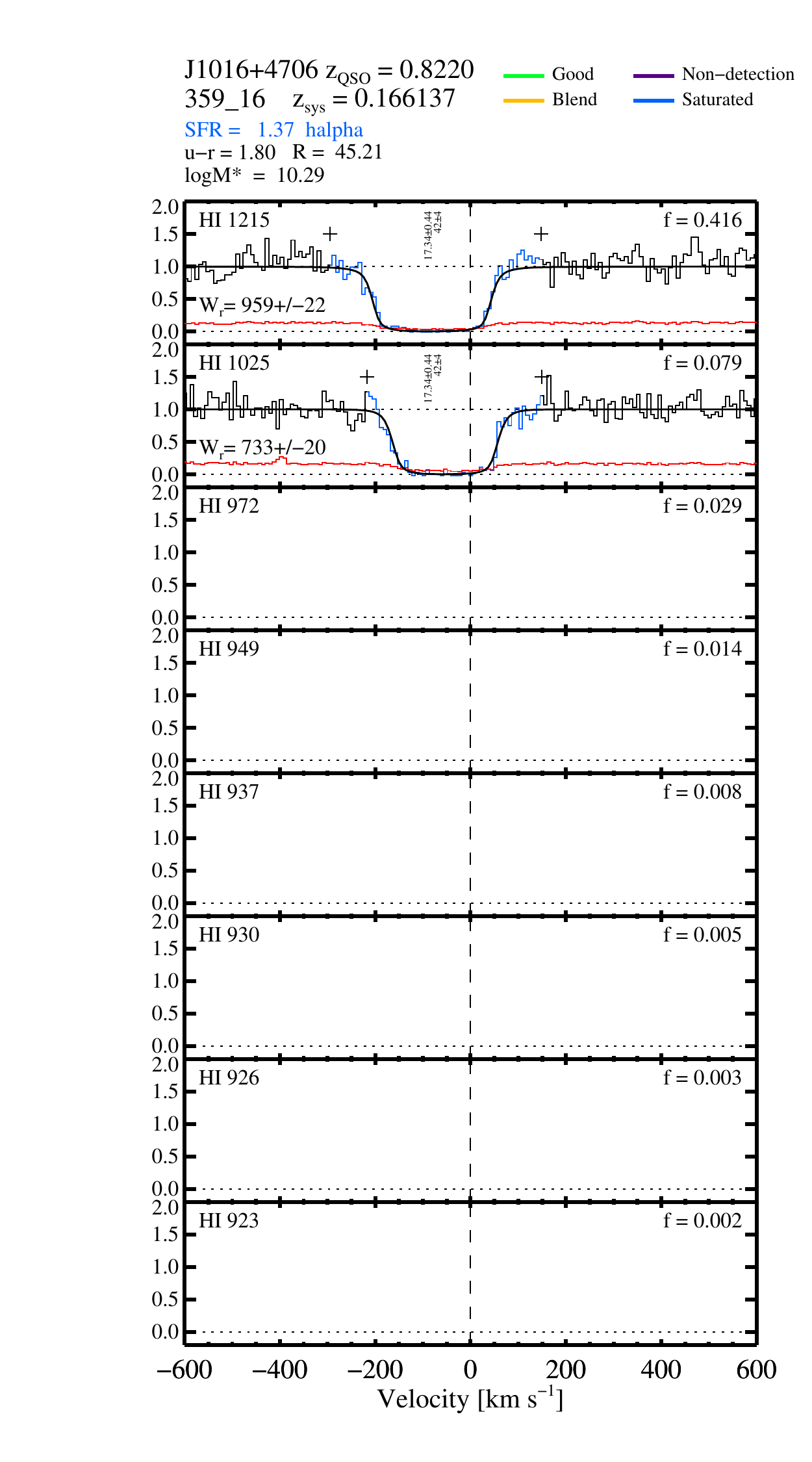}
\label{test_fig}
\end{center}
\caption{Hydrogen stack plot for system 359\_16 toward J1016$+$4706.}
\end{figure*}

\clearpage
\begin{figure*}[!h]
\begin{center}
\epsscale{0.70}
\plotone{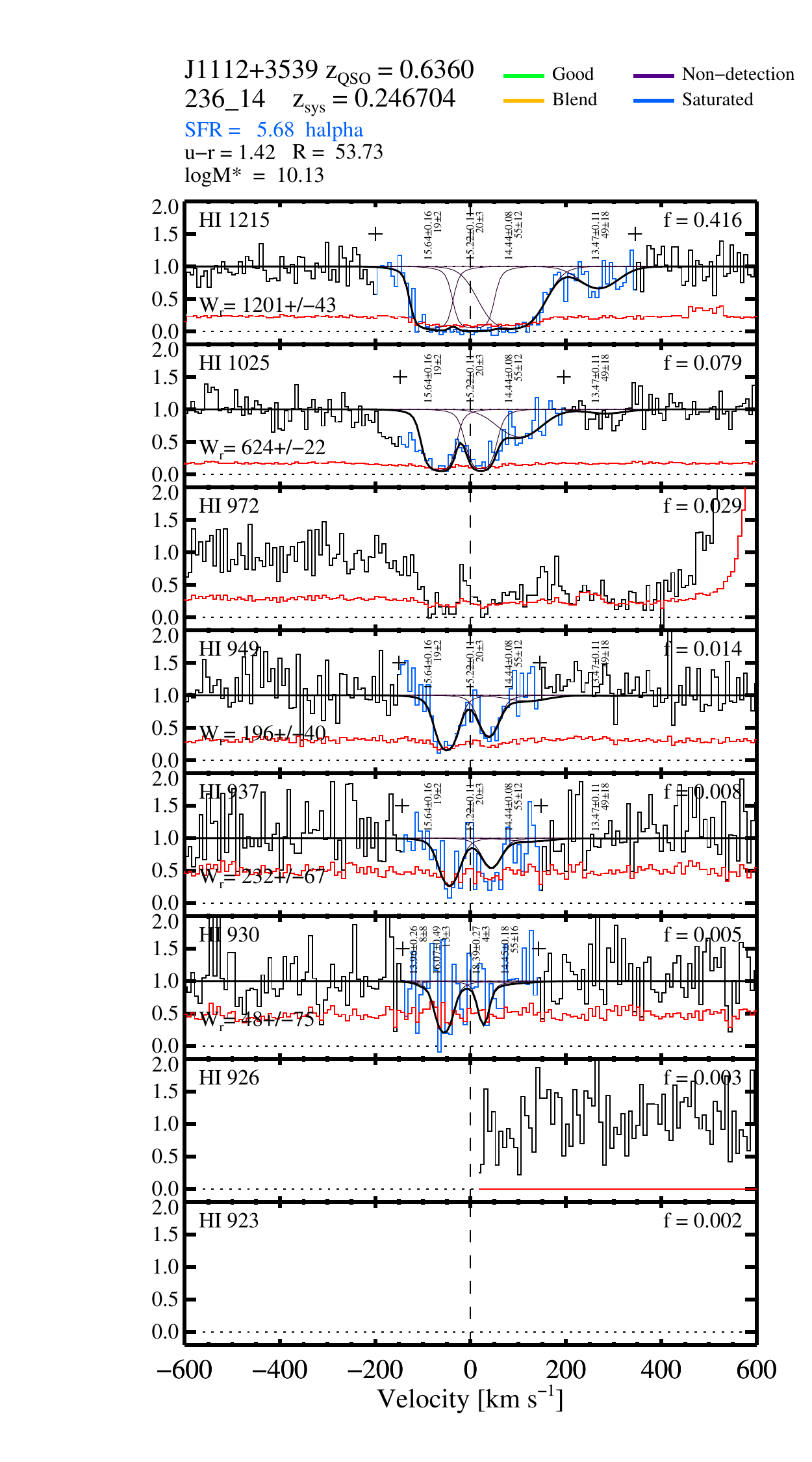} 
\label{test_fig}
\end{center}
\caption{Hydrogen stack plot for system 236\_14 toward J1112$+$3539.}
\end{figure*}

\clearpage
\begin{figure*}[!h]
\begin{center}
\epsscale{0.70}
\plotone{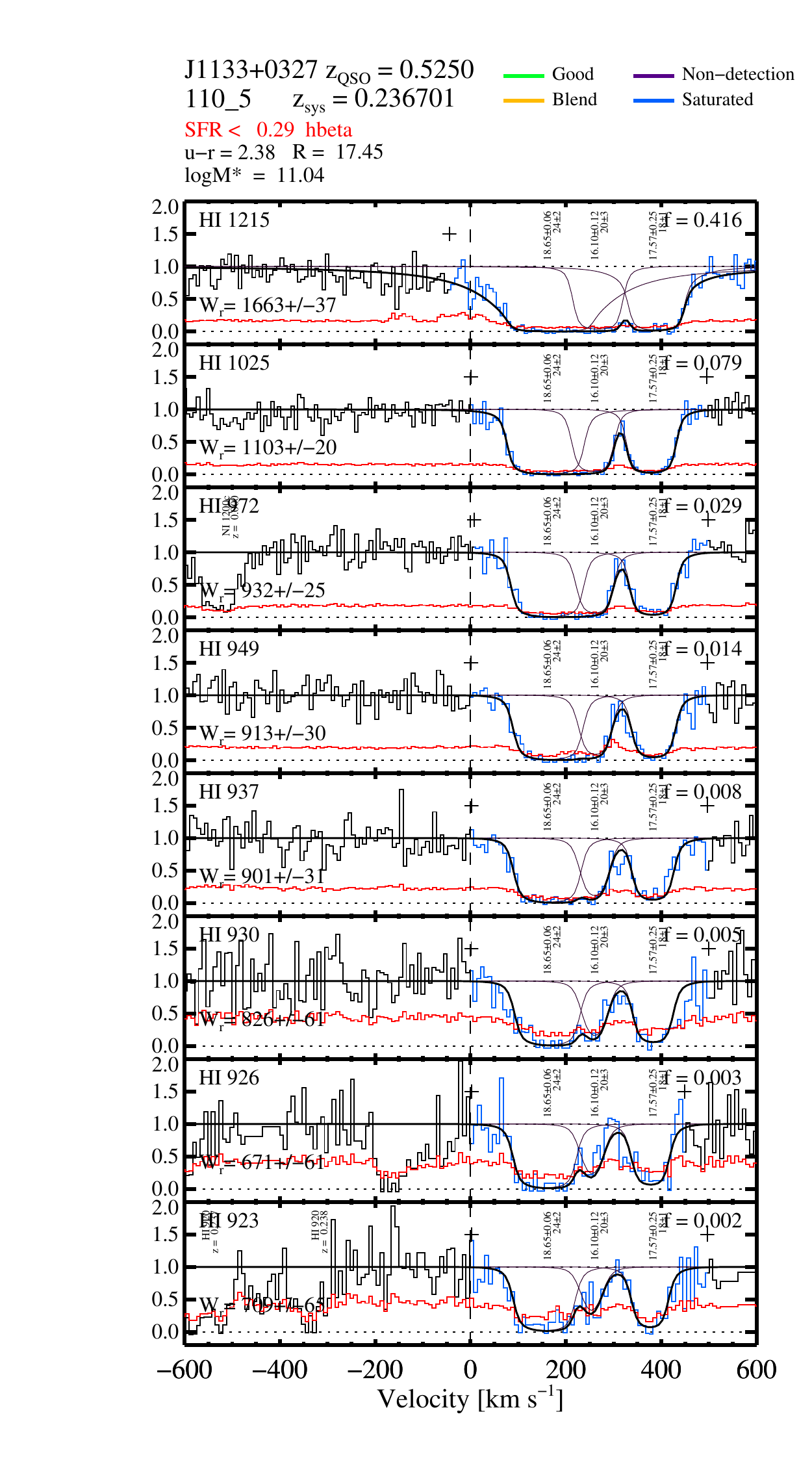} 
\label{test_fig}
\end{center}
\caption{Hydrogen stack plot for system 110\_5 toward J1133$+$0327.}
\end{figure*}
\clearpage 
\begin{figure*}[!h]
\begin{center}
\epsscale{0.70}
\plotone{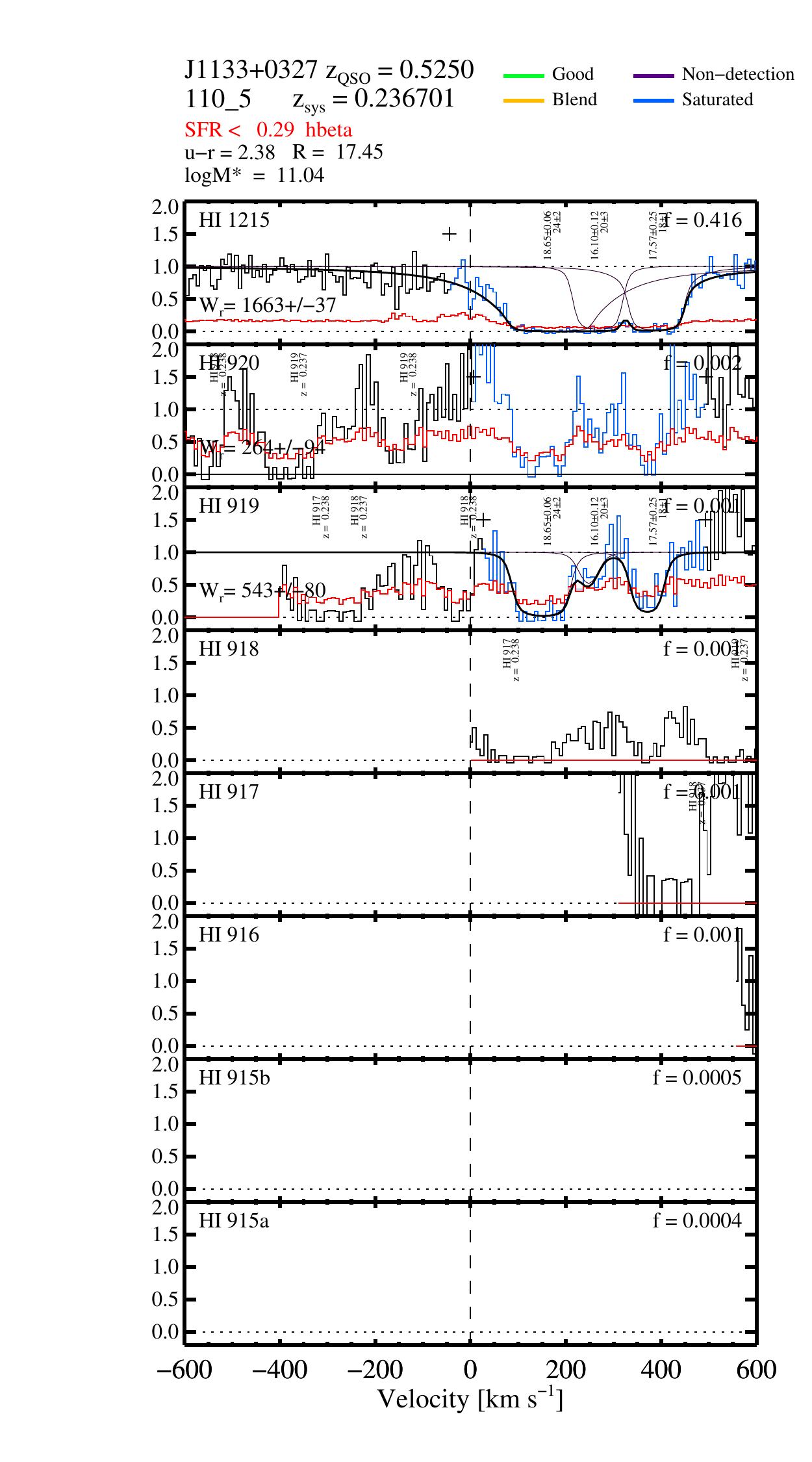} 
\label{test_fig}
\end{center}
\caption{Hydrogen stack plot for the higher Lyman lines in system 110\_5 toward J1133$+$0327.}
\end{figure*}

\clearpage
\begin{figure*}[!h]
\begin{center}
\epsscale{0.70}
\plotone{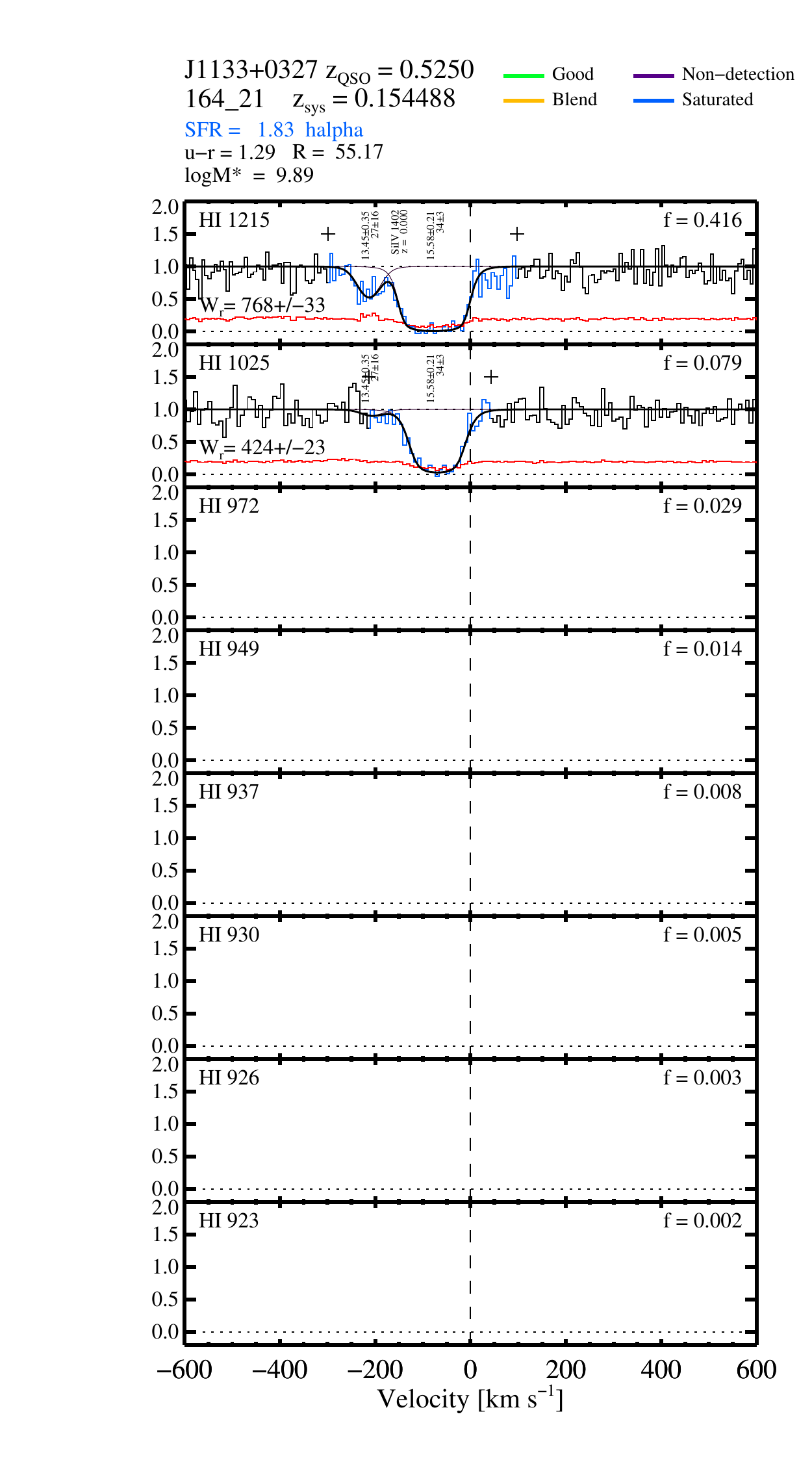} 
\label{test_fig}
\end{center}
\caption{Hydrogen stack plot for system 164\_21 toward J1133$+$0327.}
\end{figure*}

\clearpage
\begin{figure*}[!h]
\begin{center}
\epsscale{0.70}
\plotone{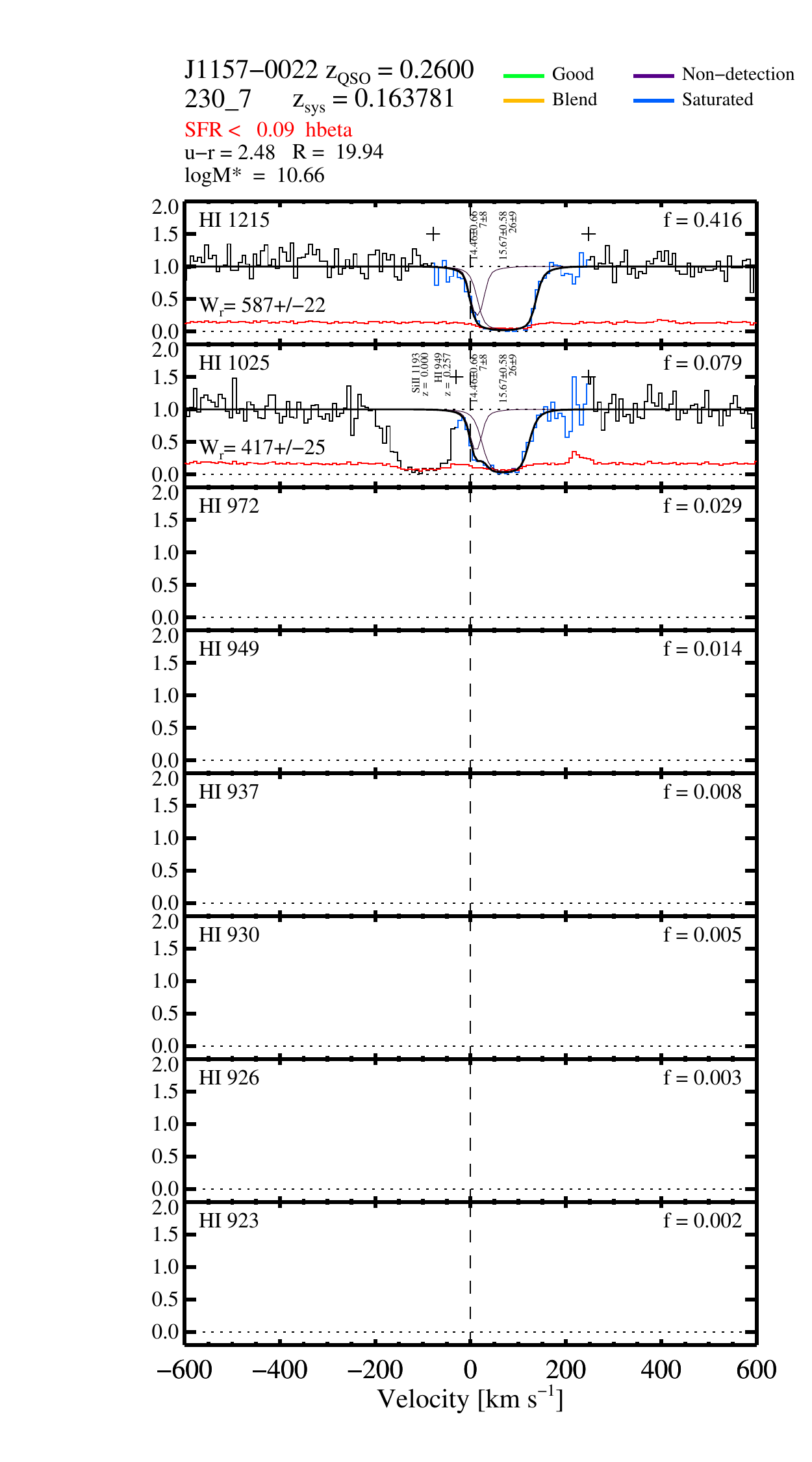} 
\label{test_fig}
\end{center}
\caption{Hydrogen stack plot for system 230\_7 toward J1157$-$0022.}
\end{figure*}

\clearpage
\begin{figure*}[!h]
\begin{center}
\epsscale{0.70}
\plotone{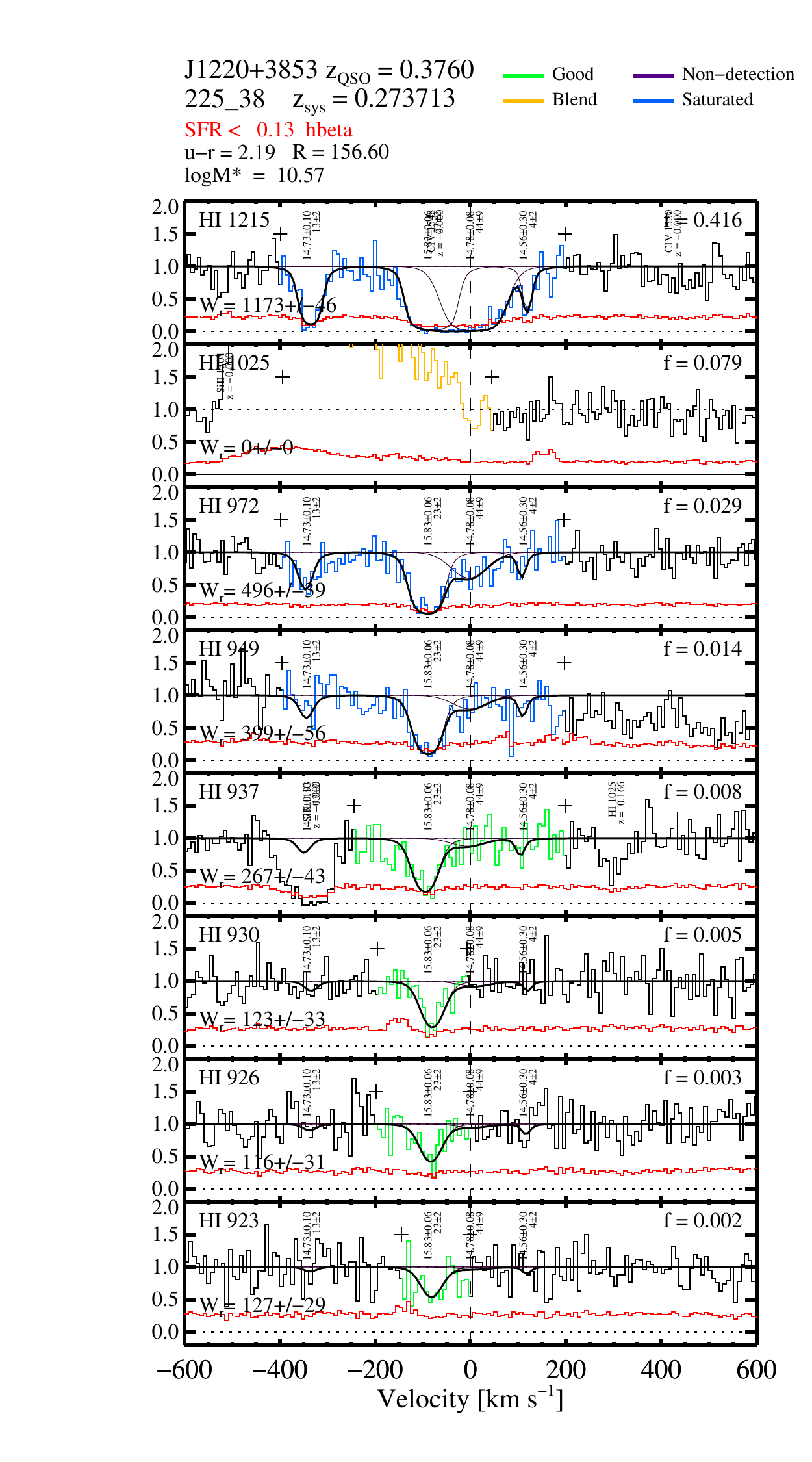} 
\label{test_fig}
\end{center}
\caption{Hydrogen stack plot for system 225\_38 toward J1220$+$3853.}
\end{figure*}

\clearpage
\begin{figure*}[!h]
\begin{center}
\epsscale{0.70}
\plotone{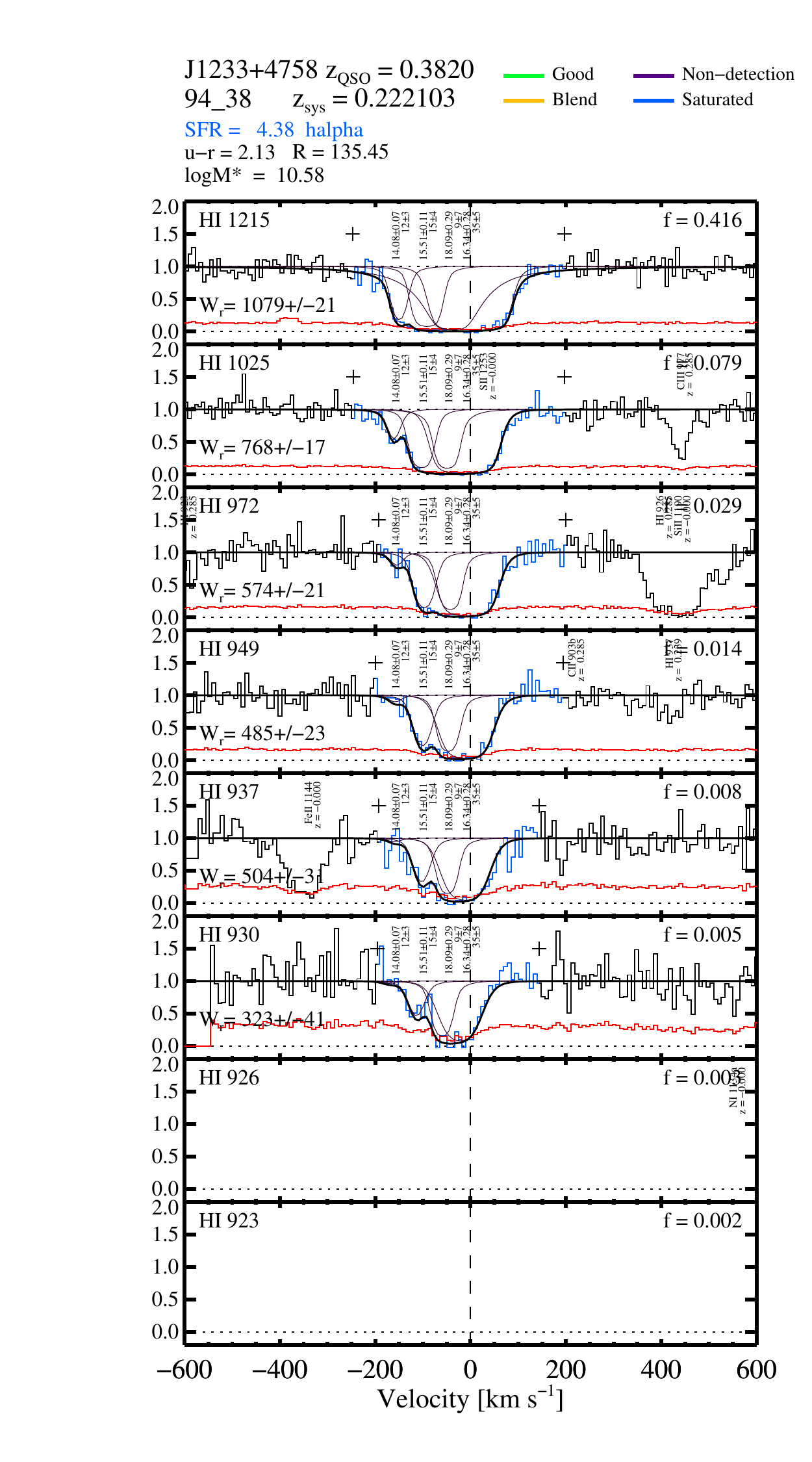} 
\label{test_fig}
\end{center}
\caption{Hydrogen stack plot for system 94\_38 toward J1233$+$4758.}
\end{figure*}

\clearpage
\begin{figure*}[!h]
\begin{center}
\epsscale{0.70}
\plotone{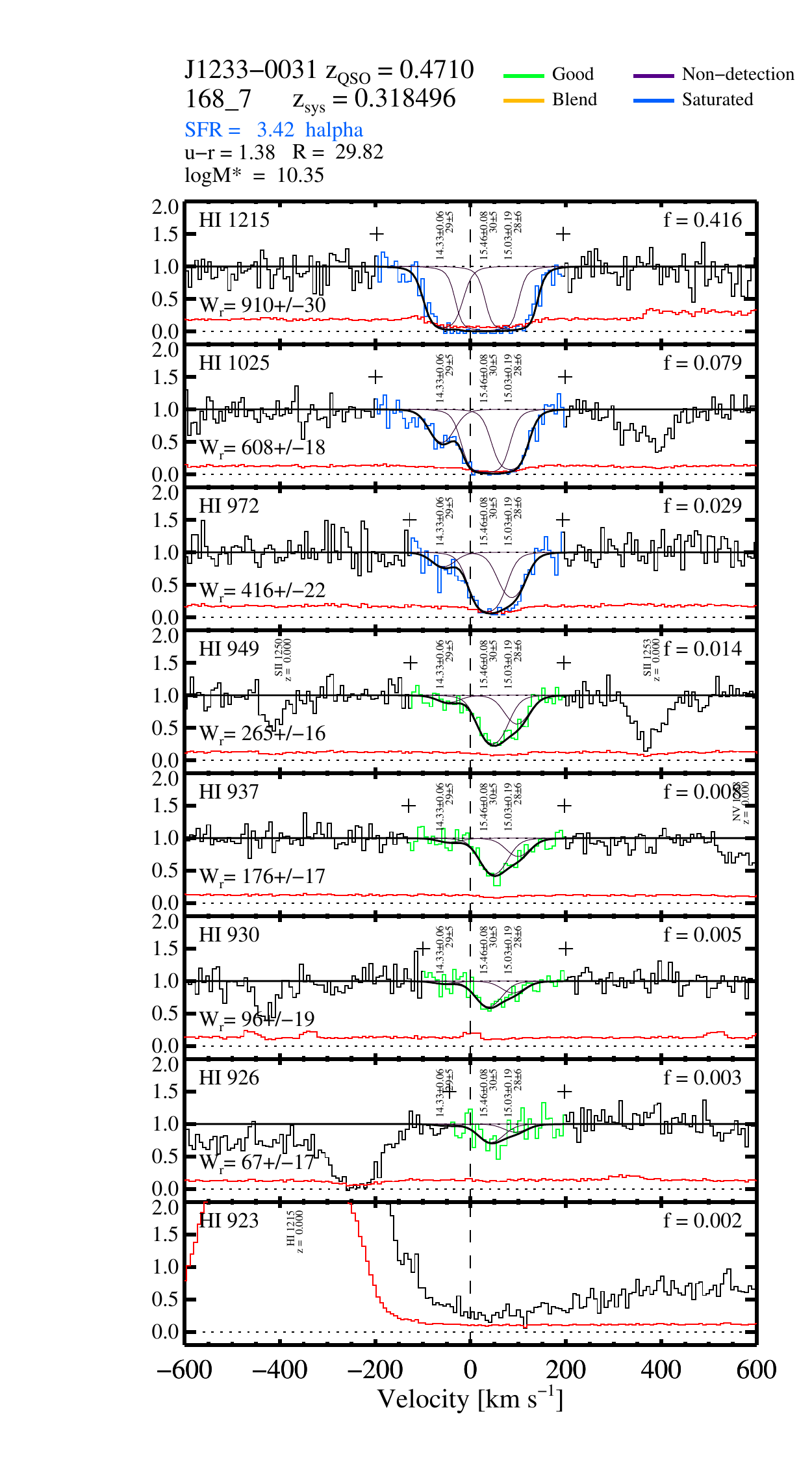} 
\label{test_fig}
\end{center}
\caption{Hydrogen stack plot for system 168\_7 toward J1233$-$0031.}
\end{figure*}

\clearpage
\begin{figure*}[!h]
\begin{center}
\epsscale{0.70}
\plotone{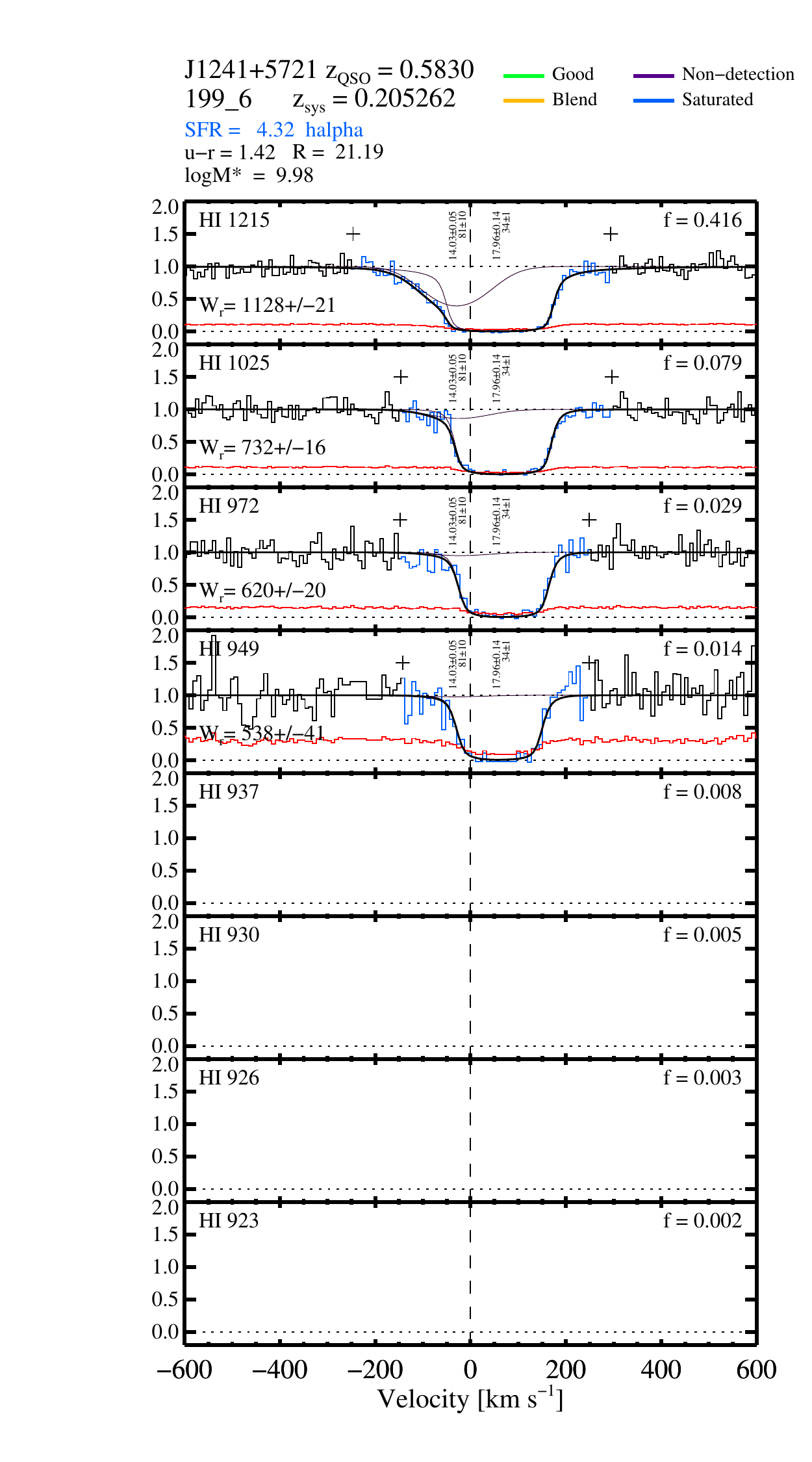} 
\label{test_fig}
\end{center}
\caption{Hydrogen stack plot for system 199\_6 toward J1241$+$5721.}
\end{figure*}

\clearpage
\begin{figure*}[!h]
\begin{center}
\epsscale{0.70}
\plotone{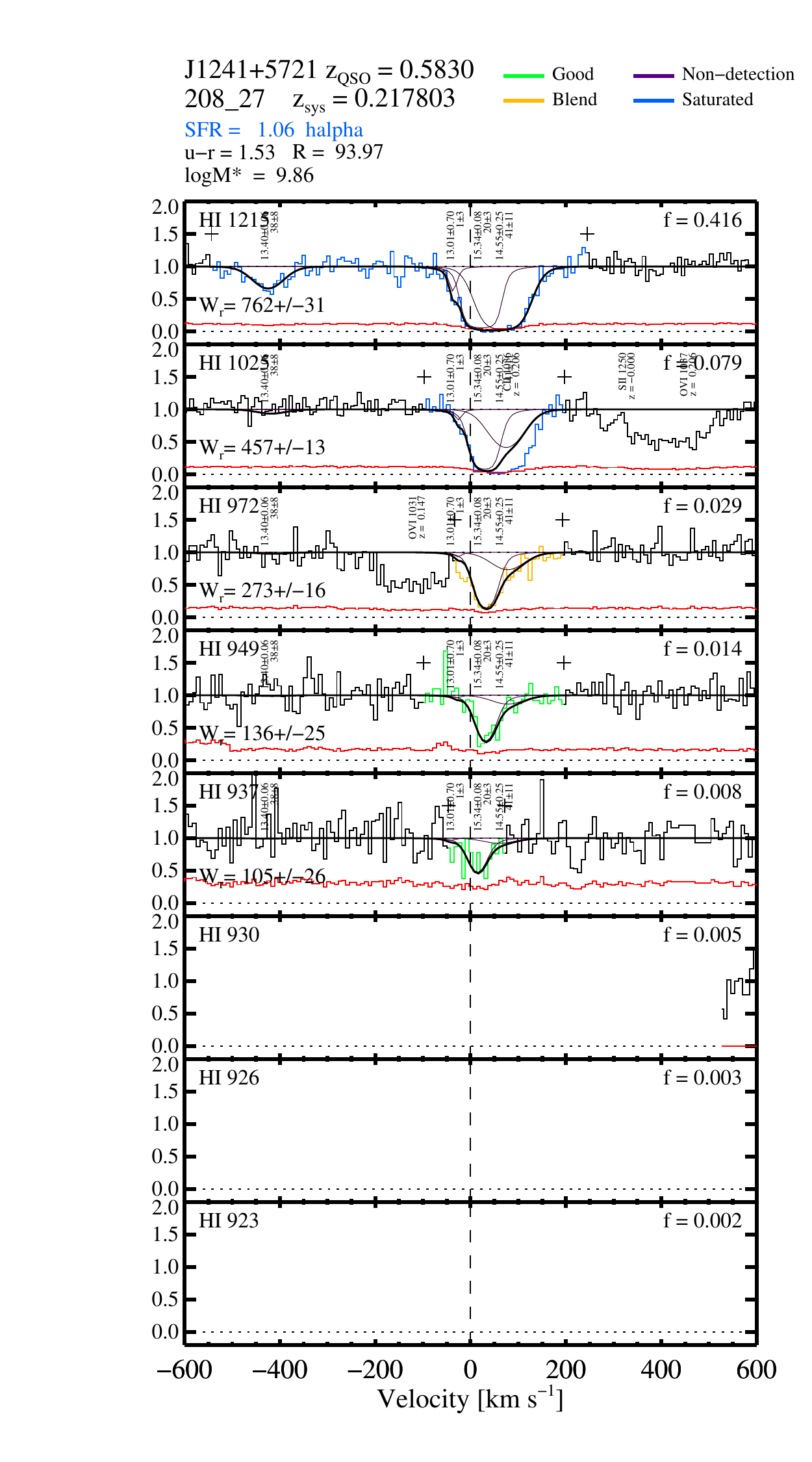} 
\label{test_fig}
\end{center}
\caption{Hydrogen stack plot for system 208\_27 toward J1241$+$5721.}
\end{figure*}

\clearpage
\begin{figure*}[!h]
\begin{center}
\epsscale{0.70}
\plotone{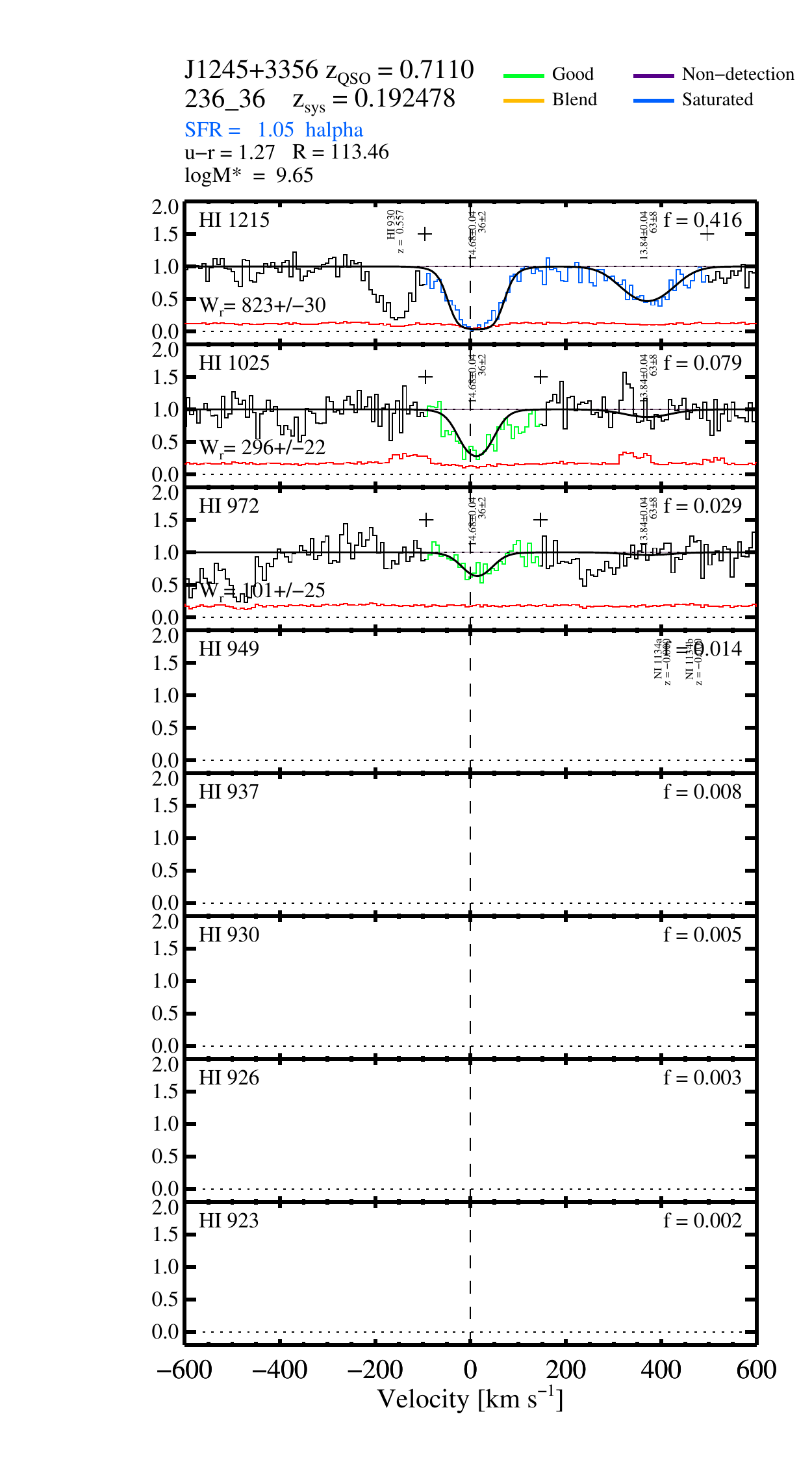} 
\label{test_fig}
\end{center}
\caption{Hydrogen stack plot for system 236\_36 toward J1245$+$3356.}
\end{figure*}

\clearpage
\begin{figure*}[!h]
\begin{center}
\epsscale{0.70}
\plotone{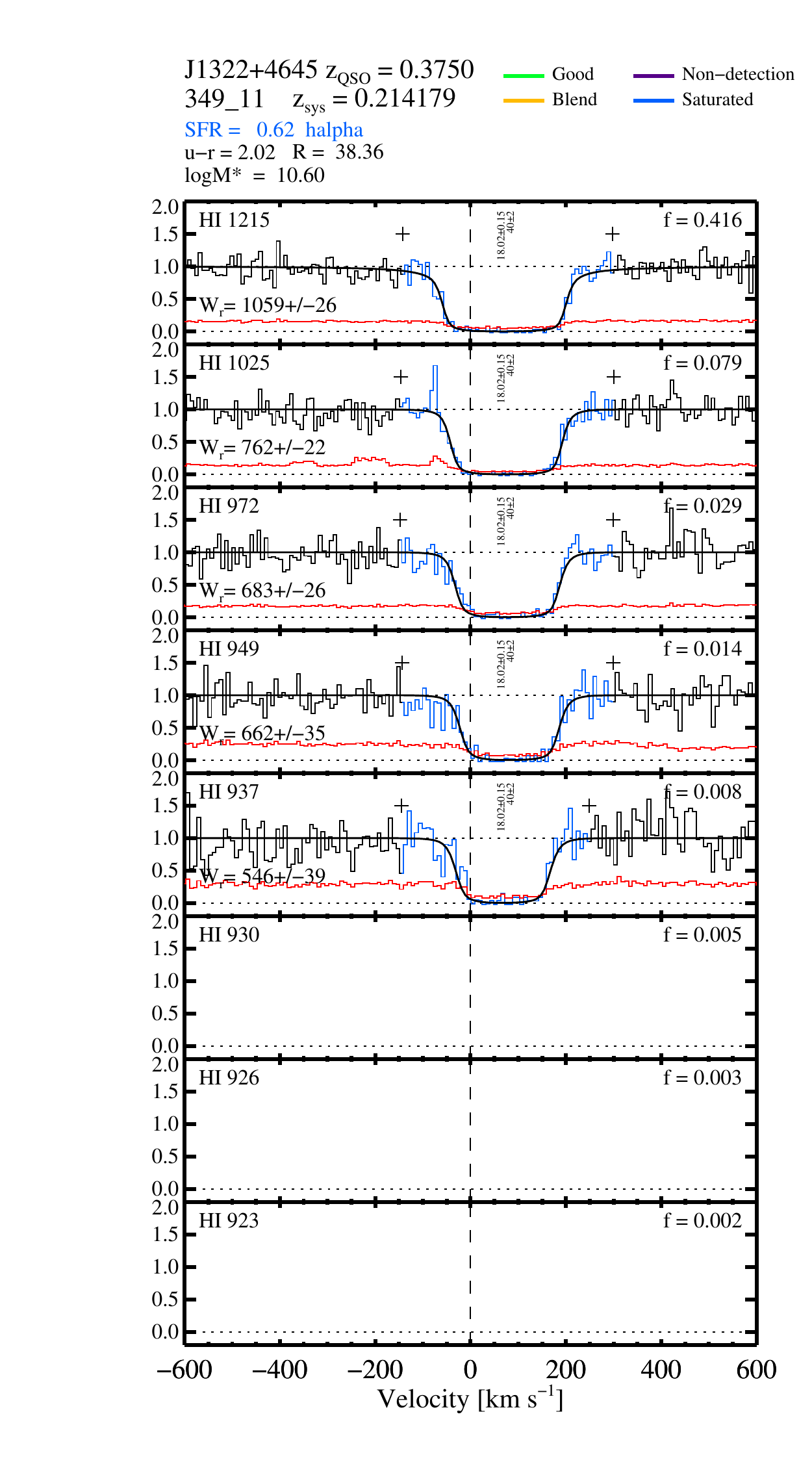} 
\label{test_fig}
\end{center}
\caption{Hydrogen stack plot for system 349\_11 toward J1322$+$4645.}
\end{figure*}

\clearpage
\begin{figure*}[!h]
\begin{center}
\epsscale{0.70}
\plotone{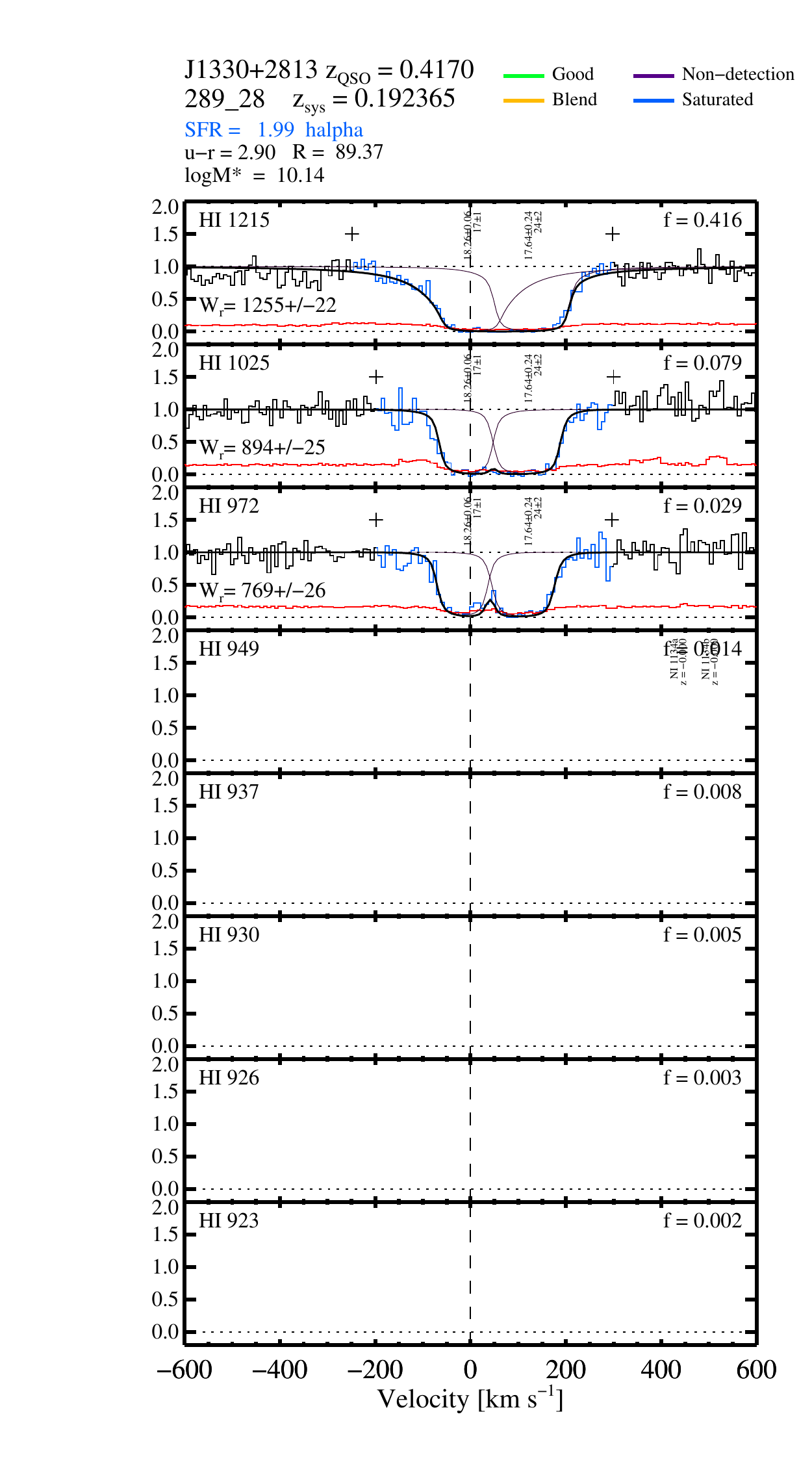} 
\label{test_fig}
\end{center}
\caption{Hydrogen stack plot for system 289\_28 toward J1330$+$2813.}
\end{figure*}

\clearpage
\begin{figure*}[!h]
\begin{center}
\epsscale{0.70}
\plotone{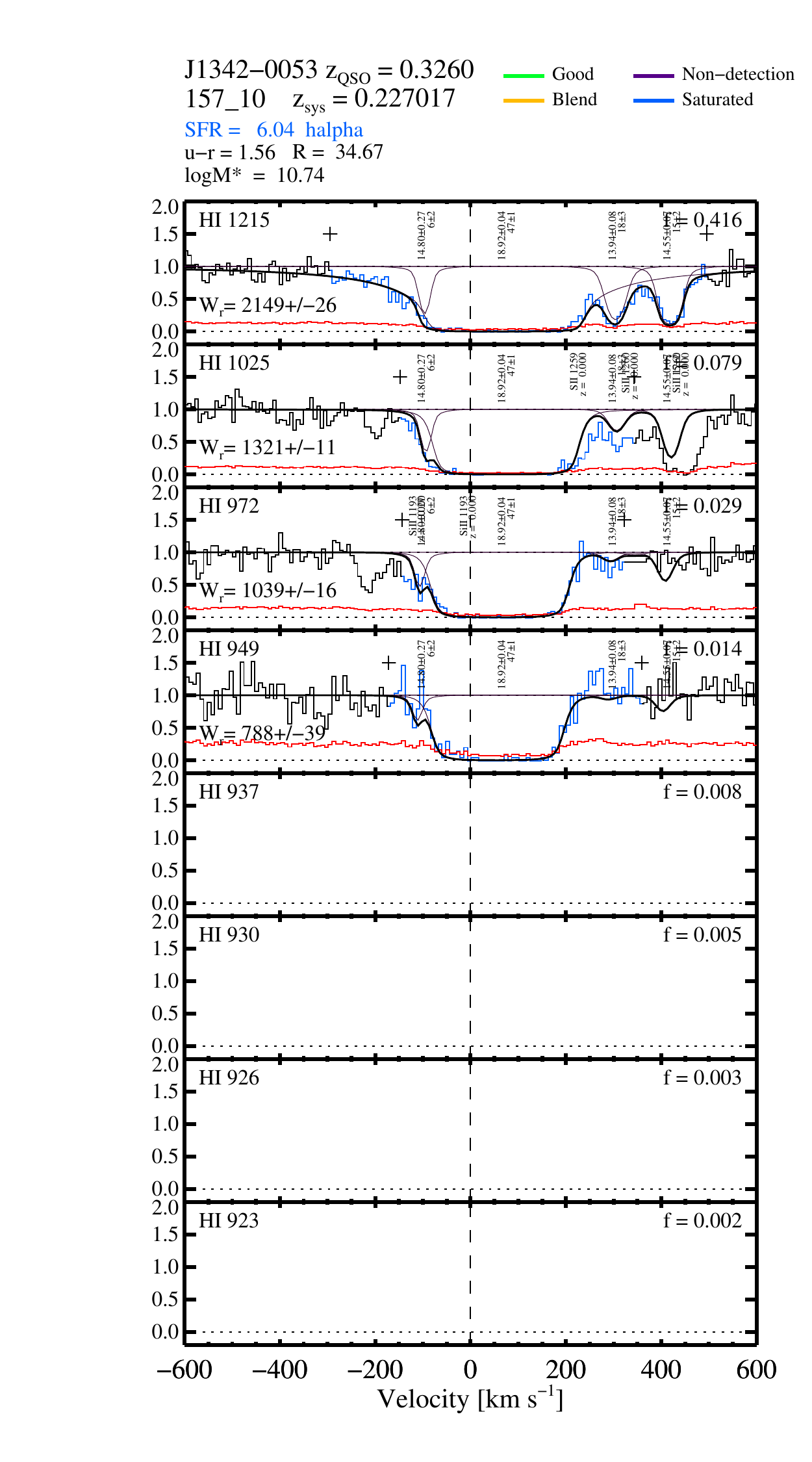} 
\label{test_fig}
\end{center}
\caption{Hydrogen stack plot for system 157\_10 toward J1342$-$0053.}
\end{figure*}

\clearpage
\begin{figure*}[!h]
\begin{center}
\epsscale{0.70}
\plotone{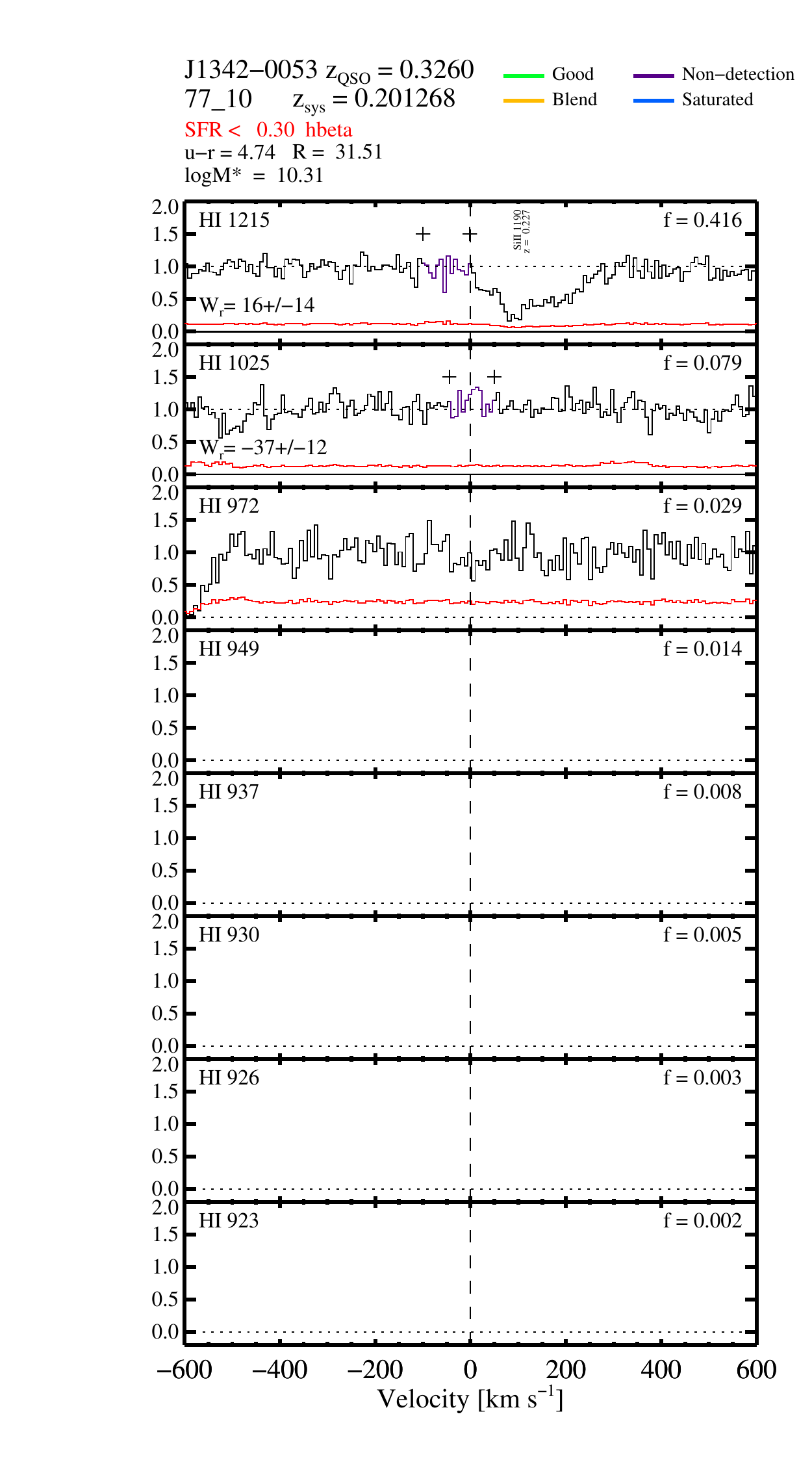} 
\label{test_fig}
\end{center}
\caption{Hydrogen stack plot for system 77\_10 toward J1342$-$0053.}
\end{figure*}

\clearpage
\begin{figure*}[!h]
\begin{center}
\epsscale{0.70}
\plotone{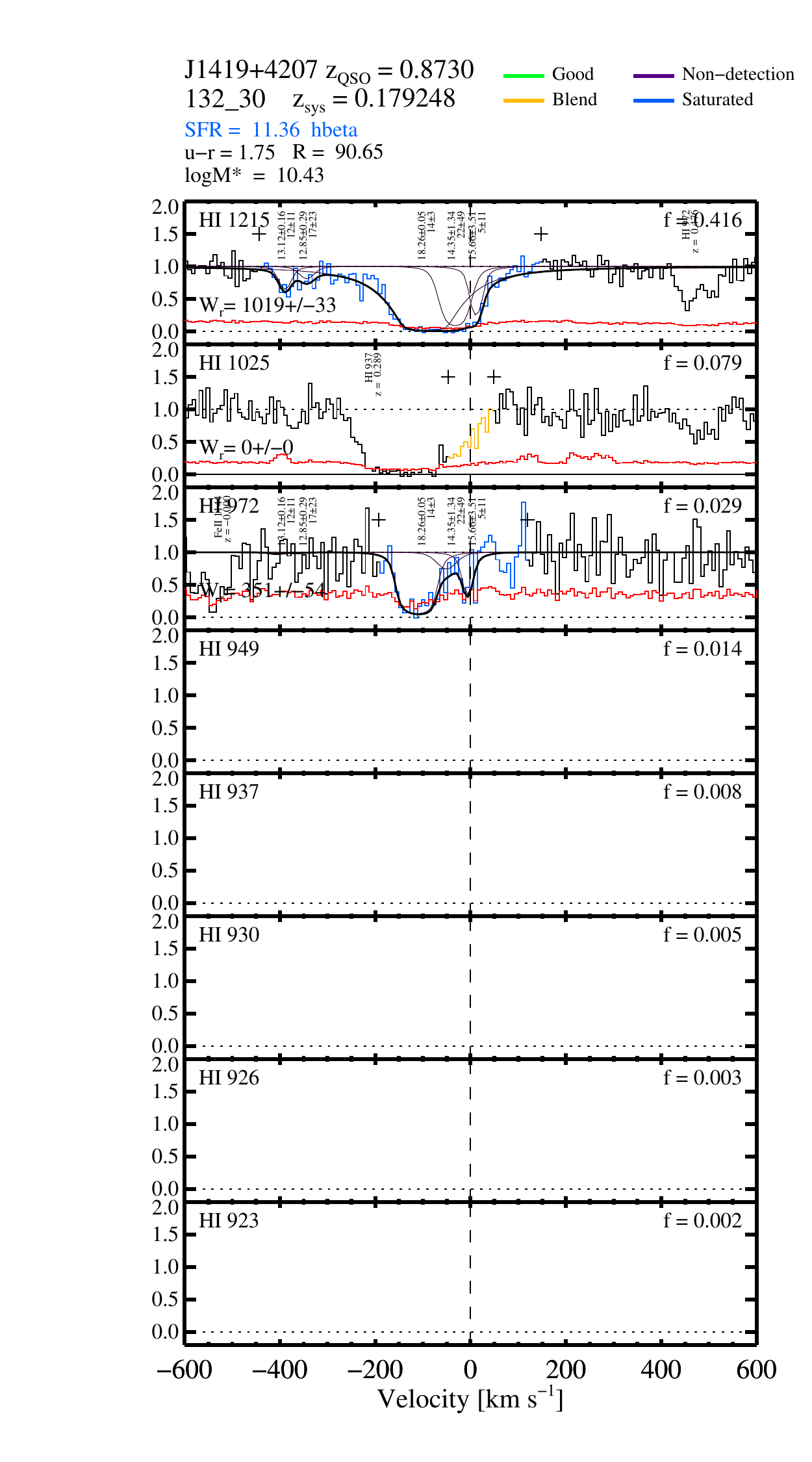}
\label{test_fig}
\end{center}
\caption{Hydrogen stack plot for system 132\_30 toward J1419$+$4207.}
\end{figure*}

\clearpage
\begin{figure*}[!h]
\begin{center}
\epsscale{0.70}
\plotone{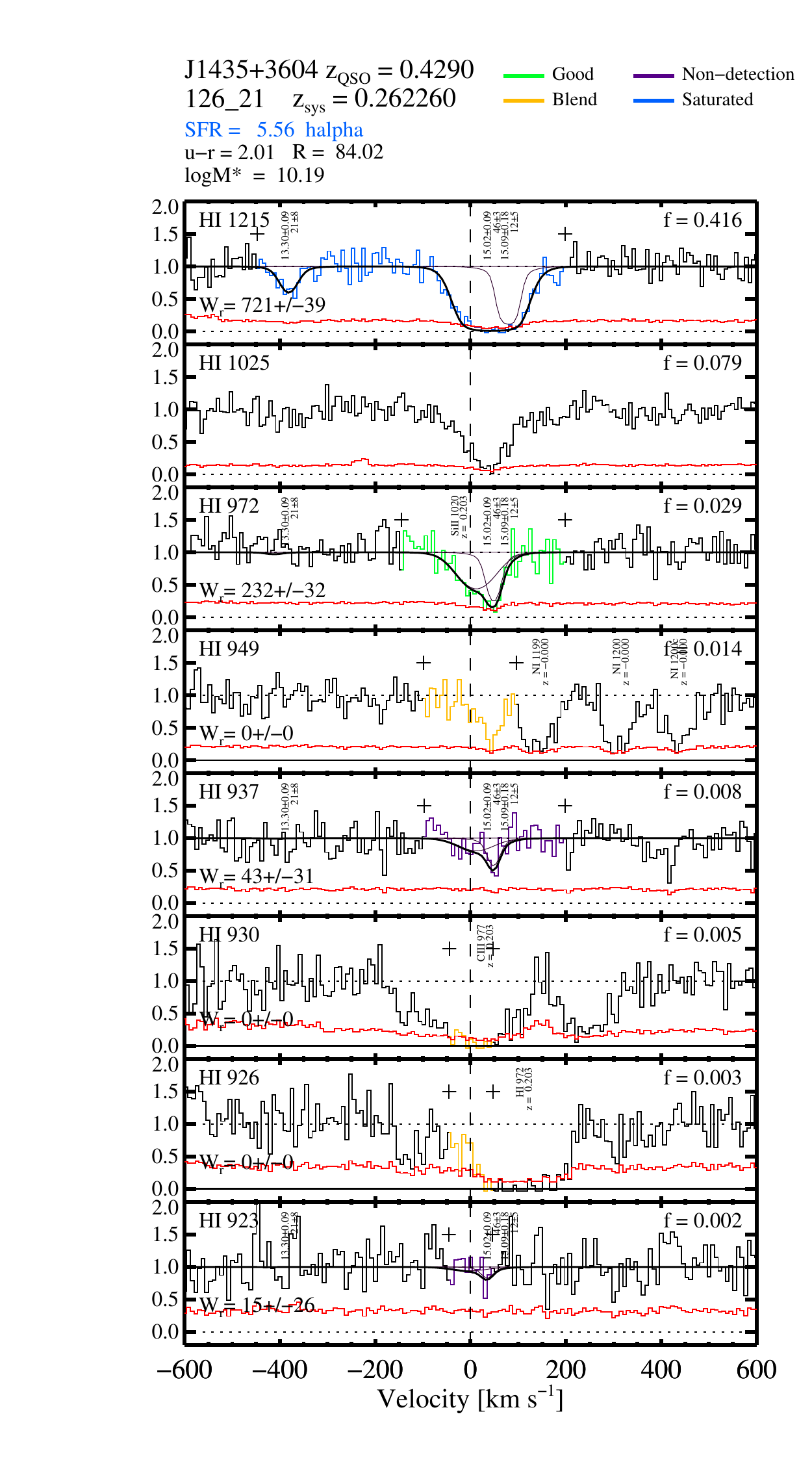}
\label{test_fig}
\end{center}
\caption{Hydrogen stack plot for system 126\_21 toward J1435$+$3604.}
\end{figure*}

\clearpage
\begin{figure*}[!h]
\begin{center}
\epsscale{0.70}
\plotone{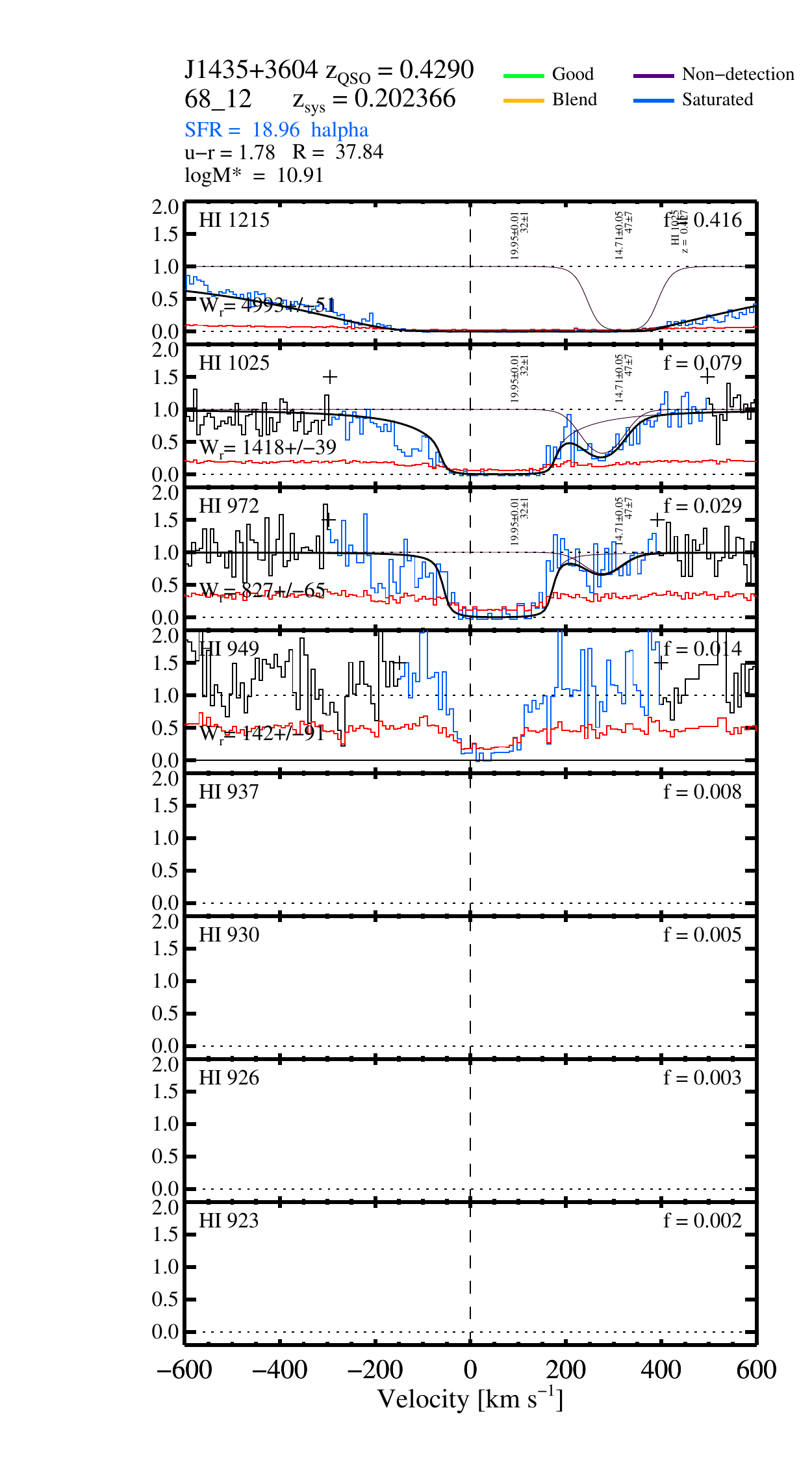}
\label{test_fig}
\end{center}
\caption{Hydrogen stack plot for system 68\_12 toward J1435$+$3604.}
\end{figure*}

\clearpage
\begin{figure*}[!h]
\begin{center}
\epsscale{0.70}
\plotone{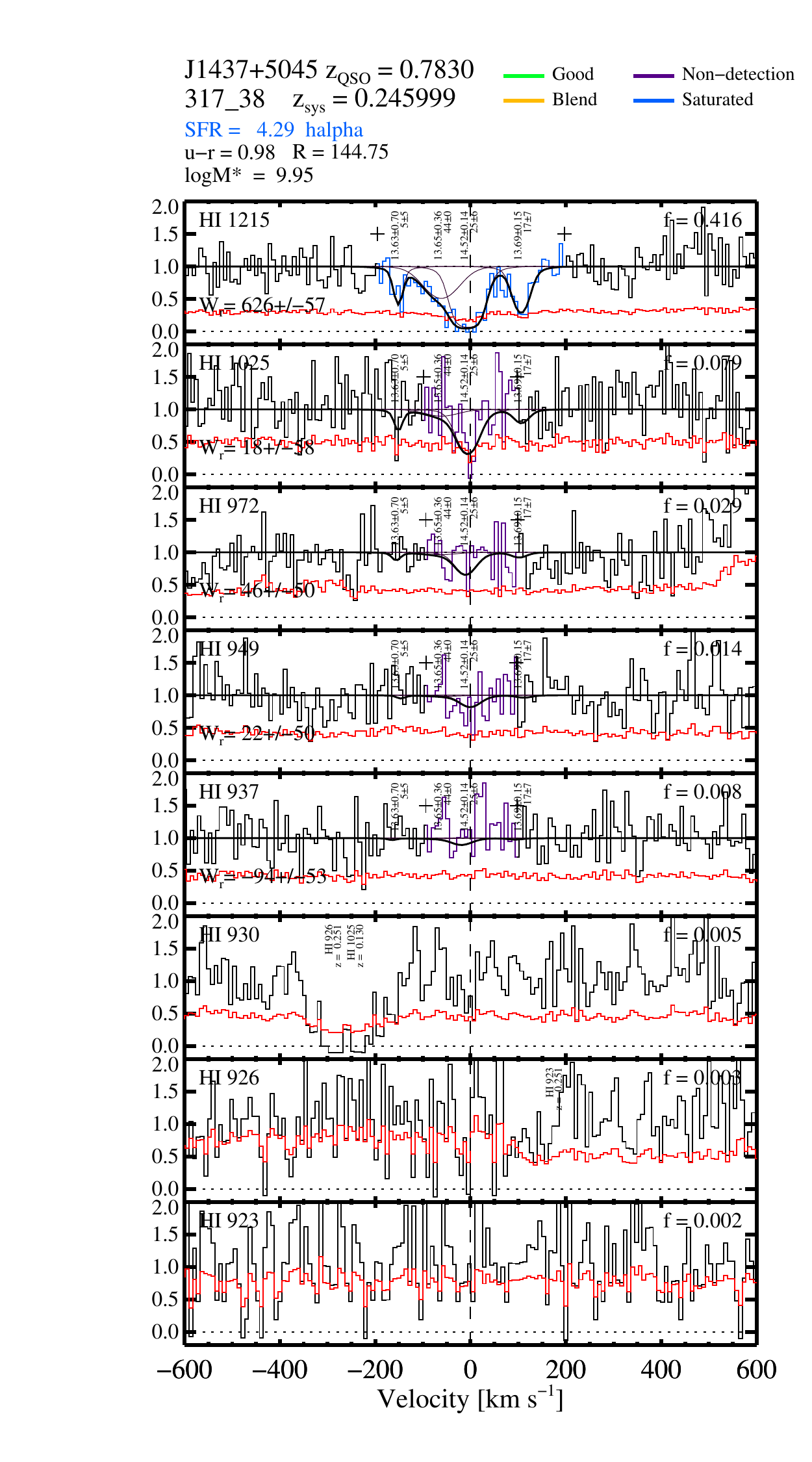}
\label{test_fig}
\end{center}
\caption{Hydrogen stack plot for system 317\_38 toward J1437$+$5045.}
\end{figure*}

\clearpage
\begin{figure*}[!h]
\begin{center}
\epsscale{0.70}
\plotone{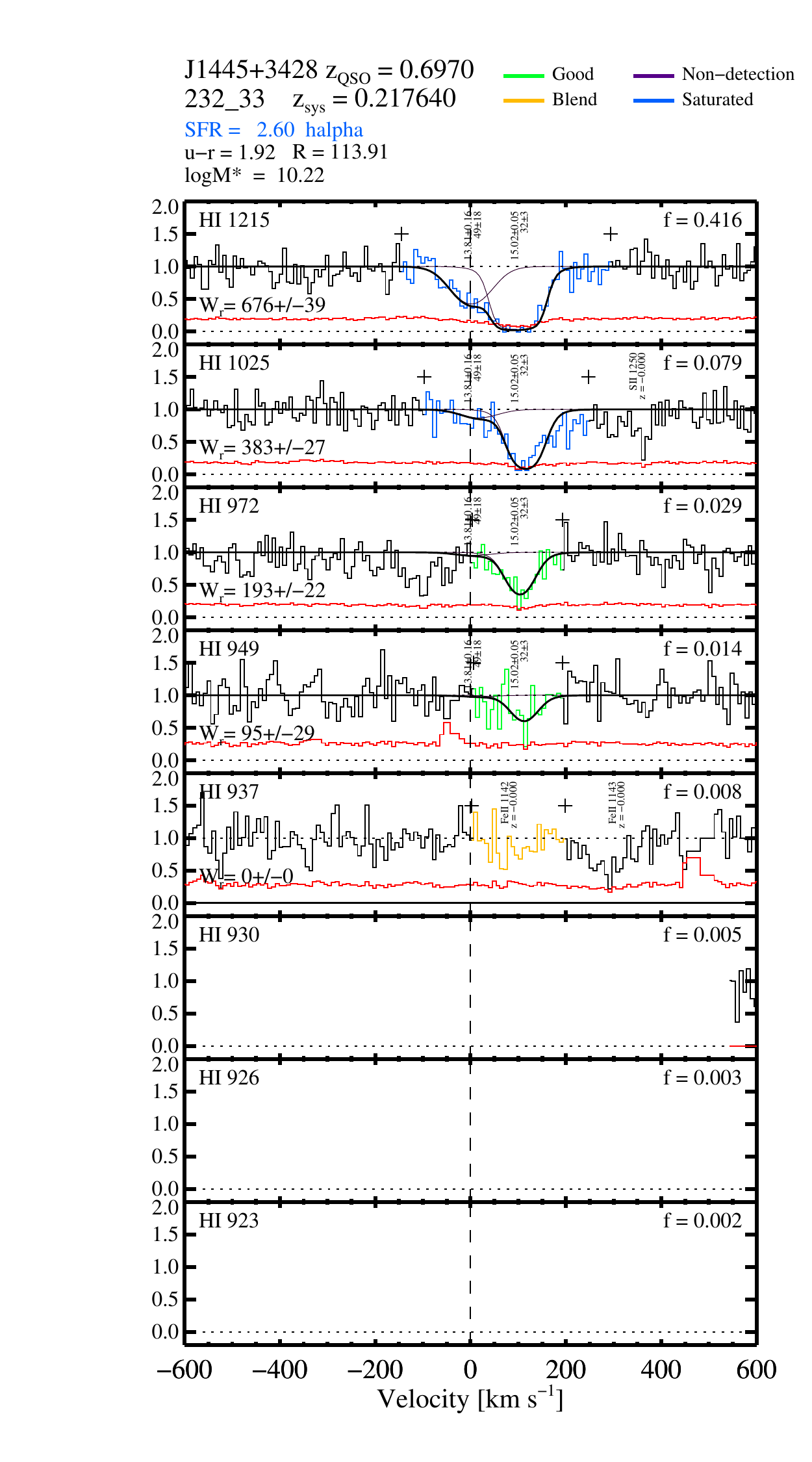}
\label{test_fig}
\end{center}
\caption{Hydrogen stack plot for system 232\_33 toward J1445$+$3428.}
\end{figure*}

\clearpage
\begin{figure*}[!h]
\begin{center}
\epsscale{0.70}
\plotone{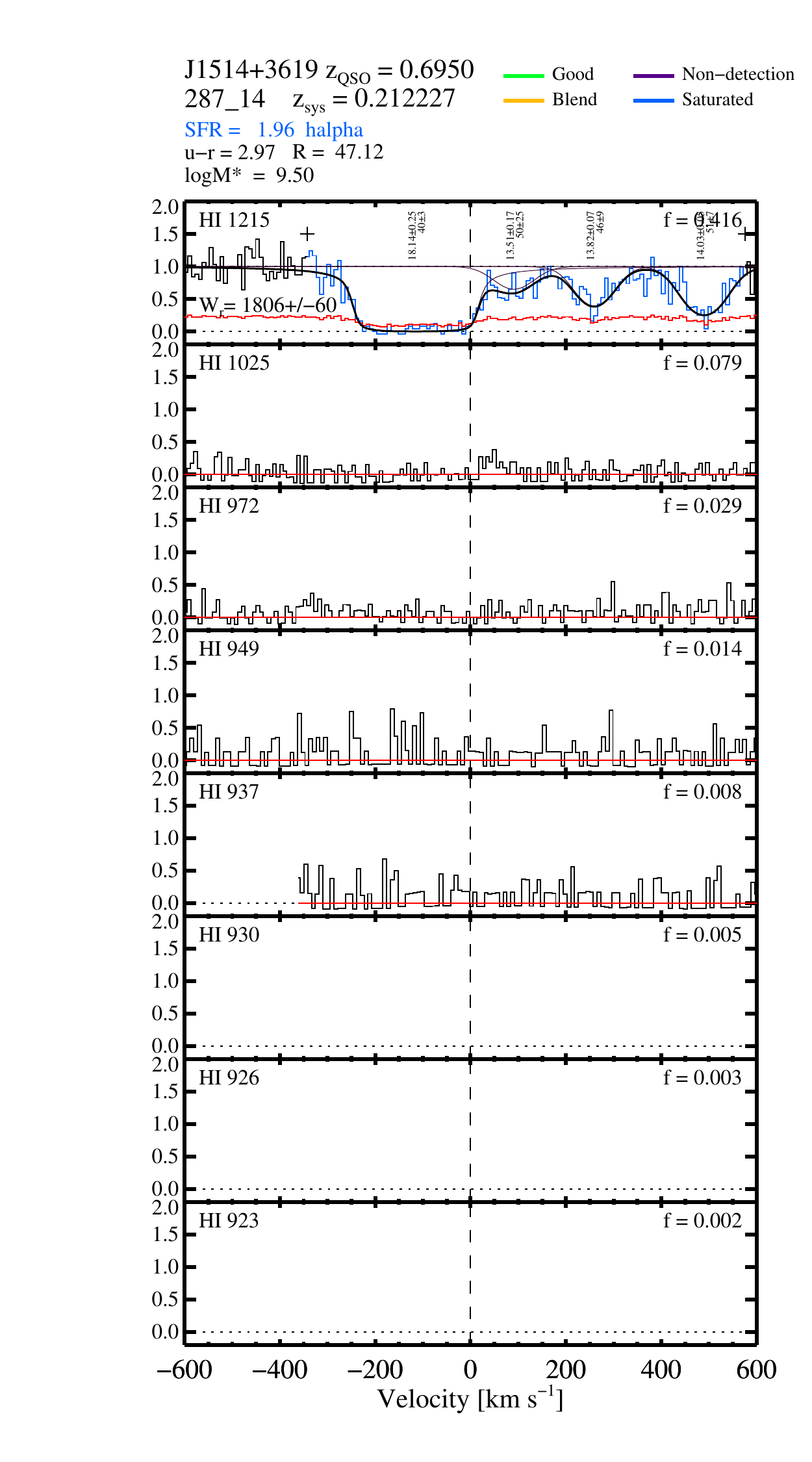}
\label{test_fig}
\end{center}
\caption{Hydrogen stack plot for system 287\_14 toward J1514$+$3619.}
\end{figure*}

\clearpage
\begin{figure*}[!h]
\begin{center}
\epsscale{0.70}
\plotone{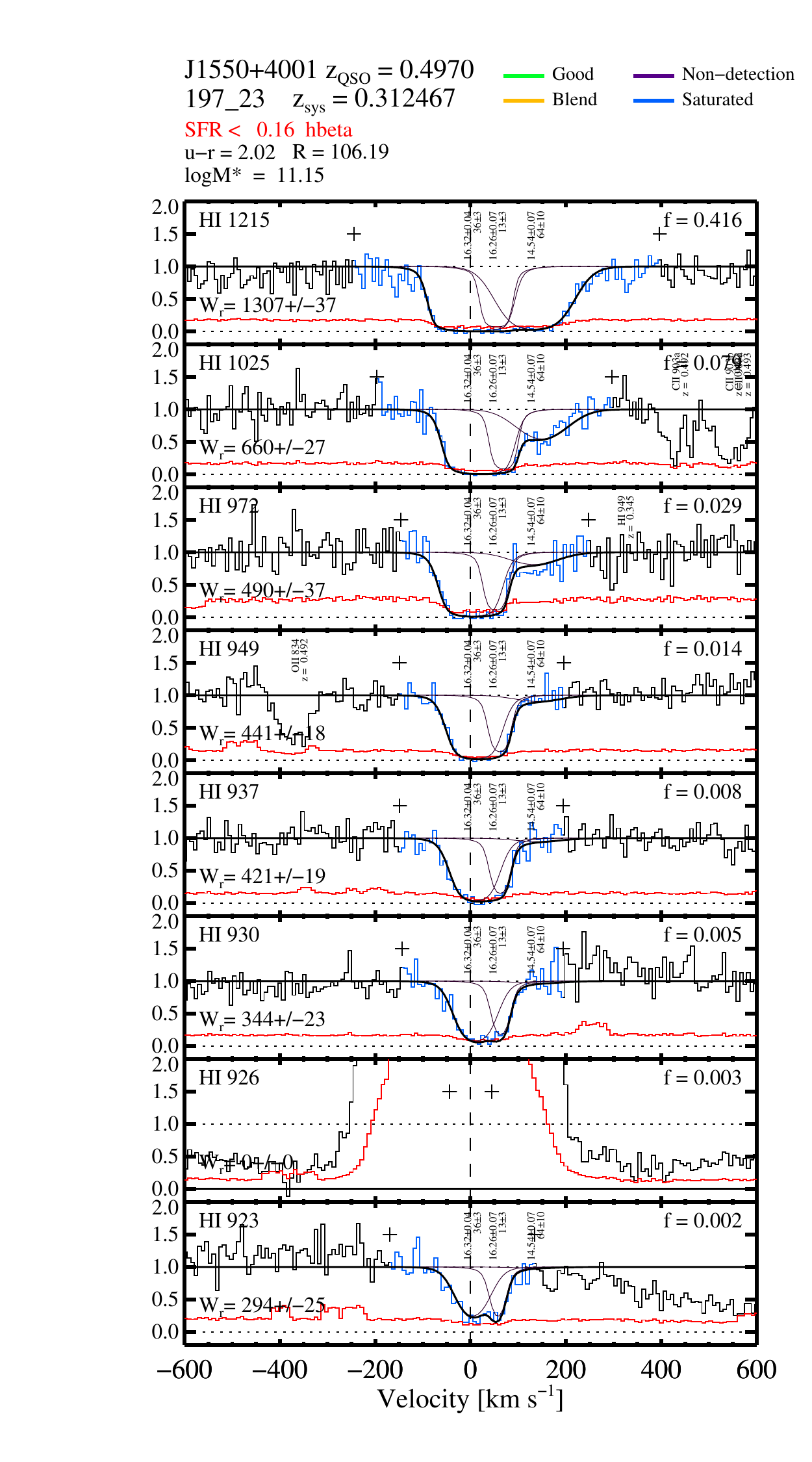}
\label{test_fig}
\end{center}
\caption{Hydrogen stack plot for system 197\_23 toward J1550$+$4001.}
\end{figure*}
\clearpage 
\begin{figure*}[!h]
\begin{center}
\epsscale{0.70}
\plotone{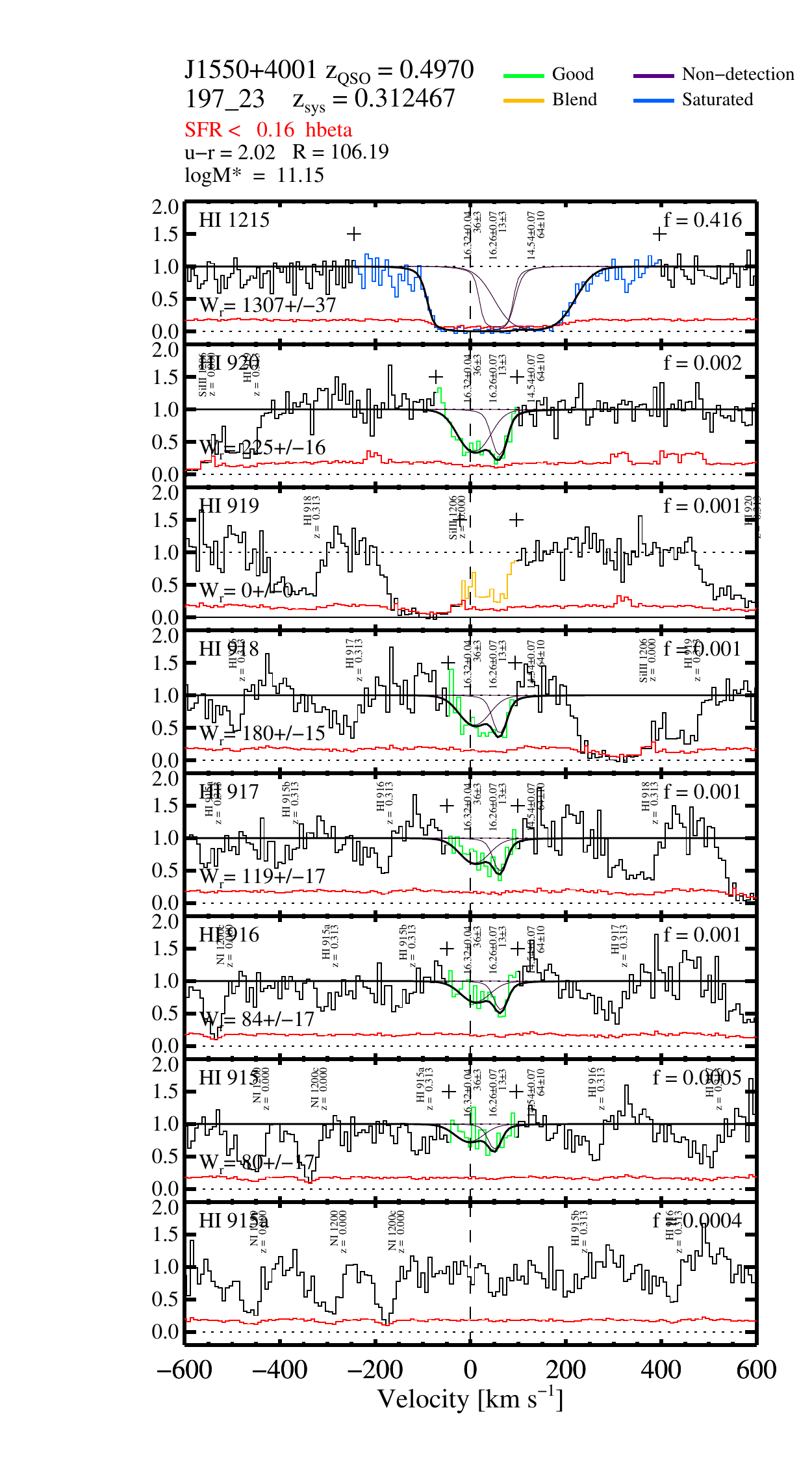}
\label{test_fig}
\end{center}
\caption{Hydrogen stack plot for the higher Lyman lines in system 197\_23 toward J1550$+$4001.}
\end{figure*}

\clearpage
\begin{figure*}[!h]
\begin{center}
\epsscale{0.70}
\plotone{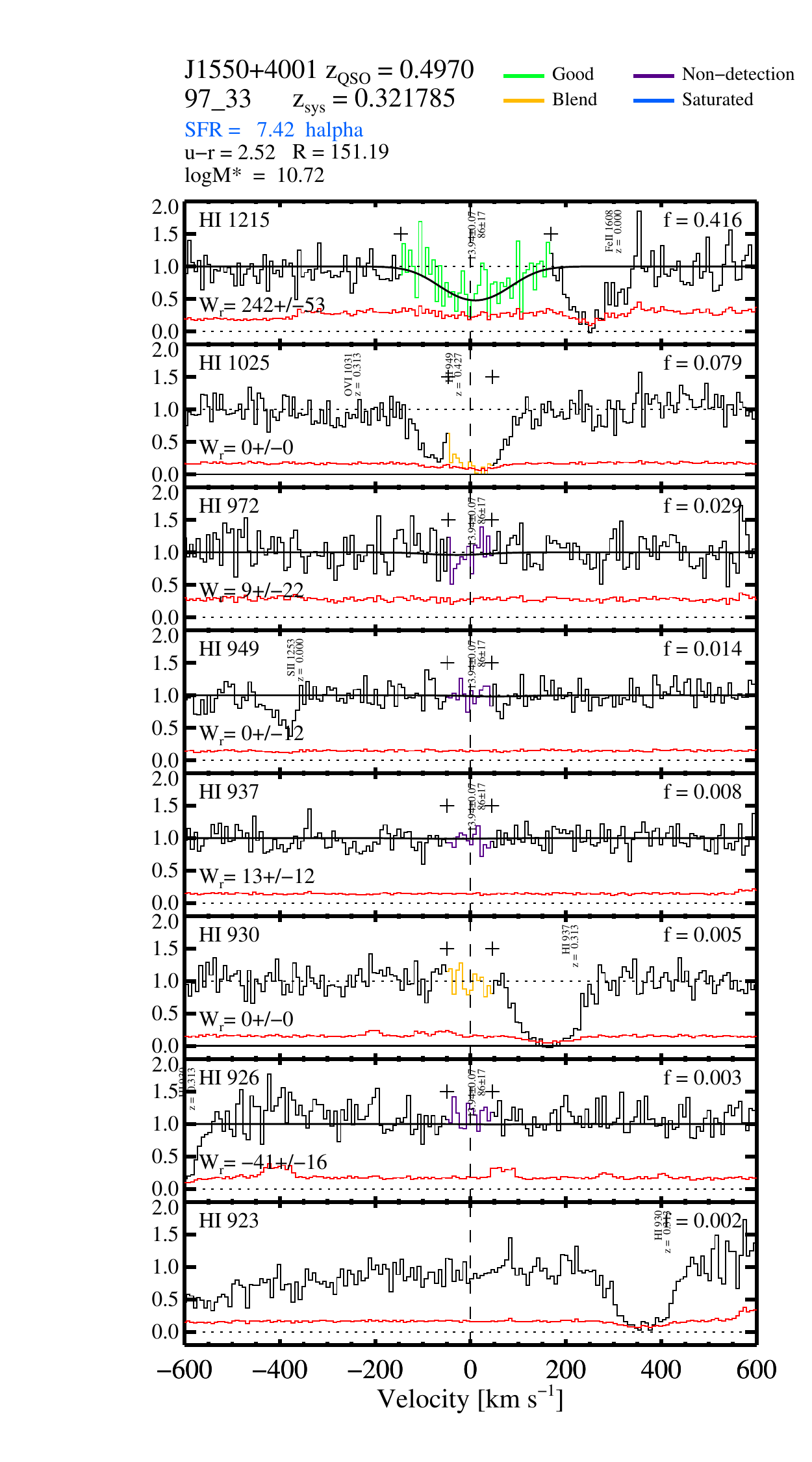}
\label{test_fig}
\end{center}
\caption{Hydrogen stack plot for system 97\_33 toward J1550$+$4001.}
\end{figure*}

\clearpage
\begin{figure*}[!h]
\begin{center}
\epsscale{0.70}
\plotone{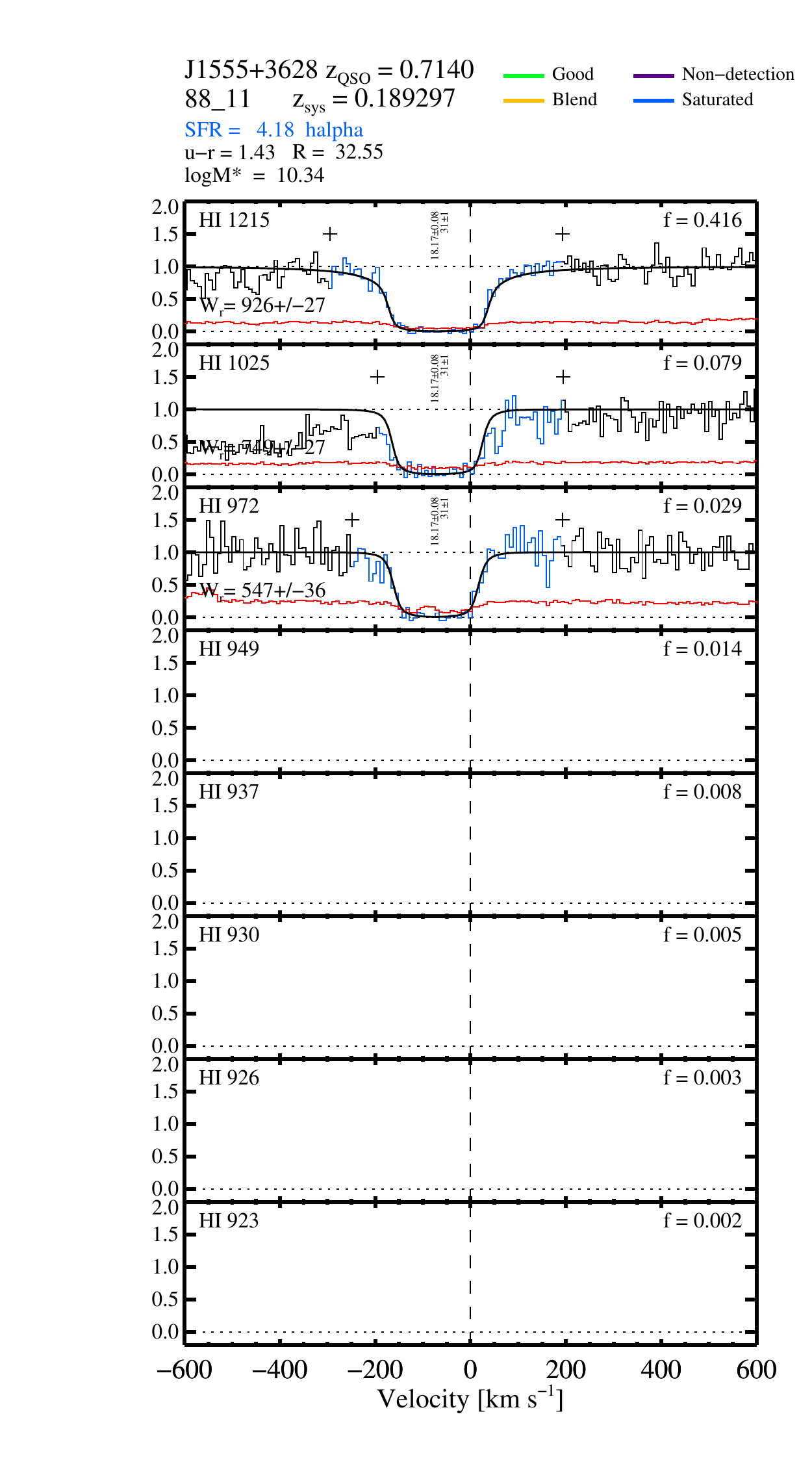}b
\label{test_fig}
\end{center}
\caption{Hydrogen stack plot for system 88\_11 toward J1555$+$3628.}
\end{figure*}

\clearpage
\begin{figure*}[!h]
\begin{center}
\epsscale{0.70}
\plotone{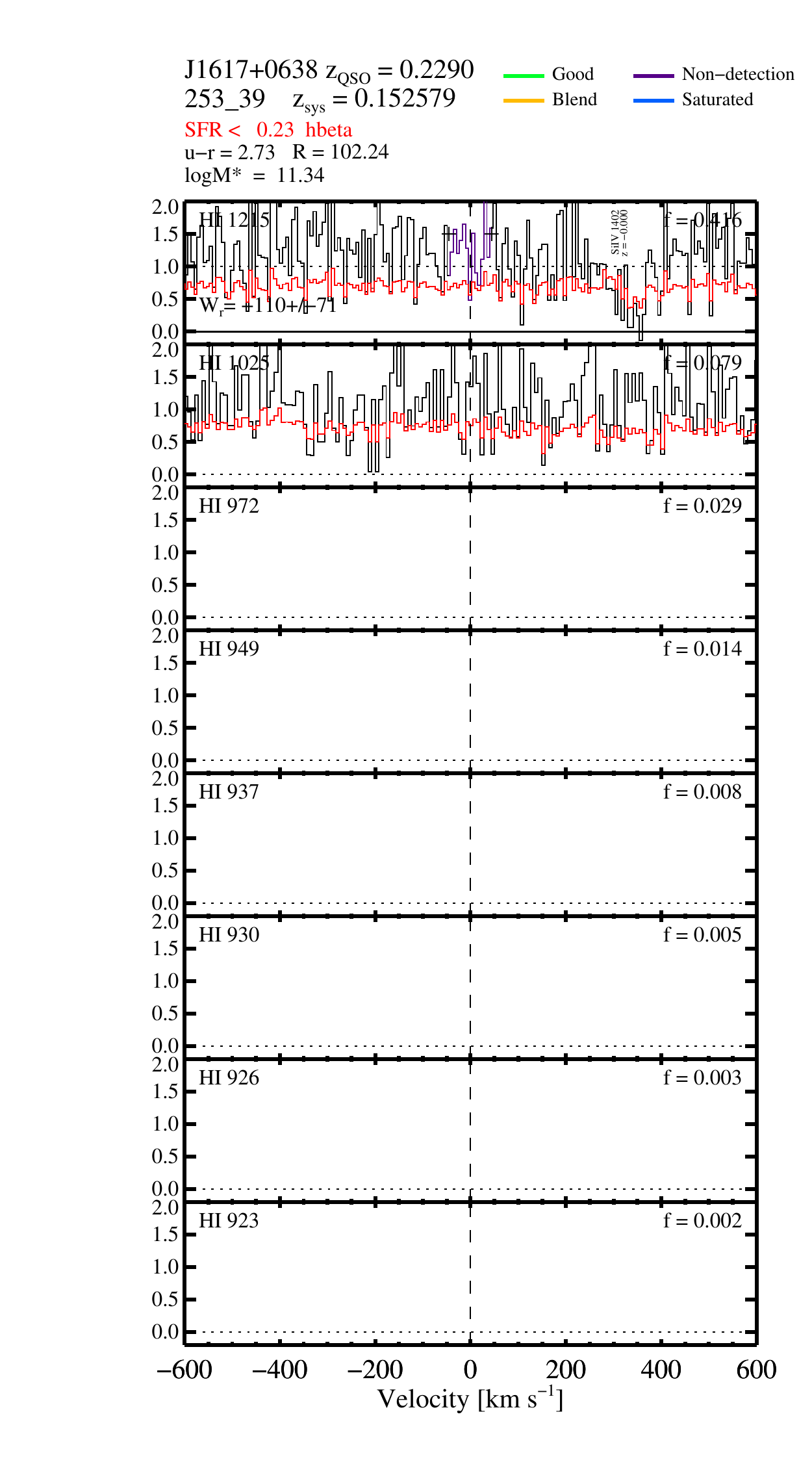}
\label{test_fig}
\end{center}
\caption{Hydrogen stack plot for system 253\_39 toward J1617$+$0638.}
\end{figure*}

\clearpage
\begin{figure*}[!h]
\begin{center}
\epsscale{0.70}
\plotone{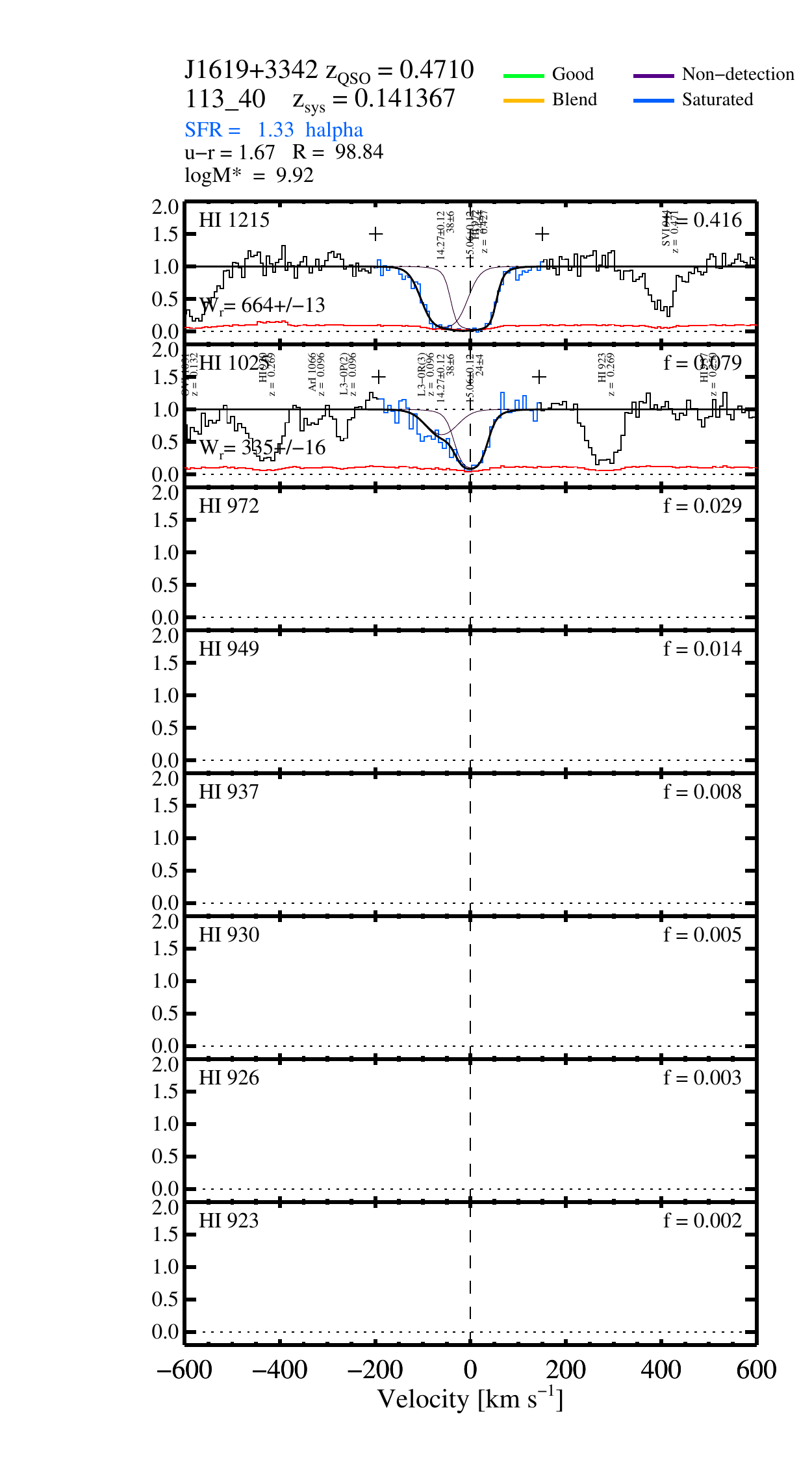}
\label{test_fig}
\end{center}
\caption{Hydrogen stack plot for system 113\_40 toward J1619$+$3342.}
\end{figure*}

\clearpage
\begin{figure*}[!h]
\begin{center}
\epsscale{0.70}
\plotone{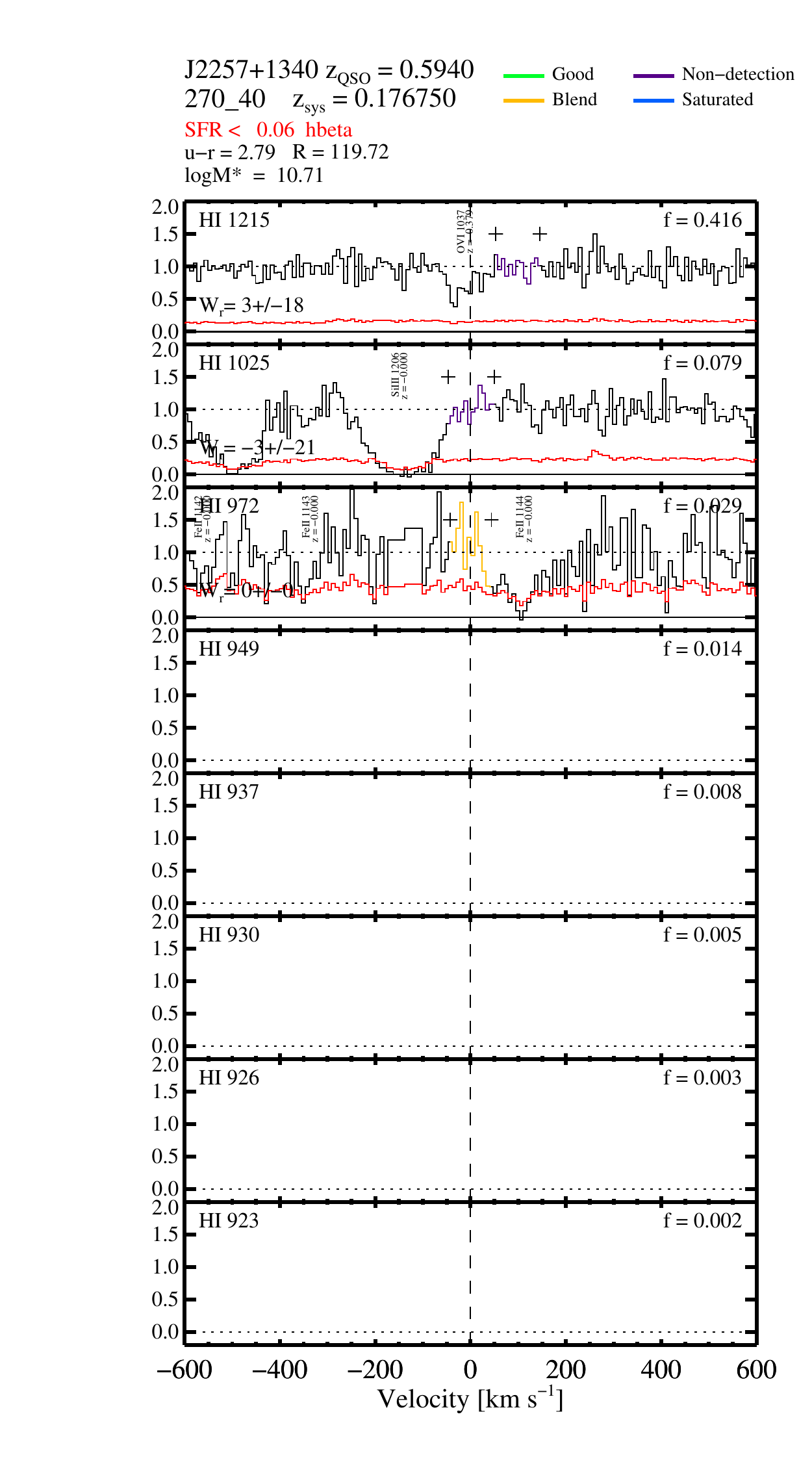}
\label{test_fig}
\end{center}
\caption{Hydrogen stack plot for system 270\_40 toward J2257$+$1340.}
\end{figure*}

\clearpage
\begin{figure*}[!h]
\begin{center}
\epsscale{0.70}
\plotone{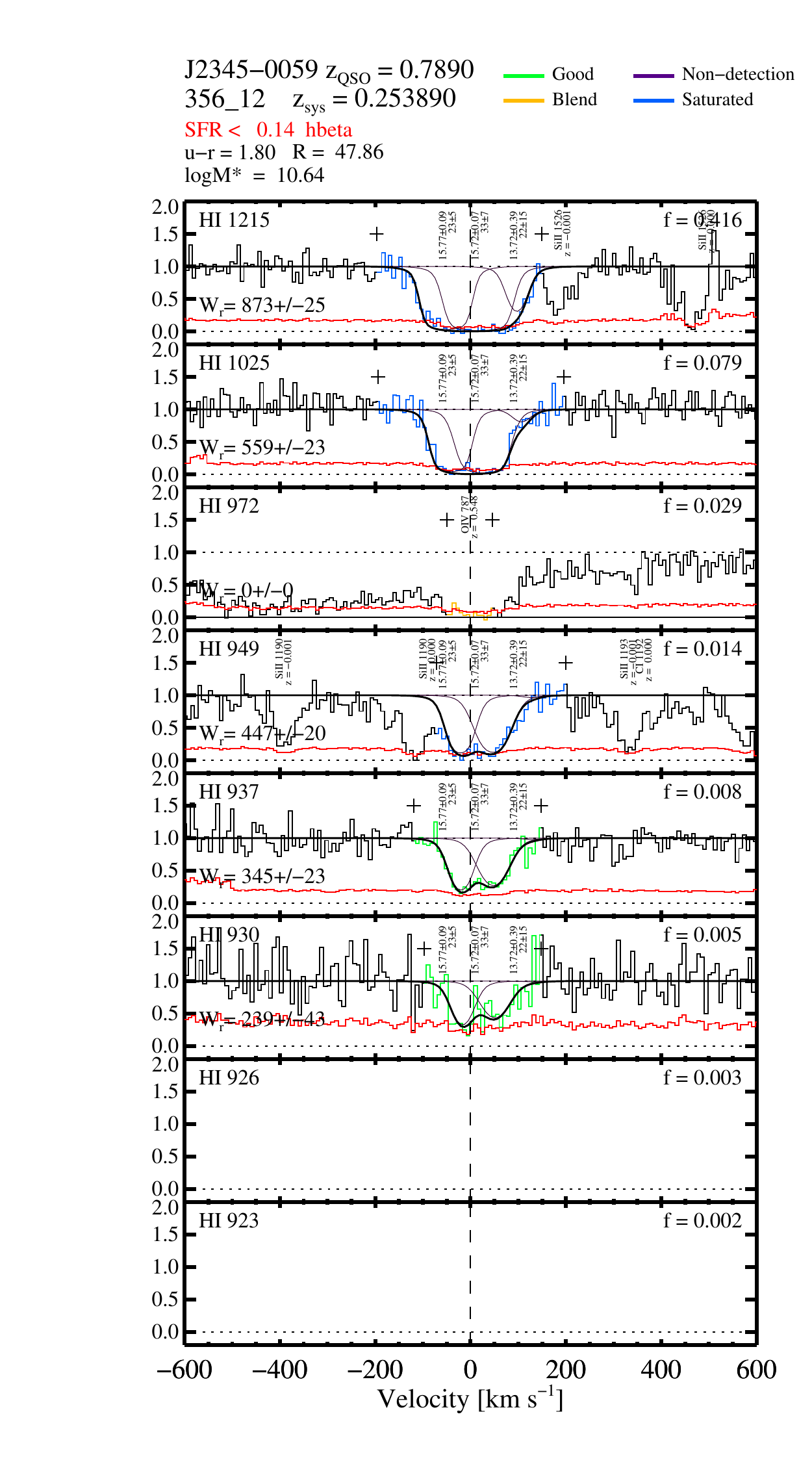}
\label{test_fig}
\end{center}
\caption{Hydrogen stack plot for system 356\_12 toward J2345$-$0059.}
\end{figure*}

\end{document}